%% file: CL18.tex
                        \newif\ifboyscout\boyscouttrue          
                        \newif\ifsubmission\submissionfalse     
                        \newif\ifblog\blogfalse 
      \boyscoutfalse                         

\documentclass[%
 aip,cha,
 reprint,%
nofootinbib, tightenlines,
nobibnotes, 
showpacs,
superscriptaddress
]{revtex4-2} 

\usepackage{amsmath,amsfonts,amssymb,amsbsy,amscd,amsgen}
\usepackage[pdftex]{graphicx}
\graphicspath{{figs/}}    

    \usepackage{xcolor} 
    \usepackage[colorlinks]{hyperref} 

\input{defsKittens}      

\makeatletter
\def\@email#1#2{%
 \endgroup
 \patchcmd{\titleblock@produce}
  {\frontmatter@RRAPformat}
  {\frontmatter@RRAPformat{\produce@RRAP{*#1\href{mailto:#2}{#2}}}\frontmatter@RRAPformat}
  {}{}
}%
\makeatother

\begin{document}

\title[Lattice field theory in 2 dimensions]
{A chaotic lattice field theory in two dimensions}      

\author{Predrag Cvitanovi{\'c}}
\author{Han Liang}%
\affiliation{ 
            School of Physics,
            Georgia Institute of Technology,
            Atlanta, GA 30332-0430
}%

    \ifboyscout\date{\today}
    \else \def\Dated@name{}
    \date
    {
    February 13, 2026 
    }
    \fi

    \begin{abstract}
\input{abstract}
    \end{abstract}



    \ifsubmission
    \else
Chaos \textbf{36} (2026), 
\HREF{https://doi.org/10.1063/5.0273642} {DOI 10.1063/5.0273642}
    \fi

\maketitle

\begin{quotation}
A temporally chaotic system is exponentially unstable with time: double 
the time, and exponentially more orbits are required to cover its strange 
attractor to the same accuracy. For a system of large spatial extent, the 
complexity of the spatial shapes also needs to be taken into account; 
double the spatial extent, and exponentially as many distinct spatial 
patterns might be required to describe the repertoire of spatial shapes to 
the same accuracy.%
    \edit{
\rf{ruelext,FMTT83,Nicolaenko86}
Based on the insight that temporal and spatial instabilities can be 
treated on equal footing, the `chronotopic' theory of 1990s offers an 
elegant description of such \spt\ chaos, with predictions of the theory 
encapsulated in an `entropy density' function.%
\rf{Grassberger89,LePoTo96,LePoTo97,LePoTo97a} 
The first decade of 2000s saw rapid progress in description of 
transitional turbulence in fluids, in terms of `{\ecs}s', 
numerically exact {\spt}ly periodic unstable solutions of {\NS} equations.%
\rf{W01,WK04,KawKida01,science04,GHCW07,WFSBC15,CPTKGS22}
This approach to turbulence aims to predict measurable observables from 
the defining (\NS) equations, without any statistical assumptions. 
But how are these solutions to be pieced together?%
\rf{focusPOT}
As a single such solution requires highly unstable time integration over 
$10^4-10^6$ discretization ODEs, estimates of `chronotopic' theory 
observables based on large spacetime volume limits of numerical 
simulations are out of question. 
In this paper we develop a new \spt\ theory of turbulence, which
assigns a new kind of a global weight, an `\Hilldet', to each {\ecs},
with a new enumeration of global solutions as sums over all {\Blatt}s. 
Our {\dft} replaces numerical extrapolations of finite spacetime 
simulations of the 1990s chronotopy by our main result, 
the {exact} \spt\ {\detZeta}, with computable cycle expansion truncations 
errors, decreasing exponentially with the \spt\ volumes of \lsts\ 
included in its evaluation.  
    } 
\end{quotation}

\section{Introduction}
\label{s:intro} 

Our goal here is to make this `{\spt} chaos' tangible and precise
(see \refsect{s:sptchaos}), 
in a series of papers that introduce its theory and its implementations. 
The companion paper I\rf{LC21} focuses on the \new{$1d$ chaotic lattice field 
theory}, and a  \new{novel treatment of time-reversal invariance}. In this paper, 
paper II, we develop the  \new{theory of 2$d$ \spt\ chaotic systems}; and in the 
companion paper III\rf{WWLAFC22} we  \new{apply the theory to several nonlinear}
field theories. As our intended audience spans many disjoint specialties, 
from fluid dynamics to quantum field theory, the exposition entails much 
pedagogical detail, so let us start by stating succinctly what the key 
novelty of our theory is. 

\begin{figure}
  \centering
{$(a)$}~~~
\includegraphics[width=0.44\textwidth]{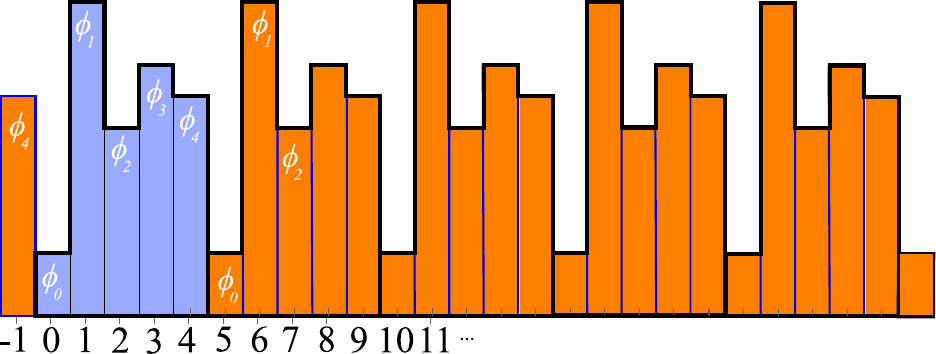}
\\\vskip 12pt
{$(b)$}~~~
\includegraphics[width=0.44\textwidth]{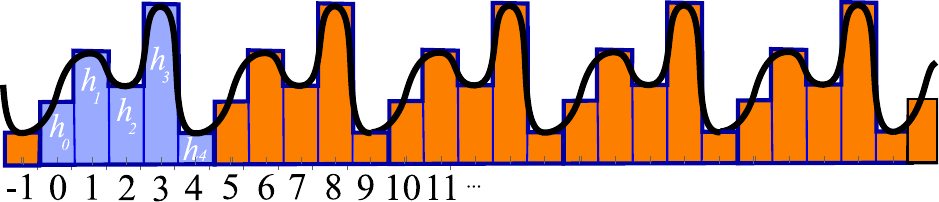}
\\\vskip 12pt
{$(c)$}~~~
\includegraphics[width=0.44\textwidth]{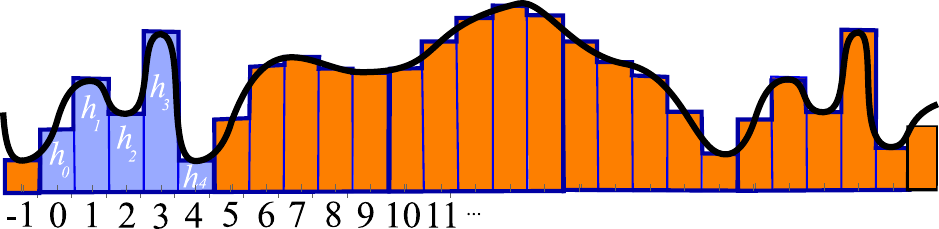}
  \caption{\label{fig:LstPerturbs} 
A one\dmn\ temporal lattice {period-5} {\lst} 
$\Xx_c=[{\ssp}_{0}\, {\ssp}_{1}\, {\ssp}_{2}\, {\ssp}_{3}\, {\ssp}_{4}]$ 
illustrated by 
(a) five repeats of the {\pcell} {\lst}. 
(b) An \emph{internal} perturbation  $h_z$ has the same periodicity as the 
{\lst}. Its spectrum,  evaluated in \refsect{s:pcellStab}, is discrete. 
(c) A \emph{transverse} perturbation $h_z$ is an arbitrary, aperiodic 
function over the infinite lattice. Its spectrum,  evaluated by the 
Floquet-Bloch theorem in \refsect{s:Bloch}, is a continuous function of 
wave number $k$. 
Horizontal: lattice sites labeled by $z\in\integers$.
Vertical: (a) value of field $\ssp_z$, (b-c) perturbation $h_z$, 
plotted as a bar centered at lattice site $z$.
Values of the field and perturbation are shown in blue 
within the {\pcell}, and in orange 
outside the {\pcell}.
          }
\end{figure}

There are two ways of studying
stabilities of translationally invariant systems,
illustrated by \reffig{fig:LstPerturbs}\,(b) and (c):

(i) In the textbook `QM-in-a-box' approach, one starts by confining a 
system to a \emph{finite} box, then takes the box size to infinity. In 
dynamical systems this point of view leads to the 
Gutzwiller-Ruelle\rf{gutbook,ruelle,ChaosBook} {\po} formulation of 
chaotic dynamics,
    \edit{
with stability of each periodic solution computed over a finite time interval. 
    } 
This approach is hampered by one simple fact that 
complicates everything: the \po\ weight is \emph{not} multiplicative for 
orbit repeats, 
\beq 
\det (\id - \jMat_p^r) 
    \neq
[\det (\id - \jMat_p)]^r 
\,. 
\ee{primCellNotMltpl} 
Much work follows,\rf{ChaosBook} with some details elaborated in 
\refsect{s:pcellStab}. 

(ii) 
A crystallographer, or field theorist starts with an \emph{infinite} 
lattice, or continuum spacetime. The approach --as we show here in 
\refsect{s:lstStab}-- yields \new{weights that are  {multiplicative}} for 
repeats of {\spt}ly periodic solutions, 
\beq
\Det \jMorb_{rp} 
    \, ``="\, {(\Det \jMorb_p)^{r}}
\ee{BravLattMltpl}
(quotation marks, as the precise statement
is in terms of {\stabexp}s rather than determinants). 
This fact simplifies everything, and yields the  \new{main result of this 
paper}, the \spt\ zeta function (\refsect{s:lattZeta}),
expressed as a product of {Euler functions} 
$\phi(t_{p})$, one for each prime \spt\ solution of weight $t_p$, 
\beq 
\zeta
        = \prod_{p}\zeta_{p} 
    \,, \qquad
1/\zeta_{p}
        = \phi(t_{p})
\,. 
\ee{intro:sptZeta2d} 

Analysis of a temporally chaotic dynamical system typically starts with 
establishing that a temporal flow (perhaps reduced to discrete time maps 
by {\PoincSec}s) is locally stretching, globally folding. Its  \statesp\  
is partitioned, the partitions labeled by an alphabet, and the 
qualitatively distinct solutions classified by their temporal symbol 
sequences.\rf{ChaosBook} 

We do not do this here: instead, we find that the natural language to 
describe `{\spt} chaos' and `turbulence' is the  \new{formalism of field 
theory}. 
        \edit{
Our work is a continuation of the 1995 chronotopic program of Politi, 
Torcini \& Lepri\rf{LePoTo96,LePoTo97,PoToLe98,GiLePo95} who, in their 
studies of propagation of {\spt} disturbances in extended systems, 
showed that the spatial stability analysis can be combined with the 
temporal stability analysis.
The Floquet-Bloch approach to stability that we deploy was introduced for 
temporal evolution in 1981-1983 by Bountis \& Helleman\rf{Bount81} and
MacKay \& Meiss,\rf{MacMei83} and, for spatially extended systems, in 
1989 by Pikovsky,\rf{Pikovsky89} who noted that for {\spt}ly chaotic 
systems space and time could be considered on the same footing in the 
sense that there are settings in which  `time'  and  `space' coordinates 
could be interchanged. 

In our formulation
the `chaos theory' is }
\new{a Euclidean \dft.}
However, as developed here, this theory looks nothing like the
textbook
expositions\rf{strogb,ottbook,deva87,ASY96,Katok95,Robinson12,ChaosBook}
of temporally chaotic, few degrees-of-freedom dynamical systems. There
one is given an {initial state}, which then \emph{evolves in time}, much
like in mechanics,  where given an {initial phase-space point}, the
integration of Hamilton's equations traces out a phase-space trajectory.
For a reader versed in fluid dynamics or atomic physics the most 
disconcerting aspect of the field-theoretic perspective is that  \new{time is 
just one of the coordinates} over which a {\fconf} is defined: each 
field-theoretic solution is a static solution over the  \new{infinite 
spacetime}. There is  \new{\emph{no} `evolution in time'},
no stable / unstable manifolds, 
no `hyperbolicity',
no `mixing'. 

Furthermore --just as the discretization of time by {\PoincSec}s 
aids analysis of temporal chaos-- we find it convenient to discretize 
both time and space. {\Spt}ly steady turbulent flows offer one physical 
motivation for considering such models: a rough approximation to such 
flows is obtained discretizing them into {\spt} cells, with each cell 
turbulent, and cells coupled to their neighbors. Lattices also arise 
naturally in many-body problems, such as many-body quantum chaos models 
studied in \refrefs{EnUrRi15a,AWGBG16,AWGG16,RiUrTo22}, see
\refsect{s:phi4}. 

\textbf{Outline}.
We start our formulation of chaotic field theory (\refsect{s:FT}) by 
defining the field theory partition sums in terms of {\spt}ly {\lsts} 
(\refsect{s:periodicCnfs}). 
A \emph{\dft} has support on the set of all solutions of systems' {\ELe}. Its
building blocks are \emph{{\lsts}}, {\spt}ly periodic 
solutions of system's defining equations (\refsect{s:DFT}). 

    \edit{
In  \refsect{s:sptFT} we introduce the field theories studied here.
In particular, the simplest of chaotic field theories, the 
\catlatt\rf{GutOsi15,GHJSC16} of \refsect{s:catlatt}, 
a discretization of the compact boson Klein-Gordon equation, 
\beq 
 (-\Box + \mu^2)\,\Xx - \Mm  =0
 \,,
\ee{intro:catlatt}
recently applied to transport in chaotic chains\rf{AlDuSh25} and to black 
holes physics,\rf{AxFlNi17,AxFlNi22} captures the essence of {\spt} 
chaos. 
\catLatt\ is a lattice of hyperbolic `anti-' or `inverted'
oscillators,\rf{Vattay1997,SHVB21} with an unstable `anti-harmonic cat' 
$\ssp_{z}$ of mass $\mu$ at each lattice site $z$, a `cat' which, when 
pushed, runs away rather than pushes back.

In \refsect{s:phi4} we motivate of choice of nonlinear field theories 
studied here by 
the 1955 {Fermi-Pasta-Ulam-Tsingou}\rf{fpu65,CaFlKi04}  model 
which illustrates how many-body `chaos theory' morphs into a Euclidian 
strong-coupling, anti-integrable \dft. 
    }

Crucial to `chaos' is the notion of stability: in \refsect{s:orbJacob} we 
describe  \new{\spt\ stability} of above field theories' {\lsts} in terms of 
their  \new{\jacobianOps}. 
{\Po} theory for a time-evolving dynamical system on a one\dmn\ temporal 
lattice is organized by grouping orbits of the same time period 
together.\rf{gutbook,ruelle,Gas97,ChaosBook,LC21} For systems 
characterized by several translational symmetries, one has to take care 
of multiple periodicities; in the language of crystallography,  \new{organize 
the \po\ sums by corresponding {\em {\Blatt}s}}, or, in the language of 
field theory, by the `sum over \emph{geometries}'. In \refsect{s:Bravais} 
we enumerate and construct spacetime geometries, or $d=2$ {\Blatt}s 
$\LTS{}{}{}$, of increasing spacetime periodicities. The classification 
of {\lsts} proceeds in two steps. On the coordinate level, periodicity is 
imposed by the hierarchy of {\Blatt}s of increasing periodicities. 
On the field-configuration level, the key to the \spt\ \po\ theory is the 
enumeration and  \new{determination of \emph{\primeOrbs}}, the basic building 
blocks of \po\ theory (\refsect{s:orbits}). 

The central idea of spatiotemporal theory 
developed here, global orbit stability, has its origins in the 1886 work 
of Hill\rf{Hill86} and {\Poincare}.\rf{Poinc1886}
The likelihood of each solution is given by the {\Hilldet}, the 
determinant of its {\spt} {\jacobianOrb}.        
Compared to the 
temporal-evolution chaos theory, the  \new{\Hilldet} (more precisely, the  
{\stabexp}) is the  \new{central innovation} of our field-theoretic formulation 
of chaotic field theory, so we return to it throughout the paper. The 
calculations are carried out on the reciprocal lattice 
(\refsect{s:Brillouin}). We discuss {\pcell} computations in 
\refsect{s:pcellStab}, as prequel  to introducing the {\stabexp} of a 
\lst\ over {\spt}ly infinite {\Blatt},  \new{computed on the reciprocal lattice} 
(\refsect{s:Bloch}). For \catlatt\ we evaluate and cross-check  
{\Hilldet}s by two methods, either on the reciprocal lattice 
(\refsect{s:rcprPcellJ}), or by the `fundamental fact' evaluation 
(\refappe{s:fact}). 

Having enumerated all {\Blatt}s (\refsect{s:Bravais}), determined {\lsts} 
over each (\refsect{s:2DprimeOrb}), computed the  \new{weight of each}  {\lst} 
(\refsect{s:Bloch}), we can now write down the  \new{{\dft} partition sum}  as a 
sum over all {\spt} solutions of the theory (\refsect{s:POT}). In 
\refsect{s:primePartSum} we reexpress the partition sum in terms of 
{\primeOrbs}, and in \refsect{s:lattZeta} we construct the  \new{{\spt} zeta 
function}.
What makes these resummations possible is the  \new{multiplicative 
property of {\Hilldet}s} announced in \refeq{BravLattMltpl}, provided by 
their evaluation over the {\spt}ly  \new{infinite {\Blatt}} 
(\refsect{s:Bloch}), the key property that is violated by 
finite-dimensional matrix  approximations that are the basis of the 
traditional Gutzwiller-Ruelle temporal {\po}s theory 
(\refsect{s:pcellStab}). 
In \refsect{s:zetaExpectV}, we explain how one computes expectation values of 
observables in deterministic chaotic field theories.

How is this global, high-dimensional orbit stability related to the 
stability of the conventional low-dimensional, forward-in-time evolution? 
The two notions of stability are related by  \new{Hill's formulas}, relations 
that rely on higher-order derivative equations being rewritten as sets of 
first order ODEs, formulas  \new{equally applicable} to energy conserving 
systems, as to viscous,  \new{dissipative systems}. We derive them in 
\refrefs{LC21,LC22}. From the field-theoretic perspective,  \new{{\Hilldet}s 
are fundamental}, while the forward-in-time evolution (a transfer matrix 
method) is merely one of the methods for computing them. 

Finally, we know that time-evolution cycle-expansions' convergence is 
accelerated by shadowing of long orbits by shorter periodic 
orbits.\rf{AACI} In \refsect{s:shadow} we check numerically that 
\catlatt\ {\lsts} that share finite {\spt} {\mosaic}s shadow each other 
to exponential precision. 
We presume (but do not show) that this shadowing property ensures that 
the predictions of the theory are dominated by the shortest period 
{\primeOrbs}. 

This completes our generalization\rf{GHJSC16,LC21,CL18,WWLAFC22} of the 
temporal-evolution deterministic chaos theory\rf{ChaosBook,Gas97} to 
\spt\ chaos / turbulence, and recasts both in the formalism of 
conventional solid state physics, field theory, and statistical 
mechanics. 
Our results are summarized and open problems discussed in 
\refsect{s:summary}. 
    \edit{
Our calculations are reported in Appendices.
    }
Icon 
\raisebox{-0.4ex}[0pt][0pt]{\includegraphics[height=1em]{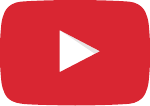}} on 
the margin links a block of text to a supplementary online video. For 
additional material, online talks and related papers, see 
\HREF{https://ChaosBook.org/overheads/spatiotemporal/} 
{ChaosBook.org/overheads/{\spt}}.\rf{CB:spatiotemp} 

\section{Lattice field theory}
\label{s:FT}   

In a $d$\dmn\ hypercubic discretization of a Euclidean space, the $d$ 
continuous Euclidean coordinates $x\in\reals^d$ are replaced by a 
hypercubic integer lattice\rf{MonMun94,MunWal00}
\beq
\lattice = 
\left\{
\sum_{j=1}^d z_j \mathbf{e}_j \mid z \in\mathbb{Z}^d
\right\}
         \,,\quad
\mathbf{e}_{j} \in
\left\{
\mathbf{e}_1, \mathbf{e}_2, \cdots, \mathbf{e}_d
\right\} \,,
\ee{SquareLatt}
spanned by a set of orthogonal 
vectors $\mathbf{e}_j$, with lattice spacing 
$a_{j}=|\mathbf{e}_{j}|=\Delta x_{j}$ along the direction of vector 
$\mathbf{e}_{j}$. We shall use lattice units, almost always setting 
\(a_j=1\) (for another, modular function parameterization choice, see 
\refeq{2tauBas}). A field $\ssp(x)$ over $d$ continuous coordinates 
$x_{j}$ is represented by a discrete array of field values over lattice 
sites 
\beq 
\ssp_z = \ssp(x) 
    \,,\quad x_j = a_j z_j= \mbox{lattice site,}
    \qquad z \in \integers^d
\,. 
\ee{LattField} 
 A lattice {\em 
{\fconf}} is a $d$\dmn\ infinite array of field values (in what follows, 
illustrative examples will be presented in one or two {\spt} dimensions) 
\beq \Xx = 
\begin{array}{ccccccc}
\cdots &\cdots &\cdots & \cdots &\cdots &\cdots & \cdots\cr
\cdots & \ssp_{-2,1} & \ssp_{-1,1} & \ssp_{0,1} & \ssp_{1,1} & \ssp_{2,1} & \cdots \cr
\cdots & \ssp_{-2,0} & \ssp_{-1,0} & \ssp_{0,0} & \ssp_{1,0} & \ssp_{2,0} & \cdots \cr
\cdots & \ssp_{-2,-1} & \ssp_{-1,-1} & \ssp_{0,-1} & \ssp_{1,-1} & \ssp_{2,-1} & \cdots \cr
\cdots &\cdots &\cdots & \cdots &\cdots &\cdots & \cdots
\end{array}
\,. 
\ee{stateSp} 
A {\fconf} \Xx\ is a \emph{point} in system's 
\emph{\statesp} 
\beq 
\pS 
  =
\left\{\Xx \mid \ssp_{z}\in\reals\,,\; z \in \integers^d \right\} 
\,, 
\ee{StateSp} 
given by all possible values of site fields, where $\ssp_z$ 
is a single scalar field, or a multiplet of real or complex fields. 

While we refer here to such discretization as a lattice field theory, the 
lattice might arise naturally from a many-body setting with the nearest 
neighbor interactions, such as many-body quantum chaos models studied in 
\refrefs{EnUrRi15a,AWGBG16,AWGG16,RiUrTo22}, 
with a multiplet of fields at every site.\rf{GutOsi15}

\subsection{Periodic {\fconfs}}
\label{s:periodicCnfs} 

\begin{figure}
\begin{center}
           \begin{minipage}[c]{0.23\textwidth}\begin{center}
\includegraphics[width=1.0\textwidth]{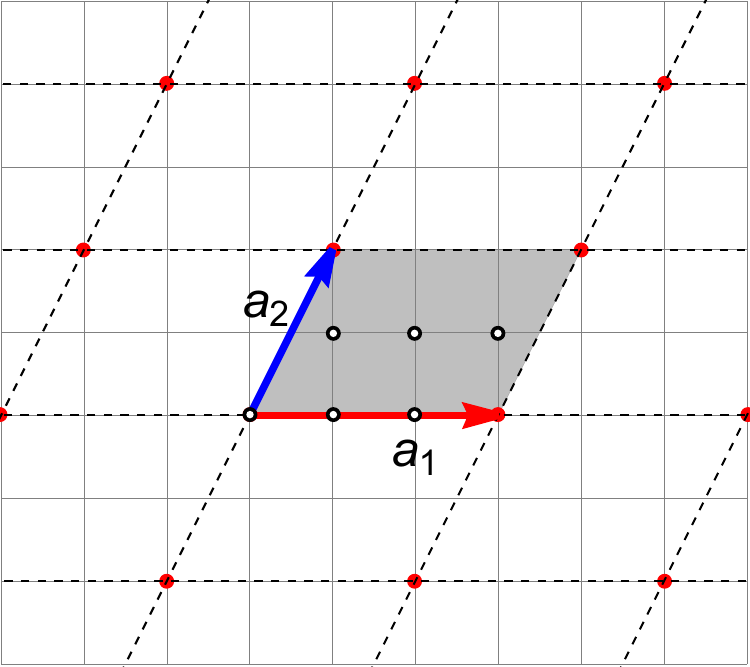}
                \\      (a)
            \end{center}\end{minipage}
                   \hskip 2ex
           \begin{minipage}[c]{0.23\textwidth}\begin{center}
\includegraphics[width=1.0\textwidth]{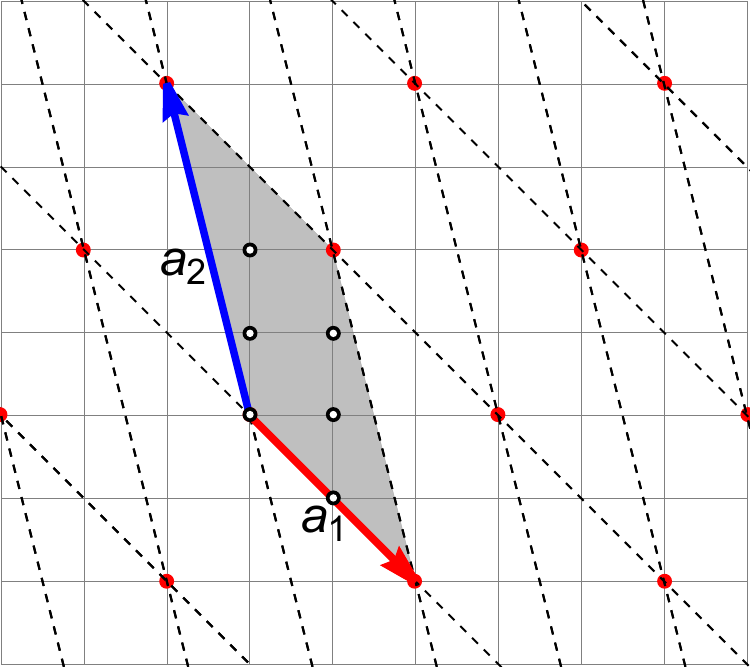}
                 \\     (b)
            \end{center}\end{minipage}
\end{center}
                \vskip -12pt     
  \caption{\label{f:BravLatt}
    The intersections of the light gray lines -lattice sites $z \in
    \integers^2$- form the integer square lattice \refeq{LattField}.
(a)
    Translations of the {\pcell} $\Bcell{A}=\BravCell{3}{2}{1}$
    spanned by primitive vectors $\mathbf{a}_1=(3,0)$ and $\mathbf{a}_2=(1,2)$
    define the  {\Blatt} $\lattice_\Bcell{A}$.
(b)
    The primitive vectors $\mathbf{a}_1=(2,-2)$ and $\mathbf{a}_2=(-1,4)$
    form a  {\pcell} ${\Bcell{A}'}$ equivalent to (a) by a
    unimodular transformation.
    The intersections (red points) of either set of dashed lines form the
    same {\Blatt} $\lattice_\Bcell{A}=\lattice_{\Bcell{A}'}$.
    The {volume} \refeq{lattVol} of either {\pcell} is $\vol_\lattice=6$,
    the number of integer lattice sites within the cell, with the tips of
    primitive vectors and tiles' outer boundaries belonging to the
    neighboring tiles.
    This will be discussed further in \reffig{f:2x1rpo}.
}
\end{figure}

    \edit{
We shall demand that {\ELe} are the same everywhere in spatially
translation-invariant directions, for all times. 
The only way to obey that is by {deterministic} solutions $\Xx_c$  
being periodic.
    }
We say that a {lattice {\fconf}} is $\lattice_\Bcell{A}$-\emph{periodic} if 
\beq 
\ssp_{z+\shift} = \ssp_z 
\ee{periodLConf}  
for any discrete translation 
$\shift=n_{1}\mathbf{a}_1+n_{2}\mathbf{a}_2+\cdots+n_{d}\mathbf{a}_d$ in 
the \emph{{\Blatt}} 
\beq 
\lattice_\Bcell{A} = \Big\{\sum_{j=1}^d n_{j} 
\mathbf{a}_j 
 \mid n_{j} \in\mathbb{Z} \Big\}
\,, 
\ee{BravLatt} 
where the $[d\!\times\!d]$ basis matrix 
\(
\Bcell{A} = 
\left[
\mathbf{a}_{1}, \mathbf{a}_{2}, \cdots,  \mathbf{a}_{d}
\right]
\) 
formed from primitive integer lattice vectors $\{\mathbf{a}_j\}$ 
defines a $d$\dmn\ \emph{\pcell}.\rf{AshMer,CoEvBu22} 
If the lattice 
spacing \refeq{LattField} is set to 1, the lattice \emph{volume}, or the 
volume of the {\pcell} 
\beq
\vol_\Bcell{A} = |\Det\Bcell{A}| 
\ee{lattVol} 
equals the number of lattice sites $z\in \Bcell{A}$ within the {\pcell}, 
see \reffig{f:BravLatt}. 

 \Pcell\ \Bcell{A} {{\fconf}}
lattice-site fields \refeq{stateSp} take values in the 
$\vol_\Bcell{A}$\dmn\ {\statesp} 
\beq
\pS_\Bcell{A} 
  =
\left\{\Xx \mid \ssp_{z}\in\reals\,,\; z\in \Bcell{A} \right\} 
\,. 
\ee{torusStatesp} 
For example, repeats of the $\vol_\Bcell{A}=15$\dmn\ 
$[5\!\times\!3]$ {\pcell} {\fconf} 
\beq 
\Xx = 
        \left[\begin{array}{ccccc}
\ssp_{-2,1} & \ssp_{-1,1} & \ssp_{0,1} & \ssp_{1,1} & \ssp_{2,1} \cr 
\ssp_{-2,0} & \ssp_{-1,0} & \ssp_{0,0} & \ssp_{1,0} & \ssp_{2,0} \cr 
\ssp_{-2,-1} & \ssp_{-1,-1} & \ssp_{0,-1} & \ssp_{1,-1} & \ssp_{2,-1} \cr 
              \end{array}\right]
\ee{stateSpTorus} 
tile periodically the doubly infinite {\fconf} 
\refeq{stateSp}. 

We focus on the one\dmn\ case for now, postpone discussion of higher\dmn\ 
{\Blatt}s to \refsect{s:Bravais}. 

\subsection{Orbits}
\label{s:orbit} 

Consider a one\dmn\ lattice $\lattice_\Bcell{A}$, defined by a single 
primitive vector $\mathbf{a}_{1}=\cl{}$ in \refeq{BravLatt}. 
One-lattice-spacing shift operator $\shift_{zz'}=\delta_{z+1,z'}$ acts on 
the {\pcell} \Bcell{A} as the {\shiftOrb} 
\beq
\shift 
= 
 \left(\begin{array}{ccccc}
0 & 1 & & & \cr 
   & 0 & 1 & & \cr
   & & \ddots & \ddots & \cr
   & & & 0 & 1 \cr
1 & & & & 0 
\end{array}\right)
\,,
\ee{hopMatrix}
a cyclic permutation matrix that translates a {\fconf} 
by one lattice site, $(\shift\Xx)_z=\ssp_{z+1}$,
\bea 
\Xx &=& 
      [{\ssp}_{0}\, {\ssp}_{1}\, {\ssp}_{2}\, {\ssp}_{3}\, \cdots \,{\ssp}_{\cl{}-1}] \,,
\continue \shift\Xx &=& 
      [{\ssp}_{1}\, {\ssp}_{2}\, {\ssp}_{3}\, \cdots \,{\ssp}_{\cl{}-1}\, {\ssp}_{0}] \,,
\continue 
                    &\cdots&
\label{1dShift}\\
\shift^{\cl{}-1}\Xx &=& 
      [{\ssp}_{\cl{}-1}\, {\ssp}_{0}\, {\ssp}_{1}\, {\ssp}_{2}\, \cdots \,{\ssp}_{\cl{}-2}]\,,
\continue \shift^{\cl{}}\Xx &=& 
      [{\ssp}_{0}\, {\ssp}_{1}\, {\ssp}_{2}\, {\ssp}_{3}\, \cdots \,{\ssp}_{\cl{}-1}] = \Xx
\,. \nnu
\eea
While each {\fconf} $\shift^{j}\Xx$ might be a distinct 
point in the {\pcell}'s \statesp\ \refeq{torusStatesp}, they are 
equivalent, in the sense that they all are the same set of lattice site 
fields $\{{\ssp}_{z}\}$, up to a cyclic relabeling of lattice sites. 

In this way actions of the {translation group} $T$ on  {\fconfs} over a 
multi-periodic {\pcell} \Bcell{A} foliate its {\statesp} into a union 
\beq \pS_\Bcell{A} = \{\Xx\} = \cup\,\pS_p 
\ee{statespFliated} of translational \emph{orbits}, \beq 
    \pS_p = \{\shift^{j} \Xx_p \mid \shift^{j} \in T\}
\ee{GroupOrbDisc} each a set of equivalent  {\fconfs}, labeled perhaps 
by $\Xx_p$, one of the configurations in the set. Each orbit is a fixed 
point of $T$, as for any translation \( \shift^{j} \pS_p = \pS_p \,. \) 
The number of distinct {\fconfs} in the orbit is \cl{p}, the period of 
the orbit. (For orbits over two\dmn\ lattices, see \refsect{s:orbits}).

\subsection{{\PrimeOrbs} over one\dmn\ {\pcell}s}
\label{s:1DprimeOrb}  

\paragraph*{Definition: {\PrimeOrb}.}
\begin{quote}
A {\pcell} {\fconf} \refeq{periodLConf} is \emph{prime} if it is not a 
repeat of a {lattice {\fconf}} of a shorter period. 
\end{quote}
The simplest example of a prime {\fconf} is a \emph{\steady} 
$\ssp_z={\ssp}$. Its {\pcell} \Bcell{A} is the unit hypercube 
\refeq{SquareLatt} of period-1 along every hypercube direction. A 
{\fconf} obtained by tiling any {\pcell} \refeq{BravLatt} by repeats of 
\steady\ ${\ssp}$ is a periodic {\fconf}, but \emph{not} a prime 
{\fconf}. 

Consider next a period-6 {\fconf} \refeq{1dShift} over a {\pcell} 
2\Bcell{A} obtained by a repeat of a {\pcell} \Bcell{A} period-3 
{\fconf}, \beq 
\begin{array}{rlrl}
\Xx_{2\Bcell{A}} &=
      [{\ssp}_{0}\, {\ssp}_{1}\, {\ssp}_{2}\, {\ssp}_{0}\, {\ssp}_{1}\, {\ssp}_{2}]
                 \,,& \;
\Xx_{\Bcell{A}}  &=
      [{\ssp}_{0}\, {\ssp}_{1}\, {\ssp}_{2}]
                 \cr
{\shift}\Xx_{2\Bcell{A}} &=
      [{\ssp}_{1}\, {\ssp}_{2}\, {\ssp}_{0}\, {\ssp}_{1}\, {\ssp}_{2}\, {\ssp}_{0}]
                 \,,& \;
{\shift}\Xx_{\Bcell{A}} &=
      [{\ssp}_{1}\, {\ssp}_{2}\, {\ssp}_{0}]
                 \cr
{\shift}^2\Xx_{2\Bcell{A}} &=
      [{\ssp}_{2}\, {\ssp}_{0}\, {\ssp}_{1}\, {\ssp}_{2}\, {\ssp}_{0}\, {\ssp}_{1}]
                 \,,& \;
{\shift}^2\Xx_{\Bcell{A}} &=
      [{\ssp}_{2}\, {\ssp}_{0}\, {\ssp}_{1}]
\end{array}
\label{1dShift3lst}
\eeq 
On the infinite {\Blatt} \refeq{stateSp}, {\fconf} $\Xx_{\Bcell{A}}$ 
and its repeat $\Xx_{2\Bcell{A}}$ are the same {\fconf}, with the same 
period-3 orbit 
$\pS_p=(\Xx_{\Bcell{A}},\shift\Xx_{\Bcell{A}},\shift^2\Xx_{\Bcell{A}})$: 
every {\Blatt} orbit is a `prime' orbit. 

The distinction arises in enumeration of {\fconfs} over a {\pcell}. The 
\pcell\ {\statesp}s \refeq{torusStatesp} are here 6-, 3\dmn, 
respectively. 
%
Both orbits depend on the same three distinct lattice site field values  
$\ssp_z$. On the {\pcell} $2\Bcell{A}$, however, the six lattice sites 
{\fconf} $\Xx_{2\Bcell{A}}$ is not \emph{prime}, it is a \emph{repeat} of 
the {\fconf} $\Xx_{\Bcell{A}}$. We elaborate this distinction in 
\refsect{s:BlattJ}. 

This is how `prime {\po}s' and their repeats work 
for the one\dmn, temporal lattice.
For a two\dmn\ square lattice the notion of `prime' is a bit trickier, so 
we postpone its discussion to \refsect{s:orbits}. 

The totality of  {\fconfs} \refeq{StateSp} can now be constructed by (i) 
enumerating all {\Blatt}s $\lattice_\Bcell{A}$, (ii) determining 
{\primeOrbs} for each {\pcell} $\Bcell{A}$, and (iii) including their 
repeats into {\fconfs} over {\pcell}s $\Bcell{AR}$. Our task is to 
identify, compute and weigh \emph{the totality} of these {\primeOrbs} for 
a given field theory.

\subsection{Observables}
\label{s:lstsExpectV} 

A goal of a physical theory is to make predictions, for example, enable 
us to evaluate the expectation value of an \emph{observable}. An 
{observable} `\obser' is a function or a set of {\fconf} functions 
$\obser_z=\obser_z[\Xx]$,
let's say temperature, measured on a \spt\ lattice site $z$.   
For a given $\lattice_\Bcell{A}$-periodic {\fconf} $\Xx$, the 
Birkhoff average of observable $\obser$ is 
given by the Birkhoff sum $\Obser$, 
\beq 
{\obser}[\Xx]_\Bcell{A} = 
\frac{1}{\vol_\Bcell{A}}\,\Obser[\Xx]_\Bcell{A} 
    \,,\qquad
\Obser[\Xx]_\Bcell{A} = \sum_{z\in\Bcell{A}} \obser_z 
\,. 
\ee{BirkhoffSum}
For example, if the observable is the field itself, 
$\obser_z=\ssp_z$, the Birkhoff average over the lattice {\fconf} $\Xx$ 
is the average `height' of the 
\edit{field $\ssp_z$ in \refeq{stateSp}}. 

In order to evaluate the \emph{expectation value} of an observable, 
\bea 
\expct{\obser}_\Bcell{A} \,&=&\, 
    \int d\Xx_\Bcell{A} \, p[\Xx]_\Bcell{A} \, \obser[\Xx]_\Bcell{A}
        \,,\quad
d\Xx_\Bcell{A} = \prod_{z \in \Bcell{A}} d\ssp_z 
\,, 
\label{expctObser} 
\eea
we need to know the \statesp\ probability density $p[\Xx]$ of {\fconf} \Xx. 
 
To understand what this probability is, and motivate the formalism that 
follows, a bit of quantum-mechanical intuition might be helpful. 
The semiclassical quantum field theory (for
a derivation, see Appendix A37 {\em WKB quantization} of \refref{appendWKB})
assigns a quantum probability amplitude to a \emph{deterministic} solution 
$\Xx_c$,\rf{VanVleck28,Verdiere07,LevSmi77,LevSmi77a}
with the partition sum
\beq 
Z_\Bcell{A}[\Source]	
    \,\approx\,  \sum_{c}
\frac{{e}^{\frac{i}{\hbar}\action[\Xx_c] + {i}\maslovInd_c + 
{i}\Xx_c\cdot\Source} } 
     {\left|\Det(\jMorb_c/\hbar)\right|^{1/2} }
\ee{n-pt-corr1} 
having support on the set of  deterministic solutions $\Xx_c$.
A deterministic solution $\Xx_c$ satisfies the stationary 
phase condition, \ie, system's Euler-Lagrange equations 
\beq 
F[\Xx_c]_z =
\frac{\delta{\action[\Xx_c]}}{\delta\ssp_z~} = 0
\ee{eqMotionLagrang} 
at every lattice site $z$
(see the sketch of \reffig{f:pde2a}\,(a)). 
In the {WKB approximation}, its weight is obtained by expanding the 
action near the \statesp\ point $\Xx_c$ to quadratic order, 
\beq
\action[\Xx] 
          \approx
\action[\Xx_c] 
       + \frac{1}{2}\transp{(\Xx-\Xx_c)}{\jMorb_c}\,(\Xx-\Xx_c) 
\,,
\ee{PathIntSaddle}
where we refer to the matrix of second derivatives
\beq (\jMorb_c)_{z'z} = \left.\frac{\delta^2 \action[\Xx]}{\delta 
\ssp_{z'}\delta \ssp_{z}} \right|_{\Xx=\Xx_c} 
\,,
\ee{SemiClJacobianOrb} 
as the {\em \jacobianOrb}. We will not return to quantum theory in this 
paper, but $\jMorb_c$ will play the central role in all that follows. 

\section{\Dft}
\label{s:DFT} 

\begin{figure}
  \setlength{\unitlength}{0.42\textwidth}
            \begin{minipage}{0.46\textwidth}
\begin{center} 
  \begin{picture}(1,1.0295231)%
    \put(0,0){\includegraphics[width=\unitlength,page=1]{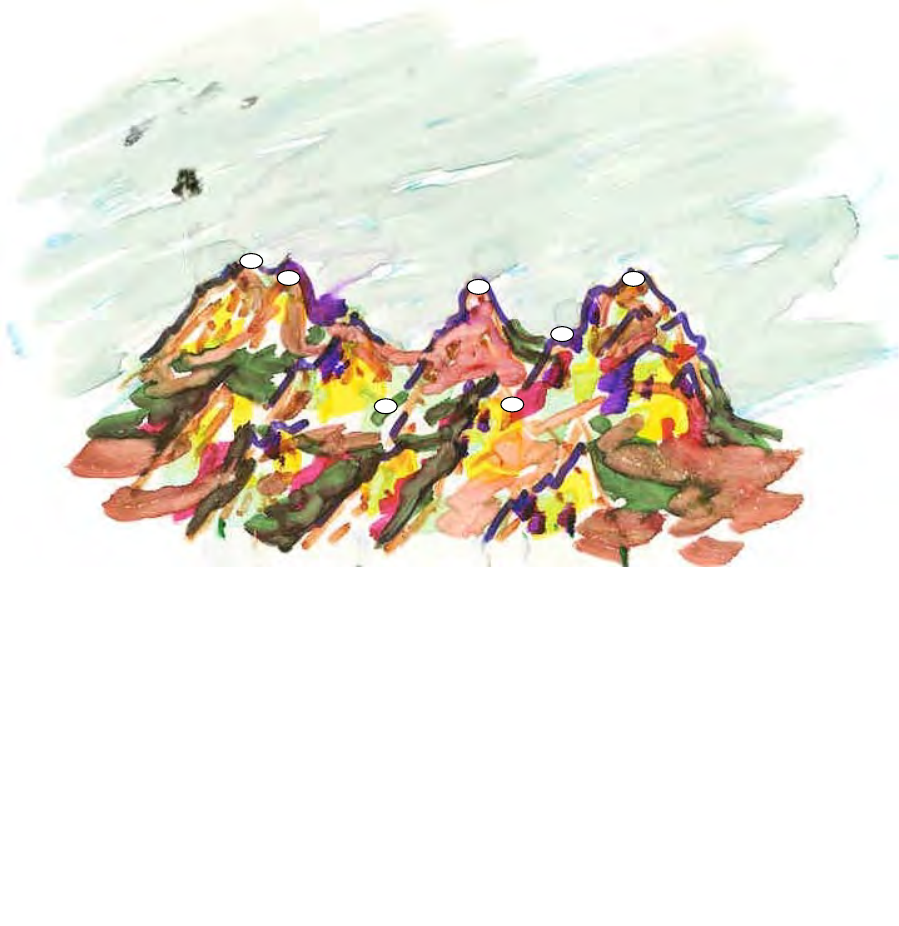}}%
    \put(0.67541654,0.76546289){\makebox(0,0)[lt]{\smash{$\action[\Xx_g]$}}}%
    \put(0.19731274,0.77757205){\makebox(0,0)[lt]{\smash{$\action[\Xx_a]$}}}%
    \put(0.30,0.92){\makebox(0,0)[lt]{\smash{$\action[\Xx]$}}}%
    \put(0,0){\includegraphics[width=\unitlength,page=2]{pde2c}}%
    \put(0.10429553,0.08068724){\rotatebox{-38.50000048}{\makebox(0,0)[lt]{\smash{$\ssp_{00}$}}}}%
    \put(0,0){\includegraphics[width=\unitlength,page=3]{pde2c}}%
    \put(0.86893778,0.06166898){\rotatebox{-24.50000011}{\makebox(0,0)[lt]{\smash{$\ssp_{zz'}$}}}}%
    \put(0.47645097,0.05512017){\rotatebox{-30.21639946}{\makebox(0,0)[lt]{\smash{$\pS_\Bcell{A}$}}}}%
    \put(0.21313753,0.15615495){\makebox(0,0)[lt]{\smash{$\Xx_a$}}}%
    \put(0.28157203,0.12175497){\makebox(0,0)[lt]{\smash{$\Xx_b$}}}%
    \put(0.3539397,0.03848268){\makebox(0,0)[lt]{\smash{$\Xx_c$}}}%
    \put(0.46794344,0.20998399){\makebox(0,0)[lt]{\smash{$\Xx_d$}}}%
    \put(0.5115623,0.14554712){\makebox(0,0)[lt]{\smash{$\Xx_e$}}}%
    \put(0.64935818,0.06326612){\makebox(0,0)[lt]{\smash{$\Xx_f$}}}%
    \put(0.71577784,0.13365105){\makebox(0,0)[lt]{\smash{$\Xx_g$}}}%
  \end{picture}%
\end{center}
        \vskip -8pt    
\[
\mbox{(a)}~~~~~~
\mbox{quantum chaos : }
\expct{\obser{}}
     \,\approx\,
 \sum_{c}^{}
\obser[\Xx_c]\, \frac{{e}^{\frac{i}{\hbar}\action[\Xx_c] + i\maslovInd_c} }
                     {\left|\Det(\jMorb_c / \hbar)\right|^{1/2}}
\] 
             \end{minipage}
        \vskip 4pt                 
            \begin{minipage}[c]{0.46\textwidth}
\begin{center} 
  \begin{picture}(1,1.0295231)%
    \put(0,0){\includegraphics[width=\unitlength,page=1]{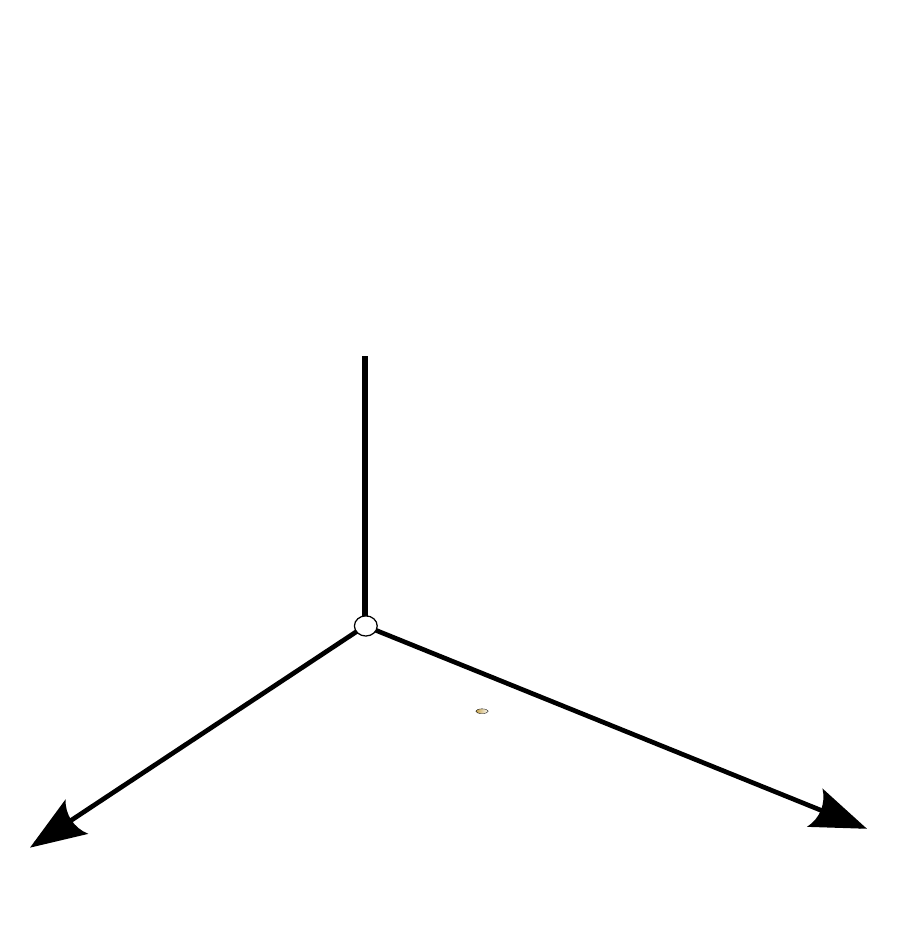}}%
    \put(0.67541654,0.76546289){\makebox(0,0)[lt]{\smash{$p[\Xx_g]$}}}%
    \put(0.19731274,0.77757205){\makebox(0,0)[lt]{\smash{$p[\Xx_a]$}}}%
    \put(0.30,0.92){\makebox(0,0)[lt]{\smash{$p[\Xx]$}}}%
    \put(0,0){\includegraphics[width=\unitlength,page=2]{pde2DFT}}%
    \put(0.10429553,0.08068724){\rotatebox{-38.50000048}{\makebox(0,0)[lt]{\smash{$\ssp_{00}$}}}}%
    \put(0.86893778,0.06166898){\rotatebox{-24.50000011}{\makebox(0,0)[lt]{\smash{$\ssp_{zz'}$}}}}%
    \put(0.47645097,0.05512017){\rotatebox{-30.21639946}{\makebox(0,0)[lt]{\smash{$\pS_\Bcell{A}$}}}}%
    \put(0.21313753,0.15615495){\makebox(0,0)[lt]{\smash{$\Xx_a$}}}%
    \put(0.28157203,0.12175497){\makebox(0,0)[lt]{\smash{$\Xx_b$}}}%
    \put(0.3539397,0.03848268){\makebox(0,0)[lt]{\smash{$\Xx_c$}}}%
    \put(0.46794344,0.20998399){\makebox(0,0)[lt]{\smash{$\Xx_d$}}}%
    \put(0.5115623,0.14554712){\makebox(0,0)[lt]{\smash{$\Xx_e$}}}%
    \put(0.64935818,0.06326612){\makebox(0,0)[lt]{\smash{$\Xx_f$}}}%
    \put(0.71577784,0.13365105){\makebox(0,0)[lt]{\smash{$\Xx_g$}}}%
  \end{picture}%
\end{center}
                \vskip -8pt     
\[
\mbox{(b)}~~~~~~
\mbox{deterministic chaos : }
\expct{\obser{}}
     \,=\,
 \sum_{c}^{}                                
\obser[\Xx_c]\, \frac{1}
                     {\left|\Det\jMorb_c\right|}
~~~~~~
\] 
             \end{minipage}
                \vskip -4pt     
    \caption{\label{f:pde2a}
A bird's eye view of the quantum action landscape over the \pcell\ 
{\statesp} $\pS_\Bcell{A}$, \refeq{torusStatesp}.
White ellipses indicate the stationary points \refeq{eqMotionLagrang}, \ie, the 
set of all deterministic solutions $\{\Xx_a,\Xx_b,\Xx_c,\cdots,\Xx_g\}$. 
They form the skeleton on which the partition sums of both quantum chaos 
and deterministic chaos / turbulence are evaluated, with the common 
deterministic backbone, but different weights. 
(a) 
For a quantum theory, the semiclassical partition sum \refeq{n-pt-corr1} 
is an approximation, with quantum probability {amplitude} phases given 
by deterministic solutions' actions, and stability weights given by 
square roots of the deterministic ones. 
(b) 
For a \dft\ the probabilities that form the partition sum 
\refeq{detPartSum} are \emph{exact}, a Dirac porcupine of delta 
function quills, a quill for each solution of {\ELe}. 
    }
\end{figure}

For chaotic (or `turbulent') systems deterministic solutions form a set 
of \statesp\ points, sketched in \reffig{f:pde2a}. 
In this paper we focus on this set of {deterministic} solutions. 
As we show here, despite vastly different appearances, the `chaos theory' of 
1970s is a deterministic Euclidean field theory (see 
\refappe{s:new}). 

For pedagogical reasons, we first 
restrict the theory to finite-dimensional \statesp\ \refeq{torusStatesp} of 
a {\pcell} $\Bcell{A}$. These finite volumes are not meant to serve as 
finite approximations to the infinite {\Blatt}s $\lattice_\Bcell{A}$: as 
is standard in solid state physics, the actual calculations 
\edit{are always}
carried out over the infinite lattice, more precisely (not standard in 
solid state physics, but necessary to describe a chaotic field theory) 
over the set of all periodic lattice {\fconfs} \refeq{periodLConf} over 
all {\Blatt}s $\lattice_\Bcell{A}$ \refeq{BravLatt}, or, in field 
theory,\rf{MalWit07,CGHMV12} as the `sum over geometries'. 

\subsection{Euclidean field theory}
\label{s:Euclidean} 

In Euclidean field theory a {\fconf} $\Xx$ over {\pcell} $\Bcell{A}$ 
occurs with {\statesp} probability density 
\beq 
p_\Bcell{A}[\Xx]\,=\, 
\frac{1}{Z_\Bcell{A}}\,{e}^{-\action[\Xx]} \,,\qquad 
Z_\Bcell{A}=Z_\Bcell{A}[0] 
\,,
\ee{ProbConf}
where $Z_\Bcell{A}$  is a normalization factor ensuring that 
probabilities add up to 1, given by the \emph{{\pcell} partition sum}, an 
integral over the \pcell\ \Bcell{A} {\statesp} \refeq{torusStatesp}, 
\beq 
Z_\Bcell{A}[\beta] = \!\!
\int\!\! d\Xx_\Bcell{A}\,{e}^{-\action[\Xx]
    + \beta\cdot\Obser_\Bcell{A}[\Xx]}
    ,\;\;
d\Xx_\Bcell{A} = \prod_{z \in \Bcell{A}} d\ssp_z 
\,. 
\ee{partFunct} 
${\Obser}_\Bcell{A}[\Xx]$, the Birkhoff sum \refeq{BirkhoffSum} of the 
observable, or a set of observables, is multiplied by a parameter, or a 
set of parameters $\beta$.
$\action[\Xx]$ is the log probability (in statistics), the Gibbs 
weight (in statistical physics), or the action (in field theory) of the 
system under consideration (for examples, see \refsect{s:sptFT}). 

What is this `action'?
If lattice site fields are not coupled, the \spt\ {\fconf} $\Xx$ 
probability density \refeq{ProbConf} is a product of lattice site 
probability densities, and the partition sum is an exponential in the 
\pcell\ volume $\vol_\Bcell{A}$. If lattice site fields are weakly 
coupled, this exponential depends on the shape of the {\pcell} 
$\Bcell{A}$, $Z_\Bcell{A}[\beta]=\exp(\vol_\Bcell{A}W_\Bcell{A}[\beta])$. 
The {expectation value} \refeq{expctObser} of observable $\obser$ can be 
extracted from  the log of the {\pcell} partition sum 
$W_\Bcell{A}[\beta]$ by application of a $d/d\beta$ derivative: 
\beq
\expct{\obser}_\Bcell{A}
     \,=\,
\left.\frac{d~}{d\beta}W_\Bcell{A}[\beta]\right|_{\beta=0}
       =
\int d\Xx_\Bcell{A}\, \obser_\Bcell{A}[\Xx] \, p_\Bcell{A}[\Xx]
\,.
\ee{exObsPcell}

In this series of papers we focus on \spt\ systems with bounded variation 
of $W_\Bcell{A}[\beta]$, 
\beq
e^{\vol_\Bcell{A}W_{min}[\beta]} <
Z_\Bcell{A}[\beta] <
e^{\vol_\Bcell{A}W_{max}[\beta]}
\,,
\ee{partFunctBounds} 
with {\fconf} independent bounds $W_{min}[\beta]$, $W_{max}[\beta]$,
to be established in \refsect{s:sptchaos}.
This requirement is the \spt\ generalization of the 
\emph{uniform hyperbolicity}\rf{HasPes08}
of time-evolving dynamical systems, with Lyapunov exponents 
strictly bounded away from 0. 

Consider a field theory over a large square \pcell\ 
$\Bcell{A}=\BravCell{\speriod{}}{\speriod{}}{}$. 
In the infinite volume $\vol_\Bcell{A}=\speriod{}^2$ limit, exponential 
bounds of \refeq{partFunctBounds} guarantee convergence to the function 
\beq
W[\beta] = \lim_{\vol_\Bcell{A}\to\infty}
    \frac{1} {\vol_\Bcell{A}} \ln Z_\Bcell{A}[\beta] 
\,,
\ee{freeEnergy}
whose derivative yields the {\em expectation value} 
\beq
\expct{\obser} \,=\, \left. \frac{d\,W[\beta]}{d\beta~~} \right|_{\beta=0}
\,.
\ee{expctObserW1}
In this limit the \pcell\ contains the full hypercubic integer lattice 
$\lattice = \integers^d$, and the averaging integral 
${\int}d\Xx\,\obser[\Xx]\,p[\Xx]$ is the integral \refeq{exObsPcell} 
evaluated over the infinite $d$\dmn\ hypercubic lattice {\statesp} 
\refeq{StateSp}, an integral which we do not know how to evaluate. 

What we actually need to evaluate is not this integral, but the derivative
$W'$. That we accomplish in \refsect{s:zetaExpectV}.

\subsection{Deterministic lattice field theory}
\label{s:latDFT} 

What these {\fconf} probabilities \refeq{ProbConf} are depends on the theory.
Here, motivated by the above semiclassical quantum field theory, we are 
led to a formulation of the \emph{\dft}, where a {\fconf} $\Xx_c$ 
contributes only if it satisfies {\ELe} 
\beq 
F[\Xx_c]_z = 0 
\ee{eqMotion} 
on every lattice site $z$ (for our examples of {\ELe}, see
Eqs.~\refeq{sptCatlatt} -- \refeq{sptPhi4}).
That is all we require, regardless of whether the system has a Lagrangian 
formulation, or not (for example, {\NS} equations). \Dft, it turns out, 
is an elegant, to a novice perhaps impenetrable, definition of what we 
already know as deterministic chaos and/or turbulence (the precise 
relation is afforded by Hill's formulas, derived in \refrefs{LC21,LC22}). 
 
\medskip

\paragraph*{Definition: {\Lst}}
\begin{quote}
 is a $\lattice_\Bcell{A}$-{periodic} set of field values 
$\Xx_c=\{\ssp_z\}$ over the $d$\dmn\ lattice $z\in\integers^d$ that 
satisfies {\ELe} on every lattice site. 
\end{quote}
As any {\fconf} $\Xx$ is a \emph{point} in $\vol_\Bcell{A}$\dmn\ 
{\statesp} \refeq{torusStatesp}, so is a {\lst} $\Xx_c$. 
Furthermore, just as a temporal evolution period $\cl{}$ periodic point 
is a fixed point of the $\cl{}$th  iterate (translation by  $\cl{}$ temporal 
lattice sites) of the dynamical time-forward map, every \lst\ is a 
\emph{fixed point} of a set of symmetries of the theory 
(\refsect{s:2DprimeOrb} and \refeq{LindZeta}). 

System's defining equations \refeq{eqMotion} are  {\ELe} everyone must 
obey: look at your left neighbor, right neighbor, remember who you were, 
make sure you fit in just right. 
The set $\{\Xx_c\}$ of all possible {\lsts} is system's `Book of Life' - 
a catalog of all possible `lives', \spt\ patterns that {\ELe} 
allow, each life a \emph{point} in system's \statesp, each 
life's likelihood given by its \Hilldet. 

A {\lst} is a fixed spacetime pattern: the `time' direction is just one 
of the coordinates. If you insist on visualizing solutions as evolving in 
time, a {\lst} is a video, not a snapshot of the system at an instant in 
time (that these are merely different visualizations is proven in 
\refref{LC22}). 

\medskip

\paragraph*{Definition: Deterministic probability density.}
\begin{quote}
For a {\dft}, the {\statesp} {probability density} is non-vanishing only 
at the {exact} solutions  of {\ELe}, 
\beq 
 p[\Xx]\,=\, \frac{1}{Z}\,\delta(F[\Xx]) 
\,, 
\ee{DiracDeltaExp} 
where the $\vol_\Bcell{A}$\dmn\ Dirac delta function 
$\delta(\cdots)$ enforces {\ELe} on every lattice site.
\end{quote}
In contrast to the quantum action landscape of  \reffig{f:pde2a}\,(a), 
for a chaotic (or `turbulent') deterministic system the probability 
density \reffig{f:pde2a}\,(b) is a Dirac porcupine, a set of delta 
function quills over the \pcell\ {\statesp} $\pS_\Bcell{A}$,
\refeq{torusStatesp}, one for each solution of {\ELe}. 
The {\pcell} {\Bcell{A}} 
\emph{\detSum} \refeq{partFunct} is given by the sum 
over all {\lsts}, here labeled `$c$',
\bea 
Z_\Bcell{A}[\beta] 	&=& 
        \sum_c \int_{\pS_c} \!\!\!\! d\Xx_\Bcell{A}\,\delta(F[\Xx])\,
        {e}^{\beta\cdot\Obser_\Bcell{A}[\Xx]}
    \continue
    &=&
        \sum_c \frac{1}{\left|\Det\!\jMorb_{{\Bcell{A}},c}\right|}\,
        {e}^{\beta\cdot{\Obser}_c}
\,, 
\label{detPartSum}  
\eea
where $\pS_c$ is an open infinitesimal neighborhood of \statesp\ point 
$\Xx_c$, and 
\beq 
{\Obser}_c \,=\, 
             \sum_{z\in\Bcell{A}} \obser_z[\Xx_c]
\ee{lst_aver}
is the Birkhoff sum \refeq{BirkhoffSum} 
of observable $\obser$ over \lst\ $\Xx_c$,
to be discussed in \refsect{s:zetaExpectV}.
    \edit{
Variants of deterministic partition sums 
had been computed, by different methods, in different contexts, by 
many,\rf{IvIzHu02,MalWit07,CaSiLo19} but \emph{always} on a given finite 
{\pcell} (geometry),  \emph{never} on the totality of infinite {\Blatt}s, 
as \new{we do here}. 
    }

We refer to 
\beq 
(\jMorb_c)_{z'z} = \frac{\delta F[\Xx_c]_{z'}}{\delta \ssp_{z}} 
\ee{jacOrb} 
    \edit{
evaluated as the $[\vol_\Bcell{A}\!\times\!\vol_\Bcell{A}]$ 
matrix over the {\pcell} $\Bcell{A}$, as the \emph{\jacobianOrb} 
$\jMorb_{\Bcell{A},c}$, to the linear operator \refeq{jacOrb}, evaluated 
over infinite {\Blatt} $\lattice_\Bcell{A}$, as the \emph{\jacobianOp} 
$\jMorb_{c}$
(also called `Hessian', a `Jacobi matrix', or a discrete 
Schr{\"o}dinger operator\rf{Bount81,Simon82,Simons97}), 
and to the determinant\rf{Hill86,Poinc1886,VanVleck28} 
$\Det\jMorb_{c}$  in \refeq{detPartSum} as the 
\emph{\Hilldet}
(also called `discriminant' or `Hill discriminant'\rf{Toda89}).
    }
We add prefix `orbit' to emphasize the distinction between the global \emph{orbit}
stability, and the  stability of a forward-in-time evolved \emph{state}. 
Our {\stabexp} $\Lyap_p$, \refeq{lnDetStabExp}, is the temporal-evolution 
sum of expanding \edit{Floquet exponents} generalized to any spacetime 
dimension. 

The {\Hilldet} is the central innovation of our formulation of \spt\ 
chaos, so we discuss at length in sections \ref{s:orbJacob}, 
\ref{s:pcellStab}, \ref{s:Bloch}, \ref{s:POT}  and 
\refappe{s:catlattComp}. 

\subsection{{\Mosaic}s}
\label{s:mosaics} 

The backbone of a \emph{deterministic} chaotic system is, thus, the set of 
all \spt\ solutions of system's defining equations \refeq{eqMotion} that we here 
refer to as \emph{{\lsts}}, or, on occasion, as (multi-)\emph{{\po}s}. 
Depending on the context, in the literature they appear under many other 
names. For example, Gutkin \&  Osipov\rf{GutOsi15} refer to a two\dmn\ 
{\lst} $\Xx_c$ as a `many-particle \po', with each lattice site field 
$\ssp_{n\zeit}$ `doubly-periodic', or `closed'. 

{\em {\Mosaic}s.~~} For a $d$\dmn\ \spt\ field theory, symbolic 
description is not a one\dmn\ temporal ``symbolic dynamics'' itinerary, 
as, for example, a symbol sequence that describes a time-evolving 
$N$-particle system. The key insight --an insight that applies to 
coupled-map lattices, and field theories modeled by 
them,\rf{BunSin88,PolTor92b,PetCorBol06,PetCorBol07,GutOsi15,GHJSC16} not 
only systems considered here-- is that a {\fconf} \( 
\Xx=\{\ssp_{z}\} 
\) over a $d$\dmn\ spacetime lattice $z\in \integers^{d}$ is labeled by 
a finite alphabet symbol lattice \( 
\Mm=\{\m_{z}\} 
\) over the same $d$\dmn\ spacetime lattice. 

\begin{figure}\begin{center}
            \begin{minipage}[c]{0.15\textwidth}\begin{center}
\includegraphics[width=1.0\textwidth]{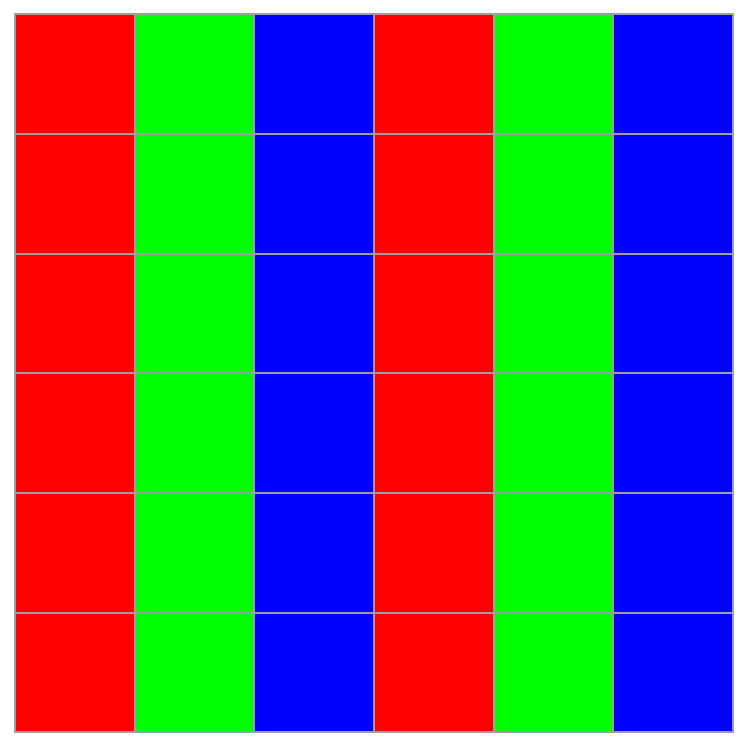}\\(a)
            \end{center}\end{minipage}
            \hskip 1.2ex
            \begin{minipage}[c]{0.15\textwidth}\begin{center}
\includegraphics[width=1.0\textwidth]{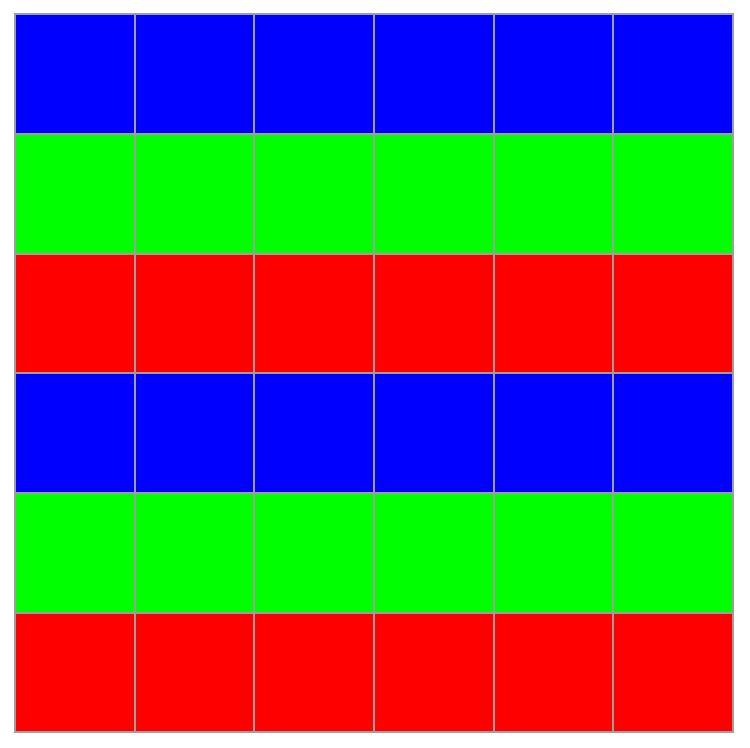}\\(b)
            \end{center}\end{minipage}
            \hskip 1.2ex
            \begin{minipage}[c]{0.15\textwidth}\begin{center}
\includegraphics[width=1.0\textwidth]{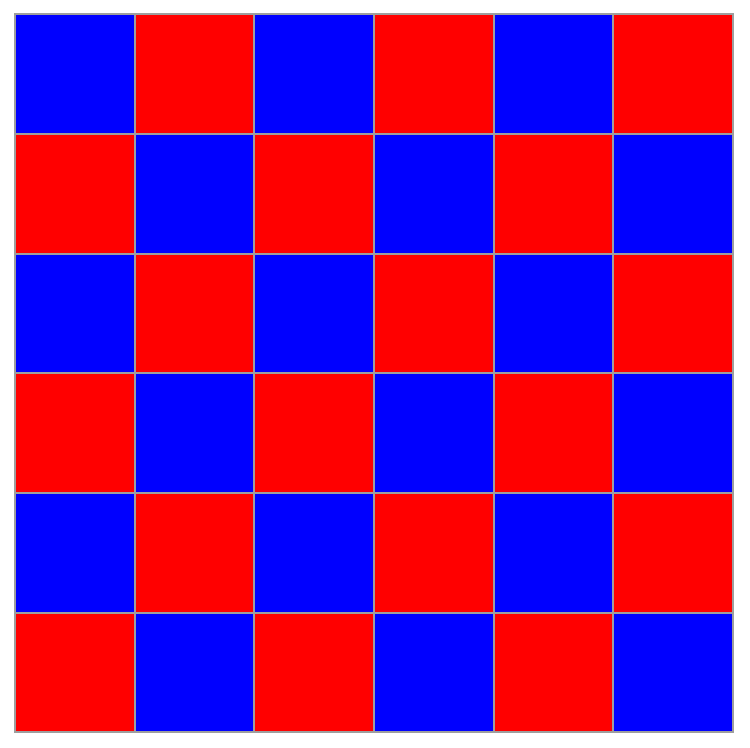}\\(c) 
            \end{center}\end{minipage}
            \\\vskip  6pt
            \begin{minipage}[c]{0.15\textwidth}\begin{center}
\includegraphics[width=1.0\textwidth]{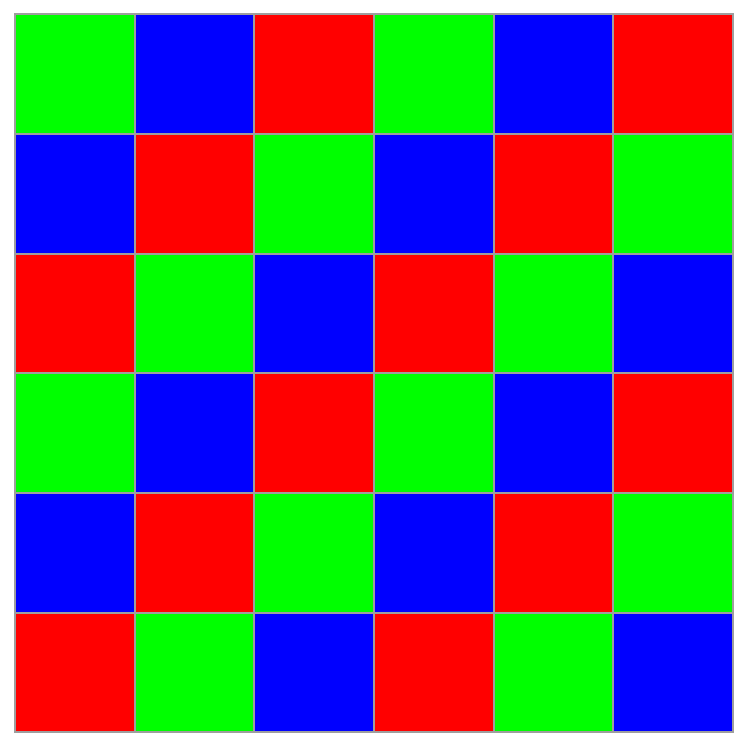}\\(d)
            \end{center}\end{minipage}
            \hskip 1.2ex
            \begin{minipage}[c]{0.15\textwidth}\begin{center}
\includegraphics[width=1.0\textwidth]{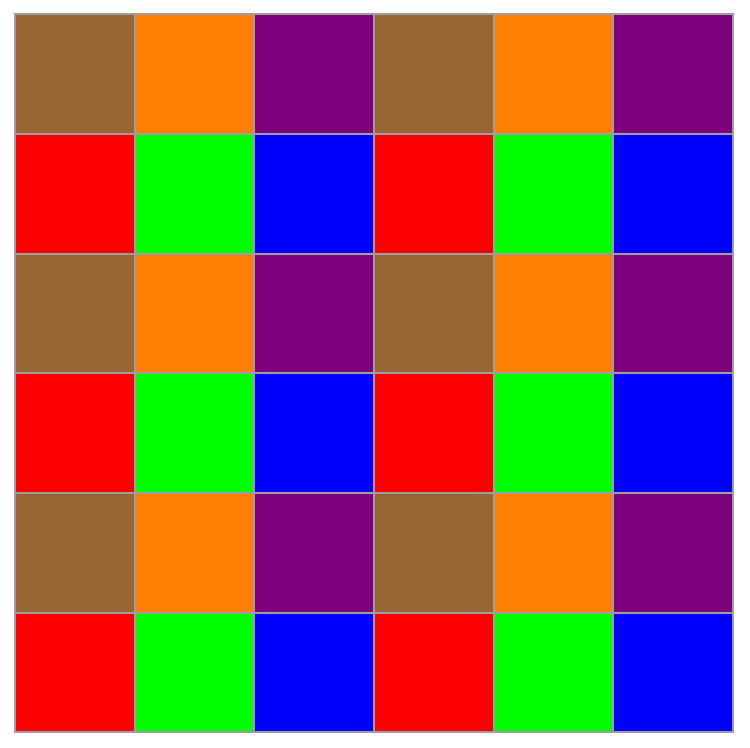}\\(e) 
            \end{center}\end{minipage}
            \hskip 1.2ex
            \begin{minipage}[c]{0.15\textwidth}\begin{center}
\includegraphics[width=1.0\textwidth]{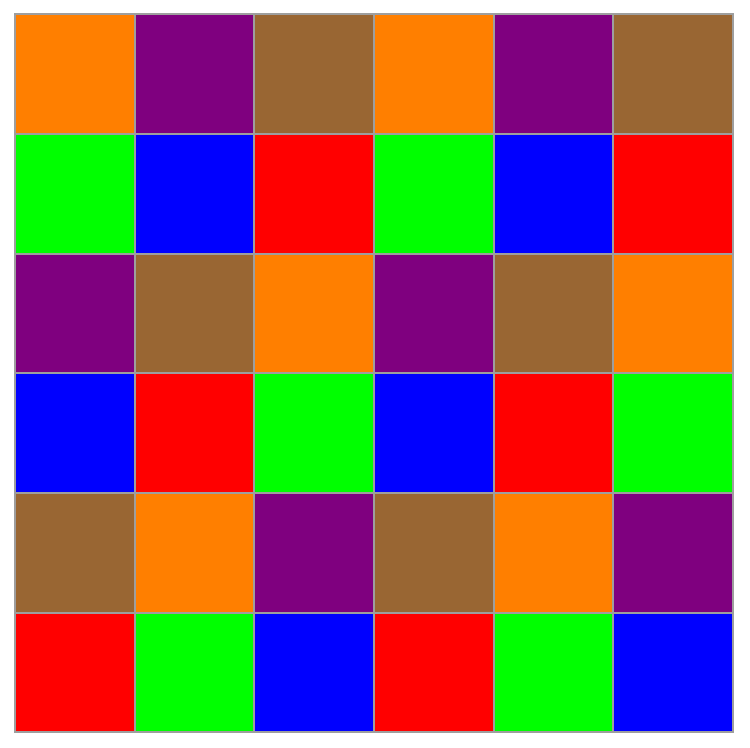}\\(f) 
            \end{center}\end{minipage}
\end{center}
            \vskip -12pt     
  \caption{\label{f:SpecialBravaisLatt}
Examples of
\spt\ {\mosaic} tilings \refeq{Mm_c} of
$\BravCell{6}{6}{0}$ {\pcell} by repeats of smaller prime
{\lsts}.
(a)
$\BravCell{3}{1}{0}$
temporally \steady.
(b)
$\BravCell{1}{3}{0}$ spatially \steady.
(c)
$\BravCell{2}{1}{1}$
relative-periodic {\primeOrb}, spatial period-2, temporal period-2;
compare with \reffig{f:2x1rpo}\,(a).
(d)
$\BravCell{3}{1}{1}$ relative-periodic {\primeOrb}, spatial period-3, temporal period-3.
(e)
$\BravCell{3}{2}{0}$ spatial period-3, temporal period-2.
(f)
$\BravCell{3}{2}{1}$ of Figs. \ref{f:2x1rpo}\,(b) and \ref{f:pcellTiling}.
It is a relative-periodic {\primeOrb}, of spatial period-3, temporal period-6.
See also \reffig{f:symmBlock} and \refappe{s:primeLatt}.
}
\end{figure}

\begin{figure}
  \centering
(a)~~~
\includegraphics[width=0.41\textwidth]{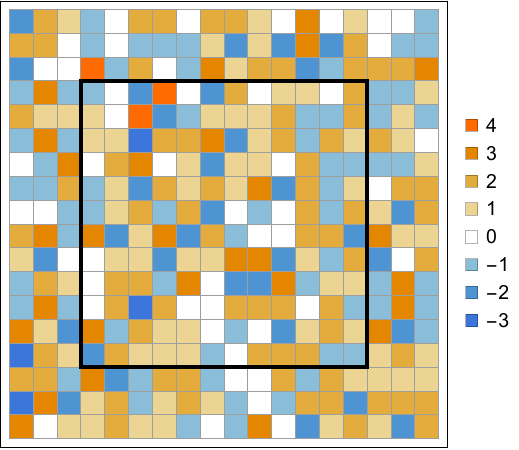}
\\\vskip 12pt
(b)~~~
\includegraphics[width=0.41\textwidth]{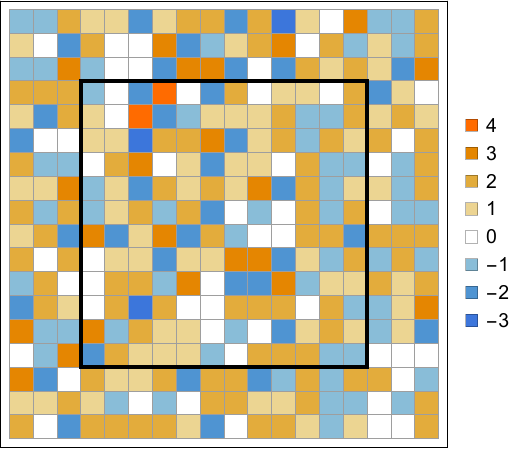}
    \caption{\label{f:symmBlock}
Mosaics \refeq{Mm_c} of two \BravCell{18}{18}{0} {\catlatt} {\lsts} which 
share the sub-{\mosaic}  within the $[12\!\times\!12]$ region enclosed by 
the black square, and have different symbols outside 
the sub-{\mosaic}. 
Color-coded 8-letter alphabet \refeq{catlatt2d}, $\mu^2=1$.
This will be discussed further in \reffig{f:18x18shadow}.
}
\end{figure}

For field theories studied here, one can partition the values of a 
lattice site field $\ssp_z$ into a set of $|\A|$ disjoint intervals, and 
label each interval by a letter $\Ssym{z}\in\A$ drawn from an alphabet 
$\A$, let us say 
\beq 
 \A  =
 \{{1},2,\cdots,|\A|\}
\,. 
\label{alphabet}
\eeq
This associates a $d$\dmn\ 
{\emph{\mosaic}} $\Mm_c$ to a  {\fconf} $\Xx_c$ over $d$\dmn\ 
lattice\rf{ChMaVa96,ChMaSh98,ChoMal95a,ChoMal95b}
\beq 
\Mm_c 
  =
\left\{\Ssym{z}\right\} \,,\quad 
   \Ssym{z}\in\A
\,, 
\ee{Mm_c} 
elsewhere called `symbolic representation';\rf{BunSin88} 
`\spt\ code';\rf{CouFer97} `symbol tensor';\rf{SterlingThesis99} `symbol 
lattice';\rf{Just05,GHJSC16} `symbol table';\rf{MacKay05} `local symbolic 
dynamics';\rf{PetCorBol07} `symbol block'.\rf{GHJSC16,LC21} A {\mosaic} 
serves both as a proxy (a `name') for the \lst\ $\Xx_c$, and its 
visualization as color-coded symbol array $\Mm_c$ (for examples, see 
\reffig{f:SpecialBravaisLatt}, \reffig{f:symmBlock}, companion paper 
III,\rf{WWLAFC22} and Lego mosaics of \refref{Perry23}). 

If there is only one, distinct {\mosaic} $\Mm_c$ for each \lst\ $\Xx_c$, 
the alphabet is said to be \emph{covering}. While each \lst\ thus gets 
assigned a unique {\mosaic} that paginates its location in the Book of 
Life, the converse is in general not true. If a given {\mosaic} $\Mm$ 
corresponds to a \lst, it is \emph{\admissible}; otherwise 
$\Mm$ has to be 
deleted from the list of {\mosaic}s. 
That's what we do in practice, see \refappe{s:catlattCount}.

In the temporal-evolution setting there is a variety of methods of 
finding grammar rules that eliminate inadmissible {\mosaic}s. While such 
rules for 2- or higher-dimensional lattice field theories remain, in 
general, not known to us, 
we are greatly helped by the observation that 
in the `anti-integrable' limit\rf{AuAb90,aub95ant,StMeiss98,Beck02,WWLAFC22} 
(also known as the `anti-continuum limit' in solid state 
physics,\rf{CaFlKi04} `large dissipation limit' in nonlinear 
dynamics,\rf{Bount81}, `weak diffusive coupling' in stochastic field 
theory\rf{Beck02}) 
finite alphabets are known, and offer good starting 
approximations\rf{WWLAFC22} to the corresponding numerically exact {\lsts}. 
For \catlatt,  see \refsect{s:open}, question~6.

\section{Examples of {\spt} lattice field theories}
\label{s:sptFT} 

We shall construct the field theory's {\detSum} 
(\refsect{s:POT}) 
by first enumerating all {\Blatt}s (geometries) $\lattice_\Bcell{A}$ 
(\refsect{s:Bravais}), determining {\primeOrbs} over each, 
computing the weight of each (\refsect{s:Bloch}), and then 
(\refsect{s:POT}) adding together the contributions of {\lsts} for each. 
The potentials may be bounded ($\ssp^4$ theory) or unbounded ($\ssp^3$ 
theory), or the system may be energy conserving or dissipative, as long 
as the set of its {\lsts} $\Xx_c$ is bounded in system's \statesp\ 
\refeq{stateSp}. To get a feel for how all this works, we illustrate the 
theory by applying it to four lattice field theories that we now 
introduce. 

A field theory is defined either by its action, for example a lattice sum 
over the Lagrangian density for a discretized scalar $d$\dmn\ Euclidean 
$\phi^k$ theory%
,\rf{FriMil89,DulMei00,LiMal04,Munster10,AnBoBa17,AnBoBa18,Anastassiou21,CvitanovicYT02b} 
\beq 
S[\Xx] = \sum_z \left\{ 
    \frac{1}{2} \sum_{\mu =1}^d
(\partial_{\mu}\ssp)_z^2 + V(\ssp_z) 
            \right\}
\,, 
\ee{actionDscr} 
    %
    %
with a {local} potential $V(\ssp)$ the same for every lattice site $z$, 
or, if lacking a variational formulation, by {\ELe} $F[\Xx]_z=0$. 

{\ELE} \refeq{eqMotion} now take the form of second-order 
difference equations 
\bea 
- \Box\,\ssp_z + V'(\ssp_z) 
    &=&
0  
\,.  
\label{1dTempFT} 
\eea 

In lattice field theory `locality' means that a field at site $z$ 
interacts only with its neighbors. To keep the exposition as simple as 
possible, we treat here the spatial and 
temporal directions on equal footing, with the graph Laplace operator%
\rf{LindMar95,Pollicott01,Cimasoni12,GodRoy13} 
\beq 
\Box\,\ssp_z = 
    \sum_{z'}^{||z'-z||=1} \!\! (\ssp_{z'} - \ssp_z)
 \quad \mbox{for all} \ z,z' \in  \integers^d
\ee{LapOp}
comparing the field on lattice site $z$ to its $2d$ nearest 
neighbors. For example, the two\dmn\ square lattice Laplace operator is 
given by 
\beq
\Box\, = \shift_1        + \shift_{2} - 4\id + 
\shift_{2}^{-1} + \shift_{1}^{-1} 
\,, 
\ee{2dLap} 
where $\shift_1,\shift_2$ {\shiftOp}s 
(see \refeq{dDimTranslGrp} for a group-theoretical  perspective) 
\beq 
(\shift_1)_{n\zeit, n'\zeit'} = \delta_{n + 1, 
n'}\,\delta_{\zeit\zeit'} 
\,, \quad (\shift_2)_{n\zeit, n'\zeit'} = 
\delta_{nn'}\,\delta_{\zeit + 1, \zeit'}
\ee{shiftOperator2d} 
translate a {\fconf} 
\[
(\shift_1\Xx)_{n\zeit}=\ssp_{n+1,\zeit}
\,,\quad
(\shift_2\Xx)_{n\zeit}=\ssp_{n,\zeit+1}
\,,
\]
by one lattice spacing \refeq{1dShift} in the spatial, temporal 
direction, respectively. 

Here, and in papers I and III\rf{LC21,WWLAFC22} we investigate {\spt}ly 
chaotic lattice field theories using as illustrative examples the 
$d$\dmn\ hypercubic lattice \refeq{LattField} discretized Klein-Gordon 
free-field theory, {\catlatt}, {\spt} {$\phi^3$} theory, and {\spt} 
{$\phi^4$} theory, defined, respectively by {\ELe} \refeq{eqMotion}
\bea 
- \Box\,\ssp_{z} \;+\;  {\mu^2}\ssp_{z} 
    &=&
0 \,, \qquad \ssp_{z} \in \reals \,, 
\label{sptTempKG}\\
- \Box\,\ssp_{z} \;+\;  {\mu^2}\ssp_{z} - \Ssym{z} 
    &=&
0 \,, \qquad \ssp_{z} \in [0,1) 
\label{sptCatlatt}\\
- \Box\,\ssp_{z} \;+\;  \mu^2\,({1}/{4}-\ssp_{z}^2) 
    &=&
0 \,, 
\label{sptHenlatt}\\
- \Box\,\ssp_{z} \;+\; \mu^2(\ssp_{z}-\,\ssp_{z}^3) 
    &=&
0 \,. 
\label{sptPhi4}
\eea 
The anti-integrable form\rf{AuAb90,aub95ant,StMeiss98} of the {\spt} 
{$\phi^3$}, \refeq{sptHenlatt}, and {\spt} {$\phi^4$}, \refeq{sptPhi4}, 
is explained in 
\refsect{s:phi4}, and the companion paper III.\rf{WWLAFC22} 
    
    \edit{
The homogeneous, free-field case,
\refeq{sptTempKG} 
is known as the discretized
\HREF{https://en.wikipedia.org/wiki/Screened_Poisson_equation}
{{\em \sPe}},\rf{Dorr70,FetWal03} with parameter $\mu$ the reciprocal
screening length in the Debye-H{\"u}ckel or Thomas-Fermi approximations.
In statistical mechanics, the related lattice discretized Helmholtz 
equation is known as the `Gaussian model',%
\rf{Kadanoff00,Fradkin13,Shankar17,Marino17}
and in field theory as the {Yukawa}
or {Klein–Gordon equation}\rf{wikiKleinGordon} for a boson of Klein-Gordon (or Yukawa) mass 
$\mu$. This free-field theory is studied by 
many, some recent examples are \refrefs{Shimizu12,CaSiLo19,CaLoSi20}. 

While the 
free-field theory teaches us much about how a field theory works, 
it is not an example of a chaotic field theory: {\ELe} 
\refeq{sptTempKG} are linear, with a single deterministic solution, the 
\steady\ $\ssp_z=0$.
    }

\subsection{\catLatt}
\label{s:catlatt} 

The {\catlatt}, \refeq{sptCatlatt}, that we derive next, is arguably the 
simplest example of a chaotic (or `turbulent') {\dft}, for which all 
local degrees of freedom are hyperbolic (anti-harmonic, `inverted 
pendula') rather than oscillatory `harmonic oscillators'.\rf{CvitanovicYT02a} We will use it 
throughout this paper to illustrate our field-theoretic formulation of 
{\spt} chaos. 

\medskip

\paragraph*{Definition: {\catLatt}}
\begin{quote}
\ELe\ are
\edit{ 
\bea
&& \left( - \Box + \mu^2 \right) \ssp_{z} \quad (\mod\; 1) = 
0 \,, \continue
&& z \in \integers^d \,, \quad \ssp_{z} \in [0,1)
\,,
\label{{catlattKGMod} }
\eea
or
}
\beq
\left( - \Box + \mu^2 \right) \ssp_{z} = 
\Ssym{z} \,, \qquad z \in \integers^d \,, \quad \ssp_{z} \in [0,1)
\,,
\ee{catlattKG} 
with the circle $\ssp_{z}\;(\mbox{mod}\;1)$ condition 
enforced by integers $\Ssym{z}$, called `winding numbers',\rf{Keating91}
or, as they shepherd stray points back into the \statesp\ unit hypercube, 
`stabilising impulses'.\rf{PerViv}
\end{quote}

\noindent
For a \pcell\ $\Bcell{A}$ we can write it in a matrix 
form,
\beq
F[\Xx_\Mm] = \jMorb_{\Bcell{A}}\Xx_\Mm-\Mm = 0 \,, \quad 
\Xx_\Mm \in [0,1)^{\vol_{\Bcell{A}}}
\,,
\ee{catlatt}
where $\jMorb_{\Bcell{A}}=-\Box+\mu^2\id$ is the {\jacobianOrb} 
\refeq{jacOrb}, and $\Mm$ a $d$\dmn\ {\mosaic}, \refeq{Mm_c}. 

    \edit{ 
Theories with lattice site field values compactified to a circle are known 
as `compact boson' or `compact scalar' theories, see for example 
\refrefs{CheSei22,FazSul23}. 
The `mod~1' in its definition makes the {\catlatt} a
discontinuous {piecewise-linear} map,
a theory that is \emph{non}linear in the sense that it is not defined 
globally by a single linear relation, such as the free-field theory 
\refeq{sptTempKG}, but by a set of distinct piecewise linear conditions, 
one for each {\mosaic}  $\Mm$. 

The \templatt, 
\beq
- \ssp_{\zeit+1} + {s}\,\ssp_{\zeit} - 
\ssp_{\zeit-1} 
    =
\Ssym{\zeit} \,, \quad \zeit\in\integers \,, \quad \ssp_{\zeit} \in 
[0,1) \,,
\ee{catMapNewt}
a one\dmn\  case of {\catlatt}
studied in companion paper I,\rf{LC21} was introduced by Percival and
Vivaldi\rf{PerViv} as a Lagrangian reformulation of the Hamiltonian
Thom-Anosov-Arnol'd-Sinai {`cat map'}\rf{ArnAve,deva87,StOtWt06}
(for a historical overview, see
{Appendix A} of companion paper I\rf{LC21}).

To derive the \catlatt, add to the
{\templatt} a ($d\!-\!1$)\dmn\
spatial lattice where each site field couples to its nearest spatial
neighbors, in addition to its nearest past and future field values.
Take the spatial coupling strength the same as the temporal
coupling strength (just a lattice constant rescaling, as in the derivation of
\refeq{CaFlKi04:1dPhi4c}). The result is the Euclidean,
space$\;\Leftrightarrow\;$time-interchange symmetric difference equation
\refeq{catlattKG}.

In two spacetime dimensions, the \catlatt\ {\ELe} \refeq{catlattKG} 
are a  five-term recurrence relation
introduced by Gutkin \& Osipov,\rf{GutOsi15,GHJSC16}
\beq 
      -\ssp_{j,\zeit+1} - \ssp_{j,\zeit-1}
+ 2{s}\,\ssp_{j\zeit} 
     - \ssp_{j+1,\zeit} - \ssp_{j-1, \zeit}
     \,=\, \Ssym{j\zeit}
\,,
\ee{CatMap2d} 
    \ifsubmission\else
\toVideo{youtube.com/embed/rTh_I0KOasY} 
    \fi
with ${\mu}^2=2({s}-2)$.\rf{CvitanovicYT02a}
    }
As in \refeq{2dLap}, the {\jacobianOp} in \refeq{catlatt} can be 
expressed in terms of {\shiftOp}s, 
\beq 
\jMorb = - 
\shift_{1} - \shift_{2} + 2{s}\id - \shift_{2}^{-1} - \shift_{1}^{-1} 
\,. 
\ee{jacOrb2d} 
We  study the one\dmn\ {\templatt} \refeq{catMapNewt} in some depth in
companion paper I.\rf{LC21} In this paper we focus on the $d=2$
{\catlatt} \refeq{CatMap2d}, with computational details relegated to
\refappe{s:catlattComp}.

    \edit{
In the Hamiltonian, forward-in-time temporal evolution formulation,
the dynamics is generated by iterations of a {piecewise linear} cat \emph{map}.
In the {\spt} formulation there is \emph{no map}, only \ELe, in the form of 
recurrence conditions, so we refer to the three-term recurrence 
\refeq{catMapNewt} as the \emph{\templatt}, and to the recurrence 
condition \refeq{catlattKG}  in higher \spt\ dimensions as the 
\emph{\catlatt}. 

How is this kind of field theory related to more familiar field theories?
Think of the discretized Helmholtz-type field theory as a
spring mattress:\rf{Zee10} you push it, and it pushes back, it oscillates.
\catLatt, on the other hand, has a `cat' (a `rotor'\rf{LC21})  at every lattice
site: you push it, and the cat runs away, but, forced by the compact boson
condition \refeq{catlattKG}, it eventually has to come back.
Chaos issues.
Our task is to herd these cats over all of the spacetime.

\subsection{A many-body example: deterministic $\phi^4$ theory}
\label{s:phi4} 

The {\sPe} \refeq{sptTempKG} is of the same form as the inhomogeneous 
Helmholtz equation, but for the sign of $\mu^2$, with the oscillatory 
$\sin$, $\cos$ solutions replaced by the hyperbolic $\sinh$, $\cosh$, and 
exponentials.\rf{GraRyz} 
To understand how dynamical systems' `chaos theory' morphs into a 
Euclidian strong-coupling, anti-integrable  \dft, consider the 1955 
{Fermi-Pasta-Ulam-Tsingou}\rf{fpu65,CaFlKi04} 
chain of molecules coupled with springs,
\beq
  \frac{d^2 \ssp_{n}}{dt^2}
{\boldsymbol{-} }
   \frac{1}{(\Delta x)^2} (\ssp_{n+1} - 2\,\ssp_{n} + \ssp_{n-1})
 - \ssp_{n} + \ssp_{n}^3
  \,=\, 0
\,,
\ee{CaFlKi04:1dPhi4} 
with spring constant $1/\Delta x$.
In absence of nonlinear terms, the solutions are oscillatory eigenmodes.
With nonlinearities they can also be breathers, intrinsic localized
modes, \etc, with perturbations that are oscillatory and bounded in
magnitude.\rf{CaFlKi04,KaRoCa24}

Next, consider Eq.~\refeq{CaFlKi04:1dPhi4} with
a ${\boldsymbol{+}}$ sign,
\[ 
  \frac{d^2 \ssp_{n}}{dt^2}
{\boldsymbol{+} }
   \frac{1}{(\Delta x)^2} (\ssp_{n+1} - 2\,\ssp_{n} + \ssp_{n-1})
 - \ssp_{n} + \ssp_{n}^3
  \,=\, 0
\] 
obtained by interpreting the \emph{imaginary} spring constant as a Klein-Gordon
mass
$\mu^2={\boldsymbol{-}}\,(\Delta x)^2$. Discretize time,
\[
  \frac{d^2 \ssp_{n}}{dt^2}
        \Rightarrow
\frac{1}{(\Delta \zeit)^2} (\ssp_{n,\zeit+1} - 2\,\ssp_{n\zeit} + \ssp_{n,\zeit-1})
\,,
\] 
rescale by $\Delta \zeit$, and combine the 2nd order derivatives into
the 2$d$ Laplacian of \refeq{2dLap},
\beq
{\boldsymbol{-}}
\Box\,\ssp_{z}
\;+\;  \mu^2(\ssp_{z}-\,\ssp_{z}^3)
  \,=\, 0
\,.
\ee{CaFlKi04:1dPhi4c}
This is the Euclidean massive scalar  Klein-Gordon $\phi^4$ field theory
\refeq{sptPhi4}, an `inverted potential' {chaotic field theory} that we study
here, and in the companion paper III,\rf{WWLAFC22} 
theory with hyperbolic instabilities and turbulence.
    } 

\subsection{Field theories that are first order in time}
\label{s:CML}

In this series of papers\rf{LC21,CL18,WWLAFC22} 
we illustrate the {\spt}ly chaotic field theory by 
discretizations of PDEs of second order both in space and time, 
such as the Euclidean $\phi^3$ and $\phi^4$ theories. 
Of equal, if not greater importance to us are `dissipative' 
PDEs which are of first order in time direction, second or higher in spatial directions, 
such as \KS\ and \NS.
Discretizations of such theories was pioneered 
by Kaneko\rf{Kaneko83,Kaneko84,Kaneko89},
whose  diffusive `coupled map lattices' are  
hypercubic spacetime lattice discretization of reaction-diffusion PDEs,
and whose study in our \spt\ field theory formulation has been initiated 
by Lippolis.\rf{Lippolis25}

While for \NS\ the coordinate $\zeit$ is the physical time, in many 
applications, such as Newton descent solution 
searches,\rf{GM00aut, CvitLanCrete02,lanCvit07} 
diffusion models,\rf{Anderson82}
normalizing flows in machine learning,\rf{TabVan10}  
Parisi-Wu stochastic quantization\rf{ParWu81}
and Beck chaotic quantization of field theories\rf{Beck95,Beck02}
$\tau$ is a  \emph{fictitious} time.   
A fascinating application of such quantization
is the 2021 Kitano, Takaura, and 
Hashimoto\rf{KiTaHa21}  lattice QED 
evaluation of the anomalous magnetic moment 
of the electron to a surprisingly good 
accuracy.\rf{Kitano24} A \spt\ \po\ reformulation of such lattice gauge 
calculations is one of the future challenges for the theory developed here.


\section{\Spt\ stability of a {\lst}}
\label{s:orbJacob}

For field theories \refeq{1dTempFT} considered here, the {\jacobianOps} 
\refeq{jacOrb} are of form \beq \jMorb_{{z}{z}'} = - \Box_{{z}{z}'} + 
V''(\ssp_z)\,\delta_{{z}{z}'} \,, \ee{ELeFT} with the free field 
\refeq{sptTempKG} and {\catlatt} \refeq{sptCatlatt}, {$\phi^3$} 
\refeq{sptHenlatt}, {$\phi^4$} \refeq{sptPhi4} {\jacobianOps} \bea 
\jMorb_{{z}{z}'} &=& 
    - \Box_{{z}{z}'} \;+\;  {\mu^2}\delta_{{z}{z}'}
\,, 
\label{ELeSptTemplatt}\\
\jMorb_{{z}{z}'} &=& - \Box_{{z}{z}'} \;-\;  2\,\mu^2 
\ssp_{z}\,\delta_{{z}{z}'} \,, 
\label{ELeSptHenlatt}\\
\jMorb_{{z}{z}'} &=& - \Box_{{z}{z}'} \;+\; 
\mu^2(1\,-\,3\,\ssp_{z}^2)\,\delta_{{z}{z}'} \,. \label{ELeSptPhi4} \eea 

Sometimes it is convenient to lump the diagonal terms of the discrete 
Laplace operator \refeq{2dLap} together with the site potential 
$V''(\ssp_z)$. In that case, the {\jacobianOp} takes the $2d+1$ banded 
form 
\bea 
\jMorb &=&  
\sum_{j=1}^d (-\shift_{j} + {\cal{D}} - \shift_{j}^{-1}) 
    \,,\continue
{\cal{D}}_{{z}{z}'} &=& \diag_z\delta_{{z}{z}'} 
    \,,\;\;
\diag_z = V''(\ssp_z)/d + 2 \,,
\label{2dLapStrech}
\eea
where $\shift_j$ {\shiftOp}s \refeq{shiftOperator2d} translate the 
{\fconf} by one lattice spacing in the $j$th hypercubic lattice 
direction, and we refer to the diagonal entry $\diag_z$ as the 
\emph{stretching factor} at lattice site $z$. For the free field and 
{\catlatt} \refeq{ELeSptTemplatt}, {$\phi^3$} \refeq{ELeSptHenlatt}, 
{$\phi^4$} \refeq{ELeSptPhi4} theories the stretching factor $\diag_z$ 
is,  respectively, \bea s &=& {\mu^2}/d + 2 \,, 
\label{stretchCatlatt}\\
\diag_z &=& -2\,\mu^2 \ssp_{z}/d+2 \,, 
\label{stretchSptHenlatt}\\
\diag_z &=& \mu^2(1\,-\,3\,\ssp_{z}^2)/d+2 \,. \label{stretchSptPhi4} 
\eea 

What can we say about the spectra of {\jacobianOps}? In the 
anti-integrable limit\rf{AuAb90,aub95ant,StMeiss98} the diagonal, 
`potential' term  in \refeq{ELeFT} dominates, and one treats the 
off-diagonal Laplacian (`kinetic energy') terms as a perturbation. For 
field theories 
\refeq{ELeSptTemplatt}-\refeq{ELeSptPhi4} considered here, in the 
anti-integrable limit, in any spacetime dimension, the eigenvalues of the 
{\jacobianOp} are proportional to the Klein-Gordon mass-squared, \beq 
\jMorb_{{z}{z}'} \,\to\, \mu^2 c_{z}\,\delta_{{z}{z}'} 
    \,,\qquad
\mu^2 \mbox{ large,} \ee{antiIntSptTemplatt} where $c_{z}$ is a 
theory-dependent constant. 
For details of  {$\phi^3$} and {$\phi^4$} field theories, see the 
companion paper III.\rf{WWLAFC22}

In what follows, it is crucial to distinguish the
$[\vol_\Bcell{A}\!\times\!\vol_\Bcell{A}]$ {\jacobianOrb}, evaluated over 
a {finite volume} {\pcell} $\Bcell{A}$, from the
    \edit{
infinite\dmn\
    } 
{\jacobianOp} \refeq{2dLapStrech} that acts on the {infinite {\Blatt}} 
$\lattice_\Bcell{A}$. 

\subsection{{\Pcell} stability}
\label{s:pcellJ} 

The {\jacobianOrb} \refeq{2dLapStrech} evaluated over a \emph{finite 
volume} {\pcell} $\Bcell{A}$ is an 
$[\vol_\Bcell{A}\!\times\!\vol_\Bcell{A}]$ \emph{matrix}, with 
$\vol_\Bcell{A}$ discrete eigenvalues. 

As an example, consider a \lst\ $c$ over the one\dmn\ \pcell\ of period 
$\cl{}$, \refsect{s:orbit}. For a {\lst} $\Xx_c$ of periodicity 
$\Bcell{A}=\cl{}$, the {\jacobianOrb} is 
\beq 
\jMorb_{c} = 
\left(\begin{array}{cccccc} 
 \diag_{0} &{-1} & 0 & \cdots & 0 &{-1} \cr
{-1} & \diag_{1} &{-1} & \cdots & 0 & 0 \cr
0 &{-1} & \diag_{2}  & \cdots & 0 & 0 \cr
\vdots & \vdots &  \vdots & \ddots & \vdots & \vdots \cr
0 & 0 & 0 & \cdots & \diag_{\cl{}-2} &{-1} \cr
{-1} & 0 & 0 & \cdots & {-1} & \diag_{\cl{}-1} 
          \end{array} \right)
\,, 
\ee{jMorb1dFT} 
where the shift operators \refeq{hopMatrix} in \refeq{2dLapStrech} are 
the off-diagonals. 

For the free field, the {\catlatt} \refeq{ELeSptTemplatt}, and any 
\steady\ (constant) solution $\ssp_z=\ssp$ of a nonlinear field theory, 
this {\jacobianOrb} is a tri-diagonal Toeplitz matrix (constant along 
each diagonal) of circulant form, 
\beq
\jMorb_{\Bcell{A}} 
  =
\left(\begin{array}{cccccc} 
  {s}&{-1}& 0 &\dots &0&{-1}\cr
{-1}&   {s}&{-1}&\dots &0&0 \cr
0 &{-1}&   {s}& \dots &0 & 0 \cr
\vdots & \vdots  & \vdots & \ddots &\vdots &\vdots\cr
0 & 0 & \dots &\dots  &  {s}&{-1}\cr
{-1}& 0 & \dots &\dots&{-1}&   {s} 
        \end{array} \right)
\,. 
\ee{steadyStab} 
In what follows, we shall refer to this type of 
stability as the \emph{\steady\ stability}. 

The \Hilldet\ \refeq{detPartSum} of a finite\dmn\  {\jacobianOrb} over a 
{\pcell} \Bcell{A} is given by the product of its eigenvalues, \beq 
\left|\Det\jMorb_{c}\right| 
    =
\prod_{j=1}^{\vol_\Bcell{A}} \left|\ExpaEig_{c,j}\right| \,. 
\ee{eigsProd} Consider such determinant in the anti-integrable limit 
\refeq{antiIntSptTemplatt}. For {\steady}s, all $\vol_{\Bcell{A}}$ 
{\jacobianOrb} eigenvalues tend to $\ExpaEig_{c,j}\simeq{\mu^2}$, so \beq 
\ln \Det\jMorb_{c} 
    =
\Tr \ln \jMorb_{c} 
    \simeq
{\vol_{\Bcell{A}} \Lyap} 
    \,, \quad
\Lyap = \ln\mu^2 \,, \ee{antiIntStabExp} where $\Lyap$ is the {\stabexp}  
{per unit-lattice-volume}, with the exact \steady\ expression given \textbf{•}in \refeq{stabExp11}. 

This suggests that we assign to each {\lst} $c$ its average 
\emph{{\stabexp}} $\Lyap_{c}$ {per unit-lattice-volume},
\beq 
\frac{1}{\left|\Det\jMorb_{c}\right|} 
    =
{e}^{-{\vol_\Bcell{A}} \Lyap_{c}} \,,\qquad \Lyap_{c} 
    =
\frac{1}{\vol_\Bcell{A}} \sum_{j=1}^{\vol_\Bcell{A}}\ln |\ExpaEig_{c,j}| 
\,, 
\ee{pcellStabExp} 
where $\Lyap_{c}$ is the Birkhoff average \refeq{BirkhoffSum} of the 
logarithms of {\jacobianOrb}'s eigenvalues. This is a generalization of 
the temporal {\po} Floquet (or `Lyapunov') {\stabexp} per unit time to 
any multi-{\lst}, in any \spt\ dimension. This will be discussed further in 
\refsect{s:steadyStab}.

\subsection{{\Blatt} stability}
\label{s:BlattJ} 

The linear orbit Jacobian \emph{operator} acts on the \emph{infinite} 
{\Blatt} $\lattice_\Bcell{A}$. For example, the {\jacobianOp} a \lst\ 
$\Xx_c$ over the one\dmn\ {\Blatt} of a period $\cl{}$,
\beq
\jMorb_c = 
\left(\begin{array}{cccccccc}
\ddots & \ddots  &\ddots & \ddots &\ddots &\ddots & \ddots & \ddots \cr
\ddots & \diag_{0} & -1    & 0      &  0     & 0     & 0           & \ddots \cr
\ddots & -1      & \diag_{1} &  -1   & 0     & 0     & 0           & \ddots \cr
\ddots & 0       & -1  &  \diag_{2}  & -1    & 0     & 0           & \ddots \cr
\vdots & \vdots   &\vdots & \ddots &\ddots &\ddots & \vdots      & \vdots \cr
\ddots & 0        & 0     & 0      & -1    & \diag_{\cl{}-2} & -1  & \ddots \cr
\ddots & 0        & 0     & 0      & 0     & -1  & \diag_{\cl{}-1} & \ddots \cr
\ddots  &\ddots   &\ddots & \ddots &\ddots &\ddots   & \ddots & \ddots 
\end{array} \right)
\,,
\ee{jMorb1dBrav}
is an infinite matrix, with the diagonal block 
$\diag_{0}\diag_{1}\cdots\diag_{\cl{}-1}$ infinitely repeated along the 
diagonal. 

The next will be an elementary but essential observation. Consider a period-3 
{\fconf} \refeq{1dShift3lst} obtained by a translation of another 
period-3 {\fconf} in its orbit. Or a period-6 {\fconf} obtained by a 
repeat of a period-3 {\fconf}. The {\jacobianOp} \refeq{jMorb1dBrav} for 
all these {\fconfs} is the \emph{same}, of period 3. So, as announced in 
\refeq{BravLattMltpl}, and elaborated in \refsect{s:Bloch}, its spectrum 
is a property of its {\em  orbit}, irrespective of whether it is 
computed over a prime {\lst} {\pcell}, or any larger {\pcell} tiled by 
repeats of a prime {\lst}. 

But what is the `\Hilldet' of an $\infty$\dmn\ linear {\Blatt} operator? 
A textbook approach to calculation of spectra of such linear operators 
(for example, quantum-mechanical Hamiltonians) is to compute them in a 
large \pcell\ $\Bcell{A}$, and then take the infinite box limit. It is 
crucial to understand that we \emph{do not} do that here. Instead, as in 
solid state physics and quantum field theory, our calculations are always 
carried out over the \emph{infinite} {\spt} 
{\Blatt},\rf{Kittel96,AshMer,Dresselhaus07} or \emph{continuous} 
spacetime,\rf{MalWit07} where one has to make sense of the 
{\Hilldet}\rf{Hill86} as a functional determinant.\rf{Poinc1886} 

As we show in \refsect{s:Bloch}, for infinite lattices the appropriate 
notion of stability is the {\stabexp} \refeq{pcellStabExp} {per 
unit-lattice-volume}, averaged over the first {Brillouin zone}, 
    \edit{
which we evaluate
        }
by means of the Floquet-Bloch theorem.


\section{{\Blatt}s}
\label{s:Bravais}

{\Po} theory of a time-evolving dynamical system on a one\dmn\ temporal 
lattice is organized by grouping orbits of the same period 
together.\rf{gutbook,ruelle,Gas97,ChaosBook,LC21} For systems 
characterized by several translational symmetries, one has to take care 
of multiple periodicities, or, in parlance of crystallography, organize 
the \po\ sums by corresponding {\em {\Blatt}s},\rf{AshMer} introduced 
here in \refsect{s:periodicCnfs}. 

The set of all transformations that overlay a lattice over itself is 
called the \emph{space group}  $\Group$. For the square lattice, the unit 
cell \refeq{SquareLatt} tiles the hypercubic lattice under action of 
translations $\shift_j$ \refeq{shiftOperator2d} in $d$ \spt\ directions, 
called `shifts' for infinite {\Blatt}s, `rotations' for finite periods 
{\pcell}s. They form the abelian {translation group} \beq 
T=\{\shift_1^{m_1}\shift_2^{m_2}\cdots\shift_d^{m_d} 
    \,|\, 
    m_j\in\integers\}
\,. 
\ee{dDimTranslGrp} 

The cosets of a space group $\Group$ by its translation subgroup $T$ form 
the group $\Group/T$, isomorphic to a {point group} $g$. For example, the 
square lattice space group $\Group=T\rtimes\Dn{4}$ is the semi-direct 
product of translations \refeq{dDimTranslGrp}, and the point group  $g$ 
of right angle rotations, time reversal, spatial reflection, and 
space-time interchanges. 
In addition, there might also be {internal} global symmetries, such as 
the invariance of \catlatt\ equations \refeq{sptCatlatt} under inversion 
of the field though the center of the $0\leq\ssp_{z}< 1$ unit interval: 
\beq {\ssp}_z \to 1 - \ssp_z \mbox{ for all } z\in\integers^d \,. 
\ee{InverCat} 

Already in the case of chaotic lattice field theory over one\dmn\ 
temporal integer lattice $\integers$ there is a sufficient amount of 
group-theoretical detail to merit the stand-alone companion paper 
I,\rf{LC21} which treats in detail the time-reversal invariance for a
$\Group=\Dn{\infty}$ dihedral space group of translations and 
reflections. 
Here we focus only on the two\dmn\ square lattice \emph{translations} 
\refeq{dDimTranslGrp}, as a full {symmetry} treatment would distract the 
reader from the main trust of the paper, the construction of the \spt\ 
zeta function (\refsect{s:POT}). 


\subsection{{\Blatt}s of the square lattice}
\label{s:Blatt} 

In crystallography, there are five {\Blatt}s over a two-dimensional space. 
The square lattice \refeq{SquareLatt} is one of them. For brevity, 
whenever we refer here to a `{\Blatt}', we mean one of the infinity of 
`full rank {sub}lattices of the square lattice'\rf{Cassels59} that we now 
describe. 

Consider a $[2\!\times\!2]$ integer basis matrix \refeq{BravLatt} \beq 
\Bcell{A} = \left[ \mathbf{a}_{1}, \mathbf{a}_{2} \right] 
           = \left[
\begin{array}{cc}
{a}_{1,1} & {a}_{2,1}\cr
{a}_{1,2} & {a}_{2,2}
\end{array}
             \right]
          \,,\quad
\mathbf{a}_{j} = \left[ \begin{array}{c} {a}_{j,1} \cr {a}_{j,2} 
\end{array} \right]
\,, 
\ee{2dPrimCell} 
formed from a pair of two\dmn\ integer lattice 
primitive vectors $\mathbf{a}_{1}$, $\mathbf{a}_{2}$. 
A two\dmn\ {\em {\Blatt}}, \reffig{f:BravLatt}, \beq \lattice_\Bcell{A} = 
\left\{\Bcell{A} \mathbf{n}  \,|\,\mathbf{n} \in \mathbb{Z}^2 \right\} 
\ee{2DBravaisLattice} 
generated by all discrete translations \( 
\Bcell{A}  \mathbf{n} 
\) is a sublattice of the integer lattice $\integers^2$. 

As in a discretized field theory the fields are defined only on the 
hypercubic integer lattice, not on a continuum, we define the {\em 
\pcell} \refeq{torusStatesp} as the set of lattice sites within the 
parallelepiped \refeq{2dPrimCell} illustrated by \reffig{f:BravLatt}. The 
tips of primitive vectors and parallelepiped's outer boundaries belong, 
by translation, to the neighboring tiles; this yields the correct lattice 
{volume} \refeq{lattVol}, the number of lattice sites $\vol_\Bcell{A}$ 
within the {\pcell} $\Bcell{A}$. 


A {\pcell} is not unique:\rf{Siegel89} the {\Blatt} 
$\lattice_{\Bcell{A}'}$ defined by basis $\Bcell{A}'$ is the same as the 
{\Blatt} $\lattice_\Bcell{A}$ defined by basis 
\( 
\Bcell{A}=\Bcell{A}'\,\Bcell{U} 
\) 
if the two are related by a $[2\!\times\!2]$ unimodular, volume 
preserving 
matrix  $\Bcell{U}\in\SLn{2}{\integers}$ 
transformation,\rf{Lang71,Cardy86,ZiLoKl99} see \reffig{f:BravLatt}\,(b). 
This equivalence underlies many of the properties of elliptic functions 
and modular forms\rf{SteSha03} (see \refeq{2tauBas}). Constructing 
\emph{all} {\Blatt}s, however, is straightforward, as each such infinite 
family of equivalent {\pcell}s contains a single, unique \emph{Hermite 
normal form} {\pcell}, with upper-triangular basis\rf{Cohen93} primitive 
vectors $\mathbf{a}_1=(\speriod{}, 0)$, 
$\mathbf{a}_2=(\tilt{},\period{})$, 
\beq 
\Bcell{A} = 
\left[\begin{array}{cc} 
\speriod{} & \tilt{} \cr
0 & \period{} 
\end{array}\right]
    \,,\qquad \vol_\Bcell{A} = \speriod{} \period{}
\,, 
\ee{2DHermiteBas} 
where $\speriod{}$, $\period{}$ are the spatial, 
temporal lattice periods, respectively, and $\vol_\Bcell{A}$ is the 
lattice {volume} \refeq{lattVol}. The tilt\rf{OKKH99} $0 \leq 
\tilt{}<\speriod{}$ imposes `relative-periodic shift' 
{\bcs}.\rf{ChaosBook} In the  literature these are also referred to as 
{`helical'},\rf{LHCLL06} {`toroidal'},\rf{IzOgCh02} 
{`screw'},\rf{Dresselhaus07} {$\tilt{}$-corkscrew,\rf{ChMaVa96}
{`twisted'}\rf{IvIzHu02} or {`twisting factor'}\rf{LHCLL06} {\bcs}. 

In the theory of elliptic functions\rf{SteSha03} the {\pcell} is 
represented by a complex modular parameter $\tau$, with spatial period 
$\speriod{}$ taken as the lattice spacing constant \refeq{LattField}, 
primitive vectors $\mathbf{a}_1=(1, 0)$, $\mathbf{a}_2=(\tau_1,\tau_2)$, 
so $\period{}\to\tau_2=\period{}/\speriod{}$, 
$\tilt{}\to\tau_1=\tilt{}/\speriod{}$, and 
\beq 
\Bcell{A} = 
\left[\begin{array}{cc} 
1 & \tau_1 \cr
0 & \tau_2 
\end{array}\right]
    \,,\qquad 
|\Det\Bcell{A}|= \tau_2 
\,, 
\ee{2tauBas} 
`Hermite normal form' 
corresponds to  the modular parameter $\tau$ values in the {fundamental 
domain}. If the corresponding torus is visualized as glueing of a unit 
square into a tube, $\tau_2$ parameterizes how the tube is stretched, and 
$\tau_1$ parameterizes how it is twisted before its edges are stitched 
together. 

Here we refer to a particular {\Blatt} by its Hermite normal form 
\refeq{2DHermiteBas}, as 
\beq 
\lattice_\Bcell{A}=\LTS{}{}{} 
\,, 
\ee{LxT_S} 
and to the set of lattice sites within the primitive 
parallelogram \Bcell{A} as its {\pcell}. Notation $\LTS{}{}{}$ refers to 
{\pcell} being a rectangle of spatial width $\speriod{}$, temporal height 
$\period{}$, with the {\pcell} above it shifted by $S$, see, for example 
the $\BravCell{3}{2}{1}$ {\pcell} shown in \reffig{f:2x1rpo}\,(b). In 
terms of lattice site fields, a {\fconf} $\ssp_{z_1 z_2}$ 
\refeq{periodLConf}, $z_1z_2\in\integers^2$, satisfies the 
$\tilt{}$-corkscrew boundary condition\rf{ChMaVa96}, 
\bea 
\mbox{horizontally: } \quad && \ssp_{z_1 z_2} \,=\, \ssp_{z_1 + 
\speriod{}, z_2} 
    \continue
\mbox{vertically: }   \quad && \ssp_{z_1 z_2} \,=\, \ssp_{z_1 + \tilt{}, 
z_2 + \period{}} 
\,, 
\label{screw} 
\eea 
see \reffig{f:2x1rpo}. 


\begin{figure}\begin{center}
           \begin{minipage}[c]{0.23\textwidth}\begin{center}
\includegraphics[width=1.0\textwidth]{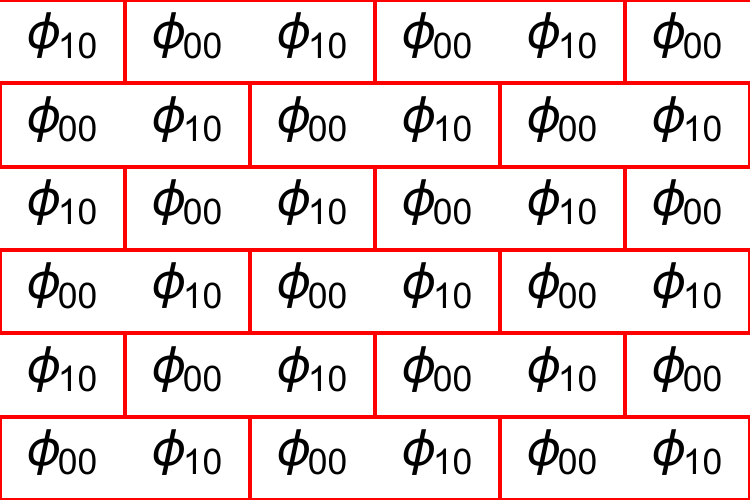}
                \\      (a)
            \end{center}\end{minipage}
                   \hskip 2ex
           \begin{minipage}[c]{0.23\textwidth}\begin{center}
\includegraphics[width=1.0\textwidth]{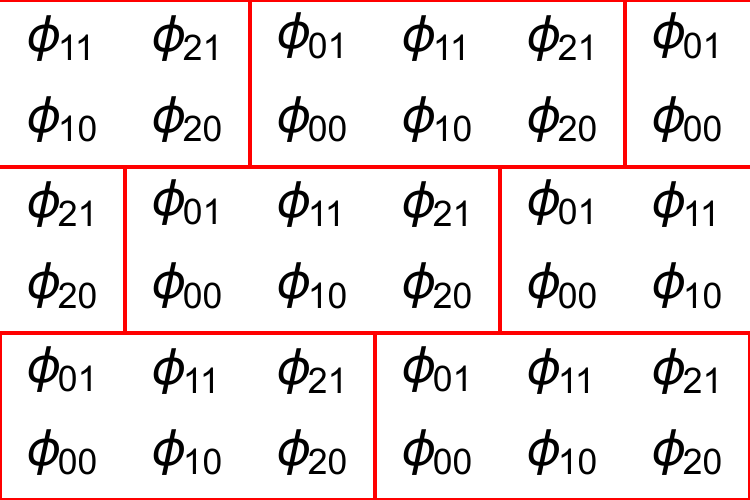} 
                 \\     (b)
            \end{center}\end{minipage}
\end{center}
                \vskip -12pt     
  \caption{\label{f:2x1rpo} 
Examples of $\LTS{}{}{}$ {\fconfs}  \refeq{2DHermiteBas}
or `bricks',
together with their \spt\ {\Blatt} tilings,
visualized as brick walls.
(a)
$\BravCell{2}{1}{1}$, primitive vectors
$\mathbf{a}_1=(2,0)$, $\mathbf{a}_2=(1,1)$;
(b)
$\BravCell{3}{2}{1}$ of \reffig{f:BravLatt}\,(a), primitive vectors
$\mathbf{a}_1=(3,0)$, $\mathbf{a}_2=(1,2)$.
Rectangles enclose the {\pcell} and its {\Blatt}
translations.
This will be discussed further in \reffig{f:pcellTiling}. 
}
\end{figure}


\section{Orbits over two\dmn\ lattices}
\label{s:orbits} 

For field theories
studied here (\refsect{s:sptFT}), the translation group $T$ 
\refeq{dDimTranslGrp} is a {symmetry}, as their defining equations \refeq{eqMotion} 
retain their form (are `{equivariant}') under lattice translations. For 
square lattice, these are 2\dmn\ translations of the form 
$\LieEl=\shift_1^{m_1}\shift_2^{m_2}$. (For symmetries other than 
translations, see remarks at the beginning of \refsect{s:Bravais}.) 

Typically a translation operation acting on {\lst} $\Xx_p$ generates an 
equivalent (up to lattice sites relabeling) but \stateDsp\ distinct 
{\lst} $\LieEl\Xx_p$. The totality of all actions of the {translation 
group} 
on  {\lsts} foliates the {\statesp} into a union \refeq{statespFliated} 
of translational \emph{orbits}
\beq 
    \pS_p = \{\LieEl\,\Xx_p \mid \LieEl \in {T}\}
\,. 
\ee{GroupOrb2d} 
{\Blatt}s $\lattice_{\Bcell{A}}$ (\refsect{s:orbit}) are infinite, and 
their translational symmetries \refeq{dDimTranslGrp} are {infinite} 
groups, but the {orbit} of a Bravais {\lst} is {finite}, generated by the 
translations of the infinite lattice curled up into a 
$\vol_{\Bcell{A}}$-site periodic {\pcell} $\Bcell{A}$.

\subsection{\PrimeOrbs\ over two\dmn\ {\pcell}s}
\label{s:2DprimeOrb}  

A \lst\  $\Xx_p$ may have all of system's symmetries, a subgroup of 
them, or have no symmetry at all. If  $\Xx_p$ has no symmetry, its 
$\speriod{p}$ horizontal translations, $\period{p}$ vertical translations 
are all distinct {\lsts}, so its orbit consists of 
$\vol_{\Bcell{A}}=\speriod{p}\period{p}$ {\lsts}. 

It is easy to check whether a one\dmn\ {\lst} $\Xx_p$ over a {\pcell} 
\Bcell{A} is prime, by comparing it to its translations, as in the 
period-6 example of \refsect{s:1DprimeOrb}. We use this test as an 
operational definition of a \emph{prime} {\lst} over a {\pcell} 
$\Bcell{A}$ for a hypercubic lattice in two (or any) dimensions. 

\medskip

\paragraph*{Definition: {\PrimeOrb}.}
\begin{quote}
A {\lst} $\Xx_p$ over {\pcell} \Bcell{A} is \emph{prime} if the number of 
distinct {\lsts} in its orbit equals $\vol_{\Bcell{A}}$, the number of 
lattice sites within its {\pcell} \Bcell{A}  \refeq{lattVol}. 
\end{quote}
This notion of a `prime' suffices to formulate our main result, the \spt\ 
zeta function (\refsect{s:lattZeta}) for field theories in two spacetime 
dimensions. 
 However, we have to emphasize that implementing this for a given theory
requires determination of \emph{all} of its {\primeOrbs}, and that is 
hard problem, in a sense all of `chaos theory'. Here we use {\catlatt} to 
test the theory, but relegate details to \refappe{s:catlattComp}, and in the 
companion paper III\rf{WWLAFC22} we apply the theory to several nonlinear 
field theories.

\subsection{Repeats of a \primeOrb\ over two\dmn\ {\pcell}s}
\label{s:repeats} 

$\ssp_z={\ssp}$, invariant under all of system's symmetries. Its {\pcell} 
$\BravCell{1}{1}{0}$ is the unit hypercube \refeq{SquareLatt} of period-1 
along every hypercube direction. 

A {\lst} obtained by tiling any larger {\pcell} by repeats of \steady\ 
${\ssp}$ is \emph{not a prime} {\lst}. There is one such for each Bravais 
sublattice constructed in \refsect{s:Blatt}, with $r_1$ copies of lattice 
site field $\ssp{}$ horizontally, $r_2$ copies vertically, and tilt $0 
\leq s < r_1$  \refeq{2DHermiteBas}, 
\beq 
\Bcell{R} = 
\left[\begin{array}{cc} 
r_1 & s \cr
0 & r_2 
\end{array}\right]
\,. 
\ee{steadyRepeats} 

Next, note that every orbit is `steady' in the sense that each orbit 
\refeq{GroupOrb2d} is a fixed point of $T$, as any translation \( 
\LieEl\, \pS_p = \pS_p 
\) only permutes the set of {\lsts} within the orbit, but leaves the set 
invariant. In particular (see \refsect{s:BlattJ}), the stability of an 
orbit is its intrinsic, translation invariant `steady' property. 

\begin{figure}
\begin{center}
           \begin{minipage}[c]{0.23\textwidth}\begin{center}
\includegraphics[width=1.0\textwidth]{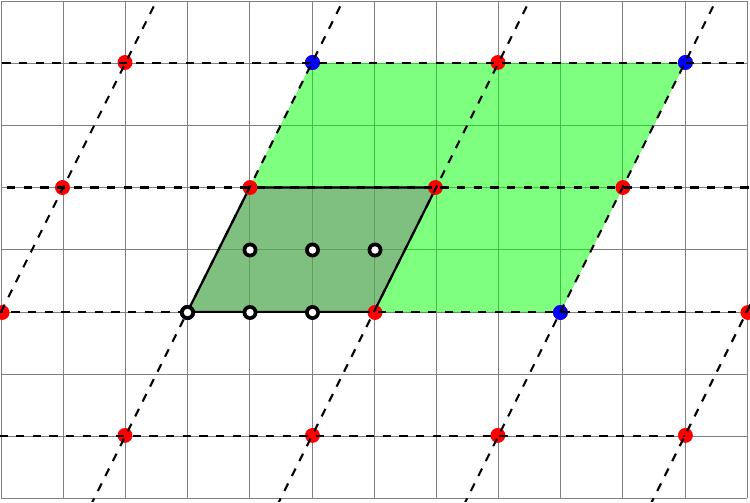}
                \\      (a)
            \end{center}\end{minipage}
                   \hskip 2ex
           \begin{minipage}[c]{0.23\textwidth}\begin{center}
\includegraphics[width=1.0\textwidth]{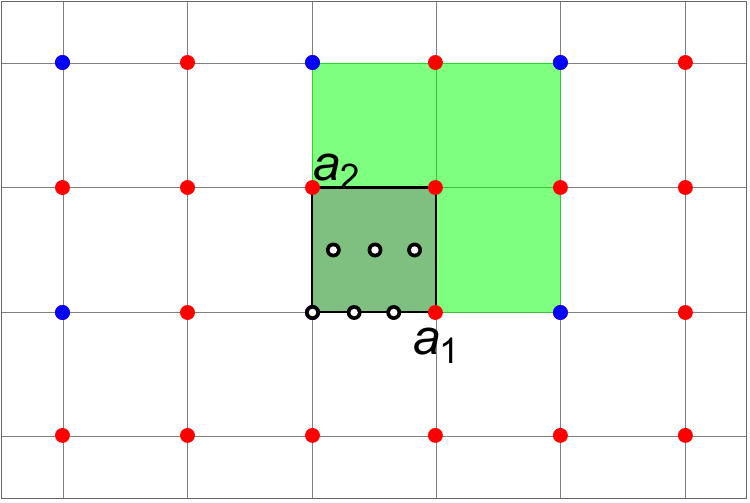}
                 \\     (b)
            \end{center}\end{minipage}
\end{center}
                \vskip -12pt     
  \caption{\label{f:pcellAsUnitSq}
    (a)
{\Blatt} $\Bcell{A}=\BravCell{6}{4}{2}$, blue dots, is a sublattice of {\Blatt}
$\Bcell{A}_p = \BravCell{3}{2}{1}$, blue and red dots.
Its {\pcell} ${\Bcell{A}}$ (green
parallelogram spanned by primitive vectors (6,0) and (2,4)) is tiled
by repeats of the {\pcell} $\Bcell{A}_p$  (gray
parallelogram spanned by primitive vectors (3,0) and (1,2)).
The primitive vectors of the two {\Blatt}s are related by
$\Bcell{A} = \Bcell{A}_p \Bcell{R}$ where $\Bcell{R} = \BravCell{2}{2}{0}$.
    (b)
Transform the {\pcell} $\Bcell{A}_p$ to the unit square of a new
square lattice, where each unit square supports a multiplet of six
fields belonging to a prime $\lattice_{\Bcell{A}_p}$-{\lst}. In
this new square lattice, the prime {\lst} is a \steady\ whose
{\pcell} is a \BravCell{1}{1}{0} unit square (gray square),
while the repeat of the prime is a
$\lattice_\Bcell{R}$-{\lst}, whose {\pcell}
is $\Bcell{R} = \BravCell{2}{2}{0}$ (green square).
}
\end{figure}

The simplest example of a prime {\lst} is a \emph{\steady} 
A way to visualize this is by multiplying the {\Blatt}  
$\lattice_{\Bcell{A}_p}$ by $\Bcell{A}_p^{-1}$, sending it into the unit 
integer lattice, as in \reffig{f:pcellAsUnitSq}\,(b): in other words, 
every  {\Blatt} is a hypercubic lattice, under an appropriate change of 
coordinates. In this new integer lattice, the {\pcell} $\Bcell{A}_p$ is 
the \emph{unit} square that supports a 
multiplet of $\vol_{\Bcell{A}}$ {\lsts} belonging to the
$\lattice_{\Bcell{A}_p}$ orbit. Under lattice translations, this 
multiplet is an $\vol_\Bcell{A}$\dmn\ \steady. 

To find all repeats of a given prime {\lst}, one only needs to find all 
{\Blatt}s $\lattice_\Bcell{R}$, which can again be accomplished using the 
Hermite normal form repeat matrix $\Bcell{R}$ \refeq{steadyRepeats}. 
Each $\Bcell{R}$ gives a non-prime {\lst} over a larger-periodicity 
Bravais sublattice $\lattice_{\Bcell{A}_p \Bcell{R}}$. 

\medskip

\paragraph*{Example:
    A repeat of \BravCell{3}{2}{1} prime {\lst}.}
Tiling of a $\lattice_\Bcell{A}=\BravCell{6}{4}{2}$ {\lst} by a repeat 
of the $\lattice_{\Bcell{A}_p}=\BravCell{3}{2}{1}$ prime {\lst} is 
shown in \reffig{f:pcellAsUnitSq}\,(a). In 
\reffig{f:pcellAsUnitSq}\,(b) the {\pcell} of the prime 
$\lattice_{\Bcell{A}_p}$-{\lst} is transformed into the unit square of 
the new integer lattice, where each unit square supports a multiplet of 
6 fields. In this new integer lattice, the {\pcell} of the repeat 
$\lattice_\Bcell{A}$-{\lst} is given by $\lattice_\Bcell{R} = 
\BravCell{2}{2}{0}$, where $\Bcell{A} = \Bcell{A}_p \Bcell{R}$. 

\medskip

\paragraph*{Example:
    A repeat of $\BravCell{3}{1}{2}$ prime {\lst}.}
\begin{figure}
\begin{center}
           \begin{minipage}[c]{0.23\textwidth}\begin{center}
\includegraphics[width=1.0\textwidth]{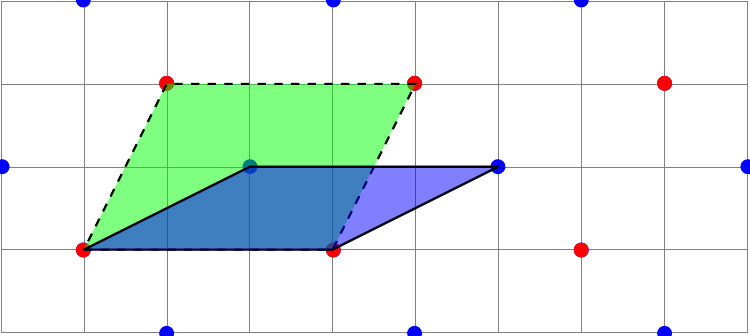}
                \\      (a)
            \end{center}\end{minipage}
                   \hskip 2ex
           \begin{minipage}[c]{0.23\textwidth}\begin{center}
\includegraphics[width=1.0\textwidth]{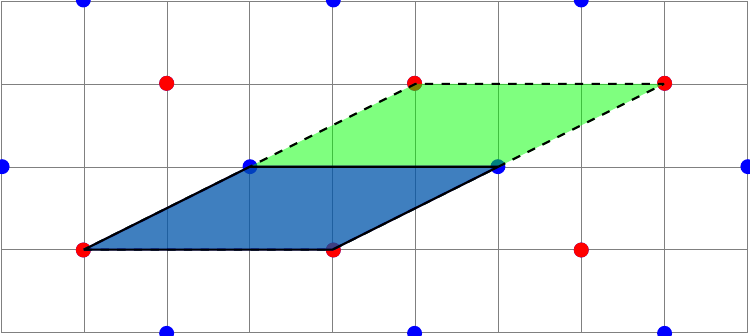}
                 \\     (b)
            \end{center}\end{minipage}
\end{center}
            \vskip -4pt     
  \caption{\label{f:pcellTiling}
    (a)
{\Blatt} $\Bcell{A}=\BravCell{3}{2}{1}$ of
\reffig{f:BravLatt}, red dots, is a sublattice of {\Blatt}
$\Bcell{A}'=\BravCell{3}{1}{2}$,
blue and red dots, even though the {\pcell} ${\Bcell{A}}$ (green
parallelogram spanned by primitive vectors (3,0) and (1,2)) does not
appear to be tiled by a repeat of the {\pcell} ${\Bcell{A}'}$  (blue
parallelogram spanned by primitive vectors (3,0) and (2,1)).
    (b)
If we shift the top edge of {\pcell} ${\Bcell{A}}$ by three lattice units,
to $\BravCell{3}{2}{4}=\BravCell{3}{2}{1}$
(green parallelogram spanned by primitive vectors (3,0) and (4,2)),
the tiling is clear.
}
\end{figure}

\emph{A priori} is not obvious that the $\BravCell{3}{1}{2}$ {\pcell} tiles 
the $\BravCell{3}{2}{1}$ {\pcell}, \reffig{f:pcellTiling}\,(a). However, if 
you stack the $\BravCell{3}{1}{2}$ {\pcell} in the shifted temporal 
direction by 2 then the left edge of the tile is shifted by 4 in the 
spatial direction. With the spatial period being 3, shift by 4 in the 
spatial direction is the same as shift by 1. Therefore, the \bcs\ of the 
$\BravCell{3}{2}{1}$ {\pcell} are satisfied by the repeat of the 
$\BravCell{3}{1}{2}$ {\pcell}. 

For further examples of {\primeOrbs}, see \refappe{s:primeLatt}. 

\medskip
In summary, to determine all {\lsts}, it suffices to enumerate all 
{\Blatt}s, then compute their {\primeOrbs} on their finite\dmn\ 
{\pcell}s. Their stabilities, however, will have to be evaluated on the 
infinite {\Blatt}s, as we shall show in \refsect{s:Bloch}. 


\section{Reciprocal lattice}
\label{s:Brillouin}  

If an operator, in case at hand the {\jacobianOp} \refeq{ELeFT}, is 
invariant under spacetime translations, its eigenvalue spectrum and 
{\Hilldet} can be efficiently computed using tools of crystallography, by 
going to the {reciprocal lattice}. 

For a $d$\dmn\ $\lattice_\Bcell{A}$-\emph{periodic} {\Blatt}, discrete 
wave vectors $\wavenum$ form a {reciprocal lattice} spanned by $d$ 
reciprocal primitive vectors which satisfy 
\beq 
\lattice_{\tilde{\Bcell{A}}}= \Big\{\wavenum = \sum_{j=1}^{d} m_{j} 
\tilde{\mathbf{a}}_j 
        \mid m_{j} \in\mathbb{Z}\Big\}
    \,,\quad
\tilde{\mathbf{a}}_i \cdot \mathbf{a}_j = 2 \pi \delta_{ij} \,. 
\ee{rcprLatt} 
Assembling the reciprocal primitive vectors 
$\{\tilde{\mathbf{a}}_j\}$ into columns of the $[d\!\times\!d]$ 
reciprocal {\pcell} matrix \( \tilde{\Bcell{A}} = 
    \left[ \tilde{\mathbf{a}}_{1}, \tilde{\mathbf{a}}_{2},
                  \cdots,          \tilde{\mathbf{a}}_{d}
    \right],
\) the reciprocity condition \refeq{rcprLatt} takes the form
 \beq 
\transp{{\tilde{\Bcell{A}}}}\Bcell{A} 
        = 2\pi\id
\,. \ee{reciprPrimTransp}

\subsection{Reciprocal {\pcell} in one and two dimensions}
\label{s:rcprLatt} 

Translation invariance of a theory suggests its reformulation in a 
discrete Fourier basis, an approach that goes all the way back to Hill's 
1886 paper.\rf{Hill86} The $\cl{}$ consecutive shifts \refeq{1dShift} 
return a period-$\cl{}$ {\fconf} to itself, so acting on a one\dmn\ 
{periodic} {\pcell}, shift operator satisfies the characteristic equation 
\beq \shift^\cl{} -\id = 
\prod_{m=0}^{\cl{}-1}(\shift-{e}^{\mathrm{i}{k}}\id) 
                   = 0
\,, \ee{shift1n} 
with eigenvalues $\{{e}^{\mathrm{i}{k}}\}$ being the $\cl{}$-th 
roots of unity, indexed by integers $m$, \beq 
    {k}= 
    \Delta k \, m
    \,,\qquad \Delta k = \frac{2\pi}{\cl{}}
    \,, \quad m=0,1,\cdots,\cl{}\!-\!1
\,, \ee{shiftEigs1n} and $\cl{}$ eigenvectors \( [\varphi(k)]_z = 
{e}^{\mathrm{i}\,{k}z} \,, \) \beq 
 [\shift\varphi(k)]_z
    =  [\varphi(k)]_{z+1}
    ={e}^{\mathrm{i}\,{k}(z+1)}
    ={e}^{\mathrm{i}{k}}[\varphi(k)]_z
\,. \ee{FrModes1d} Wave numbers $k$ 
form a one\dmn\ {reciprocal lattice} \refeq{rcprLatt}, 
\[ 
\lattice_{\tilde{\Bcell{A}}}=
\Big\{k = m\,\tilde{\mathbf{a}}_1 \mid m\in\mathbb{Z}\Big\}
    \,,\quad
\tilde{\mathbf{a}}_1 \cdot \mathbf{a}_1 = 2 \pi
\,,
\] 
with the reciprocal lattice primitive vector 
$\tilde{\mathbf{a}}_1=2\pi/\cl{}$, 
and the reciprocal {\pcell} 
--the interval $[0,2\pi)$-- that contains $n$ discrete wave numbers 
\refeq{shiftEigs1n}.

In two \spt\ dimensions, the reciprocal lattice \refeq{rcprLatt} of the 
{\Blatt} \refeq{2DHermiteBas} is given by \beq 
\lattice_{\tilde{\Bcell{A}}} = \{ \wavenum = m_1 \tilde{\mathbf{a}}_1 + 
m_2 \tilde{\mathbf{a}}_2 
  \;\vert\; m_i \in \mathbb{Z}\}
\,, \ee{2DReciprLatt} where the reciprocal lattice primitive vectors 
$\tilde{\mathbf{a}}_1 = \frac{2\pi}{\vol_\Bcell{A}} \, (\period{}, 
-\tilt{})$ and $\tilde{\mathbf{a}}_2 = \frac{2\pi}{\vol_\Bcell{A}} \, (0, 
\speriod{})$ (see \reffig{f:reciprLatt}\,(b)) satisfy the reciprocity 
condition \refeq{reciprPrimTransp}. The reciprocal {\pcell} matrix is  
also of {Hermite normal} (but lower-triangular) form, 
\beq 
\tilde{\Bcell{A}} = \frac{2\pi}{\vol_\Bcell{A}} \left[ 
\begin{array}{cc}
\period{} & 0 \cr
-\tilt{} & \speriod{}
\end{array}
\right] 
\,, 
\ee{2DHermiteRecipr} 
with the reciprocal basis condition 
\refeq{reciprPrimTransp} satisfied. The components of a reciprocal 
lattice wave vector $\wavenum$ in \refeq{2DReciprLatt} are \beq \wavenum 
= \left[ 
\begin{array}{c}
k_1 \cr
k_2
\end{array}
\right] 
            =
\frac{2\pi}{\speriod{}\period{}} \left[ 
\begin{array}{c}
m_1 \period{} \cr
- m_1 \tilt{} + m_2 \speriod{}
\end{array}
\right] \,. \ee{2DWaveVector} As in the one\dmn\ case 
\refeq{shiftEigs1n}, the wave numbers along each direction of a two\dmn\ 
square lattice can be restricted to $k_j\in[0,2\pi)$ with 
$m_1=0,1,\cdots,\speriod{}-1$, $m_2=0,1,\cdots,\period{}-1$, 
$\vol_\Bcell{A}=\speriod{}\period{}$ distinct wave vectors. This set of 
reciprocal lattice sites, indexed by integer pairs $m=m_1m_2$, forms the 
\emph{reciprocal {\pcell}} $\tilde{\Bcell{A}}$, which contains the same 
number of lattice sites $\wavenum\in\tilde{\Bcell{A}}$ as the \spt\ 
\Blatt\ {\pcell} $\Bcell{A}$ (see \reffig{f:reciprLatt}\,(b)). 

\medskip

\paragraph*{Example: 
            A \spt\ {\pcell}, reciprocal {\pcell}.}

\begin{figure}
\begin{center}
           \begin{minipage}[c]{0.23\textwidth}\begin{center}
\includegraphics[width=1.0\textwidth]{HLBravaisCell1}
                \\\vskip 13pt      (a)
            \end{center}\end{minipage}
                   \hskip 2ex
           \begin{minipage}[c]{0.23\textwidth}\begin{center}
\includegraphics[width=1.0\textwidth]{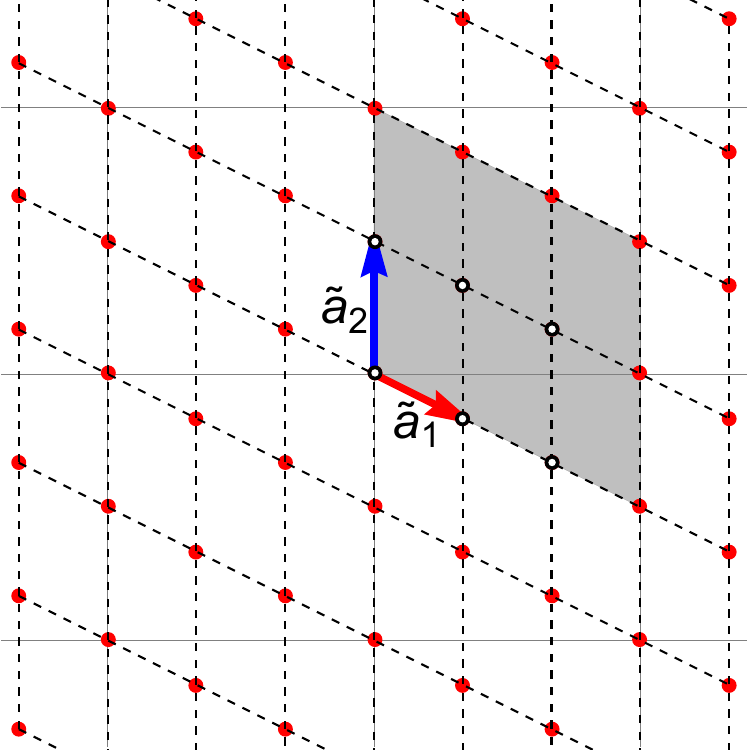}
                 \\                (b)
            \end{center}\end{minipage}
\end{center}
                \vskip -12pt     
  \caption{\label{f:reciprLatt}   
(a)
    The intersection points $z$ of the light gray lines form the integer
    square lattice \refeq{LattField}.
    The primitive vectors $\mathbf{a}_1=(3,0)$ and $\mathbf{a}_2=(1,2)$
    form the
    {\pcell} $\Bcell{A}=\BravCell{3}{2}{1}$
    (see \refeq{LxT_S} 
    and \reffig{f:BravLatt}\,(a)), 
    whose translations tile the {\Blatt} $\lattice_\Bcell{A}$ (red points).
(b)
    The intersection points $k$ of the light gray lines form the
    reciprocal square lattice.
    Translations of reciprocal primitive vectors $\tilde{\mathbf{a}}_1$ and
    $\tilde{\mathbf{a}}_2$ 
    (see \refeq{rcprLatt}, \refeq{reciprPcell321}) 
    generate the
    reciprocal lattice $\lattice_{\tilde{\Bcell{A}}}$ (red points).
    (Shaded)
    The reciprocal {\pcell} $\tilde{\Bcell{A}}$.
    A wave vector outside this region is equivalent to a wave vector
    within it by a reciprocal lattice translation.
    Note that the  number of lattice sites within the
    reciprocal {\pcell} $\tilde{\Bcell{A}}$
    equals the number of sites within the \spt\  {\pcell} $\Bcell{A}$.
}
\end{figure}

Primitive vectors $\mathbf{a}_1=(3,0)$ and $\mathbf{a}_2=(1,2)$ define 
the {\pcell} $\BravCell{3}{2}{1}$ drawn in \reffig{f:reciprLatt}\,(a), 
\beq \Bcell{A} = \left[ 
\begin{array}{cc}
3 & 1 \cr
0 & 2
\end{array}
\right] \,,\qquad \vol_\Bcell{A} = 6 
\,. \ee{pcell321} The corresponding reciprocal {\pcell} vectors (shaded 
region in \reffig{f:reciprLatt}\,(b)), \beq \tilde{\Bcell{A}} = 
\frac{2\pi}{6} \left[ 
\begin{array}{cc}
2 & 0 \cr
-1 & 3
\end{array}
\right] \,, \ee{reciprPcell321} satisfy the reciprocal bases condition 
\refeq{reciprPrimTransp}, and contain the same number of reciprocal 
lattice sites $\wavenum\in\tilde{\Bcell{A}}$ as the \Blatt\ {\pcell} 
$\Bcell{A}$ of \reffig{f:reciprLatt}\,(a). 

\medskip
\noindent The next two sections are the conceptual core of the paper: 

\medskip

\paragraph*{\refSect{s:pcellStab}~{{\Pcell} stability.}} 
As noted in the introduction, the 
textbook Gutzwiller-Ruelle {\po} theory\rf{gutbook,ruelle,ChaosBook} 
is hampered by a simple fact: its 
\po\ weight \refeq{primCellNotMltpl} is \emph{not} multiplicative for 
orbit repeats. This section recapitulates the conventional theory, in 
which all {\po} calculations are done in finite time `cells', 
as in \reffig{fig:LstPerturbs}\,(b),
with the key non-multiplicativity fact illustrated by computation of 
\refeq{nonMultHenon}. Our {\spt} theory illuminates the origin of 
this fact in several easy to grasp ways. 

\medskip

\paragraph*{\refSect{s:Bloch}~{{\Blatt} stability.}}
A crystallographer or a field theorist starts 
{\em --ab initio--} with an 
\emph{infinite} lattice or continuous spacetime,
as in \reffig{fig:LstPerturbs}\,(c).
This, we claim in the introduction, \refeq{BravLattMltpl}, is the correct 
approach which --as we show here, \refeq{lnDetStabExp}-- yields 
(multi)\lst\ weights that are  {\em multiplicative} for repeats of {\spt}ly 
periodic solutions. The {\stabexp} per unit spacetime volume is the 
spacetime generalization of the temporal {\po} Lyapunov exponent, the 
mean instability per unit time.  No matter what repeat of a prime \lst\ 
one starts with, its {\stabexp} is always given by the same  integral 
over the {\primeOrb} {Brillouin} zone. From this follows the main result 
of our paper, the \spt\ zeta function of \refsect{s:POT}. 

\section{{\Pcell} stability}
\label{s:pcellStab} 

As we now explain, it is crucial that we distinguish the \emph{finite} 
{\pcell} orbit Jacobian \emph{matrix} (finite volume \pcell\ stability, 
discussed in this section) from the \emph{infinite} orbit Jacobian 
\emph{operator} (infinite {\Blatt} stability, discussed in 
\refsect{s:Bloch}) in stability calculations. 

To the best of our knowledge, in all current implementations of the \po\ 
theory,\rf{gutbook,ruelle,Gas97,Baladi00,ChaosBook} the calculations are 
always carried out on finite {\pcell}s, so a `chaos' expert is free to 
skim over this section - it is a recapitulation of \Henon, Lorentz, \etc, 
calculations in the {\spt}, field-theoretic language. The radical 
departure takes place in \refsect{s:Bloch}. 

We start by considering the \steady\ {\jacobianOrbs}, such as 
\refeq{steadyStab}, with no lattice site dependence, $\diag_{z}= {s}$, 
which are fully diagonalized by going to the reciprocal lattice. 

\subsection{{\Pcell}  \steady\ stability in one dimension}
\label{s:pcellJtemp} 

For a one\dmn\ {\pcell} $\Bcell{A}$ of period $\cl{}$, the discrete 
Fourier transform \refeq{FrModes1d} of Laplacian 
\refeq{LapOp}, 
\bea
&& \jMorb_{\Bcell{A}} \varphi_k \,=\, (- \Box + \mu^2\id)\,\varphi_k 
                             \,=\, (p^2 + \mu^2)\,\varphi_k
    \label{tempLattEigs} \\
&& p    \,=\, 2 \sin\frac{k}{2} 
        \,,\quad 
   {k}  \,=\, \frac{2\pi}{\cl{}}m
        \,,\quad 
      m=0,1,\cdots,\cl{}-1
\,, \nnu
\eea
expresses the Fourier-diagonalized lattice 
Laplacian as the square of $p_m$, the `lattice momentum', or the 
`momentum measured in lattice units', 
\bea
(\tilde{\jMorb}_{\Bcell{A}})_{mm'}  &=& (p_{m}^2 + 
\mu^2)\,\delta_{mm'} 
    \label{Jdiag1d} \\
p_m
&=&  2\sin(\pi{m}/\cl{})    \,, 
\nnu
\eea
with $\cl{}$ eigenvalues $\ExpaEig_{m}= p_{m}^2 + 
\mu^2$ indexed by integer $m$. The cord function 
$\mathrm{crd}(\theta)=2\sin({\theta}/{2})$ was used already by Hipparchus 
cc.~130 BC in the same context, as a discretization of a circle by 
approximating $\cl{}$ arcs by $\cl{}$ cords.\rf{WikiCord,Berger1869}

\medskip

\paragraph*{Example:
            The spectrum of an \jacobianOrb\ for a \steady\ of period-3.}

The wave numbers \refeq{tempLattEigs} take values ${k} = 0, 2\pi/3, 
4\pi/3$, with lattice momentum values \( 
 p(0)       = 0 \,,\; 
 p(2\pi/3) = p(4\pi/3) = \sqrt{3} \,. \) The lattice momentum square 
${p}_{m}^2$ in \refeq{Jdiag1d} is a discrete field over the 
$\vol_\Bcell{A}=3$ lattice sites of the reciprocal {\pcell} 
$\tilde{\Bcell{A}}$, indexed by integer reciprocal lattice-site labels 
$m=0,1,2$, 
  \setlength{\unitlength}{0.11\textwidth}%
\beq 
{p}^2_{m} \;=\;   
  \raisebox{-1.6ex}[0pt][0pt]{
  \begin{picture}(1,0.67078189)%
    \put(0,0){\includegraphics[width=\unitlength]{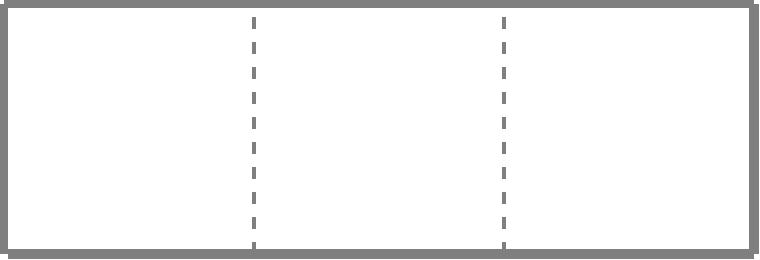}}%
    \put(0.071,0.13){\makebox(0,0)[lb]{\smash{${p}^2_{0}$}}}%
    \put(0.400,0.13){\makebox(0,0)[lb]{\smash{${p}^2_{1}$}}}%
    \put(0.730,0.13){\makebox(0,0)[lb]{\smash{${p}^2_{2}$}}}%
  \end{picture}%
                            }
\;=\;
  \raisebox{-1.6ex}[0pt][0pt]{
  \begin{picture}(1,0.67078189)%
    \put(0,0){\includegraphics[width=\unitlength]{PCrecipLatt3x1}}%
    \put(0.120,0.11){\makebox(0,0)[lb]{\smash{$0$}}}%
    \put(0.450,0.11){\makebox(0,0)[lb]{\smash{$3$}}}%
    \put(0.780,0.11){\makebox(0,0)[lb]{\smash{$3$}}}%
  \end{picture}%
                            }
\;\,, 
\ee{tempLatta3} 
The \jacobianOrb\  $\jMorb_{\Bcell{A}}$ eigenvalues 
are $ \ExpaEig_{m} = {p}^2_{m} + \mu^2 $, and the corresponding 
{\Hilldet} is the product of the three $\jMorb_{\Bcell{A}}$ 
eigenvalues. See \reftabs{tab:LxTsmuSq}{tab:LxTs=5/2} for lists of such 
computations. 

Why are eigenvalues in \refeq{tempLatta3} placed in a box? That will be 
made clearer by example \refeq{catLattaEigs3x2_1}, when we compute 
$\mathsf{p}^2_{\tilde{\Bcell{A}}}$ for a two-dimensional lattice. 

\subsection{{\Pcell}  \steady\ stability in two dimensions} 
\label{s:rcprPcellJ} 

Discrete Fourier transforms diagonalize the hypercubic {lattice} \steady\ 
{\jacobianOrb} over a periodic, `rectangular' {\pcell}  $\Bcell{A}$  in 
any \spt\ dimension $d$,
\bea
&& (\tilde{\jMorb}_{\Bcell{A}})_{mm'}  
       = (\mathsf{p}_{m}^2 + \mu^2)\,\delta_{mm'}
\label{jacOrbEigs}\\
        &&
\mathsf{p}_{m}^2 
        = \sum_{j=1}^d p_j^2
    \,,\quad
p_j = 2\sin\frac{k_j}{2} 
    \,,\quad {k}_{j}=\frac{2\pi}{\speriod{j}}\,m_j
\,, 
\nnu  
\eea 
where $p_j$  is the lattice momentum in the $j$th direction, and $\speriod{j}$ is 
the period of the {\pcell} $\Bcell{A}$ in the $j$th direction, 
with $\vol_{\Bcell{A}}$ {\jacobianOrb} eigenvalues 
$\ExpaEig_m=\mathsf{p}_{m}^2+\mu^2$ taking values on the reciprocal 
lattice sites $\wavenum$, indexed by integer multiplets 
$m=m_1m_2\cdots{m_d}$. The inverse $1/(\mathsf{p}^2 + \mu^2)$ is known as 
the free-field propagator. 

This is almost everything there is to a {\pcell} stability, except that 
the `rectangle' periodic boundary conditions \refeq{jacOrbEigs} are only 
a special case of spacetime periodicity.
Consider a {\steady} {\jacobianOrb} over a two\dmn\ integer lattice. 
For the general 
\refeq{2DHermiteBas} case, as illustrated by \reffig{f:reciprLatt}\,(b), 
the reciprocal primitive vector 
$\tilde{\mathbf{a}}_1=\frac{2\pi}{\speriod{}}\,(1,-\tilt{}/\period{})$ 
has a $0 \leq \tilt{}<\speriod{}$ tilt. Substituting wave vector 
\refeq{2DWaveVector} into the two\dmn\ plane wave (as we did for the 
one\dmn\ case, see \refeq{FrModes1d}), we find that the $k$th eigenstate 
phase evaluated on the lattice site $z$ is
\beq
[\varphi(\wavenum)]_{z} = 
{e}^{i({k}_{1} z_1 + {k}_{2} z_2)}
\ee{2dEigenvect}
where 
\bea
    {z} &=& {z_1}{z_2}    \in \integers^2 
            \,,\quad
    {\wavenum}  ={k}_{1}{k}_{2}  \in \lattice_{\tilde{\Bcell{A}}}
    \,, \quad
        m = m_1m_2
\continue
    m_1  &=&  0,1,\cdots,\speriod{}-1
            \,,\quad
    m_2  =  0,1,\cdots,\period{}-1
\continue
    {k}_{1} &=& \frac{2\pi}{\speriod{}} m_1
            \,,\quad
    {k}_{2}  = \frac{2\pi}{\period{}} (- \frac{\tilt{}}{\speriod{}} m_1 + m_2)
\,. 
\label{2dwavenum}
\eea
As illustrated by \reffig{f:reciprLatt}\,(b), there are 
$\vol_\Bcell{A}=\speriod{}\period{}$ wave 
vectors in the reciprocal {\pcell} $\tilde{\Bcell{A}}$. The \spt\ 
\jacobianOrb\ \refeq{jacOrbEigs} 
is diagonal on the reciprocal lattice, with eigenvalues
\beq 
\ExpaEig_{m_1m_2} 
     =
\mathsf{p}^2_{m_1m_2} + \mu^2 
\,.
\ee{2DjMorbEigs}
It is helpful to work out an 
example to illustrate how \refeq{2DjMorbEigs} gives us the  
{\jacobianOrb} eigenvalues.

\medskip

\paragraph*{Example: 
            The spectrum of the \steady\ \jacobianOrb, \BravCell{3}{2}{1} {\pcell}.}

Consider the {\pcell} $\BravCell{3}{2}{1}$ of example \refeq{pcell321}, 
drawn in \reffig{f:reciprLatt}\,(a). The screw boundary condition 
yields $\tilt{}/\period{}=1/2$. The wave numbers $\wavenum$ in 
\refeq{2dEigenvect} are indexed by integer pairs $m_1 = 0, 1, 2$ and 
$m_2 = 0, 1$. The $\mathsf{p}_{m_1m_2}^2$ in the reciprocal lattice 
{\jacobianOrb} \refeq{2DjMorbEigs} is 
\[
\mathsf{p}_{m_1m_2}^2 = p(k_{1})^2
 + p({k}_{2})^2
\,,
\]
where
lattice momenta 
$p(k)=2\sin({k}/{2})$ take values 
\[
 p(0)       = 0 \,,\;
 p(\pi/3)   = 1 \,,\;
 p(2\pi/3)  = \sqrt{3} \,,\;
 p(\pi)     = 2
\,.
\]
A typical reciprocal lattice site ${m_1m_2}$ evaluation: 
take $m_1=1$, $m_2=1$ in \refeq{2dwavenum}, 
\[
{p}^2_{11}
     =
p\Big(\frac{2\pi}{3}\Big)^2
 + p\Big(\pi - \frac{2\pi}{3}\frac{1}{2}\Big)^2
     = 3 + 3
\,.
\]
The values of $\mathsf{p}^2$, indexed by integer pairs $m_1m_2$, fill 
out the reciprocal lattice unit cell,

  \setlength{\unitlength}{0.11\textwidth}%
\beq 
\mathsf{p}^2_{m_1m_2} \;=\;
  \raisebox{-4.4ex}[0pt][0pt]{
  \begin{picture}(1,0.67078189)%
    \put(0,0){\includegraphics[width=\unitlength]{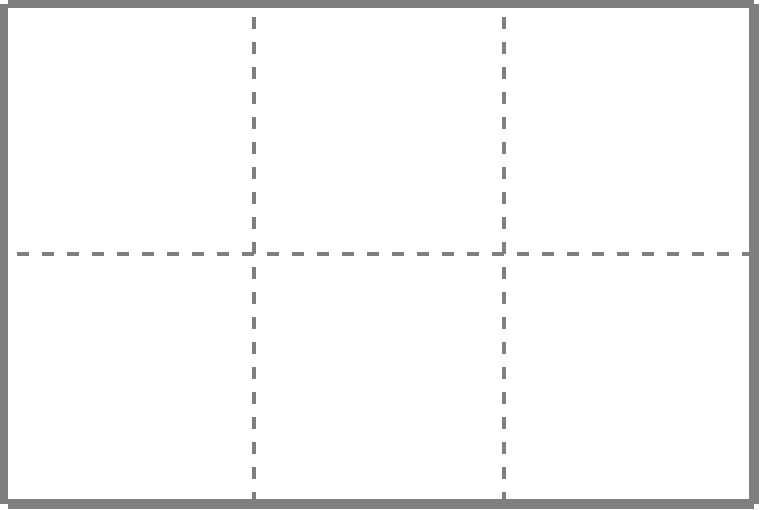}}%
    \put(0.071,0.44){\makebox(0,0)[lb]{\smash{${p}^2_{01}$}}}%
    \put(0.400,0.44){\makebox(0,0)[lb]{\smash{${p}^2_{11}$}}}%
    \put(0.730,0.44){\makebox(0,0)[lb]{\smash{${p}^2_{21}$}}}%
    \put(0.071,0.11){\makebox(0,0)[lb]{\smash{${p}^2_{00}$}}}%
    \put(0.400,0.11){\makebox(0,0)[lb]{\smash{${p}^2_{10}$}}}%
    \put(0.730,0.11){\makebox(0,0)[lb]{\smash{${p}^2_{20}$}}}%
  \end{picture}%
                            }
\;=\;\,
  \raisebox{-4.4ex}[0pt][0pt]{
  \begin{picture}(1,0.67078189)%
    \put(0,0){\includegraphics[width=\unitlength]{PCrecipLatt3x2}}%
    \put(0.120,0.44){\makebox(0,0)[lb]{\smash{$4$}}}%
    \put(0.450,0.44){\makebox(0,0)[lb]{\smash{$6$}}}%
    \put(0.780,0.44){\makebox(0,0)[lb]{\smash{$4$}}}%
    \put(0.120,0.11){\makebox(0,0)[lb]{\smash{$0$}}}%
    \put(0.450,0.11){\makebox(0,0)[lb]{\smash{$4$}}}%
    \put(0.780,0.11){\makebox(0,0)[lb]{\smash{$6$}}}%
  \end{picture}%
                            }
\;\,, 
\ee{catLattaEigs3x2_1} 
                \vskip 12pt     
\noindent
with, for example, being the $(\tilde{\jMorb}_{\Bcell{A}})_{21}$ eigenvalue $ 
\ExpaEig_{21}=4+\mu^2$, and so on.
\refFig{f:2dCatLattBand}\,(a) offers a 
perspective visualization of stability eigenvalues over such reciprocal 
cell, in that case $\lattice_\BravCell{8}{8}{0}$ {\lst}.
The corresponding {\Hilldet}s are products of the 
$\jMorb_{\Bcell{A}}$ eigenvalues, some of which are tabulated in  
\reftabs{tab:LxTsmuSq}{tab:LxTs=5/2}. 

\begin{figure}\begin{center}
(a)~~~\includegraphics[width=0.43\textwidth]{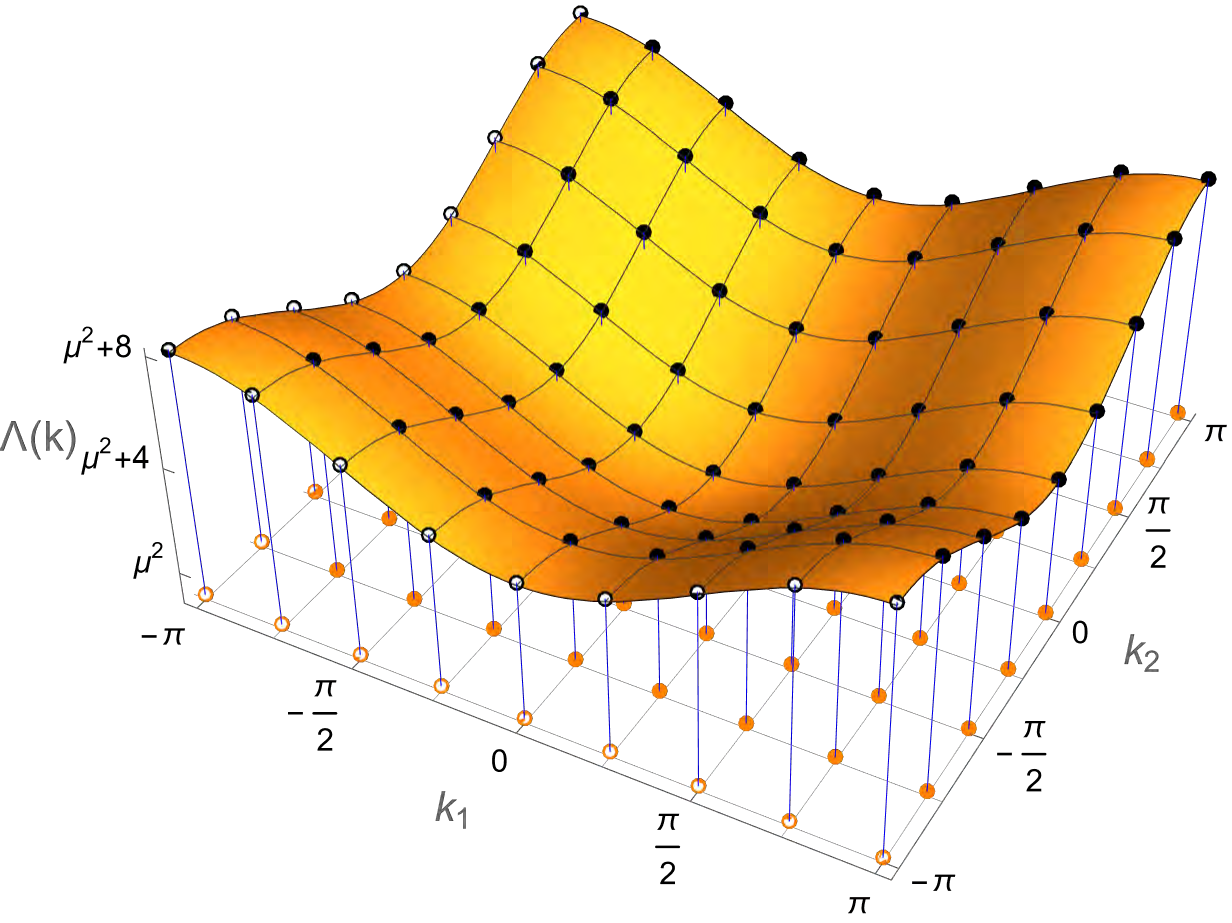}
\\\vskip 12pt
(b)~~~\includegraphics[width=0.43\textwidth]{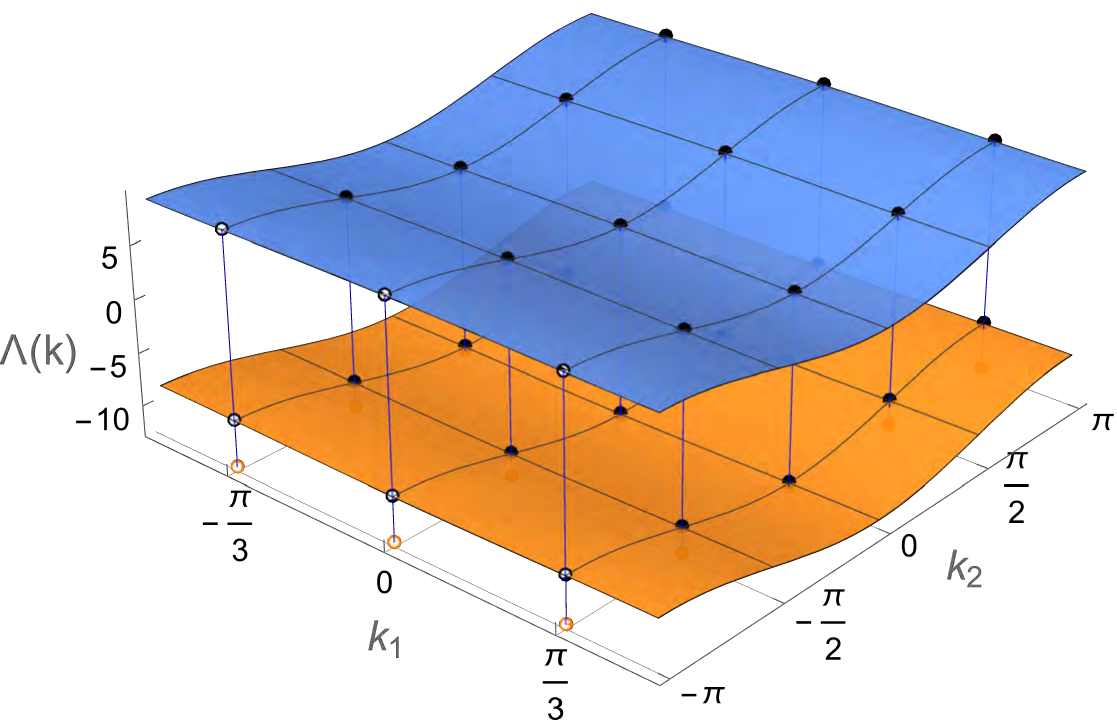}
\end{center}
                \vskip -12pt     
  \caption{\label{f:2dCatLattBand}
Square {\spt} lattice \jacobianOp\ spectra, as
functions of the wave vectors $(k_1,k_2)$. For time and space-reflection
and interchange
invariant {\lsts} the spectra are
$k_1 \to -k_1$, $k_2 \to -k_2$ and $k_1 \leftrightarrow k_2$ symmetric.
(a)
The \steady\ $\ExpaEig(\wavenum)$  Bloch band \refeq{stabExp-d} 
as a function of the wave vector $\wavenum$,
plotted for $\mu^2 = 1$ value.
Black dots are eigenvalues of the {\jacobianOrb} of {\lsts} over {\pcell}
with periodicity
\BravCell{8}{8}{0}.
(b)
The two\dmn\ $\phi^4$ lattice field theory spectra of the
{\Blatt} $\lattice_\BravCell{2}{1}{0}$ {\lst} \refeq{phi4Lst210},
plotted for $\mu^2 = 5$ value. Black
dots are eigenvalues of the {\jacobianOrb} of a \BravCell{6}{4}{0}
{\pcell} tiled by 12 repeats of a prime \BravCell{2}{1}{0} {\lst}, with
$\ExpaEig_{\pm}(\wavenum)$  Bloch bands computed in
\refappe{s:2dNonlin}.
}
\end{figure}

\begin{figure} 
    (a)~~~
\includegraphics[width=0.43\textwidth]{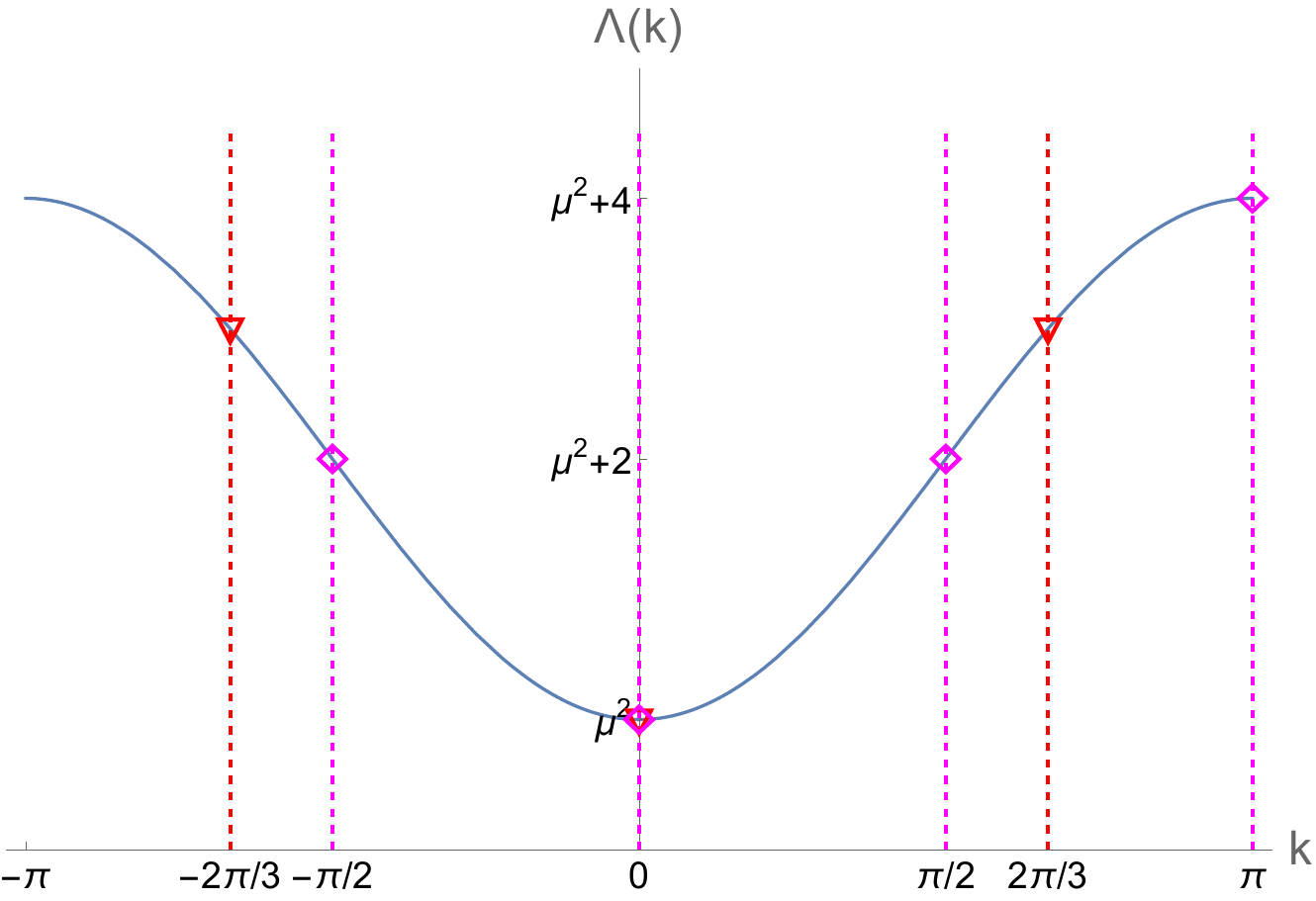}
\\\vskip 12pt
    (b)~~~
\includegraphics[width=0.43\textwidth]{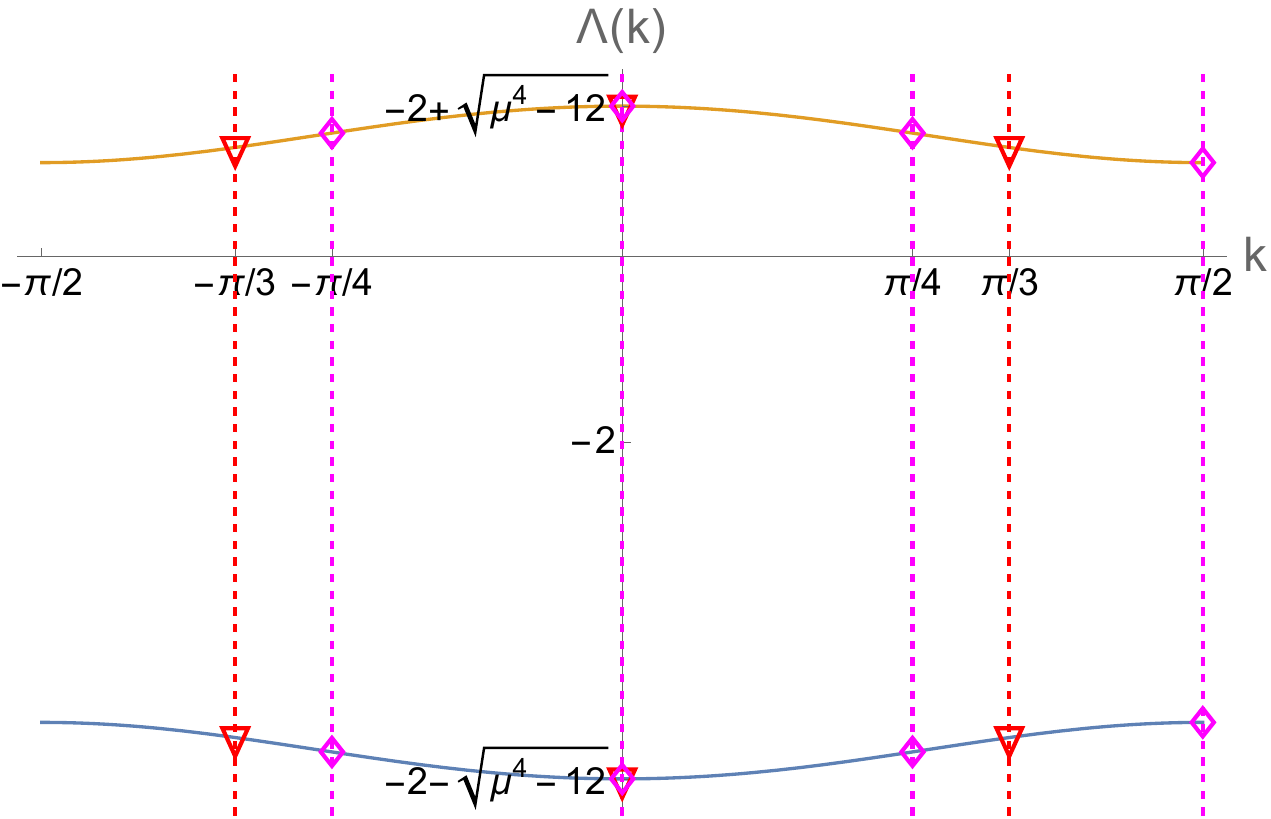}
  \caption{\label{f:recipCatCn} 
One\dmn\ lattice \jacobianOp\ spectra, as functions of the reciprocal
lattice wave number $k$. For time-reversal invariant
systems the spectra are $k\to-k$ symmetric.
(a)
The \steady\ $\ExpaEig(k)$ spectrum \refeq{tempLattEigs}, as a function of the 
first Brillouin zone wave number  $k\in(-\pi, \pi]$,
plotted for the $\mu^2 = 1$ value. 
Any period-\cl{} 
\pcell\ \refeq{steadyStab} {\jacobianOrb} spectrum consists of \cl{} 
discrete points embedded into $\ExpaEig(k)$, for example 
period-3 (red triangles) and
period-4 (magenta diamonds) eigenvalues.
(b)
The nonlinear $\phi^3$ theory $\ExpaEig_{01,\pm}(k)$ spectrum
\refeq{phi3per2Bands} 
of the {\Blatt} $\lattice_{01}$ tiled by
the period-2 {\lst}  $\Xx_{01}=\{\ssp_0,\ssp_1\}$, together with the
eigenvalues of
third repeat (red triangles) and
fourth repeat (magenta diamonds) {\pcell}s.
Plotted for the $\mu^2 = 5$ value.
See \refappe{s:1dNonlin}.
From companion paper I.\rf{LC21}
          }
\end{figure}

Note that all {\catlatt} {\Hilldet}s have a ${\mu}^2$ prefactor. This is 
due to the fact that for $\mu^2=0$ one is looking at a Laplacian, and 
Laplacian operator \refeq{LapOp}, which compares a site field to its 
neighbors, always has a zero mode for the constant eigenvector 
$\varphi_{00}$. 

The values of the lattice momentum square happen to be integers only for 
the few smallest {\pcell}s: in general their values are expressed in 
terms of Hipparchus cord functions $\mathrm{crd}(2\pi\,m_j/\speriod{j})$, 
or what we here call `lattice momenta' \refeq{Jdiag1d}. However, for 
integer values of \catlatt\ Klein-Gordon mass-square $\mu^2$, the 
{\Hilldet}s take integer values, so if we are not interested in details 
of the spectrum, their direct evaluation might be preferable, and that we do 
in \refappe{s:fact}. 

The {\Hilldet} of any {\steady}  $\ssp_z={\ssp}$ of any field theory can 
be evaluated analytically by discrete Fourier diagonalization. Its 
{\jacobianOrb} is constant along the diagonal, with eigenvalues  
evaluated in the same way as for the free-field theory and {\catlatt} 
\refeqs{2dEigenvect}{2DjMorbEigs},
\beq 
(\tilde{\jMorb}_{\Bcell{A}})_{m_1m_2} 
     =
\mathsf{p}^2_{m_1m_2} + \tilde{\mu}^2
\,,
\ee{2DjMorbEigsNonLinear}
where the 
\steady\ Klein-Gordon mass is, as explained in the companion paper 
III~\refref{WWLAFC22}, $\tilde{\mu}^2=-2\,\mu^2 \ssp$ for the {\spt} 
$\phi^3$ \refeq{ELeSptHenlatt}, and 
$\tilde{\mu}^2=\mu^2(1\,-\,3\,\ssp^2)$ for the {\spt} $\phi^4$ 
\refeq{ELeSptPhi4}.

\subsection{{\Pcell} \lst\ stability}
\label{s:pcellStabNonlin} 

Except for the \steady\ solutions discussed so far, the {\jacobianOps} of 
field theories such as the $\phi^3$ \refeq{ELeSptHenlatt} and $\phi^4$ 
\refeq{ELeSptPhi4} depend on the corresponding {\lsts}. The 
{\jacobianOrbs} evaluated over the {\pcell}s, such as \refeq{jMorb1dFT}, 
are generally not invariant under the spacetime unit-lattice spacing 
shift operator \refeq{shiftOperator2d} translations, so they are only 
block-diagonalized by Fourier transforms. 

In general,  {\lst}'s {\Hilldet}s are computed numerically, but -and that 
is basically the only exception- some period-2 {\lsts} can be worked out 
analytically. 

\medskip

\paragraph*{Example:
    One\dmn\ $\phi^3$ field theory period-2 {\lst}.
           }
The one {\spt} dimension  $\phi^3$ theory \refeq{sptHenlatt} has one 
period-2  {\primeOrb}  $\{\Xx_{01},\Xx_{10}\}$ (for details, 
see \refappe{s:1dNonlin}) with two {\jacobianOrb} \refeq{ELeSptHenlatt} 
eigenvalues 
\beq 
\ExpaEig_{01,\pm} = -2 \pm \sqrt{\mu^4 - 12} 
\,, 
\ee{HenEvLR} 
and {\Hilldet} 
\beq
\Det \jMorb_{01} = 16 - \mu^4
\,. 
\ee{HenLRHillDet} 

Consider next a period-6 {\lst} over a {\pcell} 3\Bcell{A} obtained by 
three repeats of the period-2 prime {\lst} (see 
\refeq{1dShift3lst} for another such example). The {\jacobianOrb} 
\refeq{jMorb1dFT} is a [$6\!\times\!6$] matrix, with six eigenvalues 
\bea 
\ExpaEig_{-1,\pm} &=& -2 \pm \sqrt{\mu^4 - 15}
    \continue 
\ExpaEig_{0,\pm}  &=& -2 \pm \sqrt{\mu^4 - 12}
    \continue
\ExpaEig_{1,\pm}  &=& -2 \pm \sqrt{\mu^4 - 15} 
\,, 
\label{HenEv3LR} 
\eea 
and {\Hilldet} 
\beq 
\Det \jMorb_{3*01} = (16 - \mu^4)(19 - \mu^4)^2 
\,. 
\ee{Hen3LRHillDet} 

The eigenvalues of {\jacobianOrbs} for the prime {\lst} $\Xx_{01}$ and 
its repetitions are plotted in \reffig{f:recipCatCn}\,(b). 
The bands, 
denoted by $\wavenum$ in \reffig{f:recipCatCn}\,(b), and the subscript 
of the eigenvalues \refeq{HenEv3LR} will be explained 
in \refsect{s:lstStab}. 
Two of the eigenvalues correspond to `internal' eigenstates (of the 
same periodicity as the prime \lst), so they coincide with the 
{\primeOrb} eigenvalues \refeq{HenEvLR}, while the remaining 
four correspond to `transverse' eigenstates,\rf{Pikovsky89,PikPol16} of 
periodicity of the repeat {\pcell} 3\Bcell{A}. 
 As a result, the {\Hilldet} of the third repeat is not the third power 
of the prime {\lst} {\Hilldet},
\beq 
\Det \jMorb_{3*01} \neq \left( \Det \jMorb_{01} \right)^3
\,.
\ee{nonMultHenon}
This confirms the assertion 
we had made in \refsect{s:intro}, \refeq{primCellNotMltpl}: 
{\Hilldet}s of {\pcell} {\lsts} are \emph{not multiplicative} for their 
repeats. This will be discussed further in \refsect{s:lstStab}.

\medskip

\paragraph*{Example:
    Two\dmn\ $\phi^4$ field theory \BravCell{2}{1}{0} {\lst}.
           }
In \refappe{s:2dNonlin} we work out as a further example the {\pcell} 
stability of two\dmn\ $\phi^4$ theory \refeq{sptPhi4} 
$\BravCell{2}{1}{0}$ {\lst}. 
The eigenvalues of the {\pcell} prime {\lst} and its repetition are 
plotted in \reffig{f:2dCatLattBand}\,(b). Next, note that the {\pcell} 
of {\Blatt} \BravCell{6}{4}{0} can be tiled by twelve repeats of a 
prime \BravCell{2}{1}{0} {\lst}. The eigenvalues of its {\jacobianOrb},  
plotted in \reffig{f:2dCatLattBand}\,(b), lie on the two  {\jacobianOp} 
Bloch bands, located at twelve wave vectors in the first Brillouin zone 
of \BravCell{6}{4}{0}: $k_1 = -\pi / 3, 0, \pi / 3$ and $k_2 = -\pi / 
2, 0, \pi / 2, \pi$. This will be discussed further in \refsect{s:lstStab}.

Reciprocal lattice computations of \jacobianOrb\ spectra can be 
 automated, and we have carried them out for thousands of {\Blatt}s.
Further explicit, but not particularly illuminating {\catlatt} 
calculations are relegated to \refappe{s:catlattComp}.

\section{Bravais stability} 
\label{s:Bloch}  

In \refsect{s:pcellJ} we have defined the {\stabexp} of a \lst\ over a 
finite volume {\pcell}, and in \refsect{s:pcellStab} we have explained 
how to compute them, setting the stage for the main result of section, 
the reciprocal {lattice} evaluation of the {\stabexp} for the {\em 
\jacobianOp}. 

An {\jacobianOp}  \refeq{jMorb1dBrav} acts on an infinite {\Blatt} \lst\ 
$\Xx_\Bcell{A}$ \refeq{stateSp}, so it has infinitely many eigenvalues. 
What that means in context of dynamical systems theory was first 
explained by Pikovsky:\rf{Pikovsky89} while a given \lst\ $\Xx_\Bcell{A}$ 
is $\lattice_\Bcell{A}$-periodic on its infinite {\Blatt}, its perturbations can 
have periodicity of the {\lst}, periodicity of any Bravais sublattice 
$\lattice_\Bcell{AR}$, or no periodicity at all, 
as in \reffig{fig:LstPerturbs}\,(c). 
By the Floquet-Bloch theorem \refeq{BlochThe}, the {\stabexp} $\Lyap_p$ 
then depends on continuum wave numbers $\wavenum\in\Bcell{B}$ within the 
first Brillouin zone,
\beq
k_j\in(-\pi/L_j,\pi/L_j] 
    \,,\qquad j=1,2,\cdots,d
\,. \
\ee{d-dimBrillZ} 

Continuing on the discussion of \refsect{s:pcellStabNonlin}: $k_j=0$ 
eigenvalues correspond to `internal' eigenstates, states of the same 
periodicity as the \lst\ $\Xx_\Bcell{A}$, evaluated here in 
\refsect{s:pcellStab} for the \pcell\ \Bcell{A}. The $k_j\neq0$ continuum 
corresponds to `transverse' eigenstates, perturbations that exit the 
\Bcell{A}-symmetric subspace, as in \pcell\ example 
\refeq{Hen3LRHillDet}.  

In textbook arguments leading to the Bloch theorem, 
one notes that larger and larger \spt\ {\pcell}s correspond to denser and 
denser reciprocal {\pcell}s (see, for example, 
\reffig{f:2dCatLattBand}\,(b), and the arguments of the next section), 
leading in the infinite {\pcell} limit to a parameterization by continuum 
values of wave vectors. Here we always evaluate {\stabexp}s on infinite 
{\Blatt}s, as integrals over the first Brillouin zone.

\subsection{\Steady\ stability}
\label{s:steadyStab} 

Consider the {\stabexp} \refeq{pcellStabExp} of a \steady\ $\Xx_p$, all 
lattice site fields equal,  $\ssp_z={\ssp}$, averaged over a {\pcell} 
\Bcell{A}: 
\[
{\Lyap}_{\Bcell{A}} 
=
\frac{1}{{\vol_\Bcell{A}}} \ln \Det \jMorb_\Bcell{A}
\,=\,
\frac{1}{{\vol_\Bcell{A}}} \Tr_\Bcell{A} \ln (\mathsf{p}^2 + \mu^2)
\,.
\]
The \steady\ {\jacobianOrb} \refeq{steadyStab} is translation invariant 
along each lattice direction, and thus diagonalized by discrete Fourier 
transform \refeq{FrModes1d}. 

For one\dmn\ case \refeq{Jdiag1d} 
\[
{\Lyap}_{\Bcell{A}} 
    =
\frac{1}{\cl{}}
\sum_{m=0}^{\cl{}-1}        \ln ({p}_m^2 + \mu^2)
    =
\frac{1}{2 \pi} \sum_{{k_m}} \Delta k \, \ln ({p}_m^2 + \mu^2)
\,, 
\]
where 
\[
p_m = 2\sin\frac{k_m}{2}
\,, \quad
k_m = \Delta k \, m
\,, \; 
\Delta k = \frac{2 \pi}{\cl{}}
\,.
\]
With the period $\cl{}$ of the {\pcell} ${\Bcell{A}}$ taken to infinity, 
the {\stabexp} is given by the integral over the first {Brillouin} zone, 
\beq
{\Lyap} = \frac{1}{2\pi}\int_{-\pi}^{\pi}\!\!dk\ln({p}^2 + \mu^2) 
\,,\qquad p = 2\sin\frac{k}{2} \,.
\ee{stabExp1} 

By same reasoning, for a  $d$\dmn\ hypercubic {lattice}, the \steady\ 
{\stabexp} is given by a $d$\dmn\ integral over the first {Brillouin} zone 
$\Bcell{B}$,
\bea
&&{\Lyap} \,=\, \frac{1}{(2\pi)^d}\int_{\Bcell{B}}\!\!dk^d 
\ln(\mathsf{p}^2 + \mu^2) \,,
	\continue
&&\mathsf{p}^2 = \sum_{j=1}^d     p_j^2 
    \,,\qquad
p_j = 2\sin\frac{k_j}{2}
\,,
\label{stabExp-d}
\eea
with continuous wave numbers 
restricted to $2\pi$ intervals, conventionally to the centered hypercubic 
first {Brillouin} zone
\beq
\Bcell{B} 
  =
\left\{\wavenum \mid k_1, k_2, \cdots, k_d \in(-\pi,\pi] \right\} 
\,. 
\ee{1stBrillZ} 

The one\dmn\ \steady\ integral \refeq{stabExp1} is  frequently 
encountered in solid state physics, statistical physics and field theory, 
and there are many ways of evaluating it (see, for example, {Gradshteyn 
and Ryzhik}\rf{GraRyz} Eq.~4.226 2): 
\bea
\Lyap \,&=&\,  \frac{1}{2 \pi} \int_{- \pi}^{\pi} \!\!dk \ln \left[4 
	\sin^2 \frac{k}{2} + \mu^2 \right]
	\continue
\,&=&\, \ln\mu^2 
	+2\ln\frac{1+\sqrt{1+4/\mu^2}}{2} 
\,.
\label{stabExp11}
\eea
In this, one\dmn\ 
temporal lattice example, the {\stabexp} $\Lyap$ is the cat map Lyapunov 
exponent,\rf{GHJSC16,LC21} presented here in a form that makes the 
anti-integrable limit \refeq{antiIntStabExp} explicit. 

The one\dmn\ \steady\ {\em \jacobianOp} eigenspectrum is plotted in 
\reffig{f:recipCatCn}\,(a). The discrete eigenvalues of finite\dmn\ 
{\pcell} {\em \jacobianOrbs} are points on this curve. For any finite 
period {\pcell} they only approximate the exact {\stabexp} 
\refeq{stabExp11}. 
\begin{figure}
\begin{center}
           \begin{minipage}[c]{0.23\textwidth}\begin{center}
\includegraphics[width=1.0\textwidth]{HLReciprocalLattice4}
                \\      (a)
            \end{center}\end{minipage}
                   \hskip 2ex
           \begin{minipage}[c]{0.23\textwidth}\begin{center}
\includegraphics[width=1.0\textwidth]{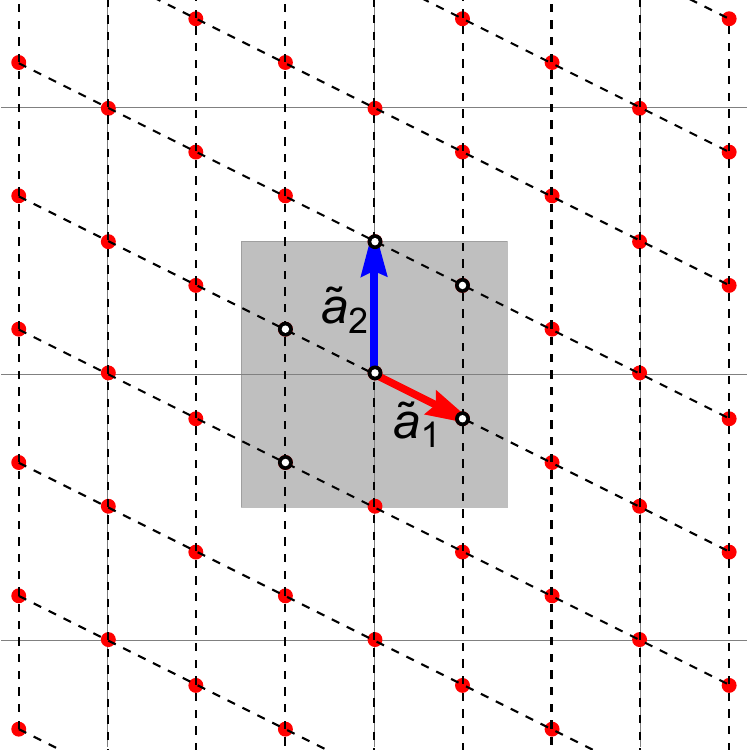}
                 \\     (b)
            \end{center}\end{minipage}
\end{center}
                \vskip -12pt     
  \caption{\label{f:BlochBrillouin} 
(a)
    As in \reffig{f:reciprLatt}\,(b):
    Translations of reciprocal primitive vectors $\tilde{\mathbf{a}}_1$ and
    $\tilde{\mathbf{a}}_2$ 
    (see \refeq{rcprLatt}, \refeq{reciprPcell321}) 
    generate the
    reciprocal lattice $\lattice_{\tilde{\Bcell{A}}}$ (red points), with
    (shaded)
    the reciprocal {\pcell} $\tilde{\Bcell{A}}$.
(b)
    By convention, one restricts the range of wave numbers to
    (shaded)
    the first {Brillouin zone} $\Bcell{B}$, with $k_1, k_2 \in(-\pi,\pi]$.
}
\end{figure}

The two\dmn\ \steady\ {\stabexp} \refeq{stabExp-d} is given by the 
integral over the square lattice two\dmn\ first {Brillouin} zone 
(conventionally a centered square, see shaded domain in 
\reffig{f:BlochBrillouin}\,(b)), 
\bea
&& \Lyap = \frac{1}{4\pi^2} 
\int_{-\pi}^{\pi}\int_{-\pi}^{\pi}\!\!\!d \wavenum \, 
\ln \left[ \mathsf{p}(\wavenum)^2 +  \mu^2 \right] \,,
	\continue
&& d \wavenum = d k_1 d k_2 \,, \quad
\mathsf{p}^2(\wavenum) = p(k_1)^2 + p({k}_{2})^2
\,.
\label{2dCatStabExp}
\eea
Spectra of the two\dmn\ \steady\ {\jacobianOps} are 
plotted in \reffig{f:2dCatLattBand}\,(a). The discrete eigenvalues of 
{\pcell} $\Bcell{A}$ {\jacobianOrbs} embedded in these spectra yield only 
finite volume {\pcell}  approximations to the exact \steady\ {\stabexp} 
\refeq{2dCatStabExp}. 

While it is possible to evaluate such \steady\ integrals analytically 
(see, for example, partition functions with twisted boundary conditions 
of Ivashkevich \etal,\rf{IvIzHu02} and 
papers\rf{Guttmann10,Martin06,Economou06} on 
{Green's function of a discrete Laplacian on a square lattice}), there 
are no analytic formulas for general \lsts, so we evaluate all such 
integrals numerically. An example is the  $\mu^2 = 1$ {\catlatt} 
{\stabexp} ${\Lyap}$ evaluated below in \refeq{catlattStabExp}. 

The \steady\ calculations are so simple, as their orbit Jacobians are 
fully diagonalized by discrete Fourier transforms. For a \steady\ the 
unit hypercube {\pcell} state is \emph{prime}, all other {\lsts} over 
larger {\pcell}s are non-prime repeats of the unit hypercube  {\lst} (see 
\refsect{s:1DprimeOrb}). 

\subsection{\Lst\ stability}
\label{s:lstStab} 

As discussed in \refsect{s:pcellStabNonlin}, the nonlinear field theory 
{\jacobianOps} typically depend on the {\lst} and cannot be diagonalized 
by discrete Fourier transforms. The eigenvalues of the {\jacobianOrbs} in 
the {\pcell}s of prime {\lsts} can only be computed numerically. 

For an arbitrary \lst, in arbitrary dimension, the {\stabexp} $\Lyap$ 
calculation is carried out with the help of the 
\HREF{https://en.wikipedia.org/wiki/Bloch\%27s_theorem} {Bloch} (or 
Floquet) theorem:\rf{Floquet1883,Brillouin30,AshMer} A linear operator 
acting on {\fconfs} with periodicity of {\Blatt} $\lattice_\Bcell{A}$ has 
continuous spectrum, with the lattice sites $z$ eigenstates of form
\beq 
[\varphi^{(\alpha)}(\wavenum)]_{z} = 
    {e}^{\mathrm{i}\wavenum\cdot{z}} [u^{(\alpha)}(\wavenum)]_{z}
\,,\quad \wavenum\in\Bcell{B} 
\,,
\ee{BlochThe}
where $u^{(\alpha)}(\wavenum)$ are band-index 
$\alpha=1,2,\cdots,\vol_\Bcell{A}$ labeled distinct 
$\lattice_\Bcell{A}$-periodic functions, and the continuous wave numbers 
$\wavenum$ are restricted to a Brillouin zone $\Bcell{B}$. In solid-state 
physics, eigenstates \refeq{BlochThe} are known as Bloch 
states.\rf{Kittel96} In mechanics they are called Floquet 
modes.\rf{MathWalk73}

For each {\pcell} {\lst} there is a corresponding prime {\lst} over the 
infinite {\Blatt}, acted upon by {\lst}'s infinite\dmn\ {\jacobianOp} 
\refeq{jMorb1dBrav}. 
We solve for the eigenvalue bands of the {\jacobianOps} as functions of 
the wave vectors $\wavenum$, using Bloch eigenstates \refeq{BlochThe}, 
\bea
{\Lyap}_p &=& 
 \frac{1}{(2\pi)^d} \int_{\Bcell{B}}\!d \mathbf{k} \,
 \ln \left| \Det \jMorb_p(\mathbf{k}) \right|
 \continue
    &=&
\frac{1}{(2\pi)^d} \sum_{\alpha}^{\vol_\Bcell{A}} \int_{\Bcell{B}}\!d 
\mathbf{k} \, 
        \ln \left| \ExpaEig_{p, \alpha}(\mathbf{k}) \right|
\,, 
\label{lnDetStabExp} 
\eea
where $\ExpaEig_{p, \alpha}(\mathbf{k})$ is the 
eigenvalue of the {\primeOrb} $p$ {\jacobianOp} on the $\alpha$-th  
eigenvalue band, 
corresponding to the eigenstate $\varphi^{(\alpha)}(\wavenum)$ in 
\refeq{BlochThe}. This {\primeOrb} {\stabexp} formula is, in the spirit 
of \refsect{s:repeats}, the  generalization of the {\steady} {\stabexp} 
\refeq{stabExp-d} to all {\lsts}. 
    
It suffices to compute the {\stabexp} ${\Lyap}_p$ for a  prime {\lst}, as 
{\stabexp} is the same for a prime {\lst} $\Xx_p=\Xx_\Bcell{A}$ and any 
of its repeats \refeq{steadyRepeats}, 
\beq
\Lyap_p = \Lyap[\Xx_\Bcell{A}] 
= \Lyap[\Xx_\Bcell{AR}] \quad \mbox{for  all $\Bcell{R}$}
\,, 
\ee{StabExpRepeat}
as explained in \refsects{s:BlattJ}{s:repeats}. 

The {Birkhoff average} \refeq{BirkhoffSum} of an observable $\obser$ over 
{\lst} $\Xx_p$ and any of its repeats is also the same, hence the weight 
of a non-prime {\lst} $\Xx_\Bcell{AR}$ contribution to {\pcell} 
{\detSum} \refeq{detPartSum} factorizes
\beq 
\frac{1}{\left|\Det\!\jMorb_\Bcell{AR}\right|}\, 
        {e}^{\vol_\Bcell{AR}\beta\cdot{\obser}_\Bcell{AR}}
    \,=\, \Big(
        {e}^{- \Lyap_p + \beta\cdot{\obser}_p}
          \Big)^{\vol_{p}r_1r_2}
\,, 
\ee{detPartSumTerm} 
with {\stabexp} $\Lyap_{p}$ 
 -and this is the central point- 
evaluated over the reciprocal lattice first {Brillouin} zone \refeq{lnDetStabExp}, 
\[
\frac{1}{\left|\Det\jMorb_{p}\right|}
    \equiv
{e}^{-{\vol_p} \Lyap_{p}}
\,.
\] 
\Pcell\ label \Bcell{A} is redundant here, as it is implicit in the 
{\lst} $p$ label: every {\lst} has its \Blatt\ $\lattice_\Bcell{A}$ 
periodicity. 

In particular, in contrast to the {\em \pcell} {\Hilldet} 
\refeq{nonMultHenon}, the {\Blatt} {\stabexp} of the third repeat 
$\Xx_{3*01}$ is thrice the {\primeOrb} $\Xx_{01}$ {\stabexp},
\beq
|\Det \jMorb_{3*01}| 
    =
\Big( 
        {e}^{- \Lyap_{01} + \beta\cdot{\obser}_{01}}
          \Big)^{3\vol_{{01}}}    
    =
|\Det \jMorb_{01}|^3 
\,.
\ee{MultHenon}
This is the precise statement of 
the  multiplicativity claim \refeq{BravLattMltpl} we had made in the 
introduction: {\Blatt} {\Hilldet}s are \emph{multiplicative} for their 
repeats. 

\medskip

\paragraph*{Example:
    One\dmn\ $\phi^3$ field theory period-2 {\lst}.}
 (This continues from \refsect{s:pcellStabNonlin}.) Consider the simplest 
not-steady solution of $\phi^3$ theory \refeq{sptHenlatt}, its period-2 
{\lst}. The {\jacobianOp} \refeq{jMorb1dBrav} of the period-2 
{\primeOrb} \refeq{henLstLR} is invariant under 
translations of period $\speriod{}=2$, so its first Brillouin zone 
\refeq{d-dimBrillZ} is 
 $(- \pi / 2, \pi / 2]$. This {\jacobianOp} has two bands (for details, 
see \refappe{s:1dNonlin}):
\bea
&& \ExpaEig_{01,\pm}(k) = -2 \pm \sqrt{\mu^4 - 12 - p(2k)^2} 
\,,
\continue
&& p(2k) = 2 \sin(k) 
\,, 
\label{phi3per2Bands} 
\eea
plotted in the first Brillouin zone in \reffig{f:recipCatCn}\,(b). 

Have a look back at the period-6 {\pcell} of 
\refsect{s:pcellStabNonlin}, 
 with the prime {\lst} $\Xx_{01}$ repeated three times. The wave 
numbers $\wavenum$ are constrained to the period-6 reciprocal lattice 
of, taking values $\wavenum=-\pi/3,0,\pi/3$ within the first Brillouin 
zone. They are embedded into the two continuous eigenvalue bands 
\refeq{phi3per2Bands}, corresponding to the \refeq{HenEv3LR}   
eigenvalues $\ExpaEig_{-1,\pm}$, $\ExpaEig_{0,\pm}$ and 
$\ExpaEig_{1,\pm}$ respectively, as illustrated in 
\reffig{f:recipCatCn}\,(b). 

\medskip

\paragraph*{Example:
    Two\dmn\ $\phi^4$ field theory \BravCell{2}{1}{0} {\lst}.}
 (This continues from \refsect{s:pcellStabNonlin}.) 
\refFig{f:2dCatLattBand}\,(b) shows the two eigenvalue bands of a 
two\dmn\ $\phi^4$ \BravCell{2}{1}{0} {\lst}, plotted over the two\dmn\ 
first Brillouin zone  \refeq{d-dimBrillZ} $k_1 \in (-\pi / 2, \pi / 
2]$, $k_2 \in (-\pi, \pi]$. For details, see \refappe{s:2dNonlin}. 

\medskip

\noindent
In summary, while a prime {\lst} and its repeats have different 
\emph{\jacobianOrbs} spectra, they share the same \emph{\jacobianOp} 
{\stabexp}, determined by integration over continuous Bloch bands. 
Inspection of \reffigs{f:recipCatCn}{f:2dCatLattBand} makes it clear what 
these different spectra are: {\jacobianOrb}  spectra are discrete 
approximations to {\jacobianOp} continuous Bloch bands, with `cords' 
approximation errors decreasing as the {\pcell} volume increases. (For 
the convergence rate of such approximations, see {shadowing} 
\refsect{s:shadow}). The wonderful property of the {\stabexp} 
${\Lyap}_{p}$ computed over the infinite spacetime (as opposed to the 
{\stabexp} computed over a finite {\pcell}) is that it is additive
for prime {\lsts} {repeated} over larger {\pcell}s. 

\section{\Po\ theory}
\label{s:POT}

Now that we know how to enumerate all {\Blatt}s
$\lattice_\Bcell{A}$ (\refsect{s:Bravais}), determine all {\lsts} over 
each (\refsect{s:orbits}), and compute the weight of each {\lst} 
(\refsect{s:Bloch}), we can combine all of that into one generating 
function sum of a very simple form, for any deterministic field theory, 
in any spacetime dimension: 

\medskip

\paragraph*{Definition: {\DetSum}.}

\begin{quote}
For integer lattices,
the {\detSum} is the sum over all  {\lsts} $\Xx_c$,
each of weight $t_c$,
\beq
 Z[\beta,{z}] \,=\, \sum_c \,t_c
\,,\quad
t_c \,=\,    \Big(
         {e}^{\beta\cdot{\obser}_c-{\Lyap}_{c}} \, z
             \Big)^{\vol_c}
\,,
\ee{detSum}
where ${\Lyap}_{c}$ is the {\stabexp} \refeq{lnDetStabExp}, $\obser_c$ 
the observable $\obser$ averaged over \lst\ $\Xx_c$, \refeq{lst_aver}, 
$\vol_c$ is the {\Blatt} $\lattice_c$ volume \refeq{lattVol}, and $z$ is 
a generating function variable. 
\end{quote}
Notation `$t_c$' is a vestige of referring to this weight in the 
time-evolution {\po} theory as the `local trace' (see 
\toChaosBook{section.18.2} {sect.~18.2}).\rf{CBcount} 
Indeed, much of the time-evolution \po\ theory developed in ChaosBook 
generalizes to the multi-periodic, \spt\ \dft, with time period \period{c} 
replaced by the \spt\ volume $\vol_c$. 
Square brackets $[\cdots]$ in quantities such as $Z[\beta,z]$ are a here
to remind us that we are dealing with \spt\
field theories,\rf{FieldThe} not with a
few degrees-of-freedom evolving forward in time (\refsect{s:DFT}).

\subsection{Chaotic field theory}
\label{s:sptchaos}

Ergodic theory of time-evolving dynamical systems is a rich subject.
In this series of papers we stay within its most robust corner that 
we refer to as the `chaotic field theory'. We say that a {\dft} is 
\emph{chaotic} if
 (1) 
all of its {\lsts} are unstable,  \ie, the 
{\stabexp}, \refeq{lnDetStabExp}, is strictly positive, ${\Lyap}_{c}>0$, 
for every deterministic solution $\Xx_c$, and 
(2) 
the number of {\lsts} $|c|_\Bcell{A}$ grows exponentially with the 
\pcell\ volume $\vol_\Bcell{A}$, with
(3)
    \edit{
{\lsts} are `shadowed' by lower volume {\lsts} (\refsect{s:shadow}). 
    }

To understand where the `generating function variable' $z$ in 
\refeq{detSum} comes from,
consider  $Z_\Bcell{A}[\beta]$, the \pcell\ \Bcell{A} partition sum
\refeq{detPartSum} over all {\lsts} $\Xx_c$ of periodicity
$\lattice_\Bcell{A}$.
Their number $|c|_{\Bcell{A}}$ is the number of admissible {\mosaic}s 
(\refsect{s:mosaics}), with the mean of the log of the number of \lsts\ 
per lattice site given by 
\(
h_{\Bcell{A}} = \frac{1} {\vol_\Bcell{A}}\ln |c|_\Bcell{A}
\,.
\)

If  $|\A|$, the number of letters in the alphabet \refeq{Mm_c}, is
bounded, there are at most $|\A|^{\vol_\Bcell{A}}$ distinct {\mosaic}s
over \pcell\ \Bcell{A}, so $|c|_{\Bcell{A}}$, the number of \spt\
solutions $\{\Xx_c\}$ of system's defining equations \refeq{eqMotion} is
bounded from above by $\exp(\vol_\Bcell{A}h_{max})$, where $h_{max}$ is
any upper bound on $h_\Bcell{A}$, for example
\beq
|c|_{\Bcell{A}} \leq {e}^{\vol_\Bcell{A} h_{max}}
    \,,\qquad
h_{max} = \ln |\A|
\,.
\ee{topEntrBound}

Now consider a system with a 2-letter alphabet (think of Ising `spins'),
with {\pcell}s \Bcell{A} accommodating very few \lsts\ $\Xx_c$, each with
almost all spins `up' or `down' (frozen phases in statistical mechanics,
Pomeau–Manneville intermittency\rf{Pomeau80} in temporal evolution
systems).
For such long correlations systems
$h_\Bcell{A}\to0$.

To guarantee chaos, we consider here only field theories for which
the number of solutions also has a strictly positive lower exponential
bound
$h_\Bcell{A} \geq h_{min}>0\,,$
with the $h_{\Bcell{A}}$
of large volume \Blatt\ bounded between  $h_{min}$ and $h_{max}$.
\beq
{e}^{\vol_\Bcell{A} h_{min}}
    \leq
{e}^{\vol_\Bcell{A} h_{\Bcell{A}}}
    \leq
{e}^{\vol_\Bcell{A} h_{max}}
\,.
\ee{topEntrPcell}
The exact value of $h_{\Bcell{A}}$ might require a calculation, and
evaluation of the expectation value of system's entropy $h$ will require
the full machinery of the \po\ theory developed here in
\refsect{s:zetaExpectV}, but all we need to ensure that the theory is
{\spt}ly chaotic is that all $ h_{\Bcell{A}}$ are
strictly positive.

Next:
a typical observable is bounded in magnitude, so its contribution to the
partition sum \refeq{detSum} weight is bounded by
$\exp(\vol_\Bcell{A}\beta\cdot{\obser}_{max})$.
And, crucially, throughout this series of papers we focus only on purely
`chaotic' systems, defined as systems for which every \lst\ is unstable
in the sense that its {\stabexp} \refeq{lnDetStabExp} is strictly
positive,
\beq
0<\Lyap_{min}
    \leq \Lyap_{c}
\,,
\ee{minStab}
as illustrated by \reffigs{f:recipCatCn}{f:2dCatLattBand} and the
anti-integrable limit calculations \refeq{antiIntStabExp} and
\refeq{stabExp11},
so:

The  {\pcell} partition sum \refeq{detPartSum} is
bounded exponentially in lattice volume $\vol_\Bcell{A}$,
\beq
Z_\Bcell{A}[\beta]	
    \,\leq\, {e}^{\vol_\Bcell{A} (\beta\cdot{\obser}_{max}
                                 -\Lyap_{min}
                                 +h_{max})}
\,.
\ee{partFunctBound}

No {\Blatt} $\lattice_\Bcell{A}$ is special,
all of them contribute, so we combine them into a generating function
\beq
\sum_{\Bcell{A}} Z_{\Bcell{A}} [\beta] \, z^{\vol_\Bcell{A}}
\,,
\ee{detSumGener}  
a sum over all `geometries'. Exponential bound \refeq{partFunctBound} ensures
that the sum is convergent for sufficiently small generating function variable $z$.
With the {\stabexp} evaluated over the reciprocal lattice, as in
\refsect{s:Bloch}, this is just the
{\detSum} \refeq{detSum} arranged as a series
in $z^{\vol}$. Our first task will be to determine the
largest $z$ for which the {\detSum} is convergent,
\refsect{s:evalZeta}.

In this paper we utilize only the translation group $T$ symmetries,
\refeq{dDimTranslGrp} (see \refsect{s:Bravais}), and
focus on the case of a two\dmn\ square lattice.
Translations stratify the {\dft}, \refeq{detSum}, into
\emph{\primeOrbs},\rf{Ruelle76a,Ruelle76,ChaosBook}
{\lsts} that are not repeats of a shorter period state
(\refsects{s:1DprimeOrb}{s:2DprimeOrb}).
To assemble the \detSum\ over all {\lsts}, we reorganize it by first
determining all {\primeOrbs} $\Xx_p$, and summing over their repeats
(\refsect{s:repeats}). Then the {\detSum} \refeq{detSum} takes form
\beq
Z[\beta,{z}] \,=\,
\sum_p Z_p
\,,
\ee{partSumPrimes}
with $(\beta,{z})$ dependent {\lsts} weights $t_c$, \refeq{detSum}.
The form of the prime partition sum $Z_p$ depends on the
spacetime dimensionality. To explain how, it suffices to consider the
simplest case: the partition sum for a theory with a single  {\primeOrb},
a {\steady}.


\subsection{{\Steady} partition sum}
\label{s:steadyPartSum}

A {\steady} $\ssp_z={\ssp}$ is a prime {\lst} whose
{\pcell} $\BravCell{1}{1}{0}$ is the unit hypercube \refeq{SquareLatt}
(\refsect{s:repeats}).
A {\lst} is then obtained by tiling any larger {\pcell} by repeats of \steady\
${\ssp}$, one such {\lst} for every Bravais sublattice.

For a one\dmn, temporal {\Blatt}, the {{\detSum}}
\refeq{detSum} is very simple.
There is a non-prime {\lst} for every repeat {\pcell} \Bcell{A} of period
$r$, all of them with orbits of period 1 (\refsect{s:1DprimeOrb}), so
thanks to repeat weights multiplicativity \refeq{detPartSumTerm}, the
contribution of $r$th repeat to the partition sum \refeq{detSum}
is
\(
t_p^{r}
    \,=\,
\left(
{e}^{\beta\cdot{\obser}_p-{\Lyap}_{p}}{z}
\right)^{r},
\) 
where ${r}$ is the volume of the period ${r}$ {\pcell}.
So, in \emph{one dimension} the {\steady} partition sum is a geometric  series,
\beq
Z_p  = \sum_{r=1}^\infty t_p^{r}
        \,=\,
\frac{t_p}{1-t_p}
\,,\qquad
t_p = {e}^{\beta\cdot{\obser}_p-{\Lyap}_{p}} {z}
\,,
\ee{partSumSteady:1d}
with
{\steady} {\stabexp} $\Lyap_{p}$ \refeq{stabExp11},
and the observable evaluated on the \steady\ $\ssp$,
$\obser_p = \obser(\ssp)$.

\medskip

Thanks to repeat weights multiplicativity \refeq{detPartSumTerm}, the
contribution of a
$\BravCell{r_1}{r_2}{s}$ \steady\ repeat
to the partition sum \refeq{detSum}
in two dimensions is
\beq
t_p^{r_1r_2}
    \,=\,
\left(
{e}^{\beta\cdot{\obser}_p-{\Lyap}_{p}} {z}
\right)^{r_1 r_2}
\,,
\ee{d2partFunctRpt} 
So for a two\dmn, \spt\ {\steady} (see \refsect{s:orbits}) there
is a non-prime {\lst} for every two\dmn\ repeat {\pcell} $\BravCell{r_1}{r_2}{s}$
constructed in \refsect{s:Blatt}, with
$r_1$ copies of lattice site field $\ssp{}$ horizontally, $r_2$ copies
vertically, with weights, as stated in \refeq{StabExpRepeat},
independent of the tilt $0 \leq s < r_1$ \refeq{steadyRepeats},
\[
Z_p \,=\, 
    \sum_{r_1=1}^\infty
    \sum_{r_2=1}^\infty r_1\,t_p^{r_1r_2}
\,,
\]
where tilts $s$ are summed over:
\[
\sum_{s=0}^{r_1-1} 1  = r_1
\,.
\]
Summing over heights  $r_2$,
the deterministic {\steady} partition sum \refeq{detSum}
in \emph{two dimensions} is thus
\beq
Z_p  \,=\,
\sum_{{n}=1}^\infty \frac{n \, t_p^{{n}}}{1-t_p^{{n}}}
\,.
\ee{partSumSteady:2d}
Its expansion in powers of variable $t_p$
\bea
Z_p
=
\sum_{\cl{}=1}^\infty \sigma(n) t_p^\cl{}
\,&=&\,
t_p+3 t_p^2+4 t_p^3+7 t_p^4+6 t_p^5
\continue
&& +12 t_p^6+8 t_p^7+\cdots
\,,
\label{EulerDivFct}
\eea
 was first studied by Euler, with $\sigma(\cl{})$ known as the Euler {sum-of-divisors function}.\rf{wikiDivisorfunc}

\subsection{{\PrimeOrb} partition sum}
\label{s:primePartSum}

As explained in \refsect{s:repeats}, every prime {\lst} is morally a
\steady, except that now the number of {\lsts} in the  {\primeOrb} is its
\pcell\ volume $\vol_p$. Hence in \emph{one dimension} the {\primeOrb} $p$
\detSum, including {\steady}s \refeq{partSumSteady:1d}
as the $\vol_p=1$ cases, is simply
\beq
Z_p  \,=\, \vol_p\,\frac{t_p}{1-t_p}
\,,\qquad
t_p =
            \left(
{e}^{\beta\cdot{\obser}_p-{\Lyap}_{p}} {z}
            \right)^{\vol_{p}}
\,,
\ee{partSumPrime:1d}
with  {\primeOrb} {\stabexp} $\Lyap_{p}$ \refeq{lnDetStabExp}, and
$\obser_p$  the {Birkhoff average}, \refeq{BirkhoffSum}, of an observable
$\obser$ over prime {\lst} $\Xx_p$.

The formula has form of the `deterministic trace formula' (see
Eq.~(21.24)} in \refref{CBtrace}) with a crucial difference:
here the weight $t_p$ is the \emph{exact}, infinite \Blatt\ weight, while
the Ruelle \dzeta\ weight $t_p$ is an \emph{approximate} weight.

The contribution of a \emph{two-dimensional} {\spt} prime {\lst} $p$ and
its repeats to the deterministic partition sum, \refeq{partSumPrimes}, is
\bea
&& Z_p  \,=\, \vol_p
\sum_{{n}=1}^\infty  \frac{n \, t_p^{{n}}}{1-t_p^{{n}}}
            \,,
\label{partSumPrime:2d} \\
&& t_p =
            \left(
{e}^{\beta\cdot{\obser}_p-{\Lyap}_{p}} {z}
            \right)^{\vol_{p}}
            \,,\;\;
        \vol_p \,=\, \speriod{p}\period{p}
\,.
\nnu
\eea

\subsection{\Spt\ zeta functions}
\label{s:lattZeta}

We have a {\detSum} \refeq{detSum} of stunning simplicity. Still, we can
do better, by taking into account symmetries  (\refsect{s:Bravais}) of
the theory.

\medskip

\paragraph*{Definition: {\DetZeta}.}
\begin{quote}
For two\dmn\ integer lattices, the  {\detZeta} is the product over all
{\primeOrbs}, of form
\beq
1/\zeta \,=\, \prod_{p} 1/\zeta_p
    \,,\qquad
1/\zeta_p \,=\, \prod_{n=1}^\infty (1 - t_p^{n})
\,.
\ee{sptZeta2d}
\end{quote}
Who ordered this `zeta'? Euler.
Euler replaced the partition sum (in
Euler's case, a weighted sum over natural numbers) by a
zeta function (in Euler's case, the product over primes formula for the
Riemann zeta function), by the logarithmic derivative relation between
the partition sum and the zeta function
\beq
    Z[\beta,z]  =
      - z \frac{\partial}{\partial z}\ln{1/\zeta[\beta,{z}]}
\ee{der-det-map}
(see, for example, \toChaosBook{section.22.3}{sect.~22.3}).\rf{CBdet}

Why?
(1) The {\detSum} \refeq{detSum} is a redundant sum over all
\emph{\lsts}, redundant as their weights depend only on
\emph{\primeOrbs}, which a zeta function counts only once per orbit.
(2)
Every \lst\ weight contributes to the \detSum\ with a positive weight.
Zeta functions are smarter, they exploit the key property of ergodic
trajectories that they are \emph{shadowed} by shorter trajectories
(\refsect{s:shadow}), with convergence of {\cycForm}s improved by
shadowing cancellations.
(3) Zeta functions have better analyticity properties, with divergence of
{\detSum} \refeq{detSumGener} corresponding to the leading zero of
{\detZeta}.

In \emph{one \spt\ dimension}, the {\detZeta} is a product over all
{\primeOrbs}, of form\rf{CvitanovicYT00}
    \ifsubmission\else
\toVideo{youtube.com/embed/_7ZNfbgJ8D4}
    \fi
\beq
1/\zeta 
    \,=\,\prod_{p} 
 \left( 1 - t_p \right)
\,,
\label{zet}
\eeq
as is easily checked by substitution into relation \refeq{der-det-map}. 
The pseudo-cycle expansion of $1/\zeta$ then leads to {averaging 
formula}s with better convergence than the {\detSum} \refeq{detSum} (see 
\toChaosBook{section.23.5}{sect.~23.5}).\rf{CBrecycle} 

In \emph{two \spt\ dimensions}, the  {\detZeta} is again
a product over all {\primeOrbs},  \refeq{sptZeta2d}.
Its correspondence to the {\detSum}, \refeq{partSumPrime:2d},
is easily checked by substitution into the relation \refeq{der-det-map}.

\subsection{Evaluation of zeta functions}
\label{s:evalZeta}

For large lattice volume $\vol$ {\pcell}s, the exponential bounds of
\refeq{partFunctBounds}, (\ref{freeEnergy}) and~(\ref{partFunctBound})
ensure convergence of high order $z^\vol$ terms in {\detSum},
\refeq{detSumGener}, to $(e^{W[\beta]}z)^{\vol}$, with sum convergent
for sufficiently small $z$.

\medskip

\paragraph*{Definition: \new{Function $W$}.} 
\begin{quote}
The largest value of $z(\beta)$ for fixed $\beta$,
\beq
z(\beta) = {e}^{-W[\beta]}
\,,
\ee{detSumAsympt}
for which the {\detSum} \refeq{detSum} converges, or, equivalently, the
value of $z(\beta)$ which is the first root of the inverse of {\detZeta},
\refeq{sptZeta2d},
\beq
Z[\beta,z(\beta)] \to \infty \,;\qquad
1/\zeta[\beta,z(\beta)] = 0
\,,
\ee{detZetaEsc}
defines system's `{\escape}'
\HLedit{$- W[0]$}.
\end{quote}

\noindent
In dynamical systems theory, the rate at which trajectories leave an open
system per unit time is called \emph{escape rate} (see \refref{KT84} and
\toChaosBook{equation.1.4.3} {Eq.~(1.3)}).
We put `\escape' into quotations here, as in \spt\ theory there is no
escape in time - the exponent is a characterization of the \nws, the
\statesp\ set formed by the deterministic solutions.
We evaluate it for the {\templatt} in \refappe{s:tempLattEscape}, and
for deterministic $\phi^3$ and $\phi^4$ theories in companion paper
III.\rf{WWLAFC22}

Much is known about the {two-\spt\ dimensions} zeta function,
\refeq{sptZeta2d}, as for each {\primeOrb} $1/\zeta_{p}$ is the {Euler function}
$\phi(t_{p})$,\rf{wikiEulerFunc}
\bea
1/\zeta_{p} &=& \phi(t_{p})=\prod_{n=1}^\infty (1-t_{p}^n)
\,,\qquad |t_{p}|<1
\,,
\label{EulerFct}
\eea
whose power series in terms of pentagonal number
powers of $z$ was given by Euler\rf{Bell05} in 1741
\bea
\phi(z) =&&
1 - z - z^2 + z^5 + z^7 - z^{12} - z^{15}
\continue
&& + z^{22} + z^{26}
- z^{35} - z^{40} + z^{51}
+ z^{57}
\continue
&& - z^{70} - z^{77}
+ z^{92} + z^{100} + \dots
\label{EulerFctExp}  
\eea
In 1996 Lind\rf{Lind96} introduced topological zeta func\-tion of
this form for two\dmn, as well as higher\dmn\ shifts.
So, while for a one\dmn\ lattice, the contribution  \refeq{zet} of a
{\primeOrb} $\Xx_p$ is simply $1/\zeta_p=1-t_{p}$, in two \spt\
dimensions the {\primeOrb} weight is a yet another `Euler function' with
an infinite power series expansion.
Presumably because of that, in our numerical work the $z$
power series expansions of two\dmn\ $1/\zeta$ do not appear to  converge
as smoothly as they do in the one\dmn, temporal settings.

\subsection{{\CycForm}}
\label{s:zetaExpectV}

While each {\pcell} probability density \refeq{ProbConf} is easily 
correctly normalized, there is no natural `overall' probability 
normalization for the generating function sum over all {\lsts}, the 
{\detSum} \refeq{detSum}. 
For one\dmn, temporally evolving systems, 
Sec.~23.5 of \refref{CBrecycle} solves this problem by the method 
of `cycle averaging'. We now generalize this method to the 
higher-dimensional, \spt\ field theories. 

The smallest value of the generating function variable $z(\beta)$ for 
which the {\detZeta} equals zero,  \refeq{detZetaEsc}, is an implicit 
equation for the root $z=z(\beta)$ satisfied on the curve $0 = 
1/\zeta[\beta,z(\beta)]$ in the $(\beta,z)$ parameters plane. 
Take the derivative of this implicit equation (for brevity, we take 
$1/\zeta=1/\zeta[\beta,z(\beta)]$): 
\[
0 = \frac{d}{d\beta} 1/\zeta
  = \frac{\pde} {\pde \beta} 1/\zeta \,+\,
    \frac{d z}{d\beta} \frac{\pde}{\pde z} 1/\zeta
\,.
\]
\refeq{detSumAsympt} relates ${d z}/{d\beta}$ to the log of partition 
function evaluated on the infinite lattice, \refeq{freeEnergy}, 
\beq
- \frac{1}{z} \frac{d z}{d\beta}
     \,=\,
\frac{dW}{d\beta}
     \,=\,
    \frac{
~\frac{\pde}{\pde \beta} 1/\zeta ~
    }{ 
z\frac{\pde}{\pde z} 1/\zeta
    }   
\,.
\ee{first-der}
The {\em expectation value} of observable 
$\obser$, \refeq{expctObserW1}, 
\[ 
\expct{\obser}
     \,=\,
\left. \frac{d~}{d\beta}W[\beta]\right|_{\beta=0}
\] 
is now given by the {\em \cycForm}\rf{CvitanovicYT00}
    \ifsubmission\else
\toVideo{youtube.com/embed/uZ4O-xDczOA} 
    \fi
\beq
\expct{\obser} \,=\,
\frac{\expct{\Obser}_\zeta}{\expct{\vol}_\zeta }
\,.
\ee{expctObserW} 
Here the weighted Birkhoff sum of the observable 
$\expct{\Obser}_\zeta$, \refeq{BirkhoffSum},
and the weighted multi-period lattice volume 
$\expct{\vol}_\zeta$, \refeq{lattVol}, are defined as 
\bea
\expct{\Obser}_\zeta &=&
- \left. \frac{\pde~}{\pde \beta} 1/\zeta[\beta,z(\beta)]
\right|_{\beta = 0, z = z(0)}
     \,,\continue 
\expct{\vol}_\zeta  &=&
- \left. z\frac{\pde~}{\pde z}  1/\zeta[\beta,z(\beta)]
\right|_{\beta = 0, z = z(0)}
\,.
\label{expVals}
\eea
where the subscript in $\expct{\cdots}_\zeta$ stands for the 
 {\detZeta}  
evaluation of such weighted sum over {\primeOrbs}.
        
{Expectation value}s  $\expct{\cdots}_\zeta$ are  evaluated by noting that all 
$(\beta,z)$ dependence of the  {\detZeta}, \refeq{sptZeta2d}, is contained 
in the {\primeOrbs} weights $t_p$, whose partial derivatives are simply 
\beq
   z \frac{\partial}{\partial z}t_p =   \vol_p t_p
    \,,\qquad
   \frac{\partial}{\partial \beta}t_p =  \Obser_p t_p
    \,.
\ee{partialDs}
As an example, we compute the expectation value of {\stabexp} of {\templatt}
in \refappe{s:HillExpPcell}.

While power series expansions in $z$ of functions such as
the Euler function, \refeq{EulerFctExp}, do not converge very well,
the theory of doubly-periodic elliptic functions suggests other,
more powerful methods to evaluate such functions.
The Euler function can be expressed as the
{Dedekind eta function} $\eta(\tau)$,\rf{wikiDedekindEta}
\beq
\phi(t_p)  \,=\, t_p^{-\frac{1}{24}} \eta(\tau_p)
    \,,\qquad
\mbox{Im}(\tau_p) > 0
\,,
\ee{DedekindEta}
where  $\tau_{p}$ is the complex phase of the Euler function
argument,  $t_{p} ={e}^{i2\pi\tau_{p}}$.
Our prime zeta function, \refeq{sptZeta2d}, complex phases
of prime {\lsts} 
follow from \refeq{partSumPrime:2d},
\beq
\tau_{p} = i \frac{{\vol_p}}{2\pi}
             ( -\beta\cdot{\obser}_p+{\Lyap}_{p}-{S})
    \,,\qquad
z        = e^{-S}
\,,
\ee{DedekindPhase}
with the \lst\ $\Xx_{p}$ probability weight having a pure positive
imaginary phase
\[
\tau_{p} =  \frac{i}{2\pi} {\vol_p}{\Lyap}_{p}
\,.
\]
The derivatives required for the evaluation of expectation values, 
\refeq{expVals}, have their own elliptic functions representations.
For example, the logarithmic derivative of Dedekind eta is known as the 
{Weierstrass zeta function},\rf{WolframDedekindEta} 
\beq
{\eta^\prime}/{\eta}=\zeta_W
\,.
\ee{derDedekEta}

The problem in evaluation of the  {\detZeta}, \refeq{sptZeta2d}, is that 
it is an \emph{infinite} product of Dedekind eta functions, and we 
currently know of no good method to \HLedit{systematically} truncate and evaluate 
such products. 

\bigskip


The {\spt} zeta function \refeq{sptZeta2d} is the main result of
this paper.
However, there are still a couple of questions of general nature that
alert reader is likely to ask.

(i) 
How is this global, high-dimensional orbit stability related to the 
stability of the conventional low-dimensional, forward-in-time evolution? 
The two notions of stability are related by Hill's formulas (also known 
as the Gel'fand-Yaglom theorem,\rf{GelYag60,Ludewig16} for continuous 
spacetime), relations that are in our formulation equally applicable to 
energy conserving systems, as to viscous, dissipative systems. We derive 
them in \refrefs{LC21,LC22}. From the field-theoretic perspective, 
{\Hilldet}s are fundamental, forward-in-time evolution is merely one of 
the methods for computing them. 

(ii) 
One might wonder why do we focus so much on computing \lsts\ over every 
\emph{small} {\pcell}, omitting none? You never see that anywhere in the 
literature. But that's what the theory of (temporal) chaos and (\spt) 
turbulence demands: the support of a deterministic field theory is on 
\emph{all} deterministic solutions, as sketched in \reffig{f:pde2a}. 

Next --the beauty of the {\po} theory of
chaos-- due to the shadowing of longer periods unstable \lsts\ by shorter
periods ones, the smallest periodicites \lsts\ dominate, the longer ones
come in only as corrections.
The convergence {\lsts}-expansions is accelerated
by shadowing of long orbits by shorter periodic orbits.\rf{AACI}
Are $d$\dmn\ tori ({\pcell}s) {\lsts} also shadowed by smaller
tori \lsts?
In \refsect{s:shadow} we check numerically that \catlatt\ {\lsts} that
share finite {\spt} {\mosaic}s indeed shadow each other to exponential
precision.    

\section{Shadowing}
\label{s:shadow}

                                                                \toCB
In ergodic theory `shadowing lemma': ``a true time-trajectory is said to 
shadow a numerical solution if it stays close to it for a time 
interval\rf{Barreto08,Palmer09}'' is often invoked to justify collecting 
statistics from numerical trajectories for integration times much longer 
than system's \HREF{https://en.wikipedia.org/wiki/Lyapunov_time} 
{Lyapunov time}.\rf{wikiLyapTime} In {\po} theory, the issue is neither 
the Lyapunov time, nor numerical accuracy: all {\po}s are `true' in the 
sense that in principle they can be computed to arbitrary 
accuracy.\rf{DingCvit14} In present context `shadowing' refers to the shortest distance 
between two orbits decreasing exponentially with the length of the 
shadowing time interval. Long orbits being shadowed by shorter ones leads 
to controllable truncations of cycle expansions,\rf{AACI} and computation 
of expectation values of observables of dynamical systems to exponential 
accuracy.\rf{ChaosBook}

{\Fconfs} are points in \statesp\ \refeq{stateSp}, with the separation of 
two {\lsts} $\Xx$, $\Xx'$ given by the \statesp\ vector $\Xx-\Xx'$, so we 
define `distance' as the average site-wise \statesp\ Euclidean 
distance-squared between {\fconfs} $\Xx$, $\Xx'$, \ie, by the Birkhoff 
average \refeq{BirkhoffSum} 
\beq
|\Xx-\Xx'|^2 = 
         \frac{1}{\vol_{\Bcell{A}}}
         \sum_{z\in\Bcell{A}} (\ssp'_z-\ssp_z)^2
\,.
\ee{lstsDist} 
This notion of distance is intrinsically \spt, it does not refer to 
time-evolving unstable trajectories separating in time. For \catlatt\ we 
have an explicit formula for pairwise separations: If two \catlatt\ 
{\lsts} $\Xx$, $\Xx'$ share a common sub-{\mosaic} \Mm, they are 
site-fields separated by
\beq
\ssp_{z}-\ssp'_{z} 
  =\sum_{z'\notin\Mm}g_{z z'}(\Ssym{}-\Ssym{}')_{z'}
  \quad
  \mod 1
\,,
\ee{catlattDist1} 
where matrix $g_{z z'}$ is the \catlatt\ Green's 
function \refeq{GreenFunc}. 

It was shown numerically by Gutkin \etal\rf{GutOsi15,GHJSC16} that 
pairs of interior alphabet \refeq{catlattIntAlph} \catlatt\ {\lsts} of a 
fixed spatial width \speriod{} that share sets of sub-{\mosaic}s, shadow 
each other when evolved forward-in-time. Here, in 
\refsect{s:catlattShadow},  we check numerically {\catlatt} shadowing for 
arbitrary \lsts, without alphabet restrictions, and without any time 
evolution. Intuitively, if two unstable {\lsts} $\Xx$, $\Xx'$ share a 
common sub-{\mosaic} \Mm\ of volume $\vol_\Mm$, they shadow each other 
with exponential accuracy of order of $\propto\exp(-\lambda \vol_\Mm)$. 
In time-evolution formulation, $\lambda$ is the leading Lyapunov 
exponent. What is it for \spt\ systems?

We first explain how the exponentially small distances follow for the 
one\dmn\ case.

\subsection{Shadowing, one\dmn\ \templatt}
\label{s:tempCatShadow} 

As the relation between the {\mosaic}s $\Mm$  and the corresponding 
{\lsts} $\Xx_\Mm$ is linear, for $\Mm$ an {\admissible} {\mosaic}, the 
corresponding {\lst} $\Xx_\Mm$ is given by the Green's function 
\beq 
\Xx_\Mm = \gd\,\Mm \,,\qquad \gd = \frac{1}{-\shift + s\id - \shift^{-1}} 
\,. 
\ee{tempCatGreen}
For an infinite one\dmn\ lattice 
$\zeit\in\integers$, the lattice field at site $\zeit$ is determined by 
the sources $\Ssym{\zeit'}$ at all sites ${\zeit'}$, by the  Green's 
function $g_{\zeit\zeit'}$ for one\dmn\ discretized heat 
equation,\rf{PerViv,varcyc}
\beq 
  \ssp_{\zeit}=\sum_{\zeit'=-\infty}^\infty g_{\zeit\zeit'} \Ssym{\zeit'}
\,, \quad 
g_{\zeit\zeit'}=\frac{1}{\ExpaEig-\ExpaEig^{-1}}\, 
                \frac{1}{\ExpaEig^{|\zeit-\zeit'|}}
\,, 
\ee{1dLatGreenFct}  
with $\ExpaEig$ the expanding cat map stability multiplier 
\refeq{tempCatMult}. 
While the {\jacobianOp} $\jMorb$ is sparse, it is not diagonal, and its 
inverse is the full matrix $\gd$, whose key feature is the matrix element 
$g_{\zeit\zeit'}$ factor $\ExpaEig^{-|t'-t|}$, which says that the 
magnitude of a matrix element falls off exponentially with its distance 
from the diagonal. This fact is essential in establishing the `shadowing' 
between {\lsts} sharing a common sub-{\mosaic} \Mm. Suppose there is a 
non-vanishing point source $\Ssym{0}\neq0$ only at the present, 
$\zeit'=0$ temporal lattice site. Its contribution to $\ssp_{\zeit}$ 
$\sim \ExpaEig^{-|\zeit|}$ decays exponentially  with the distance from 
the origin. If two {\lsts} $\Xx$, $\Xx'$ share a common sub-\mosaic\ \Mm\ 
of length \cl{}, they shadow each other with accuracy of order of 
$O(1/\ExpaEig^{\cl{}})$. 

\subsection{Shadowing,  two\dmn\ {\catlatt}}
\label{s:catlattShadow} 

Following \refrefs{GutOsi15, GHJSC16}, consider families of {\spt} orbits 
that share a sub-{\mosaic} 
region. 
    \edit{
The {\lsts} used in numerical examples of \refref{GHJSC16} 
    } 
were restricted to those whose {\mosaic}s used 
only the interior, always admissible, alphabet \refeq{catlattIntAlph}. 
Here we check numerically {\catlatt} shadowing for generic \lsts, 
with no alphabet restrictions. 

The two\dmn\  $\mu^2=1$ {\catlatt}  \refeq{catlatt} {\lsts} are labeled 
by  two\dmn\ {\mosaic}s, 8-letter alphabet \refeq{catlatt2d}, as in 
\reffig{f:symmBlock}. 

To investigate the {\catlatt} shadowing properties,
    \edit{
we considered 
{\catlatt} {\lsts} with periodicity 
\BravCell{18}{18}{0}, 
    }
all sharing the same $[12\!\times\!12]$ {sub-\mosaic}, 
see  \reffig{f:symmBlock}. 
We generated 500 such {\lsts} by 
randomly changing the lattice site symbols outside the 
{sub-\mosaic}, finding the corresponding {\lst} by 
solving the {\catlatt} defining equation \refeq{catlattKG}, and keeping 
only those admissible solutions that still contained the same 
$[12\!\times\!12]$ {sub-\mosaic}. 

The {\spt} shadowing suggests that for {\lsts} with identical 
sub-{\mosaic}s, the distances between the corresponding field 
values decrease exponentially with the size of the shared {sub-\mosaic}s. 

\begin{figure}
  \centering
(a)~~~\includegraphics[width=0.43\textwidth]{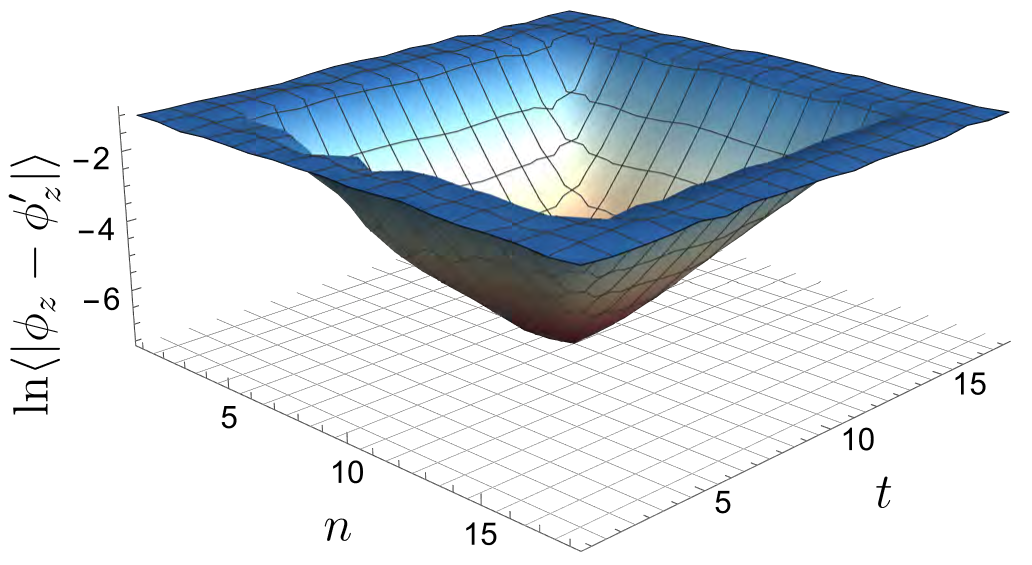}
~~~
(b)~~~\includegraphics[width=0.43\textwidth]{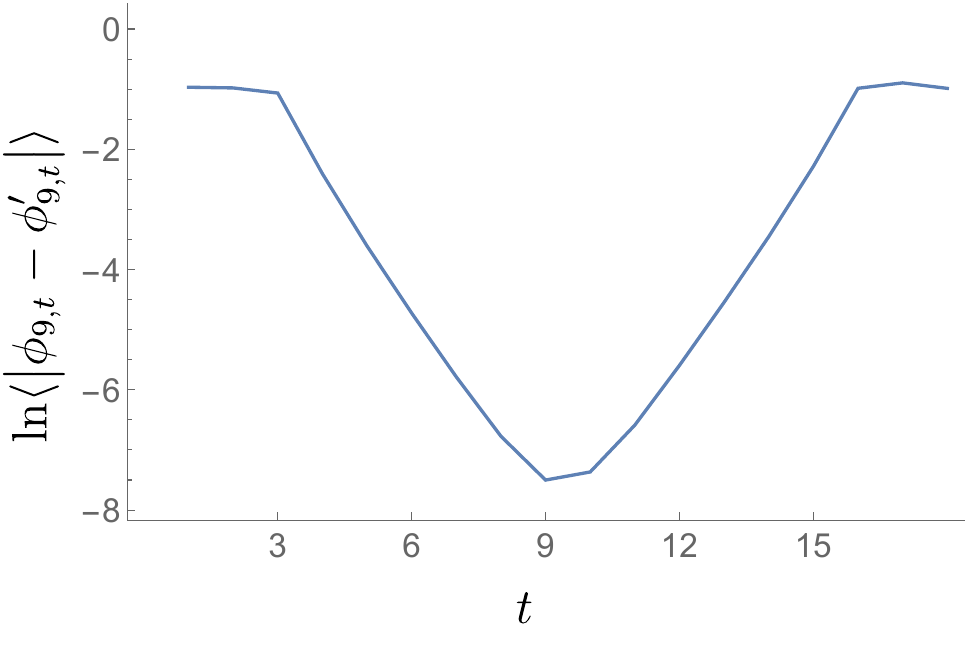}
  \caption{\label{f:18x18shadow}
  $\mu^2=1$ \catlatt.
(a)
  The log of mean of point-wise field value distances
  $|\ssp_{z}-\ssp'_{z}|$ over all lattice sites of
  $z\in[18\!\times\!18]$ {\pcell}, averaged over the 250 pairs of {\lsts},
  like the pair of \reffig{f:symmBlock}.
(b)
  The log of mean point-wise distances $|\ssp_{9,\zeit}-\ssp'_{9,\zeit}|$
  evaluated across the strip $z=(9,\zeit)$, $\zeit=1,2,\dots,18$, going
  through the center of the {\pcell}. The decrease from edge to the center
  is approximately
  linear, with slope $\approx-1.079$.
}
\end{figure}

To find the rate of decrease of distances between shadowing {\lsts}, we 
compute the mean point-wise distances of field values of the 250 pairs of 
{\lsts} over each lattice site in their {\pcell}s. The exponential 
shadowing of {\lsts} is shown in \reffig{f:18x18shadow}. The distances 
between field values of two {\lsts} $|\ssp_{z} - \ssp'_{z}|$ decrease 
exponentially as $z$ approaches the center of the common sub-{\mosaic}. 
\refFig{f:18x18shadow}\,(a) is the log plot of the mean distances. The 
logarithm of the mean distances across the center of the {\pcell} is 
plotted in \reffig{f:18x18shadow}\,(b), where the decrease is 
approximately linear, with a slope of $-1.079$. What determines this 
slope? 

\subsection{Green's function of two\dmn\ {\catlatt}}
\label{s:catlattGreen}

{\Mosaic} {\Mm} is admissible (see \refsect{s:mosaics}) if {\fconf} 
$\Xx_\Mm$ is a {\lst}, \ie, all lattice site fields are confined to 
\refeq{StateSpCatLatt}, the compact  boson hypercube \statesp\ 
$\ssp_{z}\in[0,1)$. 

The Green's function measures the correlation between two lattice sites 
in the spacetime. In our problem the distances between the shadowing 
{\lsts} can be interpreted using the Green's function, which gives 
variations of field values $\ssp_{\zeit}$ induced by a `source', in this 
example by change of a letter $\Ssym{\zeit'}$  at lattice site $z'$. The 
decrease of the differences between field values of shadowing {\lsts} is 
a result of the decay of correlations. The Green's function for massive 
free-boson on integer lattices \refeq{GreenFunc} has been extensively 
studied.%
\rf{Glaser70,BuGoNi70,Wood71,KMIHA71,KatIna71,KaInAb71,Morita71,
    MorHor71,Horiguchi71,Fiedler73,HorMor75,AndMor85,varcyc,PerViv,
    ChenM87,ChuYau00,dlLlave00,Martin06,Economou06,Asad07,BhaOst10,
    Guttmann10,Shimizu12,GHJSC16,SteGok19,CaSiLo19,CaLoSi20}
But to understand 
qualitatively the exponential falloff of spacetime correlations, it 
suffices to consider the large spacetime {\pcell} (small lattice spacing) 
continuum limit: 
\[
(-\Box + {\mu}^2)\, \ssp(x)= \Ssym{}(x)\,,\qquad x \in \reals^2
\]
whose Green's function is the radially symmetric
\beq
G(x,x') = 
\frac{1}{2 \pi} K_0 \left( \mu |x - x'| \right) 
\,, 
\ee{Han2023-07-12} 
where $K_0$ is the 
modified Bessel function of the second kind.\rf{wikiModBessel} For large spacetime 
separations, $|x - x'| \to \infty$, the asymptotic form of the Green's 
function is 
\beq 
G(x,x') \sim \sqrt{\frac{1}{8 \pi \mu r}}e^{-\mu r} 
             \,,\qquad r = |x - x'|
\,. 
\ee{ContinuumGreenAsymp}
In the numerical example of 
\refsect{s:catlattShadow}, we have set Klein-Gordon mass $\mu = 1$, so 
the Green's function of the continuum {\sPe} is a good approximation to 
the discrete  {\catlatt} Green's function, where the rate of decrease of 
correlations computed from the \reffig{f:18x18shadow}\,(b) is 
approximately $\exp(- \mu' r)$, \HLedit{with} $\mu' = -1.079$ \HLedit{the} slope 
computed from the log plot of the mean distances of field values between 
shadowing {\lsts}.

\subsection{Convergence of evaluations of observables}
\label{s:catlattStabExp} 

Computed on {\pcell}s $\Bcell{A}$ of increasing volume $\vol_\Bcell{A}$, 
the expectation value of an observable (\refsect{s:lstsExpectV}) 
converges towards the exact, infinite {\Blatt} value (\refsect{s:Bloch}). 
As the simplest case of such sequence of {\pcell} approximations, take a 
rectangular {\pcell} $\BravCell{\speriod{}}{\period{}}{0}$, and evaluate 
{\stabexp}s $\expct{\Lyap}_{\BravCell{r\speriod{}}{r\period{}}{}}$ 
(\refsect{s:rcprPcellJ}) for the sequence of {\pcell} repeats 
$\BravCell{r\speriod{}}{r\period{}}{0}$ of increasing $r$. 

That the convergence of such series of {\pcell} approximations is a 
shadowing calculation can be seen by inspection of 
\reffig{f:2dCatLattBand}. The exact {\stabexp} ${\Lyap}$ is obtained by 
integration over the bands (smooth surfaces in the figures). A shadowing 
approximation ${\Lyap}_{\BravCell{\speriod{}}{\period{}}{\tilt{}}}$ is a 
finite sum over {\pcell}s $\BravCell{\speriod{}}{\period{}}{\tilt{}}$, 
black dots in the figures, that shadows the curved surface, with 
increasing accuracy as the {\pcell} volume $\vol_\Bcell{A}$ increases. 
Here shadowing errors are Hipparchus' errors \refeq{Jdiag1d} of replacing 
arcs by cords, as in approximating $2\pi$ by the perimeter of a regular 
$n$-gon. The sense in which such shadowing or `curvature' errors are 
exponentially small for one\dmn, temporal lattice chaotic systems is 
explained in \refrefs{AACI,AACII,CBconverg}. We have not extended such 
error estimates to the \spt\ case, so here we only present numerical 
evidence that they are exponentially small.

\begin{figure}
  \begin{center} 
  \setlength{\unitlength}{0.45\textwidth}
  \begin{picture}(1,0.60833333)%
    \put(0,0){\includegraphics[width=\unitlength]{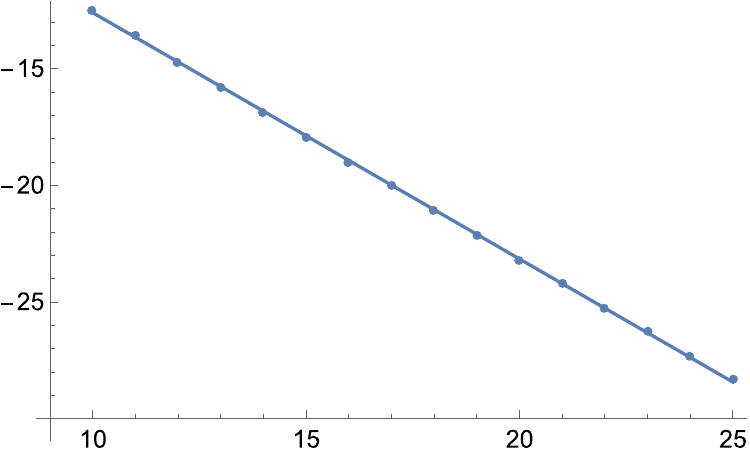}}%
    \put(0.47164459,-0.06368938){\makebox(0,0)[lb]{\smash{$L$}}}%
    \put(0.10,0.31438249){\makebox(0,0)[lb]
      {\smash{$\ln({\Lyap}-{\Lyap}_\Bcell{A})$}}}%
  \end{picture}%
\end{center}
  \caption{\label{fig:stabExpLxL}
The convergence of {\pcell}s {\stabexp}s  ${\Lyap}_\Bcell{A}$ to  
${\Lyap}$, the exact {\Blatt} value \refeq{catlattStabExp}, for square 
{\pcell}s $\BravCell{\speriod{}}{\speriod{}}{0}$ sequence 
\refeq{catlattStabexp1},
$\mu^2 = 1$. 
A linear fit of the logarithm of the distance  as a
function of the side length $\speriod{}=10,11,\cdots,25,$
with slope -1.05538.
}
\end{figure}

As a concrete example, we evaluate numerically the exact  $\mu^2 = 1$ 
{\catlatt} {\stabexp} ${\Lyap}$ for the infinite {\Blatt} {\jacobianOp} 
\refeq{2dCatStabExp}, 
\beq 
{\Lyap} = 1.507\,983\cdots 
\,, 
\ee{catlattStabExp} 
and investigate the convergence of its finite 
{\pcell} estimates ${\Lyap}_{\BravCell{r\speriod{}}{r\period{}}{0}}$. For 
the unit cell $\BravCell{1}{1}{0}$ sequence, plotted in 
\reffig{fig:stabExpLxL}, 
${\Lyap}-{\Lyap}_{\BravCell{\speriod{}}{\speriod{}}{0}}$ decreases 
linearly as the side length $\speriod{}$ increases, with a linear fit has 
slope 
\beq 
\ln({\Lyap}-{\Lyap}_{\BravCell{\speriod{}}{\speriod{}}{0}}) 
    = -2.046\,11 -1.055\,38\,\speriod{}
\,. 
\ee{catlattStabexp1} 
For various {\pcell} sequences of rectangular 
shapes $\BravCell{\speriod{}}{\period{}}{0}$, the {\stabexp}s of repeat 
{\pcell}s $\BravCell{r\speriod{}}{r\period{}}{0}$ also converge to 
${\Lyap}$ exponentially, with the same convergence rate 
$\approx1.055\cdots$. We have no theoretical estimate of this rate, but 
it appears to be close to the Klein-Gordon mass $\mu=1$, within the 
shadowing error estimates of \refsect{s:catlattShadow}.

\section{Summary and open questions}
\label{s:summary}

Gutzwiller's 1971 semiclassical quantization\rf{gutzwiller71,gutbook}
yields deep insights into the quantum behavior of low-dimensional
deterministically chaotic systems (ODEs).
In dynamical systems theory this point of view leads to the 1976 Ruelle 
{\po} formulation of chaotic dynamics.\rf{Ruelle76a,ruelle,ChaosBook}
In this series of papers\rf{LC21,CL18,WWLAFC22,LC22} we generalize the 
Ruelle temporal ODEs / iterated maps dynamical zeta functions theory
to our field-theoretic, PDEs \spt\ zeta functions formulation 
of \spt\ chaos and turbulence. 

Flows described by {\pdes} are in principle infinite dim\-ens\-ion\-al,
and, at first glance, turbulent dynamics that they exhibit might appear
hopelessly complex. However, what is actually observed in experiments and
simulations is that turbulence is dominated by repertoires of
identifiable recurrent vortices, rolls, streaks and the
like.\rf{science04} Dynamics on a low-dim\-ens\-ion\-al chaotic attractor
can be visualized as a succession of near visitations to exact unstable
periodic  solutions, interspersed by transient
interludes.\rf{ChaosBook} In the same spirit, the long-term turbulent
dynamics  of spatially   extended systems can  be thought of as a
sequence of visitations through the repertoire of {\admissible}  {\spt}
patterns (`exact coherent
structures',\rf{W01,WK04} `recurrent flows'\rf{ChaKer12}), each framed by a finite {\spt}
window. The question we address here is: can states of a strongly
nonlinear field theory  be described by such repertoires of {\admissible}
patterns, explored by turbulence?
And if yes, what is the likelihood of observing any such pattern?

These  questions have been   studied  extensively for systems of  small
spatial extent, whose inertial manifold dimension is relatively
small.\rf{Christiansen97,SCD07,CviGib10,ACHKW11,KreEck12}
Recent experimental and theoretical
advances\rf{science04,focusPOT,Suri20,CPTKGS22,CPTGS22} support such
forward-in-time dynamical vision also for spatially extended, turbulent
systems (see \refref{GibsonMovies} for a
gentle introduction).
By now thousands of such `exact coherent structures'
have been computed, always confined to small spatial
domains, while the flows of interest (pipe, channel, plane flows) are
flows on infinite spatial domains.
Going from  spatially small to spatially infinite systems
requires a radical shift in the point of view.
To describe those, we have to recast equations such as the
{\NS}  as a spacetime theory, with all infinite
translational symmetry directions treated on equal footing.
It is a bold leap, a theory of turbulence that does away with
dynamics.
However, we have no choice. For spatially extended systems evolution forward 
in time is insanely unstable.\rf{WFSBC15,GudorfThesis} Not only have 
time-evolution numerical codes%
\rf{channelflow,openpipeflow}
not worked on large domains, in retrospect 
it is clear that they never could have worked. 

Conventional numerical computations confine spatial directions to a
finite domain, then integrate forward in time, treating only time as
inherently infinite.
In contrast, our {\dft} formulation is global, in the
sense that its building blocks are \emph{orbits}, global {\fconfs} that
satisfy system's defining equations everywhere, over the infinite
spacetime. We treat all translationally invariant
directions democratically, each an infinite {`time'}. Here there
are no sketches of diverging trajectories, because in the deterministic
{\spt} field theory formulation of turbulence, there is \emph{no
evolution} in time.

The first problem that we face is \emph{global}: determining and
organizing infinities of unstable multi-\lsts\ over
$\infty$\dmn\ {\statesp}s, orbits that are presumed to form the skeleton
of turbulence.
We characterize and classify them by their shapes, captured by
corresponding \new{`{\mosaic}s'} (\refsect{s:mosaics}). 
The feel is of statistical-mechanics,
such as enumeration of Ising configurations. 
For \emph{temporal} evolution,
in the {\em large volume} $\period{}\to\infty$ limit,
our partition sum $Z$, \refeq{detSum}, 
is Ruelle's `partition function',\rf{ruelle} 
our functional $W$, \refeq{freeEnergy}, 
is Ruelle's `pressure',\rf{Ruelle72} and
our \detZeta\ $\zeta$, \refeq{zet}, 
is Ruelle's `\dzeta'\rf{Ruelle76a} 
(for nomenclature, see {remark~20.2} {\em `Pressure'}
in \refref{CBremark20-2}).
In the {\em \spt} theory,
in the \emph{large volume} $\speriod{}\period{}\to\infty$ limit
our functional $W$ is
Grassberger's `density of metric entropy'\rf{Grassberger89}
and Politi, Torcini \& Lepri's chronotopic `entropy potential function'.%
\rf{LePoTo96,LePoTo97,LePoTo97a}  
Furthermore, as had previously been done implicitly in 
cycle expansion calculations of 
observables,\rf{CGS92,Cvitanovic1995109,aver_dasbuch} 
see \refsect{s:zetaExpectV}, 
Politi \etal\ relate expectation values of observables by Legendre 
transforms of $W$ to appropriate `effective actions' or `Gibbs free 
energies'.

As we always follow Ruelle, who says this about `pressure':
``frankly the proper term ... should be free energy'',\rf{Rugh19,Rugh19a}
we provisorily settle for calling our function $W$, 
\refeq{detSumAsympt},  
the `\new{function $W$}'.

\subsection{What is \new{new}}
\label{s:new}

    \edit{
In determining the \new{totality of unstable multi-\lsts}
        }
we are helped by
working `beyond perturbation theory', in the anti-integrable, strong
coupling regime, in contrast to much of the literature that focuses on
weak coupling expansions around `ground states'.
Our calculations utilize standard optimization methods,%
\rf{ChaosBookLeastAct, 
CvitLanCrete02, 
lanVar1, 
orbithunter,
GudorfThesis, 
LFCL21, 
AzAsSc22, 
ParSch22, 
WanLan22, 
PNBK22} 
some of which precede spatiotemporal theory by decades.
They are memory costly, but as there is no evolution, neither in time nor
in space, there are no instabilities, and it suffices to determine a
solution to a modest accuracy.

The next problem is: how important is a given spacetime
configuration? Intuitively, more unstable solutions have smaller \statesp\
neighborhoods, are less likely.
Here the {\jacobianOp} \refeq{jacOrb} of a {\lst}, its {\pcell}
determinant \refeq{eigsProd}, and especially its infinite {\Blatt}
{\stabexp} \refeq{lnDetStabExp} are the most important innovations in our
theory of spatiotemporal chaos.
    
    \edit{
The \new{essential innovation} of our approach is the computation 
of \new{spatiotemporal stability exponents} over infinite Bravais lattices, 
see \reffig{fig:LstPerturbs}, \ref{f:recipCatCn} and \ref{f:2dCatLattBand},
rather than forward-in-time Floquet/Lyapunov stability of finite 
time solutions over rectangular spacetime {\pcell}s. 
We work in infinite spacetime, not a mix of modest or large spatial 
extent with either long time ergodic estimates, or compact periodic time 
solutions. 
The {\jacobianOp} \refeq{jMorb1dBrav}, \refeq{ELeFT}, the orbit 
{\stabexp}  $\Lyap_p$ 
\refeq{lnDetStabExp}, and the 
surprisingly simple exact deterministic \spt\ zeta function 
\refeq{sptZeta2d} are defined democratically over spacetime lattice 
momenta 
(Pikovsky's Bloch quasi-momenta
{\em and} Bountis' Floquet quasi-energies).

Historically, in all of the previous work on spatially extended systems, the time
and space instabilities were treated asymmetrically. While the spatial
stability was parameterized by spatial Bloch wave number $k_1$,
the time (in)stability was either estimated by forward evolution-in-time
numerical Lyapunov-Bloch exponents\rf{Pikovsky89} $\Lyap(k_1)$ (the Oseledets
multiplicative ergodic theorem),
or by means of compact, finite \pcell\ (\refsect{s:pcellStab}) periodic
time solutions. Examples of the latter are spacetime periodic solutions
of \KS\ and \NS\ PDEs studied in
\refref{Christiansen96,GHCW07,SCD07,Visw07b,DCTSCD14,GudorfThesis,WiGuOr25}.

In all previous \spt\ studies the large volume 
$\speriod{}\period{}\to\infty$ `thermodynamic' limits were estimated by  
simulations of finite volume lattices.
Our \new{\dft} is \emph{ab initio} over infinite
spacetime, there are no large volume limits to estimate.
Our \new{partition function $Z$} sum over \new{all {\Blatt}s},
our \new{functional $W$}, and our \new{\detZeta\ $\zeta$} are exact
expansions in terms of \new{exact prime orbit weights $t_p$}.

That follows from the \new{main advance} of our \new{Euclidean field theory}
of \spt\ chaos/turbulence, the functional determinant evaluations of infinite \Blatt\
\new{\stabexp\ ${\Lyap}_{c}$},
\refeq{lnDetStabExp}, 
and \new{multiplicative weight $t_c$}, which, to best of our 
knowledge, do not appear in the earlier literature.
        }
        
And, finally: now that we have a hierarchy of multi-\lsts, what is the
expectation value of any observable of the theory? The answer is given
by {\detSum}s and zeta functions of
\refsect{s:POT}.
All chaoticity is due to the intrinsic instability of \emph{deterministic}
solutions, and our \spt\ {\detSum}s \refeq{detSum}
and zeta functions \refeq{sptZeta2d} are \emph{exact}, not merely saddle
point approximations to a theory. The issues in their applications are
numerical: how many {\lsts}, evaluated to what accuracy have to be
included into a truncated zeta function, to attain a desired accuracy?

Our {\dft}
replaces numerical extrapolations of finite spacetime simulations of the 
1990s chronotopy by the \emph{exact} {\detZeta}, with computable cycle 
expansion truncations errors, decreasing exponentially with the \spt\ 
volumes of \lsts\ included in its evaluation. 
Our \new{\spt\ \detZeta}, as far as we know, is new.

\subsection{Open questions}
\label{s:open}

At the present stage of development, our \spt\ theory of chaos leaves a
number of open problems that we plan to address in future publications:

\begin{enumerate}
  \item
Can the 2- and higher- \spt\ dimension zeta function \refeq{sptZeta2d} and
expectation value 
\refeq{expctObserW} 
computations be organized into
`cycle expansions', dominated by the small spacetime volume \lsts, as is
the case for the one\dmn, temporal theory?\rf{ChaosBook}
  \item
For pedagogical reasons --in order to start out with determinants of 
finite matrices, rather than immediately grapple with functional 
determinants-- we have presented here the spacetime deterministic field 
theory in its discretized, $d$\dmn\ lattice form. 
We expect the {\po} formulation of continuum spacetime theory to be of 
essentially of the same form, 
with the spacetime periodicities ({\Blatt} primitive vectors)
defined over the continuum, $\mathbf{a}_j\in\reals^d$,
the {\stabexp} \refeq{lnDetStabExp} 
evaluated as a functional integral over its Brillouin zone, 
generating function variable $z$
replaced by a Laplace transform variable $\eigenvL$,
$z=e^{-\eigenvL}$
(see \toChaosBook{equation.21.2.20}{Eq.~(21.20)} in \refref{CBtrace}), 
and the \detSum\ \refeq{detSum} of the form
\beq
Z[\beta,\eigenvL] \,=\, \sum_c \,t_c
\,,\quad
t_c = \Big(
            {e}^{\beta\cdot{\obser}_c-{\Lyap}_{c}{-\eigenvL}}
        \Big)^{\vol_c}
\,.
\ee{detSumCont}  
We have not encountered such sums over {\Blatt}s in solid state and
mathematical physics literature. In field theory they play a key
role,\rf{MalWit07,CGHMV12} so one could refer to them as field
theorists do, as `sums over geometries'.

Show that our zeta-function  \refeq{sptZeta2d} formulation of \spt\ chaos
applies also to spacetime continuous systems, such as \KS\ and {\NS}
PDEs.

  \item
Evaluate the {\stabexp}s of a set of unstable {\KS} (or another
spatially 1\dmn\ PDE) \lsts, test the quality of zeta function predictions.
  \item
Evaluate the {\stabexp}s of a set of unstable {\NS} \lsts.
  \item
In \refsect{s:Bravais} we have assumed that the only symmetry of the
theory is the translation group $T$. However,
one needs to quotient all spacetime and internal symmetries.
For a one\dmn\ lattice field theory we have done this in companion paper
I,\rf{LC21} and derived the dihedral space group $\Group=\Dn{\infty}$
zeta function for time-reversal invariant field theories, drawing
inspiration from Lind's topological (rather than our weighted,
`dynamical') zeta func\-tion,\rf{Lind96}
\beq
\zeta_{Lind}(z) =
\exp \Big( \sum_{H} \;
            \frac{\vol_{H}}{|\Group/{H}|}z^{|\Group/H|}
      \Big)
\,,
\ee{LindZeta}
a generalization of Artin-Mazur zeta func\-tion.
In the present context, $\Group$ is the crystallographic space
group of a field theory over a hypercubic lattice $\integers^d$, $H$ a
finite-index $|\Group/{H}|$ subgroup of $\Group$, and $\vol_{H}$ is the
number of the {\lsts} that are invariant under actions of the
subgroup $H$.

  \item
Describe the admissible {\mosaic}s of the \catlatt\ \refeq{catlattKG}.
 
    \edit{ 
For one\dmn\ \templatt\ the answer is given in terms of walks on the 
{\markGraph} of 
\toChaosBook{figure.caption.276} {Fig.~14.18\,(d)}.

In two dimensions Axenides \etal\ use {Fibonacci 
polynomials}, forward in time evolution of a fixed initial spatial 
interval to generate all admissible {\mosaic}s over 
$\BravCell{\speriod{}}{\period{}}{0}$  rectangular {\pcell}s.%
\rf{AxFlNi22,AxFlNi24,AlDuSh25}
A corresponding algorithm for general \LTS{}{}{} {\Blatt}s has not been 
implemented. 
    }
  \item
Describe the admissible {\mosaic}s of a nonlinear field theory with
pruning, \ie, with couplings weaker than those topologically equivalent
to the anti-integrable limit (see \refsect{s:mosaics} and companion paper
III\rf{WWLAFC22}).

\end{enumerate}

\begin{acknowledgments}
This paper has been profoundly influenced by 
anti-integrability viewpoint of Serge Aubry,
whose 80th birthday it celebrates.
 
The work of H.~L. was fully supported, and
of P.~C. was in part supported by the family of late G. Robinson, Jr.
This research was initiated during the KITP UC Santa Barbara 2017 {\em
Recurrent Flows: The Clockwork Behind Turbulence} program, supported in
part by the National Science Foundation under Grant No. NSF PHY-1748958.
We are grateful to  S.D.V.~Williams, M.N. Gudorf and X.~Wang for many 
inspiring discussions. 
No actual cats, graduate or undergraduate, have shown interest in, or
were harmed during this research.
\end{acknowledgments}


\appendix

\section{Bravais sublattices}
\label{s:primeLatt}

When is a two\dmn\ {\Blatt} $\lattice_{\Bcell{A}}$
a sublattice of a finer {\Blatt} $\lattice_{\Bcell{A}_p}$?
Define $\lattice_{\Bcell{A}_p}$ by a pair of primitive vectors in the 
Hermite normal form \LTS{p}{p}{p}, 
\beq
\mathbf{a}^p_{1}=\left(\begin{array}{c}
  \speriod{p}\cr
  0{}
  \end{array}\right)
  \,,\qquad
\mathbf{a}^p_{2}=\left(\begin{array}{c}
  \tilt{p}\cr
  \period{p}
  \end{array}\right)
\,.
\ee{primeTile}
The sublattices $\lattice_{\Bcell{A}}$ of $\lattice_{\Bcell{A}_p}$
 have primitive vectors that are linear combinations
of $\mathbf{a}_{1}$ and $\mathbf{a}_{2}$:
\bea
\mathbf{a}_1 &=& r_1\,\mathbf{a}^p_{1} + s_2\,\mathbf{a}^p_{2}
    \continue
\mathbf{a}_2 &=& s_1\,\mathbf{a}^p_{1} + r_2\,\mathbf{a}^p_{2}
\,,
\label{subLattVec}
\eea
where $r_1$, $r_2$, $s_1$ and $s_2$ are integers, so that every lattice 
site of the sublattice $\lattice_{\Bcell{A}}$ belongs to the {\Blatt} 
$\lattice_{\Bcell{A}_p}$. If we also choose $\lattice_{\Bcell{A}}$
primitive vectors in the 
Hermite normal form \LTS{}{}{}, the relation 
\refeq{subLattVec} can be rewritten as: 
\beq
\Bcell{A} = \Bcell{A}_p \Bcell{R} \,,
\ee{sublatticeCondition1}
where
\[
\Bcell{A} =
\left[\begin{array}{cc}
\speriod{} & \tilt{} \cr
0 & \period{}
\end{array}\right]
\,, \;
\Bcell{A}_p =
\left[\begin{array}{cc}
\speriod{p} & \tilt{p} \cr
0 & \period{p}
\end{array}\right]
\,, \;
\Bcell{R} =
\left[\begin{array}{cc}
r_1 & s_1 \cr
s_2 & r_2
\end{array}\right]
\,.
\]
Then the matrix $\Bcell{R}$ is:
\beq
\Bcell{R} =
\Bcell{A}_p^{-1} \Bcell{A} =
\left[\begin{array}{cc}
\speriod{}/\speriod{p} & \tilt{}/\speriod{p} - \tilt{p} \period{} / \speriod{p} \period{p} \cr
0 & \period{}/\period{p}
\end{array}\right]
\,.
\ee{sublatticeCondition2}
Comparing \refeq{sublatticeCondition2} with \refeq{sublatticeCondition1}, we note that
$\lattice_{\Bcell{A}}$ is a sublattice of $\lattice_{\Bcell{A}_p}$ if 
$\speriod{}$ is a multiple of $\speriod{p}$,
$\period{}$ is multiple of $\period{p}$ and
\beq
\mathbf{a}_2\times\mathbf{a}^p_{2}=\tilt{}\period{p}-\period{}\tilt{p}
\ee{primeTiling}
is a multiple of the prime tile area $\speriod{p}\period{p}$.

So, given {\Blatt} $\lattice_{\Bcell{A}_p}$ with {\pcell}
$\Bcell{A}_p$, one gets all of its sublattices 
by computing
$\Bcell{A}=\Bcell{A}_p\Bcell{R}$, 
with the repeats matrix \Bcell{R} in the Hermite normal form,
\beq
\Bcell{R} =
\left[\begin{array}{cc}
r_1 & s \cr
0 & r_2
\end{array}\right] 
\,,
\ee{sublatticeCondition3}
where $r_1, r_2 > 0$ and $0 \leq s < r_1$ are integers.


\subsection{Examples of {\primeOrbs}} 
\label{s:appePrime}

\noindent
The square lattice unit {\pcell},
\beq
\Bcell{A} =
\left[
\begin{array}{cc}
1 & 0 \cr
0 & 1
\end{array}
\right]
    \,,\qquad \vol_\Bcell{A} = 1
\,,
\ee{2DUnitCell}
\BravCell{1}{1}{0}-periodic {\fconf}, or the constant lattice field
\[
\Xx =
\left[\begin{array}{c}
\ssp_{00} \cr
\end{array}\right]
\]
is the unit cell of a square $\integers^2$ integer lattice.  \\

\noindent
\BravCell{2}{1}{0}-periodic {\fconf}
\[
\Xx =
\left[\begin{array}{cc}
\ssp_{00} & \ssp_{10} \cr
\end{array}\right]
\,,
\]
\BravCell{1}{2}{0}-periodic {\fconf}
\[
\Xx =
\left[\begin{array}{c}
\ssp_{01} \cr
\ssp_{00}
\end{array}\right]
\]
have `bricks' stacked atop each other, see
mosaics of \reffig{f:SpecialBravaisLatt}\,(a)
and (b).
\BravCell{2}{1}{1}-periodic {\fconf}
\[
\Xx =
\left[\begin{array}{cc}
\ssp_{00} & \ssp_{10} \cr
\end{array}\right]
\]
has layers of `bricks' stacked atop each other, but with a
relative-periodic boundary condition, with layers shifted by
$\tilt{}=1$, as in \reffig{f:2x1rpo}\,(a).

The boundary conditions for the above three kinds of {\pcell}s can
illustrated by repeats
of the three `bricks', on top, sideways, and on top and shifted:
\[
\BravCell{2}{1}{0}\,:\,\left[\begin{array}{cc}
\ssp_{00} & \ssp_{10} \cr
\ssp_{00} & \ssp_{10} \cr
              \end{array}\right]
\, , \quad
\BravCell{1}{2}{0}\,:\,\left[\begin{array}{cc}
\ssp_{01} & \ssp_{01} \cr
\ssp_{00} & \ssp_{00} \cr
              \end{array}\right]
\]
\[
\BravCell{2}{1}{1}\,:\,\left[\begin{array}{ccc}
  & \ssp_{00} & \ssp_{10} \cr
\ssp_{00} & \ssp_{10} &   \cr
              \end{array}\right]
\,.
\]

\noindent
$\BravCell{3}{2}{1}$-periodic {\fconf}
can be presented as a field over the \emph{parallelepiped}-shaped tilted {\pcell} of
\reffig{f:BravLatt}\,(a),
\[
\BravCell{3}{2}{1} \,:\,
 \left[
 \begin{array}{cccc}
           & \ssp_{11} & \ssp_{21} & \ssp_{01} \cr
 \ssp_{00} & \ssp_{10} & \ssp_{20} &
 \end{array}
 \right]
\,,
\]
or as an $[3\!\times\!2]$ rectangular array
\beq
\Xx=
        \left[\begin{array}{ccc}
\ssp_{01} & \ssp_{11} & \ssp_{21}   \cr
\ssp_{00} & \ssp_{10} & \ssp_{20}
              \end{array}\right]
\,,
\ee{eq:block3x2}
with the {\Blatt} relative-periodicity imposed by a shift
boundary condition, as in \reffig{f:2x1rpo}\,(b) and
the {\mosaic} of \reffig{f:SpecialBravaisLatt}\,(f).

As shown above, 
an  \LTS{}{}{} {\pcell} {{\fconf}} is not prime
if it is invariant under the translations of
lattice \LTS{p}{p}{p}, and \LTS{}{}{} is a sublattice of
\LTS{p}{p}{p}.

For example, a {\fconf} over {\pcell} $\BravCell{2}{2}{0}$,
\[
\Xx =
\left[\begin{array}{ccc}
\ssp_{10} & \ssp_{00}\cr
\ssp_{00} & \ssp_{10}
\end{array}\right]
\,.
\]
is a repeat and shift of the {\fconf}
\[
\Xx_p =
\left[\begin{array}{ccc}
\ssp_{00} & \ssp_{10} \end{array}\right]
\]
over {\pcell} $\BravCell{2}{1}{1}$.
As shown in \reffig{f:2x1rpo}\,(a), the {\Blatt} \BravCell{2}{2}{0} is a sublattice
of \BravCell{2}{1}{1}.
Over the infinite spacetime $\Xx$ and $\Xx_p$ are the same {\fconf}, as
is clear by inspection of \reffig{f:SpecialBravaisLatt}\,(c).

\medskip
For further examples of orbits and their symmetries, see companion papers
I and III.\rf{LC21,WWLAFC22}

\section{Computation of {\catlatt} \lsts}
\label{s:catlattComp}

{\ELE} \refeq{catlatt} are {piecewise linear}, and, given a {\pcell} \Bcell{A}\ and a
{\mosaic} \Mm\ \refeq{Mm_c} over it,  always has a unique solution
${\Xx_\Mm}$.
We solve it by reciprocal lattice diagonalization
(\refsect{s:rcprPcellJ}), by direct determinant evaluation
(\refappe{s:fact}), or by matrix inversion,
\beq
  \ssp_{z}=\sum_{z'\in\integers^d}g_{z z'} \Ssym{z'}, \qquad  g_{z z' }
       =\left[\frac{1}{-\Box +\mu^2}\right]_{zz'}
       \,,
\ee{GreenFunc}
where $g_{zz'}$, the inverse of the {\jacobianOp}, is the Klein-Gordon
free-field \refeq{catlattKG} {Green's function}. In the literature, $g_{z z'
}$ is known as the {Green's function} for the $d$\dmn\ discretized
{\sPe}.

The solution $\Xx_\Mm$ is a \lst, and the {\mosaic} $\Mm$ is said to be
\emph{admissible}, if and only if all lattice-site field values
$\ssp_{z}$ of $\Xx_\Mm$ lie in the compact boson {\statesp}
\refeq{catlattKG}
\beq
\pS
  =
\left\{\Xx \mid \ssp_{z} \in [0,1)\,,\; z \in \integers^d \right\}
\,.
\ee{StateSpCatLatt}
So we need to define the range of permissible integers \Ssym{z}
(`covering' alphabet), and, if we are able, the grammar of admissible
{\mosaic}s $\Mm$.

\subsection{{\catLatt}  {\mosaic}s}
\label{s:catlattCount}

`Letter'  $\Ssym{z}$ is the integer part of the LHS of
{\ELe} \refeq{catlattKG}
that enforces the circle ($\mbox{mod}\;1$) condition for field $\ssp_{z}$
on lattice site $z$.
Its range depends on the Klein-Gordon mass-squared $\mu^2$,
and the lattice dimension $d$.
If all
nearest neighbor fields are as large as allowed, $\ssp_{z'}=1-\epsilon$,
in  two \spt\ dimensions the integer part of the LHS of \refeq{CatMap2d}
can be as low as $-3$, for  $\ssp_{z}=0$, or
as high as $\mu^2\!+\!3$, for $\ssp_{z}=1-\epsilon$,
hence the covering alphabet $\A=\{\Ssym{z}\}$ is
\beq
 \A  =
 \{\underline{3},\underline{2},\underline{1}
   \,;\, 0,\cdots,\mu^2 \,;\,
   \mu^2\!+\!1, \mu^2\!+\!2, \mu^2\!+\!3
 \}
\,,
\label{catlatt2d}
\eeq
where symbol $\underline{\Ssym{}_{z}}$ denotes $\Ssym{z}$ with the
negative sign, \ie, `$\underline{3}$' stands for symbol `$-3$'.
All our numerical calculations are carried out for $\mu^2=1$, with
alphabet
\beq
 \A  =
 \{\underline{3},\underline{2},\underline{1}
   \,;\, 0, 1 \,;\, 2, 3, 4
 \}
\,.
\ee{catlatt2dmu=1}
As each $\Mm$ corresponds to a unique \lst\  $\Xx_\Mm$, the \lst\ can
be visualized by its color-coded {\mosaic} $\Mm$.

Given a two\dmn\ \spt\ {\mosaic} $\Mm$, the corresponding {\lst}
can be computed using \refeq{GreenFunc},
\[
\Xx_{i_1 j_1} =
\sum_{i_2=0}^2 \sum_{j_2=0}^1 \gd_{i_1 j_1,i_2 j_2} \Mm_{i_2 j_2}
\,,
\]
provided that the correct \bcs\ are imposed on $\gd_{i_1 j_1,i_2 j_2}$.

If all nearest neighbor fields are as small as allowed, $\ssp_{z'}=0$,
the Laplacian does not contribute, and the integer part of the LHS of
\refeq{catlattKG} ranges from $0$, for $\ssp_{z}=0$, to  $\mu^2$, for
$\ssp_{z}=1$, hence the $\mu^2+7$ letter alphabet \refeq{catlatt2d} can
be divided into two subsets, the interior and the exterior alphabets
$\A_0$ and $\A_1$, respectively.
\bea
\A_0 &=& \{
0, \dots, \mu^2
\}
\,, \continue
\A_1 &=& \{
\underline{3},\underline{2},\underline{1}
\} \cup \{
\mu^2\!+\!1, \mu^2\!+\!2, \mu^2\!+\!3
\}
\,.
\label{catlattIntAlph}
\eea
If all $\Ssym{z}$ of a {\mosaic} $\Mm$ belong to the interior
alphabet $\A_0$, the {\mosaic} $\Mm$ is admissible.\rf{GHJSC16}

We have no algorithm that would generate admissible {\catlatt} {\mosaic}s 
(see question~6 of \refsect{s:open}).  
Instead, we solve the linear equation \refeq{GreenFunc} for each covering 
{\mosaic}, and then 
-for {\mosaic}s containing exterior alphabets $\A_1$ letters- discard 
those for which ${\Xx_\Mm}$ lies outside the unit hypercube 
\refeq{StateSpCatLatt}. 

For example, for $\mu^2=1$ the {\mosaic}
\[
\Mm =
 \left[
 \begin{array}{ccc}
 -1 & 1 & 0 \cr
 4 & -1 & -1
 \end{array}
 \right]
\]
over
{\pcell} $\BravCell{3}{2}{1}$ of  \reffig{f:2x1rpo}\,(b)
corresponds to the {\lst}:
\[
\Xx_\Mm =
 \left[
 \begin{array}{ccc}
 \ssp_{01} & \ssp_{11} & \ssp_{21} \cr
 \ssp_{00} & \ssp_{10} & \ssp_{20}
 \end{array}
 \right]
 =
 \frac{1}{35}
 \left[
 \begin{array}{ccc}
 5 & 17 & 6 \cr
 34 & 5 & 3
 \end{array}
 \right] \, .
\]
One can check that {\ELe} are satisfied everywhere by substituting this 
solution into \refeq{CatMap2d}. 

\subsection{{\catLatt} {\pcell}s' {\Hilldet}s}
\label{s:catlattHillexm}    

This section provides a continuation of calculations of \refsect{s:pcellStab}. 
Developing some feel for the {\Hilldet} formulas for
two\dmn\ {\catlatt} examples is now in order.
The simplest examples of {\lsts}, illustrated by \spt\ {\mosaic} tilings of
\reffig{f:SpecialBravaisLatt}, are
(i) spacetime {\steady}s over the unit cell $\BravCell{1}{1}{0}$,
(ii) spatial  {\steady}s over $\BravCell{1}{\period{}}{0}$,
(iii) temporal  {\steady}s  over $\BravCell{\speriod{}}{1}{0}$,
and
(iv) time-relative {\steady}s  over $\BravCell{\speriod{}}{1}{\tilt{}}$,
$\tilt{}\neq0$, stationary patterns in a time-reference
frame\rf{PolTor92} moving with a constant velocity $\tilt{}/\period{}$.

For explicit values of  {\Hilldet}s, we take the lowest integer value of
the Klein-Gordon mass, $\mu^2=1$,
throughout the paper.

Consider first the family of {\pcell}s of temporal period one, $\period{}=1$
in \refeq{jacOrbEigs}, 
\beq
\Det\jMorb_{\BravCell{\speriod{}}{1}{0}}
= \mu^2 \prod_{m_1=1}^{\speriod{}-1}
     \left[ p\Big(\frac{2\pi}{\speriod{}}m_1\Big)^2 + \mu^2 \right]
 \,.
\label{2Dto1DCount}
\eeq
This is the one\dmn\ \templatt\ {\Hilldet},
with calculations carried out as in \refeq{tempLatta3}.
The \steady\ {\Hilldet} is
\beq
\Det\jMorb_{\BravCell{1}{1}{0}} =  \mu^2 \Rightarrow 1
 \,,
\ee{1x1DCount}
the period-2 \lst\ {\Hilldet} is
\beq
\Det\jMorb_{\BravCell{2}{1}{0}}=\mu^2(\mu^2+4)
\Rightarrow 5
 \,,
\ee{2x1_0Count}
and so on.
However, for the simplest relative-{\lst}, with tilt
${\tilt{}}/{{\period{}}}=1$, the {\Hilldet} \refeq{2DjMorbEigs} is
already more surprising, it is larger than
$\Det\jMorb_{\BravCell{2}{1}{0}}\Rightarrow5$:
\bea
\Det\jMorb_{\BravCell{2}{1}{1}}
&=& \mu^2
     \left[
    p\left(\pi\right)^2
  + p\left(-\pi\right)^2  + \mu^2
     \right]
     \continue
&=& \mu^2(\mu^2+8)
\Rightarrow 9
 \,.
\label{2x1_1Fourier}
\eea
The \spt\ {\catlatt} calculations then proceed as in
example \refeq{catLattaEigs3x2_1},
\[
\Det\jMorb_{\BravCell{2}{2}{0}} =
\mu^2(\mu^2+4)^2(\mu^2+8) \Rightarrow 225
\,,
\]
and so on.

For example, one can check that
the  {\Hilldet} formula \refeq{jacOrbEigs}
for the
$\BravCell{3}{2}{0}$ {\lsts},
\bea
\Det\jMorb_{\BravCell{3}{2}{0}}
  &=& \prod_{m_1=0}^2\prod_{m_2=0}^1\left[
   p\Big(\frac{2\pi}{3}m_1\Big)^2
   +  p\Big(\frac{2\pi}{2}m_2\Big)^2  + \mu^2
                              \right]
\continue
   &\Rightarrow& 5120
\,,
\label{rcpr:N3times2}
\eea
is in agreement with our alternative method of its evaluation,
the fundamental fact count \refeq{N3times2} explained below.

Consider next the {\pcell}  $\BravCell{3}{2}{1}$ of
\reffig{f:BravLatt}\,(a), \reffig{f:2x1rpo}\,(b) and \reffig{f:reciprLatt}\,(a).
We have computed the eigenvalues of its
Laplacian  in \refeq{catLattaEigs3x2_1}, so the corresponding
\Hilldet\ \refeq{eigsProd} is
\beq
\Det\jMorb_\BravCell{3}{2}{1} = \mu^2(\mu^2+4)^3(\mu^2+6)^2
 = 6125
 \,.
\ee{catLattHill3x2_1}
For a list of such two\dmn\  {\catlatt} {\Hilldet}s, see
\reftab{tab:LxTsmuSq}, and
for a list of the {\catlatt} {\Hilldet}s evaluated for $\mu^2=1$,
see \reftab{tab:LxTs=5/2}.

\begin{table*}
\caption[]{\label{tab:LxTsmuSq}     \small
The numbers of \catlatt\ {\lsts} for {\pcell}s
$\Bcell{A}=\LTS{}{}{}$ up to $\BravCell{3}{3}{2}$. Here
$N_\Bcell{A}(\mu^2)$ is the number of {\lsts},
and
$M_\Bcell{A}(\mu^2)$ is the number of {\primeOrbs}.
The Klein-Gordon mass $\mu^2$ can take only integer values.
}
\begin{center}
{\small
\begin{tabular}{lll}
\\[-16pt]
$~~~\Bcell{A}$
                         & ~~~$\vol_\Bcell{A}(\mu^2)$ & $M_\Bcell{A}(\mu^2)$         \\
\hline
$\BravCell{1}{1}{0}$    &   $\mu^2$ & $\mu^2$                                        \\
$\BravCell{2}{1}{0}$    &   $\mu^2(\mu^2+4)$ & ${\mu^2}(\mu^2+3)/{2}$                \\
$\BravCell{2}{1}{1}$  &   $\mu^2(\mu^2+8)$ & ${\mu^2}(\mu^2+7)/{2}$                  \\
$\BravCell{3}{1}{0}$    &   $\mu^2(\mu^2+3)^2$ & ${\mu^2}(\mu^2+2)(\mu^2+4)/{3}$     \\
$\BravCell{3}{1}{1}$  &   $\mu^2(\mu^2+6)^2$ & ${\mu^2}(\mu^2+5)(\mu^2+7)/{3}$       \\
$\BravCell{4}{1}{0}$    &   $\mu^2 (\mu^2+2)^2(\mu^2+4)$ & ${\mu^2}(\mu^2+1)(\mu^2+3)(\mu^2+4)/{4}$ \\
$\BravCell{4}{1}{1}$  &   $\mu^2(\mu^2+4)^2(\mu^2+8)$ & ${\mu^2}(\mu^2+3)(\mu^2+4)(\mu^2+5)/{4}$    \\
$\BravCell{4}{1}{2}$  &   $\mu^2(\mu^2+4)(\mu^2+6)^2$ & ${\mu^2}(\mu^2+4)(\mu^2+5)(\mu^2+7)/{4}$    \\
$\BravCell{4}{1}{3}$  &   $\mu^2(\mu^2+4)^2(\mu^2+8)$ & ${\mu^2}(\mu^2+3)(\mu^2+5)(\mu^2+8)/{4}$    \\
$\BravCell{5}{1}{0}$    & $\mu^2(\mu^4+5\mu^2+5)^2$ & ${\mu^2}(\mu^2+1)(\mu^2+2)(\mu^2+3)(\mu^2+4)/{5}$
                                                                                                    \\
$\BravCell{5}{1}{1}$  & $\mu^2(\mu^4+10\mu^2+23)^2$ & ${\mu^2}(\mu^2+3)(\mu^2+7)(\mu^4+10\mu^2+19)/{5}$
                                                                                                    \\
$\BravCell{2}{2}{0}$    & $\mu^2(\mu^2+4)^2(\mu^2+8)$
                        & ${\mu^2}(\mu^2+3)/2\times(\mu^4 +13\mu^2 + 38)/{2}$        \\
$\BravCell{2}{2}{1}$    & $\mu^2(\mu^2+4)(\mu^2+6)^2$
                        & ${\mu^2}(\mu^2+7)/2\times(\mu^2+4)(\mu^2+5)/{2}$           \\
$\BravCell{3}{2}{0}$    & $\mu^2(\mu^2+3)^2(\mu^2+4)(\mu^2+7)^2$
	& ${\mu^2}(\mu^2+3)(\mu^2+4)(\mu^6 + 17 \mu^4 + 91 \mu^2 + 146)/{6}$
                                                                                      \\
$\BravCell{3}{2}{1}$  & $\mu^2(\mu^2+4)^3(\mu^2+6)^2$
	& ${\mu^2}(\mu^2+3) (\mu^2+5) (\mu^6 + 16 \mu^4 + 85 \mu^2 + 151)/{6}$
                                                                                      \\
$\BravCell{3}{3}{0}$    & $\mu^2(\mu^2+3)^4(\mu^2+6)^4$
                        &                                                             \\
$\BravCell{3}{3}{1}$  & $\mu^2(\mu^2+3)^2(\mu^6+15\mu^4+72\mu^2+111)^2$
	                                                                                  \\
$\BravCell{3}{3}{2}$  & $\mu^2(\mu^2+3)^2(8{s}^3+3(\mu^2+4)^2-1)^2$
\\[-18pt]
\end{tabular}
} 
\end{center}
\end{table*}

\begin{table}
\caption[]{\label{tab:LxTs=5/2}    \small
The numbers of the 
$\mu^2=1$ \catlatt\
$\LTS{}{}{}$ {\lsts}:
$N_{\LTS{}{}{}}$ is the number of {\lsts}, and
$M_{\LTS{}{}{}}$ is the number of {\primeOrbs}.
}
\begin{center}
{\small
\begin{tabular}{lrlr}
\\[-22pt]
$\LTS{}{}{}$
                  & $M$ 
                         & $N$ 
                                                     \\
\hline
$\BravCell{1}{1}{0}$  &   1  &   1                   \\
$\BravCell{2}{1}{0}$  &   2  &
  $5~=\;\;\,2\,\BravCell{2}{1}{0}+1\,\BravCell{1}{1}{0}$
                                                     \\
$\BravCell{2}{1}{1}$&   4  & 9~\;=\;\;\,$
                           {4}\,\BravCell{2}{1}{1}
                         + {1}\,\BravCell{1}{1}{0}$
                                                     \\
$\BravCell{3}{1}{0}$  &   5  & 16 =~ ${5}\,\BravCell{3}{1}{0}
                         + {1}\,\BravCell{1}{1}{0}$
                                                     \\
$\BravCell{3}{1}{1}$&   16 & $49 =16\,\BravCell{3}{1}{1}
                         + {1}\,\BravCell{1}{1}{0}$
                                                     \\
$\BravCell{4}{1}{0}$  &  10 & $45 =10\,\BravCell{4}{1}{0} 
                         + {2}\,\BravCell{2}{1}{0}
                         + {1}\,\BravCell{1}{1}{0}$
                                                    \\
$\BravCell{4}{1}{1}$&   54 & $225 =54\,\BravCell{4}{1}{1} 
			+{4}\,\BravCell{2}{1}{1}
                         + {1}\,\BravCell{1}{1}{0}$
                                                     \\
$\BravCell{4}{1}{2}$&   60 & $245 =60\,\BravCell{4}{1}{2} 
                         + {2}\,\BravCell{2}{1}{0}
                         + {1}\,\BravCell{1}{1}{0}$
                                                     \\
$\BravCell{2}{2}{0}$  &   52  & $225 =
               {52}\,\BravCell{2}{2}{0}
             + {2}\,\BravCell{2}{1}{0}
             + {2}\,\BravCell{1}{2}{0}$
                                                    \\
                      &       & $\quad\quad\;\;
             + {4}\,\BravCell{2}{1}{1}
             + {1}\,\BravCell{1}{1}{0}$
                                                    \\
$\BravCell{2}{2}{1}$&  60 & $245 = 60\,\BravCell{2}{2}{1}  
			+ {2}\,\BravCell{1}{2}{0}
                         + {1}\,\BravCell{1}{1}{0}$
                                                    \\
$\BravCell{3}{2}{0}$  & 850  &
                          $ 5\,120 =850\,\BravCell{3}{2}{0}
                          +5\,\BravCell{3}{1}{0}$
                                                     \\
                      &       & $\qquad\quad\;\,
                          +2\,\BravCell{1}{2}{0}
                          +1\,\BravCell{1}{1}{0}$
                                                     \\
$\BravCell{3}{2}{1}$& 1\,012 &
                           $ 6\,125 =1\,012\,\BravCell{3}{2}{1} 
                           +16\,\BravCell{3}{1}{2}$
                                                     \\
                      &       & $\qquad\quad~~
                           +2\,\BravCell{1}{2}{0}
                           +~\,1\,\BravCell{1}{1}{0}$
                                                     \\
$\BravCell{3}{3}{0}$  & 68\,281 &

                         $ 614\,656 = 68\,281\,\BravCell{3}{3}{0}
                         +  5\,\BravCell{3}{1}{0}$  
                                                     \\
                      &       & $\,
                         + 16\,\BravCell{3}{1}{1} +16 \,\BravCell{3}{1}{2}
                         +  5\,\BravCell{1}{3}{0}
                         + {1}\,\BravCell{1}{1}{0}$
                                                     \\
$\BravCell{3}{3}{1}$& 70\,400  &
                         $ 633\,616 =70\,400\,\BravCell{3}{3}{1}
                         + {5}\,\BravCell{1}{3}{0} 
                         + {1}\,\BravCell{1}{1}{0}$
\\[-22pt]
\end{tabular}
} 
\end{center}
\end{table}

\subsection{{\catLatt}: Fundamental fact} 
\label{s:fact}

As shown in the companion paper I,\rf{LC21} for one\dmn\ lattice
{\templatt} {\Hilldet}s count the numbers of period-\cl{} {\lsts},
\beq
N_\cl{} = |\Det\jMorb_\cl{}|
\,.
\ee{fundFact}
We now show that for a {\catlatt}, {\Hilldet} counts the number of
{\lsts} in any {\spt} dimension $d$.

{\catLatt} {\lst}  $\Xx_\Mm$ over {\pcell} $\Bcell{A}$ is a point within
the unit hypercube $[0, 1)^{\vol_\Bcell{A}}$, where $\vol_\Bcell{A}$ is
the {\pcell}  volume \refeq{lattVol}.
Visualize now what {\catlatt} defining equation \refeq{catlatt}
\[
\jMorb_{\Bcell{A}}\Xx_\Mm-\Mm = 0 
\]    
means geometrically.
The
$[\vol_\Bcell{A}\!\times\!\vol_\Bcell{A}]$ {\jacobianOrb} $\jMorb_{\Bcell{A}}$
stretches the \statesp\ unit hypercube $\Xx\in[0,1)^{\vol_\Bcell{A}}$ into an
$\vol_\Bcell{A}$\dmn\ {\em \fundPip} (or a parallelogram), and maps the 
{\lst} ${\Xx}_{\Mm}$ into a point on integer lattice 
$\integers^{\vol_\Bcell{A}}$ within it, in the $\vol_\Bcell{A}$\dmn\ 
configuration {\statesp} \refeq{torusStatesp}. 
This point is then translated by integer winding numbers $\Mm$ into the 
origin. What Baake \etal\rf{BaHePl97} call the \emph{`fundamental fact'} 
follows:\rf{CvitanovicYT02} 
    \ifsubmission\else
    %
\toVideo{youtube.com/embed/Ztt1v8uGCUE} 
    \fi
\beq
N_{\Bcell{A}} = |\Det\jMorb_{\Bcell{A}}|
\,,
\ee{fundFact2d}
the number of {\lsts} equals the number of integer lattice points within
the {\fundPip}.

For the history of `fundamental fact' see \emph{Appendix A. Historical context}
of the companion paper I.\rf{LC21}
The reader might also want to check the figures of a few {\fundPip}s there, but we
know of no good way of presenting them visually for {\pcell}s of interest
here, with $\vol_\Bcell{A}>3$.

It is a peculiarity of the {\catlatt} that it involves two
\emph{distinct} integer lattices.
(i)
The spacetime \emph{coordinates}  \refeq{LattField} are discretized
by integer lattice $\integers^d$. The {\pcell} $\Bcell{A}$
\refeq{BravLatt} is an example of a {\fundPip}, and we use the
fundamental fact when we express the volume \refeq{lattVol} of the
{\pcell}, \ie\ the determinant of the matrix $\Bcell{A}$, as the number
of lattice sites within the {\pcell}.
(ii)
For a {\catlatt}, the  lattice site  \emph{field}  $\ssp_{z}$
\refeq{catlattKG} is compactified to the unit circle $[0,1)$, imparting
an integer lattice  structure to the configuration  \emph{\statesp}
\refeq{torusStatesp}:
the {\jacobianOrb} $\jMorb_{\Bcell{A}}$ maps a {\lst} $\Xx_\Mm \in [0,
1)^{\vol_\Bcell{A}}$ to a $\integers^{\vol_\Bcell{A}}$ integer lattice site
{\Mm}.
Nothing like that, and no `fundamental fact' applies to general nonlinear
field theories of \refsect{s:sptFT}.

\medskip

\paragraph*{Example: Fundamental parallelepiped evaluation
           of a {\Hilldet}.}
\label{s:catlattRel3x2}
As a concrete example, consider {\lsts} of two\dmn\ {\catlatt} with
periodicity $\BravCell{3}{2}{0}$, \ie, space period $\speriod{} = 3$,
time period $\period{} = 2$, and tilt $\tilt{} = 0$. {\Lsts} within the
{\pcell} and their corresponding {\mosaic}s can be written as
two\dmn\ $[3\!\times\!2]$ arrays,
\bea
\Xx_{\BravCell{3}{2}{0}} &=&
 \left[\begin{array}{ccc}
 \ssp_{01} & \ssp_{11} & \ssp_{21} \cr
 \ssp_{00} & \ssp_{10} & \ssp_{20}
 \end{array}\right]
\,, \continue
\Mm_{\BravCell{3}{2}{0}} &=&
 \left[\begin{array}{ccc}
 \Ssym{01} & \Ssym{11} & \Ssym{21} \cr
 \Ssym{00} & \Ssym{10} & \Ssym{20}
 \end{array}\right]
\,.
\eea
We reshape the {\lsts} and {\mosaic}s into vectors,
\beq
\Xx_{\BravCell{3}{2}{0}} =
\left(\begin{array}{c}
 \ssp_{01} \cr
 \ssp_{00} \cr
  \hline
 \ssp_{11} \cr
 \ssp_{10} \cr
  \hline
 \ssp_{21} \cr
 \ssp_{20} \cr
      \end{array}\right)
\,,\quad
\Mm_{\BravCell{3}{2}{0}} =
\left(\begin{array}{c}
 \Ssym{01} \cr
 \Ssym{00} \cr
  \hline
 \Ssym{11} \cr
 \Ssym{10} \cr
  \hline
 \Ssym{21} \cr
 \Ssym{20} \cr
        \end{array}\right)
\,.
\ee{3times2blockVect}
The reshaped {\jacobianOrb} acting on these {\lsts} is a block matrix,
\beq
\jMorb_{\BravCell{3}{2}{0}} =
\left(
\begin{array}{cc|cc|cc}
 2 s & -2 & -1 & 0 & -1 & 0  \cr
 -2 & 2 s & 0 & -1 & 0 & -1  \cr
  \hline
 -1 & 0 & 2 s & -2 & -1 & 0  \cr
 0 & -1 & -2 & 2 s & 0 & -1  \cr
  \hline
 -1 & 0 & -1 & 0 & 2 s & -2  \cr
 0 & -1 & 0 & -1 & -2 & 2 s
\end{array}
\right)
\,.
\ee{3times2blockMat}
where the stretching factor $2s = 4+ \mu^2$.
The {\fundPip} generated by the action of {\jacobianOrb}
$\jMorb_{\BravCell{3}{2}{0}}$ on the \statesp\ unit hypercube
\refeq{catlattKG} is spanned by six primitive
vectors, the columns of the {\jacobianOrb}
\refeq{3times2blockMat}.
The `fundamental fact' now expresses the {\Hilldet}, \ie, the number of
{\lsts} within the {\fundPip}, as a polynomial
of order $\vol_\Bcell{A}$ in the
Klein-Gordon mass $\mu^2$ \refeq{stretchCatlatt},
\bea
N_{\BravCell{3}{2}{0}} &=& |\Det\jMorb_{\BravCell{3}{2}{0}}|
\continue
                   &=& \mu^2(\mu^2+3)^2(\mu^2+4)(\mu^2+7)^2
\,,
\label{N3times2}
\eea
without recourse to any explicit diagonalization, such as the reciprocal
lattice diagonalization \refeq{jacOrbEigs}. 
For $\mu^2=1$, this agrees with the reciprocal lattice evaluation
\refeq{rcpr:N3times2}.
For a list of the numbers
of {\catlatt} {\lsts} for {\pcell}s $\LTS{}{}{}$ up to
$\BravCell{3}{3}{2}$, see \reftab{tab:LxTsmuSq}.

\medskip

For the 
$\mu^2=1$ {\catlatt}
the pruning turns out to be very severe.
Only 52 of the prime
$\BravCell{2}{2}{0}$ 
{\mosaic}s are {\admissible}. As for the repeats of smaller {\mosaic}s,
there are 2 {\admissible} $\BravCell{1}{2}{0}$ {\mosaic}s repeating in time and 2
$\BravCell{2}{1}{0}$ {\mosaic}s repeating in space. There are 4 {\admissible}
$1/2$-shift periodic boundary $\BravCell{1}{2}{0}$ {\mosaic}s. And there is 1
admissible {\mosaic} which is a repeat of letter 0.
The total number of $\BravCell{2}{2}{0}$ of {\lsts} is obtained by all cyclic
permutations of \admissible\ prime {\mosaic}s,
\bea
N_{\BravCell{2}{2}{0}}
    &=&       {52}\,\BravCell{2}{2}{0}
             + {2}\,\BravCell{2}{1}{0}
             + {2}\,\BravCell{1}{2}{0}
\continue
&&        + {4}\,\BravCell{2}{1}{1}
             + {1}\,\BravCell{1}{1}{0}
     \,=\, 225
\,,
\label{[2x2]count}
\eea
summarized in \reftab{tab:LxTs=5/2}. This explicit list of \admissible\
{\primeOrbs} verifies the  {\Hilldet} formula \refeq{jacOrbEigs}. 

\subsection{Prime lattice {\fconfs}} 
\label{s:prime}




Here we show how to enumerate the total numbers of distinct periodic states
in terms of {\primeOrbs}.

The enumeration of {\catlatt} doubly {\lsts} 
proceeds in three steps:
\begin{enumerate}
  \item
Construct a hierarchy of two\dmn\ {\Blatt}s $\lattice_\Bcell{A}$, starting with
the smallest {\pcell}s, list {\Blatt}s by
increasing \LTS{}{}{}, one per each set related by translation symmetries
\refeq{dDimTranslGrp} (here we are ignoring discrete point group \Dn{4}).
  \item
For each $\lattice_\Bcell{A} = \LTS{}{}{}$ {\Blatt}, compute $N_\Bcell{A}$, the
number of doubly-periodic {\catlatt} {\lsts}, using the
`fundamental fact' $N_\Bcell{A}=\left|\Det\jMorb_\Bcell{A}\right|$.
  \item
We have defined the \emph{\primeOrb} in  \refsect{s:orbits}.
\end{enumerate}

The total number of (doubly) periodic
{\mosaic}s is the sum of all cyclic permutations of prime {\mosaic}s,
\[
N_\Bcell{A}
=
\sum_{\Bcell{A}_p|\Bcell{A}} M_{\Bcell{A}_p}\,\LTS{p}{p}{p}
\]
where the sum goes over every lattice $\lattice_{\Bcell{A}_p} = \LTS{p}{p}{p}$
which contains $\LTS{}{}{}$.

Given the number of {\lsts}, the number of $\Bcell{A}=\LTS{}{}{}$-periodic
prime orbits is computed recursively,
\beq
M_{\Bcell{A}}\,=\,\frac{1}{\speriod{}\period{}}
  \left( N_{\Bcell{A}}
          - \sum _{\Bcell{A}_p|\Bcell{A}}^{\speriod{p}\period{p}<\speriod{}\period{}}
            \speriod{p}\period{p}
                                          \, M_{\Bcell{A}_p}
  \right)
\,.
\ee{primeCount2D}

\subsection{Example: `Escape rate' of \templatt}
\label{s:tempLattEscape}

\begin{quote}
The topological zeta function of {\templatt} is,\rf{Isola90,LC21}
\bea
1/\zeta_{AM}(z) &=&
     \exp\left(- \sum_{\cl{}=1}^\infty
\frac{N_\cl{}}{\cl{}} z^\cl{}
         \right)
         \continue
         &=&
\frac{1 - (\mu^2 + 2) z + z^2} 
     {(1 - z)^2}
\,,
\label{tempCatTopoZeta}
\eea
where $\vol_{\cl{}}$ is the number of {\lsts} with
period $\cl{}$.
Due to the uniform stretching factor $\mu^2 + 2$,
the {\detZeta} of {\templatt} has the same form, up to rescaling,
\bea
&& 1/\zeta[0,{z}] =
\exp\left(- \sum_{\cl{}=1}^\infty
\frac{N_\cl{}}{\cl{}}  t^\cl{}
         \right)
         =
1/\zeta_{AM}(t)
\,,
\continue
&& t = \frac{z}{\ExpaEig}
         \,,
\label{tempCatDZeta}
\eea
where $\ExpaEig$ is the stability multiplier%
\beq
\ExpaEig = e^{\Lyap} =
         \frac{1}{2}\left(\mu^2 + 2 + \mu\sqrt{\mu^2 + 4}\right) \,.
\ee{tempCatMult}
Solving for the roots of $1/\zeta[0,{z}] = 0$, we have:
\beq
t = \ExpaEig^{\pm 1}
\to
z = 1 \mbox{ or } \ExpaEig^2 \,.
\ee{tempCatMultRoots}
The leading root is 1, so the `escape rate' is 0.
The Fredholm determinant\rf{CBdet} of {\templatt} is
\bea
F(0, z) &=& \exp\left(- \sum_{\cl{}=1}^\infty
\frac{N_\cl{} z^\cl{}}{\cl{} \left| \Det \jMorb_\cl{} \right|}
         \right) \continue
&=&
\exp\left(- \sum_{\cl{}=1}^\infty
\frac{z^\cl{}}{\cl{}}
         \right) \continue
&=&
1 - z
\,,
\label{tempCatEscRate}
\eea
where we have used the `fundamental fact' \refeq{fundFact}.
The `escape rate' is again 0, as it should be--the cat map is by construction
probability conserving.
\end{quote}

\subsection{Example:
            Expectation value of the {\stabexp} of a {\templatt}.}
\label{s:HillExpPcell}

To compute the expectation value of the {\stabexp},
take the logarithm of {\lst}'s {\pcell}
stability as the Birkhoff sum $A$, \refeq{BirkhoffSum}, {\stabexp}
observable, and compute the corresponding {\detZeta},
\beq
1/\zeta[\beta, z]
= \exp \left( - \sum_{n = 1}^\infty \frac{N_n}{n}
\frac{\exp(\beta \ln |\Det \jMorb_n|) z^n}{\ExpaEig^n}\right)
\,,
\ee{tempCatZetaStab1}
where $|\Det \jMorb_n|$ is the {\pcell} stability of period-$n$ {\lsts},
and $\ExpaEig$ is the stability multiplier, which is related to the
{\stabexp} by \refeq{tempCatMult}. Note that the number of
$n$-{\lsts} is given by the {\pcell} stability,\rf{Isola90,LC21}
\beq
N_n = |\Det \jMorb_n| = \ExpaEig^{n} + \ExpaEig^{-n} - 2 \,.
\ee{Isola90Nn}
Using \refeq{expctObserW},
the expectation value of the {\stabexp} is
\bea
\expct{\Lyap} &=&
\frac{\expct{\Obser}_\zeta}{\expct{\vol}_\zeta}
=
\left.
\frac{\partial \zeta[\beta, z]}{\partial \beta}
\middle/
\edit{z} \frac{\partial \zeta[\beta, z]}{\partial z} 
\right|_{\beta = 0, z = z(0)}
\continue
&=&
\left.
\frac{\partial \ln \zeta[\beta, z]}{\partial \beta}
\middle/ z 
\frac{\partial \ln \zeta[\beta, z]}{\partial z}
\right|_{\beta = 0, z = 1}
\,.
\label{tempCatZetaStab2}
\eea
The numerator of \refeq{tempCatZetaStab2} is
\beq
\sum_{n = 1}^\infty \frac{(\ExpaEig^{n} + \ExpaEig^{-n} - 2)
\ln(\ExpaEig^{n} + \ExpaEig^{-n} - 2)}{n \ExpaEig^{n}}
\,,
\ee{tempCatZetaStabNu}
and the denominator is
\beq
\sum_{n = 1}^\infty \frac{\ExpaEig^{n} + \ExpaEig^{-n} - 2}{\ExpaEig^{n}}
\,.
\ee{tempCatZetaStabDe}
Both the numerator and the denominator of \refeq{tempCatZetaStab2}
diverge to infinity. Using the Stolz-Ces{\`a}ro theorem,\rf{Muresan09} the
ratio of \refeq{tempCatZetaStabNu} and \refeq{tempCatZetaStabDe} equals:
\bea
\expct{\Lyap}
&=& \lim_{n \to \infty}
\frac{\ln(\ExpaEig^{n} + \ExpaEig^{-n} - 2)}{n}
\continue
&=& \ln \ExpaEig = \Lyap \,,
\label{tempCatZetaStab3}
\eea
which agrees with the fact that every {\lst} has 
the same {\stabexp} $\Lyap$.

\section{Spectra of \jacobianOps\ for nonlinear field theories}
\label{s:phi3}

The simplicity of the \catlatt\ {\jacobianOp} band spectrum
\refeq{stabExp-d}, 
plotted in \reffig{f:recipCatCn}\,(a) and
\reffig{f:2dCatLattBand}\,(a), is a bit misleading.
As explained in \refsect{s:BlattJ}, the uniform stretching factor
describes only the stability of a {\steady} solution, for any field
theory.
To get a feeling for the general case, in Sect.~10 of paper I\rf{LC21}
we compute the stability of a period-2 \lst\  for two
nonlinear field theories. Here we outline such calculations, to illustrate
the essential difference between the very special \catlatt\ case, and the
general, nonlinear case. 
For a detailed exposition, see companion paper III,\rf{WWLAFC22} where 
we evaluate stabilities of large sets of nonlinear field theories' {\lsts}. 

An analytic eigenvalue formula is feasible only for the period-2 {\lst}; 
in general, {\lsts} and the associated \jacobianOp\ spectra are evaluated 
numerically. The simplest non-constant solutions, a period-2 {\lsts}, 
suffice to illustrate the general case. 

\subsection{One\dmn\ $\phi^3$ field theory period-2 {\lst}}
\label{s:1dNonlin}

Consider the $\phi^3$ theory, \refeq{sptHenlatt},
\[
- \Box\,\ssp_{z}
\;+\;  \mu^2\,({1}/{4}-\ssp_{z}^2)
    =
0
\,.
\] 
In one \spt\ dimension, this field theory is a temporal lattice
reformulation of the forward-in-time {\Henon} map, where large numbers of
periodic solutions can be easily computed\rf{orbithunter}.
   \edit{ 
The $\phi^3$ theory, 
to which 
companion paper III\rf{WWLAFC22}  
assigns binary alphabet
\(
 \A  = \{0,1\}
\,,
\)
 \refeq{alphabet}, has one period-2  {\primeOrb} $\{\Xx_{01},\Xx_{10}\}$, 
with the 2-lattice site
{\lst}, \mosaic
    }
\beq
\Xx_{01}
=
\left[\begin{array}{c}
{\ssp_{0}} \cr
{\ssp_{1}}
\end{array}\right]
=
\left[\begin{array}{c}
\overline{\ssp} - \sqrt{\frac{1}{4}-\overline{\ssp}^2} \cr
\overline{\ssp} + \sqrt{\frac{1}{4}-\overline{\ssp}^2}
\end{array}\right]
    ,\quad
\Mm     =
\begin{array}{|c|c|}
  \hline
0 & 1 \cr
  \hline
\end{array}
\,,
\ee{henLstLR} 
where
$\overline{\ssp}=
(\ssp_{0}+\ssp_{1})/2=2/\mu^2$
is the {Birkhoff average} \refeq{BirkhoffSum} of the field
$\ssp_\zeit$.
In the anti-integrable limit \refeq{antiIntSptTemplatt} the lattice site 
field values tend to parabola 
\(
{1}/{4}-\ssp_{z}^2=0
\)
\steady\ values 
\(
\left[{\ssp_{0}},{\ssp_{1}}\right]
\,\to\,
\left[-1/2,1/2\right].
\)

The Bloch theorem \refeq{BlochThe} yields two eigenstate bands,
\beq
    \edit{ 
\ExpaEig_{\pm}(\wavenum)
     } 
                  = -2 \pm \sqrt{\mu^4 - 12 - p(2k)^2}
\,,
\ee{HenSpectPer2}
plotted  in
\reffig{f:recipCatCn}\,(b), in the $k\in(-\pi/2,\pi/2]$
{Brillouin} zone for $\mu^2 = 3.5$.
For a finite {\pcell} of even period, tiled by
$r$th repeat of the period-2 {\lst} $\Xx_p$, the eigenvalues
of its {\jacobianOrb} are $\ExpaEig_{01,\pm}(k)$ 
evaluated at $k$ restricted to a discrete set of wave vectors
$k$, multiples of $\pi/r$: an example is worked out in 
\refsect{s:pcellStabNonlin}, with
third and fourth repeats plotted in \reffig{f:recipCatCn}\,(b).

\subsection{Two\dmn\ $\phi^4$ field theory \BravCell{2}{1}{0} {\lst}}
\label{s:2dNonlin}

   \edit{ 
The {$\phi^4$} {\spt} lattice field theory, \refeq{sptPhi4},
\[
- \Box\,\ssp_{z} \;+\; \mu^2(\ssp_{z}-\,\ssp_{z}^3) = 0
\,,
\]
to which companion paper III\rf{WWLAFC22}  assigns alphabet 
\refeq{alphabet} 
\(
 \A = \{-{1},0,1\}
\,,
\)
has at most three {\steady}s. 
The two spacetime dimensions {$\phi^4$} has at most three period-2 
{\primeOrbs} of periodicity \BravCell{2}{1}{0}, with mosaics 
\(
\begin{array}{|c|c|}
  \hline
-1 & 0 \cr
  \hline
\end{array}\,,
\)
\(
\begin{array}{|c|c|}
  \hline
-1 & 1 \cr
  \hline
\end{array}\,,
\)
and
\(
\begin{array}{|c|c|}
  \hline
0 & 1 \cr
  \hline
\end{array}\,.
\)
For example, for Klein-Gordon mass-squared $\mu^2 = 5$, 
one of the period-2 {\primeOrbs} is 
    }
\beq
\Xx_{01}
=
\left[\begin{array}{c}
\sqrt{\frac{ 7 - \sqrt{33}}{10}}
\cr
\sqrt{\frac{ 7 + \sqrt{33}}{10}}
\end{array}\right]
    \,,\quad
\Mm     =
\begin{array}{|c|c|}
  \hline
0 & 1 \cr
  \hline
\end{array}
\,.
\ee{phi4Lst210}
The {\jacobianOp} has two Bloch bands,
   \edit{ 
\beq
\ExpaEig_{\pm}(\wavenum) = 
- \frac{7}{2} +  p(k_2)^2 \pm \sqrt{\frac{313}{4} - p(2k_1)^2}
\ee{phi4Lst210bands}
plotted in
\reffig{f:2dCatLattBand}\,(b). While the period-2 {\lst} \BravCell{2}{1}{0} is the same in one and two 
\spt\ dimensions, its stability, \refeq{phi4Lst210bands}, is evaluated in 
two dimensions, with transverse $k_2$, temporal direction eigenstates  
included. 
    }
For any finite {\pcell} tiled by repeats
of the {\primeOrb} $\Xx_p$, eigenstates of the {\jacobianOrb} have a
discrete set of wave vectors $k$.
As an example, eigenvalues of a \BravCell{6}{4}{0} {\lst} tiled by 12
repeats of $\Xx_p$ have wave vectors $k$ marked by black dots in
\reffig{f:2dCatLattBand}\,(b).

\section*{References}
\bibliography{../bibtex/siminos}

\end{document}

%% file: defsKittens.tex
\renewcommand{\mod}{\mbox{\rm mod}\,}

\newcommand{\fundPip}{fundamental parallelepiped}

\newcommand{\pcell}{primitive cell}              
\newcommand{\Pcell}{Primitive cell}
\newcommand{\Bcell}[1]{\ensuremath{\mathbb{#1}}} 
\newcommand{\Blatt}{Bravais lattice}
\newcommand{\stabexp}{stability exponent}

\newcommand{\SLn}[2]{\ensuremath{\textrm{SL}(#1,#2)}} 
\newcommand{\lattice}{\ensuremath{\mathcal{L}}}
\newcommand{\vol}{{N}}                          

 \newcommand{\lst}{periodic state}
 \newcommand{\Lst}{Periodic state}
 \newcommand{\lsts}{periodic states}
 \newcommand{\Lsts}{Periodic states}
 \newcommand{\primeOrb}{prime orbit}
 \newcommand{\PrimeOrb}{Prime orbit}
 \newcommand{\primeOrbs}{prime orbits}
 \newcommand{\PrimeOrbs}{Prime orbits}
 \newcommand{\fconf}{field configuration}
 \newcommand{\fconfs}{field configurations}
 
 \newcommand{\Fconfs}{Field configurations}
\newcommand{\templatt}{temporal cat}     
\newcommand{\spt}{spa\-tio\-temp\-or\-al}
\newcommand{\Spt}{Spa\-tio\-temp\-or\-al}
\newcommand{\catlatt}{spa\-tio\-temp\-or\-al cat}     
\newcommand{\catLatt}{Spa\-tio\-temp\-or\-al cat}     
\newcommand{\Henon}{H\'enon}

\newcommand{\bcs}{boundary conditions}

\newcommand{\sPe}{screened Poisson equation}

\newcommand{\jMps}{\ensuremath{J}}   
\newcommand{\jMorb}{{\ensuremath{\mathcal{\jMps}}}}
\newcommand{\diag}{d}                  
\newcommand{\jMat}{\mathbb{\jMps}}

\newcommand{\ssp}{\ensuremath{\phi}}             
\newcommand{\Mm}{\ensuremath{\mathsf{M}}}
\newcommand{\Xx}{\ensuremath{\mathsf{\Phi}}}      
\newcommand{\Source}{\mathsf{J}}

\newcommand{\A}{\ensuremath{\mathcal{A}}}       

\newcommand{\gd}{\mathsf{g}}

\newcommand{\m}{\ensuremath{m}}     
\newcommand{\reals}{\mathbb{R}}

\newcommand{\integers}{\ensuremath{\mathbb{Z}}}

\renewcommand{\det}{\mathrm{det}\,}
\newcommand{\Det}{\mathrm{Det}\,}

\newcommand{\Tr}{\mathrm{Tr}\,}

\newcommand{\pde}{\partial}

\newcommand{\mosaic}{mosaic}     
\newcommand{\Mosaic}{Mosaic}     
\newcommand{\jacobianOrb}{orbit Jacobian matrix}

\newcommand{\jacobianOrbs}{orbit Jacobian matrices}

\newcommand{\jacobianOp}{orbit Jacobian operator}

\newcommand{\jacobianOps}{orbit Jacobian operators}

\newcommand{\Hilldet}{orbit Jacobian}

\newcommand{\ecs}{exact coherent structure}

\newcommand{\HREF}[2]{{\href{#1}{#2}}}

\ifboyscout 
  \newcommand{\PC}[2]{\begin{quote}{\color{red}[#1 Predrag] \small #2}\end{quote}}
  
  \newcommand{\HL}[2]
                     {\begin{quote}\HLedit{[#1 Han] \small #2}\end{quote}}
  \newcommand{\HLedit}[1]{{\color{blue}#1}}
  \newcommand{\SVW}[2]{\begin{quote}{\color{green}[#1 Sidney] \small #2}\end{quote}} 
  \newcommand{\SVWedit}[1]{{\color{green}#1}}
  \newcommand{\BG}[2]
                     {\begin{quote}\BGedit{[#1 Boris] \small #2}\end{quote}}
  \newcommand{\BGedit}[1]{{\color{red}#1}}
  \newcommand{\AKS}[2]{$\footnotemark\footnotetext{Adrien #1: #2}$}
  
  \newcommand{\RJ}[2]
                     {\begin{quote}\RJedit{[#1 Rana] \small #2}\end{quote}}
  \newcommand{\RJedit}[1]{{\color{green}#1}}

  \newcommand{\LH}[2]
                     {\begin{quote}\LHedit{[#1 Li] \small #2}\end{quote}}
  \newcommand{\LHedit}[1]{{\color{green}#1}}
  \newcommand{\Private}[1]{{\color{blue}#1}}
  \newcommand{\toCB}{\marginpar{\footnotesize 2CB}}  
  \newcommand{\CBlibrary}[1]
             {\href{https://ChaosBook.org/library/#1.pdf} { (click here)}}
\else 
  \newcommand{\PC}[2]{}{}
  
  \newcommand{\HL}[2]{}
  \newcommand{\HLedit}[1]{#1}
  \newcommand{\SVW}[2]{}
  \newcommand{\SVWedit}[1]{}
  \newcommand{\BG}[2]{}{}
  \newcommand{\BGedit}[1]{#1}
  \newcommand{\AKS}[2]{}{}
  
  \newcommand{\RJ}[2]{}{}
  \newcommand{\RJedit}[1]{#1}
  \newcommand{\LH}[2]{}{}
  \newcommand{\LHedit}[1]{#1}
  \newcommand{\Private}[1]{}
  \newcommand{\toCB}{}
  
  \newcommand{\CBlibrary}[1]{}
\fi  

\ifsubmission 
    \newcommand{\edit}[1]{{\color{blue}#1}} 
    \newcommand{\new}[1]{{\color{red}#1}} 
\else  
    \newcommand{\edit}[1]{#1}
    \newcommand{\new}[1]{#1} 
\fi

\newsavebox{\bartName}

{\hspace*{\fill}\nolinebreak[1]\usebox{\bartName}\vspace*{1ex}\end{minipage}}

\newcommand{\beq}{\begin{equation}}
\newcommand{\continue}{\nonumber \\ }
\newcommand{\nnu}{\nonumber}
\newcommand{\eeq}{\end{equation}}
\newcommand{\ee}[1] {\label{#1} \end{equation}}
\newcommand{\bea}{\begin{eqnarray}}

\newcommand{\eea}{\end{eqnarray}}
\newcommand{\barr}{\begin{array}}
\newcommand{\earr}{\end{array}}

\newcommand{\rf}     [1] {~\cite{#1}}
\newcommand{\refref} [1] {Ref.~\onlinecite{#1}}

\newcommand{\refrefs}[1] {Refs.~\onlinecite{#1}}

\newcommand{\refeq}  [1] {Eq.~(\ref{#1})}
\newcommand{\refeqs} [2]{Eqs.~(\ref{#1}--\ref{#2})}
\newcommand{\reffig} [1] {Fig.~\ref{#1}}
\newcommand{\reffigs} [2] {Figs.~\ref{#1} and~\ref{#2}}
\newcommand{\refFig} [1] {Figure~\ref{#1}}

\newcommand{\reftab} [1] {Table~\ref{#1}}

\newcommand{\reftabs}[2] {Tables~\ref{#1} and~\ref{#2}}
\newcommand{\refsect}[1] {Sec.~\ref{#1}}
\newcommand{\refsects}[2] {sections~\ref{#1} and \ref{#2}}
\newcommand{\refSect}[1] {Sec.~\ref{#1}}

\newcommand{\refappe}[1] {Appendix~\ref{#1}}

\newcommand{\period}[1]{{\ensuremath{T_{#1}}}}         
\newcommand{\cl}[1]{{\ensuremath{n_{#1}}}}   
\newcommand{\speriod}[1]{{\ensuremath{L_{#1}}}}  
\newcommand{\tilt}[1]{{\ensuremath{S_{#1}}}}  
\newcommand{\LTS}[3]{{\ensuremath{[\speriod{#1}\!\times\!\period{#2}]_\tilt{#3}}}}
\newcommand{\BravCell}[3]{{\ensuremath{[#1\!\times\!#2]_{#3}}}}

\newcommand{\statesp}{state space}

\newcommand{\stateDsp}{state-space}

\newcommand{\escape}{re\-ject ra\-te}  
\newcommand{\eigenvL}{\ensuremath{s}} 
\newcommand{\dzeta}{dyn\-am\-ic\-al zeta func\-tion}

\newcommand{\dft}{det\-er\-min\-is\-tic field theory}
\newcommand{\Dft}{Det\-er\-min\-is\-tic field theory}

\newcommand{\detZeta}{det\-er\-min\-is\-tic zeta func\-tion}
\newcommand{\DetZeta}{Det\-er\-min\-is\-tic zeta func\-tion}
\newcommand{\detSum}{det\-er\-min\-is\-tic part\-it\-ion sum}
\newcommand{\DetSum}{Det\-er\-min\-is\-tic part\-it\-ion sum}

\newcommand{\cycForm}{{\lsts} averaging formula}
\newcommand{\CycForm}{{\Lsts} averaging formula}
\newcommand{\pdes}{partial differential equations}

\newcommand{\nws}{non--wandering set}
\newcommand{\markGraph}{transition graph} 
\newcommand{\admissible}{admissible}

\newcommand{\obser}{\ensuremath{a}}     
\newcommand{\Obser}{\ensuremath{A}}     

\newcommand{\ExpaEig}{\ensuremath{\Lambda}}
\newcommand{\Lyap}{\ensuremath{\lambda}}            

\newcommand{\Ssym}[1]{{\ensuremath{m_{#1}}}}    
\newcommand{\action}{S}               
\newcommand{\ELe}{defining equations}           
\newcommand{\ELE}{Defining equations}           
\newcommand{\shiftOrb}{shift matrix} 

\newcommand{\shiftOp}{shift operator}  

\newcommand{\maslovInd}{\ensuremath{m}}        

\newcommand{\po}{periodic orbit}
\newcommand{\Po}{Periodic orbit}

\newcommand{\steady}{steady state}
\newcommand{\Steady}{Steady state}

\newcommand{\expct}    [1]{\langle {#1} \rangle}

\newcommand{\pS}{\ensuremath{\mathcal{M}}}          
\newcommand{\Poincare}{Poincar\'e}
\newcommand{\PoincSec}{Poincar\'e section}

\newcommand{\zeit}{\ensuremath{t}}  

\newcommand{\transp}[1]{{#1}{}^\top}

\newcommand{\Group}{\ensuremath{G}}         
\newcommand{\LieEl}{\ensuremath{g}}  
\newcommand {\id}{{\ \hbox{{\rm 1}\kern-.62em\hbox{\rm 1}}}} 

\newcommand{\Dn}[1]{\ensuremath{\textrm{D}_{#1}}}              

\newcommand{\shift}{\ensuremath{\mathsf{r}}}        
\newcommand{\wavenum}{\ensuremath{\mathsf{k}}}        

\newcommand{\KS}{Ku\-ra\-mo\-to-Siva\-shin\-sky}

\newcommand{\NS}{Navier-Stokes}

\newcommand{\dmn}{-dimensional}  

\newcommand{\etc}{{etc.}}       
\newcommand{\etal}{{\em et al.}}    
\newcommand{\ie}{{i.e.}}        

\newcommand{\videoLink}[1]{    
    \raisebox{-0.4ex}[0pt][0pt]{$\!\!\!$
    \HREF{https://#1}{\includegraphics[height=1em]{YouTubeIcon}}
    $\!$             }}
\newcommand{\toVideo}[1]
{\marginpar[\videoLink{#1}]{\videoLink{#1} \hfill}}

\newcommand{\toChaosBook}[2]
{\HREF{https://ChaosBook.org/chapters/ChaosBook.pdf\##1}{ChaosBook\, #2}}
\newcommand{\toLCpdf}[2]   
    {\HREF{https://arxiv.org/pdf/2201.11325v1\##1}{I\, #2}}



%% file: abstract.tex


We describe spatiotemporally chaotic (or turbulent) field theories
discretized over d-dimensional lattices in terms of sums over their
multi-periodic orbits.
`Chaos theory' is here recast in the language of statistical mechanics,
field theory, and solid state physics, with the traditional periodic
orbits theory of low-dimensional, temporally chaotic dynamics a special,
one-dimensional case.

In the field-theoretical formulation, there is no time evolution.
Instead, treating the temporal and spatial directions on equal footing,
one determines the spatiotemporally periodic orbits that contribute to
the partition sum of the theory, each a solution of the system's defining
deterministic equations, with sums over time-periodic orbits of dynamical
systems theory replaced here by sums of d-periodic orbits over
d-dimensional spacetime, the weight of each orbit given by
the Jacobian of its spatiotemporal orbit Jacobian operator.
The weights, evaluated by application of the Bloch theorem to the
spectrum of periodic orbit's Jacobian operator, are multiplicative for
spacetime orbit repeats, leading to a spatiotemporal zeta function
formulation of the theory in terms of prime orbits.

%% file: CL18.bbl
\begin{thebibliography}{233}%
\makeatletter
\providecommand \@ifxundefined [1]{%
 \@ifx{#1\undefined}
}%
\providecommand \@ifnum [1]{%
 \ifnum #1\expandafter \@firstoftwo
 \else \expandafter \@secondoftwo
 \fi
}%
\providecommand \@ifx [1]{%
 \ifx #1\expandafter \@firstoftwo
 \else \expandafter \@secondoftwo
 \fi
}%
\providecommand \natexlab [1]{#1}%
\providecommand \enquote  [1]{``#1''}%
\providecommand \bibnamefont  [1]{#1}%
\providecommand \bibfnamefont [1]{#1}%
\providecommand \citenamefont [1]{#1}%
\providecommand \href@noop [0]{\@secondoftwo}%
\providecommand \href [0]{\begingroup \@sanitize@url \@href}%
\providecommand \@href[1]{\@@startlink{#1}\@@href}%
\providecommand \@@href[1]{\endgroup#1\@@endlink}%
\providecommand \@sanitize@url [0]{\catcode `\\12\catcode `\$12\catcode
  `\&12\catcode `\#12\catcode `\^12\catcode `\_12\catcode `\%12\relax}%
\providecommand \@@startlink[1]{}%
\providecommand \@@endlink[0]{}%
\providecommand \url  [0]{\begingroup\@sanitize@url \@url }%
\providecommand \@url [1]{\endgroup\@href {#1}{\urlprefix }}%
\providecommand \urlprefix  [0]{URL }%
\providecommand \Eprint [0]{\href }%
\providecommand \doibase [0]{https://doi.org/}%
\providecommand \selectlanguage [0]{\@gobble}%
\providecommand \bibinfo  [0]{\@secondoftwo}%
\providecommand \bibfield  [0]{\@secondoftwo}%
\providecommand \translation [1]{[#1]}%
\providecommand \BibitemOpen [0]{}%
\providecommand \bibitemStop [0]{}%
\providecommand \bibitemNoStop [0]{.\EOS\space}%
\providecommand \EOS [0]{\spacefactor3000\relax}%
\providecommand \BibitemShut  [1]{\csname bibitem#1\endcsname}%
\let\auto@bib@innerbib\@empty
\bibitem [{\citenamefont {Ruelle}(1982)}]{ruelext}%
  \BibitemOpen
  \bibfield  {author} {\bibinfo {author} {\bibfnamefont {D.}~\bibnamefont
  {Ruelle}},\ }\bibfield  {title} {\enquote {\bibinfo {title} {Large volume
  limit of the distribution of characteristic exponents in turbulence},}\
  }\href {https://doi.org/10.1007/BF01218566} {\bibfield  {journal} {\bibinfo
  {journal} {Commun. Math. Phys.}\ }\textbf {\bibinfo {volume} {87}},\ \bibinfo
  {pages} {287--302} (\bibinfo {year} {1982})}\BibitemShut {NoStop}%
\bibitem [{\citenamefont {Foias}\ \emph {et~al.}(1983)\citenamefont {Foias},
  \citenamefont {Manley}, \citenamefont {T{\'e}mam},\ and\ \citenamefont
  {Treve}}]{FMTT83}%
  \BibitemOpen
  \bibfield  {author} {\bibinfo {author} {\bibfnamefont {C.}~\bibnamefont
  {Foias}}, \bibinfo {author} {\bibfnamefont {O.~P.}\ \bibnamefont {Manley}},
  \bibinfo {author} {\bibfnamefont {R.}~\bibnamefont {T{\'e}mam}},\ and\
  \bibinfo {author} {\bibfnamefont {Y.~M.}\ \bibnamefont {Treve}},\ }\bibfield
  {title} {\enquote {\bibinfo {title} {Asymptotic analysis of the
  {Navier-Stokes} equations},}\ }\href
  {https://doi.org/10.1016/0167-2789(83)90297-X} {\bibfield  {journal}
  {\bibinfo  {journal} {Physica D}\ }\textbf {\bibinfo {volume} {9}},\ \bibinfo
  {pages} {157--188} (\bibinfo {year} {1983})}\BibitemShut {NoStop}%
\bibitem [{\citenamefont {Nicolaenko}(1986)}]{Nicolaenko86}%
  \BibitemOpen
  \bibfield  {author} {\bibinfo {author} {\bibfnamefont {B.}~\bibnamefont
  {Nicolaenko}},\ }\bibfield  {title} {\enquote {\bibinfo {title} {Some
  mathematical aspects of flame chaos and flame multiplicity},}\ }\href
  {https://doi.org/10.1016/0167-2789(86)90099-0} {\bibfield  {journal}
  {\bibinfo  {journal} {Physica D}\ }\textbf {\bibinfo {volume} {20}},\
  \bibinfo {pages} {109--121} (\bibinfo {year} {1986})}\BibitemShut {NoStop}%
\bibitem [{\citenamefont {Grassberger}(1989)}]{Grassberger89}%
  \BibitemOpen
  \bibfield  {author} {\bibinfo {author} {\bibfnamefont {P.}~\bibnamefont
  {Grassberger}},\ }\bibfield  {title} {\enquote {\bibinfo {title} {Information
  content and predictability of lumped and distributed dynamical systems},}\
  }\href {https://doi.org/10.1088/0031-8949/40/3/016} {\bibfield  {journal}
  {\bibinfo  {journal} {Phys. Scr.}\ }\textbf {\bibinfo {volume} {40}},\
  \bibinfo {pages} {346} (\bibinfo {year} {1989})}\BibitemShut {NoStop}%
\bibitem [{\citenamefont {Lepri}, \citenamefont {Politi},\ and\ \citenamefont
  {Torcini}(1996)}]{LePoTo96}%
  \BibitemOpen
  \bibfield  {author} {\bibinfo {author} {\bibfnamefont {S.}~\bibnamefont
  {Lepri}}, \bibinfo {author} {\bibfnamefont {A.}~\bibnamefont {Politi}},\ and\
  \bibinfo {author} {\bibfnamefont {A.}~\bibnamefont {Torcini}},\ }\bibfield
  {title} {\enquote {\bibinfo {title} {Chronotopic {Lyapunov} analysis. {I. A}
  detailed characterization of {1D} systems},}\ }\href
  {https://doi.org/10.1007/BF02183390} {\bibfield  {journal} {\bibinfo
  {journal} {J. Stat. Phys.}\ }\textbf {\bibinfo {volume} {82}},\ \bibinfo
  {pages} {1429--1452} (\bibinfo {year} {1996})}\BibitemShut {NoStop}%
\bibitem [{\citenamefont {Lepri}, \citenamefont {Politi},\ and\ \citenamefont
  {Torcini}(1997{\natexlab{a}})}]{LePoTo97}%
  \BibitemOpen
  \bibfield  {author} {\bibinfo {author} {\bibfnamefont {S.}~\bibnamefont
  {Lepri}}, \bibinfo {author} {\bibfnamefont {A.}~\bibnamefont {Politi}},\ and\
  \bibinfo {author} {\bibfnamefont {A.}~\bibnamefont {Torcini}},\ }\bibfield
  {title} {\enquote {\bibinfo {title} {Chronotopic {Lyapunov} analysis. {II.
  Towards} a unified approach},}\ }\href {https://doi.org/10.1007/BF02508463}
  {\bibfield  {journal} {\bibinfo  {journal} {J. Stat. Phys.}\ }\textbf
  {\bibinfo {volume} {88}},\ \bibinfo {pages} {31--45} (\bibinfo {year}
  {1997}{\natexlab{a}})}\BibitemShut {NoStop}%
\bibitem [{\citenamefont {Lepri}, \citenamefont {Politi},\ and\ \citenamefont
  {Torcini}(1997{\natexlab{b}})}]{LePoTo97a}%
  \BibitemOpen
  \bibfield  {author} {\bibinfo {author} {\bibfnamefont {S.}~\bibnamefont
  {Lepri}}, \bibinfo {author} {\bibfnamefont {A.}~\bibnamefont {Politi}},\ and\
  \bibinfo {author} {\bibfnamefont {A.}~\bibnamefont {Torcini}},\ }\bibfield
  {title} {\enquote {\bibinfo {title} {Entropy potential and {Lyapunov}
  exponents},}\ }\href {https://doi.org/10.1063/1.166268} {\bibfield  {journal}
  {\bibinfo  {journal} {Chaos}\ }\textbf {\bibinfo {volume} {7}},\ \bibinfo
  {pages} {701--709} (\bibinfo {year} {1997}{\natexlab{b}})}\BibitemShut
  {NoStop}%
\bibitem [{\citenamefont {Waleffe}(2001)}]{W01}%
  \BibitemOpen
  \bibfield  {author} {\bibinfo {author} {\bibfnamefont {F.}~\bibnamefont
  {Waleffe}},\ }\bibfield  {title} {\enquote {\bibinfo {title} {Exact coherent
  structures in channel flow},}\ }\href
  {https://doi.org/10.1017/s0022112001004189} {\bibfield  {journal} {\bibinfo
  {journal} {J. Fluid Mech.}\ }\textbf {\bibinfo {volume} {435}},\ \bibinfo
  {pages} {93--102} (\bibinfo {year} {2001})}\BibitemShut {NoStop}%
\bibitem [{\citenamefont {Wedin}\ and\ \citenamefont {Kerswell}(2004)}]{WK04}%
  \BibitemOpen
  \bibfield  {author} {\bibinfo {author} {\bibfnamefont {H.}~\bibnamefont
  {Wedin}}\ and\ \bibinfo {author} {\bibfnamefont {R.~R.}\ \bibnamefont
  {Kerswell}},\ }\bibfield  {title} {\enquote {\bibinfo {title} {Exact coherent
  structures in pipe flow},}\ }\href
  {https://doi.org/10.1017/S0022112004009346} {\bibfield  {journal} {\bibinfo
  {journal} {J. Fluid Mech.}\ }\textbf {\bibinfo {volume} {508}},\ \bibinfo
  {pages} {333--371} (\bibinfo {year} {2004})}\BibitemShut {NoStop}%
\bibitem [{\citenamefont {Kawahara}\ and\ \citenamefont
  {Kida}(2001)}]{KawKida01}%
  \BibitemOpen
  \bibfield  {author} {\bibinfo {author} {\bibfnamefont {G.}~\bibnamefont
  {Kawahara}}\ and\ \bibinfo {author} {\bibfnamefont {S.}~\bibnamefont
  {Kida}},\ }\bibfield  {title} {\enquote {\bibinfo {title} {Periodic motion
  embedded in plane {Couette} turbulence: {Regeneration} cycle and burst},}\
  }\href {https://doi.org/10.1017/s0022112001006243} {\bibfield  {journal}
  {\bibinfo  {journal} {J. Fluid Mech.}\ }\textbf {\bibinfo {volume} {449}},\
  \bibinfo {pages} {291} (\bibinfo {year} {2001})}\BibitemShut {NoStop}%
\bibitem [{\citenamefont {Hof}\ \emph {et~al.}(2004)\citenamefont {Hof},
  \citenamefont {van Doorne}, \citenamefont {Westerweel}, \citenamefont
  {Nieuwstadt}, \citenamefont {Faisst}, \citenamefont {Eckhardt}, \citenamefont
  {Wedin}, \citenamefont {Kerswell},\ and\ \citenamefont
  {Waleffe}}]{science04}%
  \BibitemOpen
  \bibfield  {author} {\bibinfo {author} {\bibfnamefont {B.}~\bibnamefont
  {Hof}}, \bibinfo {author} {\bibfnamefont {C.~W.~H.}\ \bibnamefont {van
  Doorne}}, \bibinfo {author} {\bibfnamefont {J.}~\bibnamefont {Westerweel}},
  \bibinfo {author} {\bibfnamefont {F.~T.~M.}\ \bibnamefont {Nieuwstadt}},
  \bibinfo {author} {\bibfnamefont {H.}~\bibnamefont {Faisst}}, \bibinfo
  {author} {\bibfnamefont {B.}~\bibnamefont {Eckhardt}}, \bibinfo {author}
  {\bibfnamefont {H.}~\bibnamefont {Wedin}}, \bibinfo {author} {\bibfnamefont
  {R.~R.}\ \bibnamefont {Kerswell}},\ and\ \bibinfo {author} {\bibfnamefont
  {F.}~\bibnamefont {Waleffe}},\ }\bibfield  {title} {\enquote {\bibinfo
  {title} {Experimental observation of nonlinear traveling waves in turbulent
  pipe flow},}\ }\href {https://doi.org/10.1126/science.1100393} {\bibfield
  {journal} {\bibinfo  {journal} {Science}\ }\textbf {\bibinfo {volume}
  {305}},\ \bibinfo {pages} {1594--1598} (\bibinfo {year} {2004})}\BibitemShut
  {NoStop}%
\bibitem [{\citenamefont {Gibson}, \citenamefont {Halcrow},\ and\ \citenamefont
  {Cvitanovi{\'c}}(2008)}]{GHCW07}%
  \BibitemOpen
  \bibfield  {author} {\bibinfo {author} {\bibfnamefont {J.~F.}\ \bibnamefont
  {Gibson}}, \bibinfo {author} {\bibfnamefont {J.}~\bibnamefont {Halcrow}},\
  and\ \bibinfo {author} {\bibfnamefont {P.}~\bibnamefont {Cvitanovi{\'c}}},\
  }\bibfield  {title} {\enquote {\bibinfo {title} {Visualizing the geometry of
  state-space in plane {Couette} flow},}\ }\href
  {https://doi.org/10.1017/S002211200800267X} {\bibfield  {journal} {\bibinfo
  {journal} {J. Fluid Mech.}\ }\textbf {\bibinfo {volume} {611}},\ \bibinfo
  {pages} {107--130} (\bibinfo {year} {2008})}\BibitemShut {NoStop}%
\bibitem [{\citenamefont {Budanur}\ \emph {et~al.}(2017)\citenamefont
  {Budanur}, \citenamefont {Short}, \citenamefont {Farazmand}, \citenamefont
  {Willis},\ and\ \citenamefont {Cvitanovi{\'c}}}]{WFSBC15}%
  \BibitemOpen
  \bibfield  {author} {\bibinfo {author} {\bibfnamefont {N.~B.}\ \bibnamefont
  {Budanur}}, \bibinfo {author} {\bibfnamefont {K.~Y.}\ \bibnamefont {Short}},
  \bibinfo {author} {\bibfnamefont {M.}~\bibnamefont {Farazmand}}, \bibinfo
  {author} {\bibfnamefont {A.~P.}\ \bibnamefont {Willis}},\ and\ \bibinfo
  {author} {\bibfnamefont {P.}~\bibnamefont {Cvitanovi{\'c}}},\ }\bibfield
  {title} {\enquote {\bibinfo {title} {Relative periodic orbits form the
  backbone of turbulent pipe flow},}\ }\href
  {https://doi.org/10.1017/jfm.2017.699} {\bibfield  {journal} {\bibinfo
  {journal} {J. Fluid Mech.}\ }\textbf {\bibinfo {volume} {833}},\ \bibinfo
  {pages} {274--301} (\bibinfo {year} {2017})}\BibitemShut {NoStop}%
\bibitem [{\citenamefont {Crowley}\ \emph
  {et~al.}(2022{\natexlab{a}})\citenamefont {Crowley}, \citenamefont
  {Pughe-Sanford}, \citenamefont {Toler}, \citenamefont {Krygier},
  \citenamefont {Grigoriev},\ and\ \citenamefont {Schatz}}]{CPTKGS22}%
  \BibitemOpen
  \bibfield  {author} {\bibinfo {author} {\bibfnamefont {C.~J.}\ \bibnamefont
  {Crowley}}, \bibinfo {author} {\bibfnamefont {J.~L.}\ \bibnamefont
  {Pughe-Sanford}}, \bibinfo {author} {\bibfnamefont {W.}~\bibnamefont
  {Toler}}, \bibinfo {author} {\bibfnamefont {M.~C.}\ \bibnamefont {Krygier}},
  \bibinfo {author} {\bibfnamefont {R.~O.}\ \bibnamefont {Grigoriev}},\ and\
  \bibinfo {author} {\bibfnamefont {M.~F.}\ \bibnamefont {Schatz}},\ }\bibfield
   {title} {\enquote {\bibinfo {title} {Turbulence tracks recurrent
  solutions},}\ }\href {https://doi.org/10.1073/pnas.2120665119} {\bibfield
  {journal} {\bibinfo  {journal} {Proc. Natl. Acad. Sci.}\ }\textbf {\bibinfo
  {volume} {119}},\ \bibinfo {pages} {120665119} (\bibinfo {year}
  {2022}{\natexlab{a}})}\BibitemShut {NoStop}%
\bibitem [{\citenamefont {Cvitanovi{\'c}}(2013)}]{focusPOT}%
  \BibitemOpen
  \bibfield  {author} {\bibinfo {author} {\bibfnamefont {P.}~\bibnamefont
  {Cvitanovi{\'c}}},\ }\bibfield  {title} {\enquote {\bibinfo {title}
  {Recurrent flows: {The} clockwork behind turbulence},}\ }\href
  {https://doi.org/10.1017/jfm.2013.198} {\bibfield  {journal} {\bibinfo
  {journal} {J. Fluid Mech. Focus Fluids}\ }\textbf {\bibinfo {volume} {726}},\
  \bibinfo {pages} {1--4} (\bibinfo {year} {2013})}\BibitemShut {NoStop}%
\bibitem [{\citenamefont {Liang}\ and\ \citenamefont
  {Cvitanovi{\'c}}(2022)}]{LC21}%
  \BibitemOpen
  \bibfield  {author} {\bibinfo {author} {\bibfnamefont {H.}~\bibnamefont
  {Liang}}\ and\ \bibinfo {author} {\bibfnamefont {P.}~\bibnamefont
  {Cvitanovi{\'c}}},\ }\bibfield  {title} {\enquote {\bibinfo {title} {A
  chaotic lattice field theory in one dimension},}\ }\href
  {https://doi.org/10.1088/1751-8121/ac76f8} {\bibfield  {journal} {\bibinfo
  {journal} {J. Phys. A}\ }\textbf {\bibinfo {volume} {55}},\ \bibinfo {pages}
  {304002} (\bibinfo {year} {2022})}\BibitemShut {NoStop}%
\bibitem [{\citenamefont {Williams}\ \emph {et~al.}(2025)\citenamefont
  {Williams}, \citenamefont {Wang}, \citenamefont {Liang},\ and\ \citenamefont
  {Cvitanovi{\'c}}}]{WWLAFC22}%
  \BibitemOpen
  \bibfield  {author} {\bibinfo {author} {\bibfnamefont {S.~D.~V.}\
  \bibnamefont {Williams}}, \bibinfo {author} {\bibfnamefont {X.}~\bibnamefont
  {Wang}}, \bibinfo {author} {\bibfnamefont {H.}~\bibnamefont {Liang}},\ and\
  \bibinfo {author} {\bibfnamefont {P.}~\bibnamefont {Cvitanovi{\'c}}},\ }\href
  {https://ChaosBook.org/overheads/spatiotemporal/} {\enquote {\bibinfo {title}
  {Nonlinear chaotic lattice field theory},}\ } (\bibinfo {year} {2025}),\
  \bibinfo {note} {in preparation}\BibitemShut {NoStop}%
\bibitem [{\citenamefont {Gutzwiller}(1990)}]{gutbook}%
  \BibitemOpen
  \bibfield  {author} {\bibinfo {author} {\bibfnamefont {M.~C.}\ \bibnamefont
  {Gutzwiller}},\ }\href@noop {} {\emph {\bibinfo {title} {{Chaos in Classical
  and Quantum Mechanics}}}}\ (\bibinfo  {publisher} {Springer},\ \bibinfo
  {address} {New York},\ \bibinfo {year} {1990})\BibitemShut {NoStop}%
\bibitem [{\citenamefont {Ruelle}(2004)}]{ruelle}%
  \BibitemOpen
  \bibfield  {author} {\bibinfo {author} {\bibfnamefont {D.}~\bibnamefont
  {Ruelle}},\ }\href {https://doi.org/10.1017/cbo9780511617546} {\emph
  {\bibinfo {title} {{Thermodynamic Formalism: The Mathematical Structure of
  Equilibrium Statistical Mechanics}}}},\ \bibinfo {edition} {2nd}\ ed.\
  (\bibinfo  {publisher} {Cambridge Univ. Press},\ \bibinfo {address}
  {Cambridge},\ \bibinfo {year} {2004})\BibitemShut {NoStop}%
\bibitem [{\citenamefont {Cvitanovi{\'c}}\ \emph
  {et~al.}(2025{\natexlab{a}})\citenamefont {Cvitanovi{\'c}}, \citenamefont
  {Artuso}, \citenamefont {Mainieri}, \citenamefont {Tanner},\ and\
  \citenamefont {Vattay}}]{ChaosBook}%
  \BibitemOpen
  \bibfield  {author} {\bibinfo {author} {\bibfnamefont {P.}~\bibnamefont
  {Cvitanovi{\'c}}}, \bibinfo {author} {\bibfnamefont {R.}~\bibnamefont
  {Artuso}}, \bibinfo {author} {\bibfnamefont {R.}~\bibnamefont {Mainieri}},
  \bibinfo {author} {\bibfnamefont {G.}~\bibnamefont {Tanner}},\ and\ \bibinfo
  {author} {\bibfnamefont {G.}~\bibnamefont {Vattay}},\ }\href
  {https://ChaosBook.org/} {\emph {\bibinfo {title} {{Chaos: Classical and
  Quantum}}}}\ (\bibinfo  {publisher} {Niels Bohr Inst.},\ \bibinfo {address}
  {Copenhagen},\ \bibinfo {year} {2025})\BibitemShut {NoStop}%
\bibitem [{\citenamefont {Politi}, \citenamefont {Torcini},\ and\ \citenamefont
  {Lepri}(1998)}]{PoToLe98}%
  \BibitemOpen
  \bibfield  {author} {\bibinfo {author} {\bibfnamefont {A.}~\bibnamefont
  {Politi}}, \bibinfo {author} {\bibfnamefont {A.}~\bibnamefont {Torcini}},\
  and\ \bibinfo {author} {\bibfnamefont {S.}~\bibnamefont {Lepri}},\ }\bibfield
   {title} {\enquote {\bibinfo {title} {{Lyapunov} exponents from node-counting
  arguments},}\ }\href {https://doi.org/10.1051/jp4:1998636} {\bibfield
  {journal} {\bibinfo  {journal} {J. Phys. IV}\ }\textbf {\bibinfo {volume}
  {8}},\ \bibinfo {pages} {263} (\bibinfo {year} {1998})}\BibitemShut {NoStop}%
\bibitem [{\citenamefont {Giacomelli}, \citenamefont {Lepri},\ and\
  \citenamefont {Politi}(1995)}]{GiLePo95}%
  \BibitemOpen
  \bibfield  {author} {\bibinfo {author} {\bibfnamefont {G.}~\bibnamefont
  {Giacomelli}}, \bibinfo {author} {\bibfnamefont {S.}~\bibnamefont {Lepri}},\
  and\ \bibinfo {author} {\bibfnamefont {A.}~\bibnamefont {Politi}},\
  }\bibfield  {title} {\enquote {\bibinfo {title} {Statistical properties of
  bidimensional patterns generated from delayed and extended maps},}\ }\href
  {https://doi.org/10.1103/PhysRevE.51.3939} {\bibfield  {journal} {\bibinfo
  {journal} {Phys. Rev. E}\ }\textbf {\bibinfo {volume} {51}},\ \bibinfo
  {pages} {3939--3944} (\bibinfo {year} {1995})}\BibitemShut {NoStop}%
\bibitem [{\citenamefont {Bountis}\ and\ \citenamefont
  {Helleman}(1981)}]{Bount81}%
  \BibitemOpen
  \bibfield  {author} {\bibinfo {author} {\bibfnamefont {T.}~\bibnamefont
  {Bountis}}\ and\ \bibinfo {author} {\bibfnamefont {R.~H.~G.}\ \bibnamefont
  {Helleman}},\ }\bibfield  {title} {\enquote {\bibinfo {title} {On the
  stability of periodic orbits of two-dimensional mappings},}\ }\href
  {https://doi.org/10.1063/1.525159} {\bibfield  {journal} {\bibinfo  {journal}
  {J. Math. Phys.}\ }\textbf {\bibinfo {volume} {22}},\ \bibinfo {pages}
  {1867--1877} (\bibinfo {year} {1981})}\BibitemShut {NoStop}%
\bibitem [{\citenamefont {MacKay}\ and\ \citenamefont
  {Meiss}(1983)}]{MacMei83}%
  \BibitemOpen
  \bibfield  {author} {\bibinfo {author} {\bibfnamefont {R.~S.}\ \bibnamefont
  {MacKay}}\ and\ \bibinfo {author} {\bibfnamefont {J.~D.}\ \bibnamefont
  {Meiss}},\ }\bibfield  {title} {\enquote {\bibinfo {title} {Linear stability
  of periodic orbits in {Lagrangian} systems},}\ }\href
  {https://doi.org/10.1016/0375-9601(83)90735-1} {\bibfield  {journal}
  {\bibinfo  {journal} {Phys. Lett. A}\ }\textbf {\bibinfo {volume} {98}},\
  \bibinfo {pages} {92--94} (\bibinfo {year} {1983})}\BibitemShut {NoStop}%
\bibitem [{\citenamefont {Pikovsky}(1989)}]{Pikovsky89}%
  \BibitemOpen
  \bibfield  {author} {\bibinfo {author} {\bibfnamefont {A.~S.}\ \bibnamefont
  {Pikovsky}},\ }\bibfield  {title} {\enquote {\bibinfo {title} {Spatial
  development of chaos in nonlinear media},}\ }\href
  {https://doi.org/10.1016/0375-9601(89)90096-0} {\bibfield  {journal}
  {\bibinfo  {journal} {Phys. Lett. A}\ }\textbf {\bibinfo {volume} {137}},\
  \bibinfo {pages} {121--127} (\bibinfo {year} {1989})}\BibitemShut {NoStop}%
\bibitem [{\citenamefont {Strogatz}(2014)}]{strogb}%
  \BibitemOpen
  \bibfield  {author} {\bibinfo {author} {\bibfnamefont {S.~H.}\ \bibnamefont
  {Strogatz}},\ }\href@noop {} {\emph {\bibinfo {title} {{Nonlinear Dynamics
  and Chaos}}}}\ (\bibinfo  {publisher} {Westview Press},\ \bibinfo {address}
  {Boulder, CO},\ \bibinfo {year} {2014})\BibitemShut {NoStop}%
\bibitem [{\citenamefont {Ott}(2002)}]{ottbook}%
  \BibitemOpen
  \bibfield  {author} {\bibinfo {author} {\bibfnamefont {E.}~\bibnamefont
  {Ott}},\ }\href {https://doi.org/10.1017/cbo9780511803260} {\emph {\bibinfo
  {title} {{Chaos and Dynamical Systems}}}}\ (\bibinfo  {publisher} {Cambridge
  Univ. Press},\ \bibinfo {address} {Cambridge},\ \bibinfo {year}
  {2002})\BibitemShut {NoStop}%
\bibitem [{\citenamefont {Devaney}(2008)}]{deva87}%
  \BibitemOpen
  \bibfield  {author} {\bibinfo {author} {\bibfnamefont {R.~L.}\ \bibnamefont
  {Devaney}},\ }\href@noop {} {\emph {\bibinfo {title} {{An Introduction to
  Chaotic Dynamical systems}}}},\ \bibinfo {edition} {2nd}\ ed.\ (\bibinfo
  {publisher} {Westview Press},\ \bibinfo {address} {Cambridge, Mass},\
  \bibinfo {year} {2008})\BibitemShut {NoStop}%
\bibitem [{\citenamefont {Alligood}, \citenamefont {Sauer},\ and\ \citenamefont
  {Yorke}(1996)}]{ASY96}%
  \BibitemOpen
  \bibfield  {author} {\bibinfo {author} {\bibfnamefont {K.~T.}\ \bibnamefont
  {Alligood}}, \bibinfo {author} {\bibfnamefont {T.~D.}\ \bibnamefont
  {Sauer}},\ and\ \bibinfo {author} {\bibfnamefont {J.~A.}\ \bibnamefont
  {Yorke}},\ }\href {https://doi.org/10.1007/b97589} {\emph {\bibinfo {title}
  {{Chaos, An Introduction to Dynamical Systems}}}}\ (\bibinfo  {publisher}
  {Springer},\ \bibinfo {address} {New York},\ \bibinfo {year}
  {1996})\BibitemShut {NoStop}%
\bibitem [{\citenamefont {Katok}\ and\ \citenamefont
  {Hasselblatt}(1995)}]{Katok95}%
  \BibitemOpen
  \bibfield  {author} {\bibinfo {author} {\bibfnamefont {A.}~\bibnamefont
  {Katok}}\ and\ \bibinfo {author} {\bibfnamefont {B.}~\bibnamefont
  {Hasselblatt}},\ }\href {https://doi.org/10.1017/cbo9780511809187} {\emph
  {\bibinfo {title} {{Introduction to the Modern Theory of Dynamical
  Systems}}}}\ (\bibinfo  {publisher} {Cambridge Univ. Press},\ \bibinfo
  {address} {Cambridge},\ \bibinfo {year} {1995})\BibitemShut {NoStop}%
\bibitem [{\citenamefont {Robinson}(2012)}]{Robinson12}%
  \BibitemOpen
  \bibfield  {author} {\bibinfo {author} {\bibfnamefont {R.~C.}\ \bibnamefont
  {Robinson}},\ }\href@noop {} {\emph {\bibinfo {title} {{An Introduction to
  Dynamical Systems: Continuous and Discrete}}}}\ (\bibinfo  {publisher} {Amer.
  Math. Soc.},\ \bibinfo {address} {New York},\ \bibinfo {year}
  {2012})\BibitemShut {NoStop}%
\bibitem [{\citenamefont {Engl}, \citenamefont {Urbina},\ and\ \citenamefont
  {Richter}(2015)}]{EnUrRi15a}%
  \BibitemOpen
  \bibfield  {author} {\bibinfo {author} {\bibfnamefont {T.}~\bibnamefont
  {Engl}}, \bibinfo {author} {\bibfnamefont {J.~D.}\ \bibnamefont {Urbina}},\
  and\ \bibinfo {author} {\bibfnamefont {K.}~\bibnamefont {Richter}},\
  }\bibfield  {title} {\enquote {\bibinfo {title} {Periodic mean-field
  solutions and the spectra of discrete bosonic fields: {Trace} formula for
  {Bose-Hubbard} models},}\ }\href {https://doi.org/10.1103/physreve.92.062907}
  {\bibfield  {journal} {\bibinfo  {journal} {Phys. Rev. E}\ }\textbf {\bibinfo
  {volume} {92}},\ \bibinfo {pages} {062907} (\bibinfo {year}
  {2015})}\BibitemShut {NoStop}%
\bibitem [{\citenamefont {Akila}\ \emph {et~al.}(2017)\citenamefont {Akila},
  \citenamefont {Waltner}, \citenamefont {Gutkin}, \citenamefont {Braun},\ and\
  \citenamefont {Guhr}}]{AWGBG16}%
  \BibitemOpen
  \bibfield  {author} {\bibinfo {author} {\bibfnamefont {M.}~\bibnamefont
  {Akila}}, \bibinfo {author} {\bibfnamefont {D.}~\bibnamefont {Waltner}},
  \bibinfo {author} {\bibfnamefont {B.}~\bibnamefont {Gutkin}}, \bibinfo
  {author} {\bibfnamefont {P.}~\bibnamefont {Braun}},\ and\ \bibinfo {author}
  {\bibfnamefont {T.}~\bibnamefont {Guhr}},\ }\bibfield  {title} {\enquote
  {\bibinfo {title} {Semiclassical identification of periodic orbits in a
  quantum many-body system},}\ }\href
  {https://doi.org/10.1103/physrevlett.118.164101} {\bibfield  {journal}
  {\bibinfo  {journal} {Phys. Rev. Lett.}\ }\textbf {\bibinfo {volume} {118}},\
  \bibinfo {pages} {164101} (\bibinfo {year} {2017})}\BibitemShut {NoStop}%
\bibitem [{\citenamefont {Akila}\ \emph {et~al.}(2016)\citenamefont {Akila},
  \citenamefont {Waltner}, \citenamefont {Gutkin},\ and\ \citenamefont
  {Guhr}}]{AWGG16}%
  \BibitemOpen
  \bibfield  {author} {\bibinfo {author} {\bibfnamefont {M.}~\bibnamefont
  {Akila}}, \bibinfo {author} {\bibfnamefont {D.}~\bibnamefont {Waltner}},
  \bibinfo {author} {\bibfnamefont {B.}~\bibnamefont {Gutkin}},\ and\ \bibinfo
  {author} {\bibfnamefont {T.}~\bibnamefont {Guhr}},\ }\bibfield  {title}
  {\enquote {\bibinfo {title} {Particle-time duality in the kicked {Ising} spin
  chain},}\ }\href {https://doi.org/10.1088/1751-8113/49/37/375101} {\bibfield
  {journal} {\bibinfo  {journal} {J. Phys. A}\ }\textbf {\bibinfo {volume}
  {49}},\ \bibinfo {pages} {375101} (\bibinfo {year} {2016})}\BibitemShut
  {NoStop}%
\bibitem [{\citenamefont {Richter}, \citenamefont {Urbina},\ and\ \citenamefont
  {Tomsovic}(2022)}]{RiUrTo22}%
  \BibitemOpen
  \bibfield  {author} {\bibinfo {author} {\bibfnamefont {K.}~\bibnamefont
  {Richter}}, \bibinfo {author} {\bibfnamefont {J.~D.}\ \bibnamefont
  {Urbina}},\ and\ \bibinfo {author} {\bibfnamefont {S.}~\bibnamefont
  {Tomsovic}},\ }\bibfield  {title} {\enquote {\bibinfo {title} {Semiclassical
  roots of universality in many-body quantum chaos},}\ }\href
  {https://doi.org/10.1088/1751-8121/ac9e4e} {\bibfield  {journal} {\bibinfo
  {journal} {J. Phys. A}\ }\textbf {\bibinfo {volume} {55}},\ \bibinfo {pages}
  {453001} (\bibinfo {year} {2022})}\BibitemShut {NoStop}%
\bibitem [{\citenamefont {Gutkin}\ and\ \citenamefont
  {Osipov}(2016)}]{GutOsi15}%
  \BibitemOpen
  \bibfield  {author} {\bibinfo {author} {\bibfnamefont {B.}~\bibnamefont
  {Gutkin}}\ and\ \bibinfo {author} {\bibfnamefont {V.}~\bibnamefont
  {Osipov}},\ }\bibfield  {title} {\enquote {\bibinfo {title} {Classical
  foundations of many-particle quantum chaos},}\ }\href
  {https://doi.org/10.1088/0951-7715/29/2/325} {\bibfield  {journal} {\bibinfo
  {journal} {Nonlinearity}\ }\textbf {\bibinfo {volume} {29}},\ \bibinfo
  {pages} {325--356} (\bibinfo {year} {2016})}\BibitemShut {NoStop}%
\bibitem [{\citenamefont {Gutkin}\ \emph {et~al.}(2021)\citenamefont {Gutkin},
  \citenamefont {Han}, \citenamefont {Jafari}, \citenamefont {Saremi},\ and\
  \citenamefont {Cvitanovi{\'c}}}]{GHJSC16}%
  \BibitemOpen
  \bibfield  {author} {\bibinfo {author} {\bibfnamefont {B.}~\bibnamefont
  {Gutkin}}, \bibinfo {author} {\bibfnamefont {L.}~\bibnamefont {Han}},
  \bibinfo {author} {\bibfnamefont {R.}~\bibnamefont {Jafari}}, \bibinfo
  {author} {\bibfnamefont {A.~K.}\ \bibnamefont {Saremi}},\ and\ \bibinfo
  {author} {\bibfnamefont {P.}~\bibnamefont {Cvitanovi{\'c}}},\ }\bibfield
  {title} {\enquote {\bibinfo {title} {Linear encoding of the spatiotemporal
  cat map},}\ }\href {https://doi.org/10.1088/1361-6544/abd7c8} {\bibfield
  {journal} {\bibinfo  {journal} {Nonlinearity}\ }\textbf {\bibinfo {volume}
  {34}},\ \bibinfo {pages} {2800--2836} (\bibinfo {year} {2021})}\BibitemShut
  {NoStop}%
\bibitem [{\citenamefont {Alderson}, \citenamefont {Dubertrand},\ and\
  \citenamefont {Shudo}(2025)}]{AlDuSh25}%
  \BibitemOpen
  \bibfield  {author} {\bibinfo {author} {\bibfnamefont {W.}~\bibnamefont
  {Alderson}}, \bibinfo {author} {\bibfnamefont {R.}~\bibnamefont
  {Dubertrand}},\ and\ \bibinfo {author} {\bibfnamefont {A.}~\bibnamefont
  {Shudo}},\ }\bibfield  {title} {\enquote {\bibinfo {title} {Classical
  transport in a maximally chaotic chain},}\ }\href
  {https://doi.org/10.1088/1751-8121/adc217} {\bibfield  {journal} {\bibinfo
  {journal} {J. Phys. A}\ }\textbf {\bibinfo {volume} {58}},\ \bibinfo {pages}
  {165201} (\bibinfo {year} {2025})}\BibitemShut {NoStop}%
\bibitem [{\citenamefont {Axenides}, \citenamefont {Floratos},\ and\
  \citenamefont {Nicolis}(2018)}]{AxFlNi17}%
  \BibitemOpen
  \bibfield  {author} {\bibinfo {author} {\bibfnamefont {M.}~\bibnamefont
  {Axenides}}, \bibinfo {author} {\bibfnamefont {E.}~\bibnamefont {Floratos}},\
  and\ \bibinfo {author} {\bibfnamefont {S.}~\bibnamefont {Nicolis}},\
  }\bibfield  {title} {\enquote {\bibinfo {title} {The quantum cat map on the
  modular discretization of extremal black hole horizons},}\ }\href
  {https://doi.org/10.1140/epjc/s10052-018-5850-9} {\bibfield  {journal}
  {\bibinfo  {journal} {Eur. Phys. J. C}\ }\textbf {\bibinfo {volume} {78}},\
  \bibinfo {pages} {412} (\bibinfo {year} {2018})}\BibitemShut {NoStop}%
\bibitem [{\citenamefont {Axenides}, \citenamefont {Floratos},\ and\
  \citenamefont {Nicolis}(2023)}]{AxFlNi22}%
  \BibitemOpen
  \bibfield  {author} {\bibinfo {author} {\bibfnamefont {M.}~\bibnamefont
  {Axenides}}, \bibinfo {author} {\bibfnamefont {E.}~\bibnamefont {Floratos}},\
  and\ \bibinfo {author} {\bibfnamefont {S.}~\bibnamefont {Nicolis}},\
  }\bibfield  {title} {\enquote {\bibinfo {title} {Arnol’d cat map
  lattices},}\ }\href {https://doi.org/10.1103/physreve.107.064206} {\bibfield
  {journal} {\bibinfo  {journal} {Phys. Rev. E}\ }\textbf {\bibinfo {volume}
  {107}},\ \bibinfo {pages} {064206} (\bibinfo {year} {2023})}\BibitemShut
  {NoStop}%
\bibitem [{\citenamefont {Vattay}(1997)}]{Vattay1997}%
  \BibitemOpen
  \bibfield  {author} {\bibinfo {author} {\bibfnamefont {G.}~\bibnamefont
  {Vattay}},\ }\bibfield  {title} {\enquote {\bibinfo {title} {Noise and
  quantum corrections to trace formulas},}\ }in\ \href
  {https://ChaosBook.org/paper.shtml#trace} {\emph {\bibinfo {booktitle}
  {{Chaos: Classical and Quantum}}}},\ \bibinfo {editor} {edited by\ \bibinfo
  {editor} {\bibfnamefont {P.}~\bibnamefont {Cvitanovi{\'c}}}, \bibinfo
  {editor} {\bibfnamefont {R.}~\bibnamefont {Artuso}}, \bibinfo {editor}
  {\bibfnamefont {R.}~\bibnamefont {Mainieri}}, \bibinfo {editor}
  {\bibfnamefont {G.}~\bibnamefont {Tanner}},\ and\ \bibinfo {editor}
  {\bibfnamefont {G.}~\bibnamefont {Vattay}}}\ (\bibinfo  {publisher} {Niels
  Bohr Inst.},\ \bibinfo {address} {Copenhagen},\ \bibinfo {year}
  {1997})\BibitemShut {NoStop}%
\bibitem [{\citenamefont {Subramanyan}\ \emph {et~al.}(2021)\citenamefont
  {Subramanyan}, \citenamefont {Hegde}, \citenamefont {Vishveshwara},\ and\
  \citenamefont {Bradlyn}}]{SHVB21}%
  \BibitemOpen
  \bibfield  {author} {\bibinfo {author} {\bibfnamefont {V.}~\bibnamefont
  {Subramanyan}}, \bibinfo {author} {\bibfnamefont {S.~S.}\ \bibnamefont
  {Hegde}}, \bibinfo {author} {\bibfnamefont {S.}~\bibnamefont
  {Vishveshwara}},\ and\ \bibinfo {author} {\bibfnamefont {B.}~\bibnamefont
  {Bradlyn}},\ }\bibfield  {title} {\enquote {\bibinfo {title} {Physics of the
  inverted harmonic oscillator: {From} the lowest {Landau} level to event
  horizons},}\ }\href {https://doi.org/10.1016/j.aop.2021.168470} {\bibfield
  {journal} {\bibinfo  {journal} {Ann. Phys.}\ }\textbf {\bibinfo {volume}
  {435}},\ \bibinfo {pages} {168470} (\bibinfo {year} {2021})}\BibitemShut
  {NoStop}%
\bibitem [{\citenamefont {Fermi}, \citenamefont {Pasta},\ and\ \citenamefont
  {Ulam}(1965)}]{fpu65}%
  \BibitemOpen
  \bibfield  {author} {\bibinfo {author} {\bibfnamefont {E.}~\bibnamefont
  {Fermi}}, \bibinfo {author} {\bibfnamefont {J.}~\bibnamefont {Pasta}},\ and\
  \bibinfo {author} {\bibfnamefont {S.}~\bibnamefont {Ulam}},\ }\bibfield
  {title} {\enquote {\bibinfo {title} {Studies of nonlinear problems},}\ }in\
  \href
  {https://archive.org/details/collectedpapersn0000ferm_r2f8/page/n1113/mode/2up}
  {\emph {\bibinfo {booktitle} {{E. Fermi: Note e Memorie (Collected
  Papers)}}}},\ Vol.~\bibinfo {volume} {{II}}\ (\bibinfo  {publisher}
  {Accademia Nazionale dei Lincei, Roma, and The University of Chicago Press,
  Chicago},\ \bibinfo {year} {1965})\ pp.\ \bibinfo {pages}
  {977--988}\BibitemShut {NoStop}%
\bibitem [{\citenamefont {Campbell}, \citenamefont {Flach},\ and\ \citenamefont
  {Kivshar}(2004)}]{CaFlKi04}%
  \BibitemOpen
  \bibfield  {author} {\bibinfo {author} {\bibfnamefont {D.~K.}\ \bibnamefont
  {Campbell}}, \bibinfo {author} {\bibfnamefont {S.}~\bibnamefont {Flach}},\
  and\ \bibinfo {author} {\bibfnamefont {Y.~S.}\ \bibnamefont {Kivshar}},\
  }\bibfield  {title} {\enquote {\bibinfo {title} {Localizing energy through
  nonlinearity and discreteness},}\ }\href {https://doi.org/10.1063/1.1650069}
  {\bibfield  {journal} {\bibinfo  {journal} {Phys. Today}\ }\textbf {\bibinfo
  {volume} {57}},\ \bibinfo {pages} {43--49} (\bibinfo {year}
  {2004})}\BibitemShut {NoStop}%
\bibitem [{\citenamefont {Gaspard}(1997)}]{Gas97}%
  \BibitemOpen
  \bibfield  {author} {\bibinfo {author} {\bibfnamefont {P.}~\bibnamefont
  {Gaspard}},\ }\href@noop {} {\emph {\bibinfo {title} {{Chaos, Scattering and
  Statistical Mechanics}}}}\ (\bibinfo  {publisher} {Cambridge Univ. Press},\
  \bibinfo {address} {Cambridge},\ \bibinfo {year} {1997})\BibitemShut
  {NoStop}%
\bibitem [{\citenamefont {Hill}(1886)}]{Hill86}%
  \BibitemOpen
  \bibfield  {author} {\bibinfo {author} {\bibfnamefont {G.~W.}\ \bibnamefont
  {Hill}},\ }\bibfield  {title} {\enquote {\bibinfo {title} {On the part of the
  motion of the lunar perigee which is a function of the mean motions of the
  sun and moon},}\ }\href {https://doi.org/10.1007/bf02417081} {\bibfield
  {journal} {\bibinfo  {journal} {Acta Math.}\ }\textbf {\bibinfo {volume}
  {8}},\ \bibinfo {pages} {1--36} (\bibinfo {year} {1886})}\BibitemShut
  {NoStop}%
\bibitem [{\citenamefont {Poincar{\'e}}(1886)}]{Poinc1886}%
  \BibitemOpen
  \bibfield  {author} {\bibinfo {author} {\bibfnamefont {H.}~\bibnamefont
  {Poincar{\'e}}},\ }\bibfield  {title} {\enquote {\bibinfo {title} {Sur les
  d{\'e}terminants d'ordre infini},}\ }\href {http://eudml.org/doc/85600}
  {\bibfield  {journal} {\bibinfo  {journal} {Bull. Soc. Math. France}\
  }\textbf {\bibinfo {volume} {14}},\ \bibinfo {pages} {77--90} (\bibinfo
  {year} {1886})}\BibitemShut {NoStop}%
\bibitem [{\citenamefont {Liang}\ and\ \citenamefont
  {Cvitanovi{\'c}}(2025)}]{LC22}%
  \BibitemOpen
  \bibfield  {author} {\bibinfo {author} {\bibfnamefont {H.}~\bibnamefont
  {Liang}}\ and\ \bibinfo {author} {\bibfnamefont {P.}~\bibnamefont
  {Cvitanovi{\'c}}},\ }\href {https://ChaosBook.org/overheads/spatiotemporal/}
  {\enquote {\bibinfo {title} {A derivation of {Hill}'s formulas},}\ }
  (\bibinfo {year} {2025}),\ \bibinfo {note} {in preparation}\BibitemShut
  {NoStop}%
\bibitem [{\citenamefont {Artuso}, \citenamefont {Aurell},\ and\ \citenamefont
  {Cvitanovi\'c}(1990{\natexlab{a}})}]{AACI}%
  \BibitemOpen
  \bibfield  {author} {\bibinfo {author} {\bibfnamefont {R.}~\bibnamefont
  {Artuso}}, \bibinfo {author} {\bibfnamefont {E.}~\bibnamefont {Aurell}},\
  and\ \bibinfo {author} {\bibfnamefont {P.}~\bibnamefont {Cvitanovi\'c}},\
  }\bibfield  {title} {\enquote {\bibinfo {title} {Recycling of strange sets:
  {I. Cycle} expansions},}\ }\href {https://doi.org/10.1088/0951-7715/3/2/005}
  {\bibfield  {journal} {\bibinfo  {journal} {Nonlinearity}\ }\textbf {\bibinfo
  {volume} {3}},\ \bibinfo {pages} {325--359} (\bibinfo {year}
  {1990}{\natexlab{a}})}\BibitemShut {NoStop}%
\bibitem [{\citenamefont {Cvitanovi{\'c}}\ and\ \citenamefont
  {Liang}(2026)}]{CL18}%
  \BibitemOpen
  \bibfield  {author} {\bibinfo {author} {\bibfnamefont {P.}~\bibnamefont
  {Cvitanovi{\'c}}}\ and\ \bibinfo {author} {\bibfnamefont {H.}~\bibnamefont
  {Liang}},\ }\bibfield  {title} {\enquote {\bibinfo {title} {A chaotic lattice
  field theory in two dimensions},}\ }\href {https://doi.org/10.1063/5.0273642}
  {\bibfield  {journal} {\bibinfo  {journal} {Chaos}\ }\textbf {\bibinfo
  {volume} {36}} (\bibinfo {year} {2026}),\ 10.1063/5.0273642}\BibitemShut
  {NoStop}%
\bibitem [{\citenamefont {Cvitanovi{\'c}}(2026{\natexlab{a}})}]{CB:spatiotemp}%
  \BibitemOpen
  \bibfield  {author} {\bibinfo {author} {\bibfnamefont {P.}~\bibnamefont
  {Cvitanovi{\'c}}},\ }\href {https://ChaosBook.org/overheads/spatiotemporal}
  {\enquote {\bibinfo {title} {Turbulence in spacetime},}\ } (\bibinfo {year}
  {2026}{\natexlab{a}})\BibitemShut {NoStop}%
\bibitem [{\citenamefont {Montvay}\ and\ \citenamefont
  {M{\"u}nster}(1994)}]{MonMun94}%
  \BibitemOpen
  \bibfield  {author} {\bibinfo {author} {\bibfnamefont {I.}~\bibnamefont
  {Montvay}}\ and\ \bibinfo {author} {\bibfnamefont {G.}~\bibnamefont
  {M{\"u}nster}},\ }\href {https://doi.org/10.1017/cbo9780511470783} {\emph
  {\bibinfo {title} {{Quantum Fields on a Lattice}}}}\ (\bibinfo  {publisher}
  {Cambridge Univ. Press},\ \bibinfo {address} {Cambridge},\ \bibinfo {year}
  {1994})\BibitemShut {NoStop}%
\bibitem [{\citenamefont {M{\"u}nster}\ and\ \citenamefont
  {Walzl}(2000)}]{MunWal00}%
  \BibitemOpen
  \bibfield  {author} {\bibinfo {author} {\bibfnamefont {G.}~\bibnamefont
  {M{\"u}nster}}\ and\ \bibinfo {author} {\bibfnamefont {M.}~\bibnamefont
  {Walzl}},\ }\href {https://arxiv.org/abs/hep-lat/0012005} {\enquote {\bibinfo
  {title} {Lattice gauge theory - {A} short primer},}\ } (\bibinfo {year}
  {2000})\BibitemShut {NoStop}%
\bibitem [{\citenamefont {Ashcroft}\ and\ \citenamefont
  {Mermin}(1976)}]{AshMer}%
  \BibitemOpen
  \bibfield  {author} {\bibinfo {author} {\bibfnamefont {N.~W.}\ \bibnamefont
  {Ashcroft}}\ and\ \bibinfo {author} {\bibfnamefont {N.~D.}\ \bibnamefont
  {Mermin}},\ }\href {https://doi.org/10.1063/1.3037370} {\emph {\bibinfo
  {title} {{Solid State Physics}}}}\ (\bibinfo  {publisher} {Holt, Rinehart and
  Winston},\ \bibinfo {year} {1976})\BibitemShut {NoStop}%
\bibitem [{\citenamefont {Couchman}, \citenamefont {Evans},\ and\ \citenamefont
  {Bush}(2022)}]{CoEvBu22}%
  \BibitemOpen
  \bibfield  {author} {\bibinfo {author} {\bibfnamefont {M.~M.~P.}\
  \bibnamefont {Couchman}}, \bibinfo {author} {\bibfnamefont {D.~J.}\
  \bibnamefont {Evans}},\ and\ \bibinfo {author} {\bibfnamefont {J.~W.~M.}\
  \bibnamefont {Bush}},\ }\bibfield  {title} {\enquote {\bibinfo {title} {The
  stability of a hydrodynamic {Bravais} lattice},}\ }\href
  {https://doi.org/10.3390/sym14081524} {\bibfield  {journal} {\bibinfo
  {journal} {Symmetry}\ }\textbf {\bibinfo {volume} {14}},\ \bibinfo {pages}
  {1524} (\bibinfo {year} {2022})}\BibitemShut {NoStop}%
\bibitem [{\citenamefont {Cvitanovi{\'c}}\ and\ \citenamefont
  {Liang}(2025)}]{appendWKB}%
  \BibitemOpen
  \bibfield  {author} {\bibinfo {author} {\bibfnamefont {P.}~\bibnamefont
  {Cvitanovi{\'c}}}\ and\ \bibinfo {author} {\bibfnamefont {H.}~\bibnamefont
  {Liang}},\ }\bibfield  {title} {\enquote {\bibinfo {title} {Appendix {WKB}
  quantization},}\ }in\ \href {https://ChaosBook.org/paper.shtml#appendWKB}
  {\emph {\bibinfo {booktitle} {{Chaos: Classical and Quantum}}}},\ \bibinfo
  {editor} {edited by\ \bibinfo {editor} {\bibfnamefont {P.}~\bibnamefont
  {Cvitanovi{\'c}}}, \bibinfo {editor} {\bibfnamefont {R.}~\bibnamefont
  {Artuso}}, \bibinfo {editor} {\bibfnamefont {R.}~\bibnamefont {Mainieri}},
  \bibinfo {editor} {\bibfnamefont {G.}~\bibnamefont {Tanner}},\ and\ \bibinfo
  {editor} {\bibfnamefont {G.}~\bibnamefont {Vattay}}}\ (\bibinfo  {publisher}
  {Niels Bohr Inst.},\ \bibinfo {address} {Copenhagen},\ \bibinfo {year}
  {2025})\BibitemShut {NoStop}%
\bibitem [{\citenamefont {Van~Vleck}(1928)}]{VanVleck28}%
  \BibitemOpen
  \bibfield  {author} {\bibinfo {author} {\bibfnamefont {J.~H.}\ \bibnamefont
  {Van~Vleck}},\ }\bibfield  {title} {\enquote {\bibinfo {title} {The
  correspondence principle in the statistical interpretation of quantum
  mechanics},}\ }\href {https://doi.org/10.1073/pnas.14.2.178} {\bibfield
  {journal} {\bibinfo  {journal} {Proc. Natl. Acad. Sci.}\ }\textbf {\bibinfo
  {volume} {14}},\ \bibinfo {pages} {178--188} (\bibinfo {year}
  {1928})}\BibitemShut {NoStop}%
\bibitem [{\citenamefont {Colin~de Verdi{\`{e}}re}(2007)}]{Verdiere07}%
  \BibitemOpen
  \bibfield  {author} {\bibinfo {author} {\bibfnamefont {Y.}~\bibnamefont
  {Colin~de Verdi{\`{e}}re}},\ }\bibfield  {title} {\enquote {\bibinfo {title}
  {Spectrum of the {Laplace} operator and periodic geodesics: thirty years
  after},}\ }\href {https://doi.org/10.5802/aif.2339} {\bibfield  {journal}
  {\bibinfo  {journal} {Ann. Inst. Fourier}\ }\textbf {\bibinfo {volume}
  {57}},\ \bibinfo {pages} {2429--2463} (\bibinfo {year} {2007})}\BibitemShut
  {NoStop}%
\bibitem [{\citenamefont {Levit}\ and\ \citenamefont
  {Smilansky}(1977{\natexlab{a}})}]{LevSmi77}%
  \BibitemOpen
  \bibfield  {author} {\bibinfo {author} {\bibfnamefont {S.}~\bibnamefont
  {Levit}}\ and\ \bibinfo {author} {\bibfnamefont {U.}~\bibnamefont
  {Smilansky}},\ }\bibfield  {title} {\enquote {\bibinfo {title} {A theorem on
  infinite products of eigenvalues of {Sturm-Liouville} type operators},}\
  }\href {https://doi.org/10.1090/s0002-9939-1977-0457836-8} {\bibfield
  {journal} {\bibinfo  {journal} {Proc. Amer. Math. Soc.}\ }\textbf {\bibinfo
  {volume} {65}},\ \bibinfo {pages} {299--299} (\bibinfo {year}
  {1977}{\natexlab{a}})}\BibitemShut {NoStop}%
\bibitem [{\citenamefont {Levit}\ and\ \citenamefont
  {Smilansky}(1977{\natexlab{b}})}]{LevSmi77a}%
  \BibitemOpen
  \bibfield  {author} {\bibinfo {author} {\bibfnamefont {S.}~\bibnamefont
  {Levit}}\ and\ \bibinfo {author} {\bibfnamefont {U.}~\bibnamefont
  {Smilansky}},\ }\bibfield  {title} {\enquote {\bibinfo {title} {A new
  approach to {Gaussian} path integrals and the evaluation of the semiclassical
  propagator},}\ }\href {https://doi.org/10.1016/0003-4916(77)90269-x}
  {\bibfield  {journal} {\bibinfo  {journal} {Ann. Phys.}\ }\textbf {\bibinfo
  {volume} {103}},\ \bibinfo {pages} {198--207} (\bibinfo {year}
  {1977}{\natexlab{b}})}\BibitemShut {NoStop}%
\bibitem [{\citenamefont {Maloney}\ and\ \citenamefont
  {Witten}(2010)}]{MalWit07}%
  \BibitemOpen
  \bibfield  {author} {\bibinfo {author} {\bibfnamefont {A.}~\bibnamefont
  {Maloney}}\ and\ \bibinfo {author} {\bibfnamefont {E.}~\bibnamefont
  {Witten}},\ }\bibfield  {title} {\enquote {\bibinfo {title} {Quantum gravity
  partition functions in three dimensions},}\ }\href
  {https://doi.org/10.1007/jhep02(2010)029} {\bibfield  {journal} {\bibinfo
  {journal} {J. High Energy Phys.}\ }\textbf {\bibinfo {volume} {2010}},\
  \bibinfo {pages} {029} (\bibinfo {year} {2010})}\BibitemShut {NoStop}%
\bibitem [{\citenamefont {Castro}\ \emph {et~al.}(2012)\citenamefont {Castro},
  \citenamefont {Gaberdiel}, \citenamefont {Hartman}, \citenamefont {Maloney},\
  and\ \citenamefont {Volpato}}]{CGHMV12}%
  \BibitemOpen
  \bibfield  {author} {\bibinfo {author} {\bibfnamefont {A.}~\bibnamefont
  {Castro}}, \bibinfo {author} {\bibfnamefont {M.~R.}\ \bibnamefont
  {Gaberdiel}}, \bibinfo {author} {\bibfnamefont {T.}~\bibnamefont {Hartman}},
  \bibinfo {author} {\bibfnamefont {A.}~\bibnamefont {Maloney}},\ and\ \bibinfo
  {author} {\bibfnamefont {R.}~\bibnamefont {Volpato}},\ }\bibfield  {title}
  {\enquote {\bibinfo {title} {Gravity dual of the {Ising} model},}\ }\href
  {https://doi.org/10.1103/physrevd.85.024032} {\bibfield  {journal} {\bibinfo
  {journal} {Phys. Rev. D}\ }\textbf {\bibinfo {volume} {85}},\ \bibinfo
  {pages} {024032} (\bibinfo {year} {2012})}\BibitemShut {NoStop}%
\bibitem [{\citenamefont {Hasselblatt}\ and\ \citenamefont
  {Pesin}(2008)}]{HasPes08}%
  \BibitemOpen
  \bibfield  {author} {\bibinfo {author} {\bibfnamefont {B.}~\bibnamefont
  {Hasselblatt}}\ and\ \bibinfo {author} {\bibfnamefont {Y.~B.}\ \bibnamefont
  {Pesin}},\ }\bibfield  {title} {\enquote {\bibinfo {title} {Hyperbolic
  dynamics},}\ }\href {https://doi.org/10.4249/scholarpedia.2208} {\bibfield
  {journal} {\bibinfo  {journal} {Scholarpedia}\ }\textbf {\bibinfo {volume}
  {3}},\ \bibinfo {pages} {2208} (\bibinfo {year} {2008})}\BibitemShut
  {NoStop}%
\bibitem [{\citenamefont {Ivashkevich}, \citenamefont {Izmailian},\ and\
  \citenamefont {Hu}(2002)}]{IvIzHu02}%
  \BibitemOpen
  \bibfield  {author} {\bibinfo {author} {\bibfnamefont {E.~V.}\ \bibnamefont
  {Ivashkevich}}, \bibinfo {author} {\bibfnamefont {N.~S.}\ \bibnamefont
  {Izmailian}},\ and\ \bibinfo {author} {\bibfnamefont {C.-K.}\ \bibnamefont
  {Hu}},\ }\bibfield  {title} {\enquote {\bibinfo {title} {Kronecker's double
  series and exact asymptotic expansions for free models of statistical
  mechanics on torus},}\ }\href {https://doi.org/10.1088/0305-4470/35/27/302}
  {\bibfield  {journal} {\bibinfo  {journal} {J. Phys. A}\ }\textbf {\bibinfo
  {volume} {35}},\ \bibinfo {pages} {5543--5561} (\bibinfo {year}
  {2002})}\BibitemShut {NoStop}%
\bibitem [{\citenamefont {Campos}, \citenamefont {Sierra},\ and\ \citenamefont
  {L{\'o}pez}(2019)}]{CaSiLo19}%
  \BibitemOpen
  \bibfield  {author} {\bibinfo {author} {\bibfnamefont {M.}~\bibnamefont
  {Campos}}, \bibinfo {author} {\bibfnamefont {G.}~\bibnamefont {Sierra}},\
  and\ \bibinfo {author} {\bibfnamefont {E.}~\bibnamefont {L{\'o}pez}},\
  }\bibfield  {title} {\enquote {\bibinfo {title} {Tensor renormalization group
  in bosonic field theory},}\ }\href
  {https://doi.org/10.1103/physrevb.100.195106} {\bibfield  {journal} {\bibinfo
   {journal} {Phys. Rev. B}\ }\textbf {\bibinfo {volume} {100}},\ \bibinfo
  {pages} {195106} (\bibinfo {year} {2019})}\BibitemShut {NoStop}%
\bibitem [{\citenamefont {Simon}(1982)}]{Simon82}%
  \BibitemOpen
  \bibfield  {author} {\bibinfo {author} {\bibfnamefont {B.}~\bibnamefont
  {Simon}},\ }\bibfield  {title} {\enquote {\bibinfo {title} {Almost periodic
  {Schr{\"o}dinger} operators: {A review}},}\ }\href
  {https://doi.org/10.1016/s0196-8858(82)80018-3} {\bibfield  {journal}
  {\bibinfo  {journal} {Adv. Appl. Math.}\ }\textbf {\bibinfo {volume} {3}},\
  \bibinfo {pages} {463--490} (\bibinfo {year} {1982})}\BibitemShut {NoStop}%
\bibitem [{\citenamefont {Simons}(1997)}]{Simons97}%
  \BibitemOpen
  \bibfield  {author} {\bibinfo {author} {\bibfnamefont {S.}~\bibnamefont
  {Simons}},\ }\bibfield  {title} {\enquote {\bibinfo {title} {Analytical
  inversion of a particular type of banded matrix},}\ }\href
  {https://doi.org/10.1088/0305-4470/30/2/034} {\bibfield  {journal} {\bibinfo
  {journal} {J. Phys. A}\ }\textbf {\bibinfo {volume} {30}},\ \bibinfo {pages}
  {755} (\bibinfo {year} {1997})}\BibitemShut {NoStop}%
\bibitem [{\citenamefont {Toda}(1989)}]{Toda89}%
  \BibitemOpen
  \bibfield  {author} {\bibinfo {author} {\bibfnamefont {M.}~\bibnamefont
  {Toda}},\ }\href {https://doi.org/10.1007/978-3-642-83219-2} {\emph {\bibinfo
  {title} {{Theory of Nonlinear Lattices}}}}\ (\bibinfo  {publisher}
  {Springer},\ \bibinfo {address} {Berlin},\ \bibinfo {year}
  {1989})\BibitemShut {NoStop}%
\bibitem [{\citenamefont {Bunimovich}\ and\ \citenamefont
  {Sinai}(1988)}]{BunSin88}%
  \BibitemOpen
  \bibfield  {author} {\bibinfo {author} {\bibfnamefont {L.~A.}\ \bibnamefont
  {Bunimovich}}\ and\ \bibinfo {author} {\bibfnamefont {Y.~G.}\ \bibnamefont
  {Sinai}},\ }\bibfield  {title} {\enquote {\bibinfo {title} {Spacetime chaos
  in coupled map lattices},}\ }\href
  {https://doi.org/10.1088/0951-7715/1/4/001} {\bibfield  {journal} {\bibinfo
  {journal} {Nonlinearity}\ }\textbf {\bibinfo {volume} {1}},\ \bibinfo {pages}
  {491} (\bibinfo {year} {1988})}\BibitemShut {NoStop}%
\bibitem [{\citenamefont {Politi}\ and\ \citenamefont
  {Torcini}(1992{\natexlab{a}})}]{PolTor92b}%
  \BibitemOpen
  \bibfield  {author} {\bibinfo {author} {\bibfnamefont {A.}~\bibnamefont
  {Politi}}\ and\ \bibinfo {author} {\bibfnamefont {A.}~\bibnamefont
  {Torcini}},\ }\bibfield  {title} {\enquote {\bibinfo {title} {Towards a
  statistical mechanics of spatiotemporal chaos},}\ }\href
  {https://doi.org/10.1103/PhysRevLett.69.3421} {\bibfield  {journal} {\bibinfo
   {journal} {Phys. Rev. Lett.}\ }\textbf {\bibinfo {volume} {69}},\ \bibinfo
  {pages} {3421--3424} (\bibinfo {year} {1992}{\natexlab{a}})}\BibitemShut
  {NoStop}%
\bibitem [{\citenamefont {Pethel}, \citenamefont {Corron},\ and\ \citenamefont
  {Bollt}(2006)}]{PetCorBol06}%
  \BibitemOpen
  \bibfield  {author} {\bibinfo {author} {\bibfnamefont {S.~D.}\ \bibnamefont
  {Pethel}}, \bibinfo {author} {\bibfnamefont {N.~J.}\ \bibnamefont {Corron}},\
  and\ \bibinfo {author} {\bibfnamefont {E.}~\bibnamefont {Bollt}},\ }\bibfield
   {title} {\enquote {\bibinfo {title} {Symbolic dynamics of coupled map
  lattices},}\ }\href {https://doi.org/10.1103/PhysRevLett.96.034105}
  {\bibfield  {journal} {\bibinfo  {journal} {Phys. Rev. Lett.}\ }\textbf
  {\bibinfo {volume} {96}},\ \bibinfo {pages} {034105} (\bibinfo {year}
  {2006})}\BibitemShut {NoStop}%
\bibitem [{\citenamefont {Pethel}, \citenamefont {Corron},\ and\ \citenamefont
  {Bollt}(2007)}]{PetCorBol07}%
  \BibitemOpen
  \bibfield  {author} {\bibinfo {author} {\bibfnamefont {S.~D.}\ \bibnamefont
  {Pethel}}, \bibinfo {author} {\bibfnamefont {N.~J.}\ \bibnamefont {Corron}},\
  and\ \bibinfo {author} {\bibfnamefont {E.}~\bibnamefont {Bollt}},\ }\bibfield
   {title} {\enquote {\bibinfo {title} {Deconstructing spatiotemporal chaos
  using local symbolic dynamics},}\ }\href
  {https://doi.org/10.1103/PhysRevLett.99.214101} {\bibfield  {journal}
  {\bibinfo  {journal} {Phys. Rev. Lett.}\ }\textbf {\bibinfo {volume} {99}},\
  \bibinfo {pages} {214101} (\bibinfo {year} {2007})}\BibitemShut {NoStop}%
\bibitem [{\citenamefont {Chow}, \citenamefont {Mallet-Paret},\ and\
  \citenamefont {Van~Vleck}(1996)}]{ChMaVa96}%
  \BibitemOpen
  \bibfield  {author} {\bibinfo {author} {\bibfnamefont {S.-N.}\ \bibnamefont
  {Chow}}, \bibinfo {author} {\bibfnamefont {J.}~\bibnamefont {Mallet-Paret}},\
  and\ \bibinfo {author} {\bibfnamefont {E.~S.}\ \bibnamefont {Van~Vleck}},\
  }\bibfield  {title} {\enquote {\bibinfo {title} {Dynamics of lattice
  differential equations},}\ }\href {https://doi.org/10.1142/s0218127496000977}
  {\bibfield  {journal} {\bibinfo  {journal} {Int. J. Bifurcat. Chaos}\
  }\textbf {\bibinfo {volume} {06}},\ \bibinfo {pages} {1605--1621} (\bibinfo
  {year} {1996})}\BibitemShut {NoStop}%
\bibitem [{\citenamefont {Chow}, \citenamefont {Mallet-Paret},\ and\
  \citenamefont {Shen}(1998)}]{ChMaSh98}%
  \BibitemOpen
  \bibfield  {author} {\bibinfo {author} {\bibfnamefont {S.-N.}\ \bibnamefont
  {Chow}}, \bibinfo {author} {\bibfnamefont {J.}~\bibnamefont {Mallet-Paret}},\
  and\ \bibinfo {author} {\bibfnamefont {W.}~\bibnamefont {Shen}},\ }\bibfield
  {title} {\enquote {\bibinfo {title} {Traveling waves in lattice dynamical
  systems},}\ }\href {https://doi.org/10.1006/jdeq.1998.3478} {\bibfield
  {journal} {\bibinfo  {journal} {J. Diff. Equ.}\ }\textbf {\bibinfo {volume}
  {149}},\ \bibinfo {pages} {248--291} (\bibinfo {year} {1998})}\BibitemShut
  {NoStop}%
\bibitem [{\citenamefont {Mallet-Paret}\ and\ \citenamefont
  {Chow}(1995{\natexlab{a}})}]{ChoMal95a}%
  \BibitemOpen
  \bibfield  {author} {\bibinfo {author} {\bibfnamefont {J.}~\bibnamefont
  {Mallet-Paret}}\ and\ \bibinfo {author} {\bibfnamefont {S.-N.}\ \bibnamefont
  {Chow}},\ }\bibfield  {title} {\enquote {\bibinfo {title} {Pattern formation
  and spatial chaos in lattice dynamical systems. {I}},}\ }\href
  {https://doi.org/10.1109/81.473583} {\bibfield  {journal} {\bibinfo
  {journal} {IEEE Trans. Circuits Systems I Fund. Theory Appl.}\ }\textbf
  {\bibinfo {volume} {42}},\ \bibinfo {pages} {746--751} (\bibinfo {year}
  {1995}{\natexlab{a}})}\BibitemShut {NoStop}%
\bibitem [{\citenamefont {Mallet-Paret}\ and\ \citenamefont
  {Chow}(1995{\natexlab{b}})}]{ChoMal95b}%
  \BibitemOpen
  \bibfield  {author} {\bibinfo {author} {\bibfnamefont {J.}~\bibnamefont
  {Mallet-Paret}}\ and\ \bibinfo {author} {\bibfnamefont {S.-N.}\ \bibnamefont
  {Chow}},\ }\bibfield  {title} {\enquote {\bibinfo {title} {Pattern formation
  and spatial chaos in lattice dynamical systems. {II}},}\ }\href
  {https://doi.org/10.1109/81.473584} {\bibfield  {journal} {\bibinfo
  {journal} {IEEE Trans. Circuits Systems I Fund. Theory Appl.}\ }\textbf
  {\bibinfo {volume} {42}},\ \bibinfo {pages} {752--756} (\bibinfo {year}
  {1995}{\natexlab{b}})}\BibitemShut {NoStop}%
\bibitem [{\citenamefont {Coutinho}\ and\ \citenamefont
  {Fernandez}(1997)}]{CouFer97}%
  \BibitemOpen
  \bibfield  {author} {\bibinfo {author} {\bibfnamefont {R.}~\bibnamefont
  {Coutinho}}\ and\ \bibinfo {author} {\bibfnamefont {B.}~\bibnamefont
  {Fernandez}},\ }\bibfield  {title} {\enquote {\bibinfo {title} {Extended
  symbolic dynamics in bistable {CML: Existence} and stability of fronts},}\
  }\href {https://doi.org/10.1016/s0167-2789(97)82005-2} {\bibfield  {journal}
  {\bibinfo  {journal} {Physica D}\ }\textbf {\bibinfo {volume} {108}},\
  \bibinfo {pages} {60--80} (\bibinfo {year} {1997})}\BibitemShut {NoStop}%
\bibitem [{\citenamefont {Sterling}(1999)}]{SterlingThesis99}%
  \BibitemOpen
  \bibfield  {author} {\bibinfo {author} {\bibfnamefont {D.~G.}\ \bibnamefont
  {Sterling}},\ }\emph {\bibinfo {title} {{Anti-integrable Continuation and the
  Destruction of Chaos}}},\ \href
  {https://search.proquest.com/dissertations-theses/anti-integrable-continuation-destruction-chaos/docview/304508605/se-2?accountid=11107}
  {Ph.D. thesis},\ \bibinfo  {school} {Univ. Colorado}, \bibinfo {address}
  {Boulder, CO} (\bibinfo {year} {1999})\BibitemShut {NoStop}%
\bibitem [{\citenamefont {Just}(2005)}]{Just05}%
  \BibitemOpen
  \bibfield  {author} {\bibinfo {author} {\bibfnamefont {W.}~\bibnamefont
  {Just}},\ }\bibfield  {title} {\enquote {\bibinfo {title} {On symbolic
  dynamics of space-time chaotic models},}\ }in\ \href
  {https://doi.org/10.1007/3-540-26869-3_15} {\emph {\bibinfo {booktitle}
  {{Collective Dynamics of Nonlinear and Disordered Systems}}}}\ (\bibinfo
  {publisher} {Springer},\ \bibinfo {year} {2005})\ pp.\ \bibinfo {pages}
  {339--357}\BibitemShut {NoStop}%
\bibitem [{\citenamefont {MacKay}(2005)}]{MacKay05}%
  \BibitemOpen
  \bibfield  {author} {\bibinfo {author} {\bibfnamefont {R.~S.}\ \bibnamefont
  {MacKay}},\ }\bibfield  {title} {\enquote {\bibinfo {title} {Indecomposable
  coupled map lattices with non-unique phase},}\ }in\ \href
  {https://doi.org/10.1007/11360810_4} {\emph {\bibinfo {booktitle} {{{Dynamics
  of Coupled Map Lattices and of Related Spatially Extended Systems}}}}},\
  \bibinfo {editor} {edited by\ \bibinfo {editor} {\bibfnamefont {J.-R.}\
  \bibnamefont {Chazottes}}\ and\ \bibinfo {editor} {\bibfnamefont
  {B.}~\bibnamefont {Fernandez}}}\ (\bibinfo  {publisher} {Springer},\ \bibinfo
  {year} {2005})\ pp.\ \bibinfo {pages} {65--94}\BibitemShut {NoStop}%
\bibitem [{\citenamefont {Perry}(2023)}]{Perry23}%
  \BibitemOpen
  \bibfield  {author} {\bibinfo {author} {\bibfnamefont {K.}~\bibnamefont
  {Perry}},\ }\href
  {https://www.washingtonpost.com/parenting/interactive/2023/lego-bricks-colors-history/}
  {\enquote {\bibinfo {title} {How {Lego} bricks went from five colors to
  nearly 200},}\ } (\bibinfo {year} {2023})\BibitemShut {NoStop}%
\bibitem [{\citenamefont {Aubry}\ and\ \citenamefont
  {Abramovici}(1990)}]{AuAb90}%
  \BibitemOpen
  \bibfield  {author} {\bibinfo {author} {\bibfnamefont {S.}~\bibnamefont
  {Aubry}}\ and\ \bibinfo {author} {\bibfnamefont {G.}~\bibnamefont
  {Abramovici}},\ }\bibfield  {title} {\enquote {\bibinfo {title} {Chaotic
  trajectories in the standard map. {The} concept of anti-integrability},}\
  }\href {https://doi.org/10.1016/0167-2789(90)90133-A} {\bibfield  {journal}
  {\bibinfo  {journal} {Physica D}\ }\textbf {\bibinfo {volume} {43}},\
  \bibinfo {pages} {199--219} (\bibinfo {year} {1990})}\BibitemShut {NoStop}%
\bibitem [{\citenamefont {Aubry}(1995)}]{aub95ant}%
  \BibitemOpen
  \bibfield  {author} {\bibinfo {author} {\bibfnamefont {S.}~\bibnamefont
  {Aubry}},\ }\bibfield  {title} {\enquote {\bibinfo {title}
  {Anti-integrability in dynamical and variational problems},}\ }\href
  {https://doi.org/10.1016/0167-2789(95)00109-h} {\bibfield  {journal}
  {\bibinfo  {journal} {Physica D}\ }\textbf {\bibinfo {volume} {86}},\
  \bibinfo {pages} {284--296} (\bibinfo {year} {1995})}\BibitemShut {NoStop}%
\bibitem [{\citenamefont {Sterling}\ and\ \citenamefont
  {Meiss}(1998)}]{StMeiss98}%
  \BibitemOpen
  \bibfield  {author} {\bibinfo {author} {\bibfnamefont {D.}~\bibnamefont
  {Sterling}}\ and\ \bibinfo {author} {\bibfnamefont {J.~D.}\ \bibnamefont
  {Meiss}},\ }\bibfield  {title} {\enquote {\bibinfo {title} {Computing
  periodic orbits using the anti-integrable limit},}\ }\href
  {https://doi.org/10.1016/s0375-9601(98)00094-2} {\bibfield  {journal}
  {\bibinfo  {journal} {Phys. Lett. A}\ }\textbf {\bibinfo {volume} {241}},\
  \bibinfo {pages} {46--52} (\bibinfo {year} {1998})}\BibitemShut {NoStop}%
\bibitem [{\citenamefont {Beck}(2002)}]{Beck02}%
  \BibitemOpen
  \bibfield  {author} {\bibinfo {author} {\bibfnamefont {C.}~\bibnamefont
  {Beck}},\ }\href {https://doi.org/10.1142/9789812778239} {\emph {\bibinfo
  {title} {{Spatio-Temporal Chaos and Vacuum Fluctuations of Quantized
  Fields}}}}\ (\bibinfo  {publisher} {World Scientific},\ \bibinfo {address}
  {Singapore},\ \bibinfo {year} {2002})\BibitemShut {NoStop}%
\bibitem [{\citenamefont {Friedland}\ and\ \citenamefont
  {Milnor}(1989)}]{FriMil89}%
  \BibitemOpen
  \bibfield  {author} {\bibinfo {author} {\bibfnamefont {S.}~\bibnamefont
  {Friedland}}\ and\ \bibinfo {author} {\bibfnamefont {J.}~\bibnamefont
  {Milnor}},\ }\bibfield  {title} {\enquote {\bibinfo {title} {Dynamical
  properties of plane polynomial automorphisms},}\ }\href
  {https://doi.org/10.1017/s014338570000482x} {\bibfield  {journal} {\bibinfo
  {journal} {Ergodic Theory Dynam. Systems}\ }\textbf {\bibinfo {volume} {9}},\
  \bibinfo {pages} {67--99} (\bibinfo {year} {1989})}\BibitemShut {NoStop}%
\bibitem [{\citenamefont {Dullin}\ and\ \citenamefont
  {Meiss}(2000)}]{DulMei00}%
  \BibitemOpen
  \bibfield  {author} {\bibinfo {author} {\bibfnamefont {H.~R.}\ \bibnamefont
  {Dullin}}\ and\ \bibinfo {author} {\bibfnamefont {J.~D.}\ \bibnamefont
  {Meiss}},\ }\bibfield  {title} {\enquote {\bibinfo {title} {Generalized
  {H{{\'e}}non} maps: the cubic diffeomorphisms of the plane},}\ }\href
  {https://doi.org/10.1016/s0167-2789(00)00105-6} {\bibfield  {journal}
  {\bibinfo  {journal} {Physica D}\ }\textbf {\bibinfo {volume} {143}},\
  \bibinfo {pages} {262--289} (\bibinfo {year} {2000})}\BibitemShut {NoStop}%
\bibitem [{\citenamefont {Li}\ and\ \citenamefont {Malkin}(2004)}]{LiMal04}%
  \BibitemOpen
  \bibfield  {author} {\bibinfo {author} {\bibfnamefont {M.-C.}\ \bibnamefont
  {Li}}\ and\ \bibinfo {author} {\bibfnamefont {M.}~\bibnamefont {Malkin}},\
  }\bibfield  {title} {\enquote {\bibinfo {title} {Bounded nonwandering sets
  for polynomial mappings},}\ }\href
  {https://doi.org/10.1023/b:jods.0000034436.39278.37} {\bibfield  {journal}
  {\bibinfo  {journal} {J. Dynam. Control Systems}\ }\textbf {\bibinfo {volume}
  {10}},\ \bibinfo {pages} {377--389} (\bibinfo {year} {2004})}\BibitemShut
  {NoStop}%
\bibitem [{\citenamefont {M{\"u}nster}(2010)}]{Munster10}%
  \BibitemOpen
  \bibfield  {author} {\bibinfo {author} {\bibfnamefont {G.}~\bibnamefont
  {M{\"u}nster}},\ }\bibfield  {title} {\enquote {\bibinfo {title} {Lattice
  quantum field theory},}\ }\href {https://doi.org/10.4249/scholarpedia.8613}
  {\bibfield  {journal} {\bibinfo  {journal} {Scholarpedia}\ }\textbf {\bibinfo
  {volume} {5}},\ \bibinfo {pages} {8613} (\bibinfo {year} {2010})}\BibitemShut
  {NoStop}%
\bibitem [{\citenamefont {Anastassiou}, \citenamefont {Bountis},\ and\
  \citenamefont {B{\"a}cker}(2017)}]{AnBoBa17}%
  \BibitemOpen
  \bibfield  {author} {\bibinfo {author} {\bibfnamefont {S.}~\bibnamefont
  {Anastassiou}}, \bibinfo {author} {\bibfnamefont {A.}~\bibnamefont
  {Bountis}},\ and\ \bibinfo {author} {\bibfnamefont {A.}~\bibnamefont
  {B{\"a}cker}},\ }\bibfield  {title} {\enquote {\bibinfo {title} {Homoclinic
  points of {2D} and {4D} maps via the parametrization method},}\ }\href
  {https://doi.org/10.1088/1361-6544/aa7e9b} {\bibfield  {journal} {\bibinfo
  {journal} {Nonlinearity}\ }\textbf {\bibinfo {volume} {30}},\ \bibinfo
  {pages} {3799--3820} (\bibinfo {year} {2017})}\BibitemShut {NoStop}%
\bibitem [{\citenamefont {Anastassiou}, \citenamefont {Bountis},\ and\
  \citenamefont {B{\"a}cker}(2018)}]{AnBoBa18}%
  \BibitemOpen
  \bibfield  {author} {\bibinfo {author} {\bibfnamefont {S.}~\bibnamefont
  {Anastassiou}}, \bibinfo {author} {\bibfnamefont {A.}~\bibnamefont
  {Bountis}},\ and\ \bibinfo {author} {\bibfnamefont {A.}~\bibnamefont
  {B{\"a}cker}},\ }\bibfield  {title} {\enquote {\bibinfo {title} {Recent
  results on the dynamics of higher-dimensional {H{{\'e}}non} maps},}\ }\href
  {https://doi.org/10.1134/s156035471802003x} {\bibfield  {journal} {\bibinfo
  {journal} {Regul. Chaotic Dyn.}\ }\textbf {\bibinfo {volume} {23}},\ \bibinfo
  {pages} {161--177} (\bibinfo {year} {2018})}\BibitemShut {NoStop}%
\bibitem [{\citenamefont {Anastassiou}(2021)}]{Anastassiou21}%
  \BibitemOpen
  \bibfield  {author} {\bibinfo {author} {\bibfnamefont {S.}~\bibnamefont
  {Anastassiou}},\ }\bibfield  {title} {\enquote {\bibinfo {title} {Complicated
  behavior in cubic {H{{\'e}}non} maps},}\ }\href
  {https://doi.org/10.1134/s0040577921050032} {\bibfield  {journal} {\bibinfo
  {journal} {Theoret. Math. Phys.}\ }\textbf {\bibinfo {volume} {207}},\
  \bibinfo {pages} {572--578} (\bibinfo {year} {2021})}\BibitemShut {NoStop}%
\bibitem [{\citenamefont
  {Cvitanovi{\'c}}(2022{\natexlab{a}})}]{CvitanovicYT02b}%
  \BibitemOpen
  \bibfield  {author} {\bibinfo {author} {\bibfnamefont {P.}~\bibnamefont
  {Cvitanovi{\'c}}},\ }\href {https://youtu.be/F-iOrF-G-1M} {\enquote {\bibinfo
  {title} {Strong coupling field theory lives on a fractal},}\ } (\bibinfo
  {year} {2022}{\natexlab{a}})\BibitemShut {NoStop}%
\bibitem [{\citenamefont {Lind}\ and\ \citenamefont
  {Marcus}(1995)}]{LindMar95}%
  \BibitemOpen
  \bibfield  {author} {\bibinfo {author} {\bibfnamefont {D.~A.}\ \bibnamefont
  {Lind}}\ and\ \bibinfo {author} {\bibfnamefont {B.}~\bibnamefont {Marcus}},\
  }\href {https://doi.org/10.1017/cbo9780511626302} {\emph {\bibinfo {title}
  {{An Introduction to Symbolic Dynamics and Coding}}}}\ (\bibinfo  {publisher}
  {Cambridge Univ. Press},\ \bibinfo {address} {Cambridge},\ \bibinfo {year}
  {1995})\BibitemShut {NoStop}%
\bibitem [{\citenamefont {Pollicott}(2001)}]{Pollicott01}%
  \BibitemOpen
  \bibfield  {author} {\bibinfo {author} {\bibfnamefont {M.}~\bibnamefont
  {Pollicott}},\ }\bibfield  {title} {\enquote {\bibinfo {title} {Dynamical
  zeta functions},}\ }in\ \href {https://doi.org/10.1090/pspum/069} {\emph
  {\bibinfo {booktitle} {{Smooth Ergodic Theory and Its Applications}}}},\
  Vol.~\bibinfo {volume} {69},\ \bibinfo {editor} {edited by\ \bibinfo {editor}
  {\bibfnamefont {A.}~\bibnamefont {Katok}}, \bibinfo {editor} {\bibfnamefont
  {R.}~\bibnamefont {de~la Llave}}, \bibinfo {editor} {\bibfnamefont {Y.~B.}\
  \bibnamefont {Pesin}},\ and\ \bibinfo {editor} {\bibfnamefont
  {H.}~\bibnamefont {Weiss}}}\ (\bibinfo  {publisher} {Amer. Math. Soc.},\
  \bibinfo {address} {Providence RI},\ \bibinfo {year} {2001})\ pp.\ \bibinfo
  {pages} {409--428}\BibitemShut {NoStop}%
\bibitem [{\citenamefont {Cimasoni}(2012)}]{Cimasoni12}%
  \BibitemOpen
  \bibfield  {author} {\bibinfo {author} {\bibfnamefont {D.}~\bibnamefont
  {Cimasoni}},\ }\bibfield  {title} {\enquote {\bibinfo {title} {The critical
  {Ising} model via {Kac-Ward} matrices},}\ }\href
  {https://doi.org/10.1007/s00220-012-1575-z} {\bibfield  {journal} {\bibinfo
  {journal} {Commun. Math. Phys.}\ }\textbf {\bibinfo {volume} {316}},\
  \bibinfo {pages} {99--126} (\bibinfo {year} {2012})}\BibitemShut {NoStop}%
\bibitem [{\citenamefont {Godsil}\ and\ \citenamefont
  {Royle}(2013)}]{GodRoy13}%
  \BibitemOpen
  \bibfield  {author} {\bibinfo {author} {\bibfnamefont {C.}~\bibnamefont
  {Godsil}}\ and\ \bibinfo {author} {\bibfnamefont {G.~F.}\ \bibnamefont
  {Royle}},\ }\href {https://doi.org/10.1007/978-1-4613-0163-9} {\emph
  {\bibinfo {title} {{Algebraic Graph Theory}}}}\ (\bibinfo  {publisher}
  {Springer},\ \bibinfo {address} {New York},\ \bibinfo {year}
  {2013})\BibitemShut {NoStop}%
\bibitem [{\citenamefont {Dorr}(1970)}]{Dorr70}%
  \BibitemOpen
  \bibfield  {author} {\bibinfo {author} {\bibfnamefont {F.~W.}\ \bibnamefont
  {Dorr}},\ }\bibfield  {title} {\enquote {\bibinfo {title} {The direct
  solution of the discrete {Poisson} equation on a rectangle},}\ }\href
  {https://doi.org/10.1137/1012045} {\bibfield  {journal} {\bibinfo  {journal}
  {SIAM Rev.}\ }\textbf {\bibinfo {volume} {12}},\ \bibinfo {pages} {248--263}
  (\bibinfo {year} {1970})}\BibitemShut {NoStop}%
\bibitem [{\citenamefont {Fetter}\ and\ \citenamefont
  {Walecka}(2003)}]{FetWal03}%
  \BibitemOpen
  \bibfield  {author} {\bibinfo {author} {\bibfnamefont {A.~L.}\ \bibnamefont
  {Fetter}}\ and\ \bibinfo {author} {\bibfnamefont {J.~D.}\ \bibnamefont
  {Walecka}},\ }\href@noop {} {\emph {\bibinfo {title} {{Theoretical Mechanics
  of Particles and Continua}}}}\ (\bibinfo  {publisher} {Dover},\ \bibinfo
  {address} {New York},\ \bibinfo {year} {2003})\BibitemShut {NoStop}%
\bibitem [{\citenamefont {Kadanoff}(2000)}]{Kadanoff00}%
  \BibitemOpen
  \bibfield  {author} {\bibinfo {author} {\bibfnamefont {L.~P.}\ \bibnamefont
  {Kadanoff}},\ }\href {https://doi.org/10.1142/4016} {\emph {\bibinfo {title}
  {{Statistical Physics: Statics, Dynamics and Renormalization}}}}\ (\bibinfo
  {publisher} {World Scientific},\ \bibinfo {address} {Singapore},\ \bibinfo
  {year} {2000})\BibitemShut {NoStop}%
\bibitem [{\citenamefont {Fradkin}(2013)}]{Fradkin13}%
  \BibitemOpen
  \bibfield  {author} {\bibinfo {author} {\bibfnamefont {E.}~\bibnamefont
  {Fradkin}},\ }\href {https://doi.org/10.1017/cbo9781139015509} {\emph
  {\bibinfo {title} {{Field Theories of Condensed Matter Physics}}}}\ (\bibinfo
   {publisher} {Cambridge Univ. Press},\ \bibinfo {address} {Cambridge UK},\
  \bibinfo {year} {2013})\BibitemShut {NoStop}%
\bibitem [{\citenamefont {Shankar}(2017)}]{Shankar17}%
  \BibitemOpen
  \bibfield  {author} {\bibinfo {author} {\bibfnamefont {R.}~\bibnamefont
  {Shankar}},\ }\href {https://doi.org/10.1017/9781139044349} {\emph {\bibinfo
  {title} {{Quantum Field Theory and Condensed Matter}}}}\ (\bibinfo
  {publisher} {Cambridge Univ. Press},\ \bibinfo {address} {Cambridge UK},\
  \bibinfo {year} {2017})\BibitemShut {NoStop}%
\bibitem [{\citenamefont {Marino}(2017)}]{Marino17}%
  \BibitemOpen
  \bibfield  {author} {\bibinfo {author} {\bibfnamefont {E.~C.}\ \bibnamefont
  {Marino}},\ }\href {https://doi.org/10.1017/9781139696548} {\emph {\bibinfo
  {title} {{Quantum Field Theory Approach to Condensed Matter Physics}}}}\
  (\bibinfo  {publisher} {Cambridge Univ. Press},\ \bibinfo {address}
  {Cambridge UK},\ \bibinfo {year} {2017})\BibitemShut {NoStop}%
\bibitem [{\citenamefont {{Wikipedia
  contributors}}(2025{\natexlab{a}})}]{wikiKleinGordon}%
  \BibitemOpen
  \bibfield  {author} {\bibinfo {author} {\bibnamefont {{Wikipedia
  contributors}}},\ }\href
  {https://en.wikipedia.org/wiki/Klein%E2%80%93Gordon_equation} {\enquote
  {\bibinfo {title} {Klein–gordon equation --- {Wikipedia, The Free
  Encyclopedia}},}\ } (\bibinfo {year} {2025}{\natexlab{a}})\BibitemShut
  {NoStop}%
\bibitem [{\citenamefont {Shimizu}(2012)}]{Shimizu12}%
  \BibitemOpen
  \bibfield  {author} {\bibinfo {author} {\bibfnamefont {Y.}~\bibnamefont
  {Shimizu}},\ }\bibfield  {title} {\enquote {\bibinfo {title} {Tensor
  renormalization group approach to a lattice boson model},}\ }\href
  {https://doi.org/10.1142/s0217732312500356} {\bibfield  {journal} {\bibinfo
  {journal} {Mod. Phys. Lett. A}\ }\textbf {\bibinfo {volume} {27}},\ \bibinfo
  {pages} {1250035} (\bibinfo {year} {2012})}\BibitemShut {NoStop}%
\bibitem [{\citenamefont {Campos}, \citenamefont {L{\'o}pez},\ and\
  \citenamefont {Sierra}(2021)}]{CaLoSi20}%
  \BibitemOpen
  \bibfield  {author} {\bibinfo {author} {\bibfnamefont {M.}~\bibnamefont
  {Campos}}, \bibinfo {author} {\bibfnamefont {E.}~\bibnamefont {L{\'o}pez}},\
  and\ \bibinfo {author} {\bibfnamefont {G.}~\bibnamefont {Sierra}},\
  }\bibfield  {title} {\enquote {\bibinfo {title} {Integrability and scattering
  of the boson field theory on a lattice},}\ }\href
  {https://doi.org/10.1088/1751-8121/abd5c7} {\bibfield  {journal} {\bibinfo
  {journal} {J. Phys. A}\ }\textbf {\bibinfo {volume} {54}},\ \bibinfo {pages}
  {055001} (\bibinfo {year} {2021})}\BibitemShut {NoStop}%
\bibitem [{\citenamefont
  {Cvitanovi{\'c}}(2022{\natexlab{b}})}]{CvitanovicYT02a}%
  \BibitemOpen
  \bibfield  {author} {\bibinfo {author} {\bibfnamefont {P.}~\bibnamefont
  {Cvitanovi{\'c}}},\ }\href {https://youtu.be/rTh_I0KOasY} {\enquote {\bibinfo
  {title} {A chaotic field theory: {Spatiotemporal} cat},}\ } (\bibinfo {year}
  {2022}{\natexlab{b}})\BibitemShut {NoStop}%
\bibitem [{\citenamefont {Keating}(1991)}]{Keating91}%
  \BibitemOpen
  \bibfield  {author} {\bibinfo {author} {\bibfnamefont {J.~P.}\ \bibnamefont
  {Keating}},\ }\bibfield  {title} {\enquote {\bibinfo {title} {The cat maps:
  quantum mechanics and classical motion},}\ }\href
  {https://doi.org/10.1088/0951-7715/4/2/006} {\bibfield  {journal} {\bibinfo
  {journal} {Nonlinearity}\ }\textbf {\bibinfo {volume} {4}},\ \bibinfo {pages}
  {309--341} (\bibinfo {year} {1991})}\BibitemShut {NoStop}%
\bibitem [{\citenamefont {Percival}\ and\ \citenamefont
  {Vivaldi}(1987)}]{PerViv}%
  \BibitemOpen
  \bibfield  {author} {\bibinfo {author} {\bibfnamefont {I.}~\bibnamefont
  {Percival}}\ and\ \bibinfo {author} {\bibfnamefont {F.}~\bibnamefont
  {Vivaldi}},\ }\bibfield  {title} {\enquote {\bibinfo {title} {A linear code
  for the sawtooth and cat maps},}\ }\href
  {https://doi.org/10.1016/0167-2789(87)90037-6} {\bibfield  {journal}
  {\bibinfo  {journal} {Physica D}\ }\textbf {\bibinfo {volume} {27}},\
  \bibinfo {pages} {373--386} (\bibinfo {year} {1987})}\BibitemShut {NoStop}%
\bibitem [{\citenamefont {Cheng}\ and\ \citenamefont
  {Seiberg}(2023)}]{CheSei22}%
  \BibitemOpen
  \bibfield  {author} {\bibinfo {author} {\bibfnamefont {M.}~\bibnamefont
  {Cheng}}\ and\ \bibinfo {author} {\bibfnamefont {N.}~\bibnamefont
  {Seiberg}},\ }\bibfield  {title} {\enquote {\bibinfo {title}
  {{Lieb-Schultz-Mattis, Luttinger, and 't Hooft} -- anomaly matching in
  lattice systems},}\ }\href {https://doi.org/10.21468/SciPostPhys.15.2.051}
  {\bibfield  {journal} {\bibinfo  {journal} {SciPost Phys.}\ }\textbf
  {\bibinfo {volume} {15}},\ \bibinfo {pages} {051} (\bibinfo {year}
  {2023})}\BibitemShut {NoStop}%
\bibitem [{\citenamefont {Fazza}\ and\ \citenamefont
  {Sulejmanpasic}(2023)}]{FazSul23}%
  \BibitemOpen
  \bibfield  {author} {\bibinfo {author} {\bibfnamefont {L.}~\bibnamefont
  {Fazza}}\ and\ \bibinfo {author} {\bibfnamefont {T.}~\bibnamefont
  {Sulejmanpasic}},\ }\bibfield  {title} {\enquote {\bibinfo {title} {Lattice
  quantum {Villain Hamiltonians}: compact scalars, {U(1)} gauge theories,
  fracton models and quantum {Ising} model dualities},}\ }\href
  {https://doi.org/10.1007/jhep05(2023)017} {\bibfield  {journal} {\bibinfo
  {journal} {J. High Energy Phys.}\ }\textbf {\bibinfo {volume} {2023}},\
  \bibinfo {pages} {17} (\bibinfo {year} {2023})}\BibitemShut {NoStop}%
\bibitem [{\citenamefont {Arnol'd}\ and\ \citenamefont {Avez}(1989)}]{ArnAve}%
  \BibitemOpen
  \bibfield  {author} {\bibinfo {author} {\bibfnamefont {V.~I.}\ \bibnamefont
  {Arnol'd}}\ and\ \bibinfo {author} {\bibfnamefont {A.}~\bibnamefont {Avez}},\
  }\href@noop {} {\emph {\bibinfo {title} {{Ergodic Problems of Classical
  Mechanics}}}}\ (\bibinfo  {publisher} {Addison-Wesley},\ \bibinfo {address}
  {Redwood City},\ \bibinfo {year} {1989})\BibitemShut {NoStop}%
\bibitem [{\citenamefont {Sturman}, \citenamefont {Ottino},\ and\ \citenamefont
  {Wiggins}(2006)}]{StOtWt06}%
  \BibitemOpen
  \bibfield  {author} {\bibinfo {author} {\bibfnamefont {R.}~\bibnamefont
  {Sturman}}, \bibinfo {author} {\bibfnamefont {J.~M.}\ \bibnamefont
  {Ottino}},\ and\ \bibinfo {author} {\bibfnamefont {S.}~\bibnamefont
  {Wiggins}},\ }\href {https://doi.org/10.1017/cbo9780511618116} {\emph
  {\bibinfo {title} {{The Mathematical Foundations of Mixing}}}}\ (\bibinfo
  {publisher} {Cambridge Univ. Press},\ \bibinfo {year} {2006})\BibitemShut
  {NoStop}%
\bibitem [{\citenamefont {Zee}(2010)}]{Zee10}%
  \BibitemOpen
  \bibfield  {author} {\bibinfo {author} {\bibfnamefont {A.}~\bibnamefont
  {Zee}},\ }\href@noop {} {\emph {\bibinfo {title} {{Quantum Field Theory in a
  Nutshell}}}},\ \bibinfo {edition} {2nd}\ ed.\ (\bibinfo  {publisher}
  {Princeton Univ. Press},\ \bibinfo {address} {Princeton NJ},\ \bibinfo {year}
  {2010})\BibitemShut {NoStop}%
\bibitem [{\citenamefont {Gradshteyn}\ and\ \citenamefont
  {Ryzhik}(2014)}]{GraRyz}%
  \BibitemOpen
  \bibfield  {author} {\bibinfo {author} {\bibfnamefont {I.~S.}\ \bibnamefont
  {Gradshteyn}}\ and\ \bibinfo {author} {\bibfnamefont {I.~M.}\ \bibnamefont
  {Ryzhik}},\ }\href@noop {} {\emph {\bibinfo {title} {{Tables of Integrals,
  Series and Products}}}},\ \bibinfo {edition} {8th}\ ed.\ (\bibinfo
  {publisher} {Elsevier LTD, Oxford},\ \bibinfo {address} {New York},\ \bibinfo
  {year} {2014})\BibitemShut {NoStop}%
\bibitem [{\citenamefont {Karve}, \citenamefont {Rose},\ and\ \citenamefont
  {Campbell}(2024)}]{KaRoCa24}%
  \BibitemOpen
  \bibfield  {author} {\bibinfo {author} {\bibfnamefont {N.}~\bibnamefont
  {Karve}}, \bibinfo {author} {\bibfnamefont {N.}~\bibnamefont {Rose}},\ and\
  \bibinfo {author} {\bibfnamefont {D.}~\bibnamefont {Campbell}},\ }\bibfield
  {title} {\enquote {\bibinfo {title} {Periodic orbits in
  {Fermi-Pasta-Ulam-Tsingou} systems},}\ }\href
  {https://doi.org/https://doi.org/10.1063/5.0223767} {\bibfield  {journal}
  {\bibinfo  {journal} {Chaos: An Interdisciplinary Journal of Nonlinear
  Science}\ }\textbf {\bibinfo {volume} {34}},\ \bibinfo {pages} {093117}
  (\bibinfo {year} {2024})}\BibitemShut {NoStop}%
\bibitem [{\citenamefont {Kaneko}(1983)}]{Kaneko83}%
  \BibitemOpen
  \bibfield  {author} {\bibinfo {author} {\bibfnamefont {K.}~\bibnamefont
  {Kaneko}},\ }\bibfield  {title} {\enquote {\bibinfo {title} {Transition from
  torus to chaos accompanied by frequency lockings with symmetry breaking: {In}
  connection with the coupled-logistic map},}\ }\href
  {https://doi.org/10.1143/PTP.69.1427} {\bibfield  {journal} {\bibinfo
  {journal} {Prog. Theor. Phys.}\ }\textbf {\bibinfo {volume} {69}},\ \bibinfo
  {pages} {1427--1442} (\bibinfo {year} {1983})}\BibitemShut {NoStop}%
\bibitem [{\citenamefont {Kaneko}(1984)}]{Kaneko84}%
  \BibitemOpen
  \bibfield  {author} {\bibinfo {author} {\bibfnamefont {K.}~\bibnamefont
  {Kaneko}},\ }\bibfield  {title} {\enquote {\bibinfo {title} {Period-doubling
  of kink-antikink patterns, quasiperiodicity in antiferro-like structures and
  spatial intermittency in coupled logistic lattice: {Towards} a prelude of a
  ``field theory of chaos''},}\ }\href {https://doi.org/10.1143/PTP.72.480}
  {\bibfield  {journal} {\bibinfo  {journal} {Prog. Theor. Phys.}\ }\textbf
  {\bibinfo {volume} {72}},\ \bibinfo {pages} {480--486} (\bibinfo {year}
  {1984})}\BibitemShut {NoStop}%
\bibitem [{\citenamefont {Kaneko}(1989)}]{Kaneko89}%
  \BibitemOpen
  \bibfield  {author} {\bibinfo {author} {\bibfnamefont {K.}~\bibnamefont
  {Kaneko}},\ }\bibfield  {title} {\enquote {\bibinfo {title} {Spatiotemporal
  chaos in one- and two-dimensional coupled map lattices},}\ }\href
  {https://doi.org/https://doi.org/10.1016/0167-2789(89)90117-6} {\bibfield
  {journal} {\bibinfo  {journal} {Physica D}\ }\textbf {\bibinfo {volume}
  {37}},\ \bibinfo {pages} {60--82} (\bibinfo {year} {1989})}\BibitemShut
  {NoStop}%
\bibitem [{\citenamefont {Lippolis}(2025)}]{Lippolis25}%
  \BibitemOpen
  \bibfield  {author} {\bibinfo {author} {\bibfnamefont {D.}~\bibnamefont
  {Lippolis}},\ }\href {https://doi.org/10.48550/arXiv.2510.12532} {\enquote
  {\bibinfo {title} {Spatiotemporal stability of synchronized coupled map
  lattice states},}\ } (\bibinfo {year} {2025})\BibitemShut {NoStop}%
\bibitem [{\citenamefont {Guckenheimer}\ and\ \citenamefont
  {Meloon}(2000)}]{GM00aut}%
  \BibitemOpen
  \bibfield  {author} {\bibinfo {author} {\bibfnamefont {J.}~\bibnamefont
  {Guckenheimer}}\ and\ \bibinfo {author} {\bibfnamefont {B.}~\bibnamefont
  {Meloon}},\ }\bibfield  {title} {\enquote {\bibinfo {title} {Computing
  periodic orbits and their bifurcations with automatic differentiation},}\
  }\href {https://doi.org/10.1137/s1064827599359278} {\bibfield  {journal}
  {\bibinfo  {journal} {SIAM J. Sci. Comput.}\ }\textbf {\bibinfo {volume}
  {22}},\ \bibinfo {pages} {951--985} (\bibinfo {year} {2000})}\BibitemShut
  {NoStop}%
\bibitem [{\citenamefont {Cvitanovi{\'c}}\ and\ \citenamefont
  {Lan}(2003)}]{CvitLanCrete02}%
  \BibitemOpen
  \bibfield  {author} {\bibinfo {author} {\bibfnamefont {P.}~\bibnamefont
  {Cvitanovi{\'c}}}\ and\ \bibinfo {author} {\bibfnamefont {Y.}~\bibnamefont
  {Lan}},\ }\bibfield  {title} {\enquote {\bibinfo {title} {Turbulent fields
  and their recurrences},}\ }in\ \href
  {https://doi.org/10.1142/9789812704641_0032} {\emph {\bibinfo {booktitle}
  {{Correlations and Fluctuations in QCD : Proceedings of 10. International
  Workshop on Multiparticle Production}}}},\ \bibinfo {editor} {edited by\
  \bibinfo {editor} {\bibfnamefont {N.}~\bibnamefont {Antoniou}}}\ (\bibinfo
  {publisher} {World Scientific},\ \bibinfo {address} {Singapore},\ \bibinfo
  {year} {2003})\ pp.\ \bibinfo {pages} {313--325}\BibitemShut {NoStop}%
\bibitem [{\citenamefont {Lan}\ and\ \citenamefont
  {Cvitanovi{\'c}}(2008)}]{lanCvit07}%
  \BibitemOpen
  \bibfield  {author} {\bibinfo {author} {\bibfnamefont {Y.}~\bibnamefont
  {Lan}}\ and\ \bibinfo {author} {\bibfnamefont {P.}~\bibnamefont
  {Cvitanovi{\'c}}},\ }\bibfield  {title} {\enquote {\bibinfo {title} {Unstable
  recurrent patterns in {Kuramoto-Sivashinsky} dynamics},}\ }\href
  {https://doi.org/10.1103/PhysRevE.78.026208} {\bibfield  {journal} {\bibinfo
  {journal} {Phys. Rev. E}\ }\textbf {\bibinfo {volume} {78}},\ \bibinfo
  {pages} {026208} (\bibinfo {year} {2008})}\BibitemShut {NoStop}%
\bibitem [{\citenamefont {Anderson}(1982)}]{Anderson82}%
  \BibitemOpen
  \bibfield  {author} {\bibinfo {author} {\bibfnamefont {B.~D.~O.}\
  \bibnamefont {Anderson}},\ }\bibfield  {title} {\enquote {\bibinfo {title}
  {Reverse-time diffusion equation models},}\ }\href
  {https://doi.org/10.1016/0304-4149(82)90051-5} {\bibfield  {journal}
  {\bibinfo  {journal} {Stochastic Process. Appl.}\ }\textbf {\bibinfo {volume}
  {12}},\ \bibinfo {pages} {313--326} (\bibinfo {year} {1982})}\BibitemShut
  {NoStop}%
\bibitem [{\citenamefont {Tabak}\ and\ \citenamefont
  {Vanden-Eijnden}(2010)}]{TabVan10}%
  \BibitemOpen
  \bibfield  {author} {\bibinfo {author} {\bibfnamefont {E.~G.}\ \bibnamefont
  {Tabak}}\ and\ \bibinfo {author} {\bibfnamefont {E.}~\bibnamefont
  {Vanden-Eijnden}},\ }\bibfield  {title} {\enquote {\bibinfo {title} {Density
  estimation by dual ascent of the log-likelihood},}\ }\href
  {https://doi.org/10.4310/cms.2010.v8.n1.a11} {\bibfield  {journal} {\bibinfo
  {journal} {Commun. Math. Sci.}\ }\textbf {\bibinfo {volume} {8}},\ \bibinfo
  {pages} {217--233} (\bibinfo {year} {2010})}\BibitemShut {NoStop}%
\bibitem [{\citenamefont {Parisi}\ and\ \citenamefont {Wu}(1981)}]{ParWu81}%
  \BibitemOpen
  \bibfield  {author} {\bibinfo {author} {\bibfnamefont {G.}~\bibnamefont
  {Parisi}}\ and\ \bibinfo {author} {\bibfnamefont {Y.~S.}\ \bibnamefont
  {Wu}},\ }\bibfield  {title} {\enquote {\bibinfo {title} {Perturbation-theory
  without gauge fixing},}\ }\href
  {https://doi.org/10.15161/oar.it/1447948233.36} {\bibfield  {journal}
  {\bibinfo  {journal} {Scientia Sinica}\ }\textbf {\bibinfo {volume} {24}},\
  \bibinfo {pages} {483--496} (\bibinfo {year} {1981})}\BibitemShut {NoStop}%
\bibitem [{\citenamefont {Beck}(1995)}]{Beck95}%
  \BibitemOpen
  \bibfield  {author} {\bibinfo {author} {\bibfnamefont {C.}~\bibnamefont
  {Beck}},\ }\bibfield  {title} {\enquote {\bibinfo {title} {Chaotic
  quantization of field theories},}\ }\href
  {https://doi.org/10.1088/0951-7715/8/3/008} {\bibfield  {journal} {\bibinfo
  {journal} {Nonlinearity}\ }\textbf {\bibinfo {volume} {8}},\ \bibinfo {pages}
  {423--441} (\bibinfo {year} {1995})}\BibitemShut {NoStop}%
\bibitem [{\citenamefont {Kitano}, \citenamefont {Takaura},\ and\ \citenamefont
  {Hashimoto}(2021)}]{KiTaHa21}%
  \BibitemOpen
  \bibfield  {author} {\bibinfo {author} {\bibfnamefont {R.}~\bibnamefont
  {Kitano}}, \bibinfo {author} {\bibfnamefont {H.}~\bibnamefont {Takaura}},\
  and\ \bibinfo {author} {\bibfnamefont {S.}~\bibnamefont {Hashimoto}},\
  }\bibfield  {title} {\enquote {\bibinfo {title} {Stochastic computation of
  g-2 in {QED}},}\ }\href {https://doi.org/10.1007/jhep05(2021)119} {\bibfield
  {journal} {\bibinfo  {journal} {J. High Energy Phys.}\ }\textbf {\bibinfo
  {volume} {2021}},\ \bibinfo {pages} {199} (\bibinfo {year}
  {2021})}\BibitemShut {NoStop}%
\bibitem [{\citenamefont {Kitano}(2024)}]{Kitano24}%
  \BibitemOpen
  \bibfield  {author} {\bibinfo {author} {\bibfnamefont {R.}~\bibnamefont
  {Kitano}},\ }\bibfield  {title} {\enquote {\bibinfo {title} {{QED} five-loop
  on the lattice},}\ }\href {https://doi.org/10.1093/ptep/ptae194} {\bibfield
  {journal} {\bibinfo  {journal} {Prog. Theor. Exp. Phys.}\ }\textbf {\bibinfo
  {volume} {2025}},\ \bibinfo {pages} {013B02} (\bibinfo {year}
  {2024})}\BibitemShut {NoStop}%
\bibitem [{\citenamefont {Kittel}(2004)}]{Kittel96}%
  \BibitemOpen
  \bibfield  {author} {\bibinfo {author} {\bibfnamefont {C.}~\bibnamefont
  {Kittel}},\ }\href {https://doi.org/10.1063/1.3060399} {\emph {\bibinfo
  {title} {{Introduction to Solid State Physics}}}},\ \bibinfo {edition} {8th}\
  ed.\ (\bibinfo  {publisher} {Wiley},\ \bibinfo {year} {2004})\BibitemShut
  {NoStop}%
\bibitem [{\citenamefont {Dresselhaus}, \citenamefont {Dresselhaus},\ and\
  \citenamefont {Jorio}(2007)}]{Dresselhaus07}%
  \BibitemOpen
  \bibfield  {author} {\bibinfo {author} {\bibfnamefont {M.~S.}\ \bibnamefont
  {Dresselhaus}}, \bibinfo {author} {\bibfnamefont {G.}~\bibnamefont
  {Dresselhaus}},\ and\ \bibinfo {author} {\bibfnamefont {A.}~\bibnamefont
  {Jorio}},\ }\href {https://doi.org/10.1007/978-3-540-32899-5} {\emph
  {\bibinfo {title} {{Group Theory: Application to the Physics of Condensed
  Matter}}}}\ (\bibinfo  {publisher} {Springer},\ \bibinfo {address} {New
  York},\ \bibinfo {year} {2007})\BibitemShut {NoStop}%
\bibitem [{\citenamefont {Cassels}(1959)}]{Cassels59}%
  \BibitemOpen
  \bibfield  {author} {\bibinfo {author} {\bibfnamefont {J.~W.~S.}\
  \bibnamefont {Cassels}},\ }\href {https://doi.org/10.1007/978-3-642-62035-5}
  {\emph {\bibinfo {title} {{An Introduction to the Geometry of Numbers}}}}\
  (\bibinfo  {publisher} {Springer},\ \bibinfo {address} {Berlin},\ \bibinfo
  {year} {1959})\BibitemShut {NoStop}%
\bibitem [{\citenamefont {Siegel}\ and\ \citenamefont
  {Chandrasekharan}(1989)}]{Siegel89}%
  \BibitemOpen
  \bibfield  {author} {\bibinfo {author} {\bibfnamefont {C.~L.}\ \bibnamefont
  {Siegel}}\ and\ \bibinfo {author} {\bibfnamefont {K.}~\bibnamefont
  {Chandrasekharan}},\ }\href {https://doi.org/10.1007/978-3-662-08287-4}
  {\emph {\bibinfo {title} {{Lectures on the Geometry of Numbers}}}}\ (\bibinfo
   {publisher} {Springer Berlin Heidelberg},\ \bibinfo {address} {Berlin,
  Heidelberg},\ \bibinfo {year} {1989})\BibitemShut {NoStop}%
\bibitem [{\citenamefont {Lang}(1987)}]{Lang71}%
  \BibitemOpen
  \bibfield  {author} {\bibinfo {author} {\bibfnamefont {S.}~\bibnamefont
  {Lang}},\ }\href {https://doi.org/10.1007/978-1-4757-1949-9} {\emph {\bibinfo
  {title} {{Linear Algebra}}}}\ (\bibinfo  {publisher} {Addison-Wesley},\
  \bibinfo {address} {Reading, MA},\ \bibinfo {year} {1987})\BibitemShut
  {NoStop}%
\bibitem [{\citenamefont {Cardy}(1986)}]{Cardy86}%
  \BibitemOpen
  \bibfield  {author} {\bibinfo {author} {\bibfnamefont {J.~L.}\ \bibnamefont
  {Cardy}},\ }\bibfield  {title} {\enquote {\bibinfo {title} {Operator content
  of two-dimensional conformally invariant theories},}\ }\href
  {https://doi.org/10.1016/0550-3213(86)90552-3} {\bibfield  {journal}
  {\bibinfo  {journal} {Nucl. Phys. B}\ }\textbf {\bibinfo {volume} {270}},\
  \bibinfo {pages} {186--204} (\bibinfo {year} {1986})}\BibitemShut {NoStop}%
\bibitem [{\citenamefont {Ziff}, \citenamefont {Lorenz},\ and\ \citenamefont
  {Kleban}(1999)}]{ZiLoKl99}%
  \BibitemOpen
  \bibfield  {author} {\bibinfo {author} {\bibfnamefont {R.~M.}\ \bibnamefont
  {Ziff}}, \bibinfo {author} {\bibfnamefont {C.~D.}\ \bibnamefont {Lorenz}},\
  and\ \bibinfo {author} {\bibfnamefont {P.}~\bibnamefont {Kleban}},\
  }\bibfield  {title} {\enquote {\bibinfo {title} {Shape-dependent universality
  in percolation},}\ }\href {https://doi.org/10.1016/s0378-4371(98)00569-x}
  {\bibfield  {journal} {\bibinfo  {journal} {Physica A}\ }\textbf {\bibinfo
  {volume} {266}},\ \bibinfo {pages} {17--26} (\bibinfo {year}
  {1999})}\BibitemShut {NoStop}%
\bibitem [{\citenamefont {Stein}\ and\ \citenamefont
  {Shakarchi}(2003)}]{SteSha03}%
  \BibitemOpen
  \bibfield  {author} {\bibinfo {author} {\bibfnamefont {E.~M.}\ \bibnamefont
  {Stein}}\ and\ \bibinfo {author} {\bibfnamefont {R.}~\bibnamefont
  {Shakarchi}},\ }\href@noop {} {\emph {\bibinfo {title} {{Complex
  Analysis}}}}\ (\bibinfo  {publisher} {Princeton Univ. Press},\ \bibinfo
  {address} {Princeton},\ \bibinfo {year} {2003})\BibitemShut {NoStop}%
\bibitem [{\citenamefont {Cohen}(1993)}]{Cohen93}%
  \BibitemOpen
  \bibfield  {author} {\bibinfo {author} {\bibfnamefont {H.}~\bibnamefont
  {Cohen}},\ }\href {https://doi.org/10.1007/978-3-662-02945-9} {\emph
  {\bibinfo {title} {{A Course in Computational Algebraic Number Theory}}}}\
  (\bibinfo  {publisher} {Springer},\ \bibinfo {address} {Berlin},\ \bibinfo
  {year} {1993})\BibitemShut {NoStop}%
\bibitem [{\citenamefont {Okabe}\ \emph {et~al.}(1999)\citenamefont {Okabe},
  \citenamefont {Kaneda}, \citenamefont {Kikuchi},\ and\ \citenamefont
  {Hu}}]{OKKH99}%
  \BibitemOpen
  \bibfield  {author} {\bibinfo {author} {\bibfnamefont {Y.}~\bibnamefont
  {Okabe}}, \bibinfo {author} {\bibfnamefont {K.}~\bibnamefont {Kaneda}},
  \bibinfo {author} {\bibfnamefont {M.}~\bibnamefont {Kikuchi}},\ and\ \bibinfo
  {author} {\bibfnamefont {C.-K.}\ \bibnamefont {Hu}},\ }\bibfield  {title}
  {\enquote {\bibinfo {title} {Universal finite-size scaling functions for
  critical systems with tilted boundary conditions},}\ }\href
  {https://doi.org/10.1103/physreve.59.1585} {\bibfield  {journal} {\bibinfo
  {journal} {Phys. Rev. E}\ }\textbf {\bibinfo {volume} {59}},\ \bibinfo
  {pages} {1585--1588} (\bibinfo {year} {1999})}\BibitemShut {NoStop}%
\bibitem [{\citenamefont {Liaw}\ \emph {et~al.}(2006)\citenamefont {Liaw},
  \citenamefont {Huang}, \citenamefont {Chou}, \citenamefont {Lin},\ and\
  \citenamefont {Li}}]{LHCLL06}%
  \BibitemOpen
  \bibfield  {author} {\bibinfo {author} {\bibfnamefont {T.~M.}\ \bibnamefont
  {Liaw}}, \bibinfo {author} {\bibfnamefont {M.~C.}\ \bibnamefont {Huang}},
  \bibinfo {author} {\bibfnamefont {Y.~L.}\ \bibnamefont {Chou}}, \bibinfo
  {author} {\bibfnamefont {S.~C.}\ \bibnamefont {Lin}},\ and\ \bibinfo {author}
  {\bibfnamefont {F.~Y.}\ \bibnamefont {Li}},\ }\bibfield  {title} {\enquote
  {\bibinfo {title} {Partition functions and finite-size scalings of {Ising}
  model on helical tori},}\ }\href {https://doi.org/10.1103/physreve.73.055101}
  {\bibfield  {journal} {\bibinfo  {journal} {Phys. Rev. E}\ }\textbf {\bibinfo
  {volume} {73}},\ \bibinfo {pages} {041118} (\bibinfo {year}
  {2006})}\BibitemShut {NoStop}%
\bibitem [{\citenamefont {Izmailian}, \citenamefont {Oganesyan},\ and\
  \citenamefont {Hu}(2002)}]{IzOgCh02}%
  \BibitemOpen
  \bibfield  {author} {\bibinfo {author} {\bibfnamefont {N.~S.}\ \bibnamefont
  {Izmailian}}, \bibinfo {author} {\bibfnamefont {K.~B.}\ \bibnamefont
  {Oganesyan}},\ and\ \bibinfo {author} {\bibfnamefont {C.-K.}\ \bibnamefont
  {Hu}},\ }\bibfield  {title} {\enquote {\bibinfo {title} {Exact finite-size
  corrections for the square-lattice {Ising} model with {Brascamp-Kunz}
  boundary conditions},}\ }\href {https://doi.org/10.1103/physreve.65.056132}
  {\bibfield  {journal} {\bibinfo  {journal} {Phys. Rev. E}\ }\textbf {\bibinfo
  {volume} {65}},\ \bibinfo {pages} {056132} (\bibinfo {year}
  {2002})}\BibitemShut {NoStop}%
\bibitem [{\citenamefont {Baladi}(2000)}]{Baladi00}%
  \BibitemOpen
  \bibfield  {author} {\bibinfo {author} {\bibfnamefont {V.}~\bibnamefont
  {Baladi}},\ }\href {https://doi.org/10.1142/9789812813633} {\emph {\bibinfo
  {title} {{Positive transfer operators and decay of correlations}}}}\
  (\bibinfo  {publisher} {World Scientific},\ \bibinfo {address} {Singapore},\
  \bibinfo {year} {2000})\BibitemShut {NoStop}%
\bibitem [{\citenamefont {{Wikipedia
  contributors}}(2023{\natexlab{a}})}]{WikiCord}%
  \BibitemOpen
  \bibfield  {author} {\bibinfo {author} {\bibnamefont {{Wikipedia
  contributors}}},\ }\href {https://en.wikipedia.org/wiki/Chord_(geometry)}
  {\enquote {\bibinfo {title} {Chord (geometry) --- {Wikipedia, The Free
  Encyclopedia}},}\ } (\bibinfo {year} {2023}{\natexlab{a}})\BibitemShut
  {NoStop}%
\bibitem [{\citenamefont {Berger}(1869)}]{Berger1869}%
  \BibitemOpen
  \bibfield  {author} {\bibinfo {author} {\bibfnamefont {E.~H.}\ \bibnamefont
  {Berger}},\ }\href@noop {} {\emph {\bibinfo {title} {{Die geographischen
  Fragmente des Hipparch}}}}\ (\bibinfo  {publisher} {Teubner},\ \bibinfo
  {year} {1869})\BibitemShut {NoStop}%
\bibitem [{\citenamefont {Pikovsky}\ and\ \citenamefont
  {Politi}(2016)}]{PikPol16}%
  \BibitemOpen
  \bibfield  {author} {\bibinfo {author} {\bibfnamefont {A.}~\bibnamefont
  {Pikovsky}}\ and\ \bibinfo {author} {\bibfnamefont {A.}~\bibnamefont
  {Politi}},\ }\href {https://doi.org/10.1017/cbo9781139343473} {\emph
  {\bibinfo {title} {{Lyapunov Exponents: A Tool to Explore Complex
  Dynamics}}}}\ (\bibinfo  {publisher} {Cambridge Univ. Press},\ \bibinfo
  {address} {Cambridge},\ \bibinfo {year} {2016})\BibitemShut {NoStop}%
\bibitem [{\citenamefont {Guttmann}(2010)}]{Guttmann10}%
  \BibitemOpen
  \bibfield  {author} {\bibinfo {author} {\bibfnamefont {A.~J.}\ \bibnamefont
  {Guttmann}},\ }\bibfield  {title} {\enquote {\bibinfo {title} {Lattice
  {Green's} functions in all dimensions},}\ }\href
  {https://doi.org/10.1088/1751-8113/43/30/305205} {\bibfield  {journal}
  {\bibinfo  {journal} {J. Phys. A}\ }\textbf {\bibinfo {volume} {43}},\
  \bibinfo {pages} {305205} (\bibinfo {year} {2010})}\BibitemShut {NoStop}%
\bibitem [{\citenamefont {Martin}(2006)}]{Martin06}%
  \BibitemOpen
  \bibfield  {author} {\bibinfo {author} {\bibfnamefont {P.~A.}\ \bibnamefont
  {Martin}},\ }\bibfield  {title} {\enquote {\bibinfo {title} {Discrete
  scattering theory: {Green}'s function for a square lattice},}\ }\href
  {https://doi.org/10.1016/j.wavemoti.2006.05.006} {\bibfield  {journal}
  {\bibinfo  {journal} {Wave Motion}\ }\textbf {\bibinfo {volume} {43}},\
  \bibinfo {pages} {619--629} (\bibinfo {year} {2006})}\BibitemShut {NoStop}%
\bibitem [{\citenamefont {Economou}(2006)}]{Economou06}%
  \BibitemOpen
  \bibfield  {author} {\bibinfo {author} {\bibfnamefont {E.~N.}\ \bibnamefont
  {Economou}},\ }\href {https://doi.org/10.1007/3-540-28841-4} {\emph {\bibinfo
  {title} {{Green's Functions in Quantum Physics}}}}\ (\bibinfo  {publisher}
  {Springer},\ \bibinfo {address} {Berlin},\ \bibinfo {year}
  {2006})\BibitemShut {NoStop}%
\bibitem [{\citenamefont {Floquet}(1883)}]{Floquet1883}%
  \BibitemOpen
  \bibfield  {author} {\bibinfo {author} {\bibfnamefont {G.}~\bibnamefont
  {Floquet}},\ }\bibfield  {title} {\enquote {\bibinfo {title} {Sur les
  {\'e}quations diff{\'e}rentielles lin{\'e}aires \`a coefficients
  p{\'e}riodiques},}\ }\href {http://eudml.org/doc/80895} {\bibfield  {journal}
  {\bibinfo  {journal} {Ann. Sci. Ec. Norm. S{\'e}r}\ }\textbf {\bibinfo
  {volume} {12}},\ \bibinfo {pages} {47--88} (\bibinfo {year}
  {1883})}\BibitemShut {NoStop}%
\bibitem [{\citenamefont {Brillouin}(1930)}]{Brillouin30}%
  \BibitemOpen
  \bibfield  {author} {\bibinfo {author} {\bibfnamefont {L.}~\bibnamefont
  {Brillouin}},\ }\bibfield  {title} {\enquote {\bibinfo {title} {Les
  {{\'e}}lectrons libres dans les m{{\'e}}taux et le role des r{{\'e}}flexions
  de {Bragg}},}\ }\href {https://doi.org/10.1051/jphysrad:01930001011037700}
  {\bibfield  {journal} {\bibinfo  {journal} {J. Phys. Radium}\ }\textbf
  {\bibinfo {volume} {1}},\ \bibinfo {pages} {377--400} (\bibinfo {year}
  {1930})}\BibitemShut {NoStop}%
\bibitem [{\citenamefont {Mathews}\ and\ \citenamefont
  {Walker}(1973)}]{MathWalk73}%
  \BibitemOpen
  \bibfield  {author} {\bibinfo {author} {\bibfnamefont {J.}~\bibnamefont
  {Mathews}}\ and\ \bibinfo {author} {\bibfnamefont {R.~L.}\ \bibnamefont
  {Walker}},\ }\href {https://doi.org/10.2307/2316002} {\emph {\bibinfo {title}
  {{Mathematical Methods of Physics}}}}\ (\bibinfo  {publisher}
  {Addison-Wesley},\ \bibinfo {address} {Reading, MA},\ \bibinfo {year}
  {1973})\BibitemShut {NoStop}%
\bibitem [{\citenamefont {Cvitanovi{\'c}}(2025{\natexlab{a}})}]{CBcount}%
  \BibitemOpen
  \bibfield  {author} {\bibinfo {author} {\bibfnamefont {P.}~\bibnamefont
  {Cvitanovi{\'c}}},\ }\bibfield  {title} {\enquote {\bibinfo {title}
  {Counting},}\ }in\ \href {https://ChaosBook.org/paper.shtml#count} {\emph
  {\bibinfo {booktitle} {{Chaos: Classical and Quantum}}}},\ \bibinfo {editor}
  {edited by\ \bibinfo {editor} {\bibfnamefont {P.}~\bibnamefont
  {Cvitanovi{\'c}}}, \bibinfo {editor} {\bibfnamefont {R.}~\bibnamefont
  {Artuso}}, \bibinfo {editor} {\bibfnamefont {R.}~\bibnamefont {Mainieri}},
  \bibinfo {editor} {\bibfnamefont {G.}~\bibnamefont {Tanner}},\ and\ \bibinfo
  {editor} {\bibfnamefont {G.}~\bibnamefont {Vattay}}}\ (\bibinfo  {publisher}
  {Niels Bohr Inst.},\ \bibinfo {address} {Copenhagen},\ \bibinfo {year}
  {2025})\BibitemShut {NoStop}%
\bibitem [{\citenamefont {Cvitanovi{\'c}}(1983)}]{FieldThe}%
  \BibitemOpen
  \bibfield  {author} {\bibinfo {author} {\bibfnamefont {P.}~\bibnamefont
  {Cvitanovi{\'c}}},\ }\href {https://ChaosBook.org/FieldTheory} {\emph
  {\bibinfo {title} {{Field Theory}}}}\ (\bibinfo  {publisher} {Nordita},\
  \bibinfo {address} {Copenhagen},\ \bibinfo {year} {1983})\ \bibinfo {note}
  {notes prepared by E. Gyldenkerne}\BibitemShut {NoStop}%
\bibitem [{\citenamefont {Pomeau}\ and\ \citenamefont
  {Manneville}(1980)}]{Pomeau80}%
  \BibitemOpen
  \bibfield  {author} {\bibinfo {author} {\bibfnamefont {Y.}~\bibnamefont
  {Pomeau}}\ and\ \bibinfo {author} {\bibfnamefont {P.}~\bibnamefont
  {Manneville}},\ }\bibfield  {title} {\enquote {\bibinfo {title} {Intermittent
  transition to turbulence in dissipative dynamical systems},}\ }\href
  {https://doi.org/10.1007/BF01197757} {\bibfield  {journal} {\bibinfo
  {journal} {Commun. Math. Phys.}\ }\textbf {\bibinfo {volume} {74}},\ \bibinfo
  {pages} {189} (\bibinfo {year} {1980})}\BibitemShut {NoStop}%
\bibitem [{\citenamefont {Ruelle}(1976{\natexlab{a}})}]{Ruelle76a}%
  \BibitemOpen
  \bibfield  {author} {\bibinfo {author} {\bibfnamefont {D.}~\bibnamefont
  {Ruelle}},\ }\bibfield  {title} {\enquote {\bibinfo {title} {Generalized
  zeta-functions for {Axiom A} basic sets},}\ }\href
  {https://doi.org/10.1090/S0002-9904-1976-14003-7} {\bibfield  {journal}
  {\bibinfo  {journal} {Bull. Amer. Math. Soc}\ }\textbf {\bibinfo {volume}
  {82}},\ \bibinfo {pages} {153--157} (\bibinfo {year}
  {1976}{\natexlab{a}})}\BibitemShut {NoStop}%
\bibitem [{\citenamefont {Ruelle}(1976{\natexlab{b}})}]{Ruelle76}%
  \BibitemOpen
  \bibfield  {author} {\bibinfo {author} {\bibfnamefont {D.}~\bibnamefont
  {Ruelle}},\ }\bibfield  {title} {\enquote {\bibinfo {title} {Zeta-functions
  for expanding maps and {Anosov} flows},}\ }\href
  {https://doi.org/10.1007/BF01403069} {\bibfield  {journal} {\bibinfo
  {journal} {Inv. Math.}\ }\textbf {\bibinfo {volume} {34}},\ \bibinfo {pages}
  {231--242} (\bibinfo {year} {1976}{\natexlab{b}})}\BibitemShut {NoStop}%
\bibitem [{\citenamefont {{Wikipedia
  contributors}}(2025{\natexlab{b}})}]{wikiDivisorfunc}%
  \BibitemOpen
  \bibfield  {author} {\bibinfo {author} {\bibnamefont {{Wikipedia
  contributors}}},\ }\href {https://en.wikipedia.org/wiki/Divisor_function}
  {\enquote {\bibinfo {title} {Divisor function --- {Wikipedia, The Free
  Encyclopedia}},}\ } (\bibinfo {year} {2025}{\natexlab{b}})\BibitemShut
  {NoStop}%
\bibitem [{\citenamefont {Cvitanovi{\'c}}(2025{\natexlab{b}})}]{CBtrace}%
  \BibitemOpen
  \bibfield  {author} {\bibinfo {author} {\bibfnamefont {P.}~\bibnamefont
  {Cvitanovi{\'c}}},\ }\bibfield  {title} {\enquote {\bibinfo {title} {Trace
  formulas},}\ }in\ \href {https://ChaosBook.org/paper.shtml#trace} {\emph
  {\bibinfo {booktitle} {{Chaos: Classical and Quantum}}}},\ \bibinfo {editor}
  {edited by\ \bibinfo {editor} {\bibfnamefont {P.}~\bibnamefont
  {Cvitanovi{\'c}}}, \bibinfo {editor} {\bibfnamefont {R.}~\bibnamefont
  {Artuso}}, \bibinfo {editor} {\bibfnamefont {R.}~\bibnamefont {Mainieri}},
  \bibinfo {editor} {\bibfnamefont {G.}~\bibnamefont {Tanner}},\ and\ \bibinfo
  {editor} {\bibfnamefont {G.}~\bibnamefont {Vattay}}}\ (\bibinfo  {publisher}
  {Niels Bohr Inst.},\ \bibinfo {address} {Copenhagen},\ \bibinfo {year}
  {2025})\BibitemShut {NoStop}%
\bibitem [{\citenamefont {Cvitanovi{\'c}}(2025{\natexlab{c}})}]{CBdet}%
  \BibitemOpen
  \bibfield  {author} {\bibinfo {author} {\bibfnamefont {P.}~\bibnamefont
  {Cvitanovi{\'c}}},\ }\bibfield  {title} {\enquote {\bibinfo {title} {Spectral
  determinants},}\ }in\ \href {https://ChaosBook.org/paper.shtml#det} {\emph
  {\bibinfo {booktitle} {{Chaos: Classical and Quantum}}}},\ \bibinfo {editor}
  {edited by\ \bibinfo {editor} {\bibfnamefont {P.}~\bibnamefont
  {Cvitanovi{\'c}}}, \bibinfo {editor} {\bibfnamefont {R.}~\bibnamefont
  {Artuso}}, \bibinfo {editor} {\bibfnamefont {R.}~\bibnamefont {Mainieri}},
  \bibinfo {editor} {\bibfnamefont {G.}~\bibnamefont {Tanner}},\ and\ \bibinfo
  {editor} {\bibfnamefont {G.}~\bibnamefont {Vattay}}}\ (\bibinfo  {publisher}
  {Niels Bohr Inst.},\ \bibinfo {address} {Copenhagen},\ \bibinfo {year}
  {2025})\BibitemShut {NoStop}%
\bibitem [{\citenamefont {Cvitanovi{\'c}}(2020)}]{CvitanovicYT00}%
  \BibitemOpen
  \bibfield  {author} {\bibinfo {author} {\bibfnamefont {P.}~\bibnamefont
  {Cvitanovi{\'c}}},\ }\href {https://youtu.be/_7ZNfbgJ8D4} {\enquote {\bibinfo
  {title} {Spectral determinants, zeta functions: why and what?}}\ } (\bibinfo
  {year} {2020})\BibitemShut {NoStop}%
\bibitem [{\citenamefont {Cvitanovi{\'c}}\ \emph
  {et~al.}(2025{\natexlab{b}})\citenamefont {Cvitanovi{\'c}}, \citenamefont
  {Artuso}, \citenamefont {Rondoni},\ and\ \citenamefont
  {Spiegel}}]{CBrecycle}%
  \BibitemOpen
  \bibfield  {author} {\bibinfo {author} {\bibfnamefont {P.}~\bibnamefont
  {Cvitanovi{\'c}}}, \bibinfo {author} {\bibfnamefont {R.}~\bibnamefont
  {Artuso}}, \bibinfo {author} {\bibfnamefont {L.}~\bibnamefont {Rondoni}},\
  and\ \bibinfo {author} {\bibfnamefont {E.~A.}\ \bibnamefont {Spiegel}},\
  }\bibfield  {title} {\enquote {\bibinfo {title} {Cycle expansions},}\ }in\
  \href {https://ChaosBook.org/paper.shtml#recycle} {\emph {\bibinfo
  {booktitle} {{Chaos: Classical and Quantum}}}},\ \bibinfo {editor} {edited
  by\ \bibinfo {editor} {\bibfnamefont {P.}~\bibnamefont {Cvitanovi{\'c}}},
  \bibinfo {editor} {\bibfnamefont {R.}~\bibnamefont {Artuso}}, \bibinfo
  {editor} {\bibfnamefont {R.}~\bibnamefont {Mainieri}}, \bibinfo {editor}
  {\bibfnamefont {G.}~\bibnamefont {Tanner}},\ and\ \bibinfo {editor}
  {\bibfnamefont {G.}~\bibnamefont {Vattay}}}\ (\bibinfo  {publisher} {Niels
  Bohr Inst.},\ \bibinfo {address} {Copenhagen},\ \bibinfo {year}
  {2025})\BibitemShut {NoStop}%
\bibitem [{\citenamefont {Kadanoff}\ and\ \citenamefont {Tang}(1984)}]{KT84}%
  \BibitemOpen
  \bibfield  {author} {\bibinfo {author} {\bibfnamefont {L.}~\bibnamefont
  {Kadanoff}}\ and\ \bibinfo {author} {\bibfnamefont {C.}~\bibnamefont
  {Tang}},\ }\bibfield  {title} {\enquote {\bibinfo {title} {Escape rate from
  strange repellers},}\ }\href {https://doi.org/10.1073/pnas.81.4.1276}
  {\bibfield  {journal} {\bibinfo  {journal} {Proc. Natl. Acad. Sci.}\ }\textbf
  {\bibinfo {volume} {81}},\ \bibinfo {pages} {1276--1279} (\bibinfo {year}
  {1984})}\BibitemShut {NoStop}%
\bibitem [{\citenamefont {{Wikipedia
  contributors}}(2025{\natexlab{c}})}]{wikiEulerFunc}%
  \BibitemOpen
  \bibfield  {author} {\bibinfo {author} {\bibnamefont {{Wikipedia
  contributors}}},\ }\href {https://en.wikipedia.org/wiki/Euler_function}
  {\enquote {\bibinfo {title} {Euler function --- {Wikipedia, The Free
  Encyclopedia}},}\ } (\bibinfo {year} {2025}{\natexlab{c}})\BibitemShut
  {NoStop}%
\bibitem [{\citenamefont {Bell}(2005)}]{Bell05}%
  \BibitemOpen
  \bibfield  {author} {\bibinfo {author} {\bibfnamefont {J.}~\bibnamefont
  {Bell}},\ }\href {https://doi.org/10.48550/ARXIV.MATH/0510054} {\enquote
  {\bibinfo {title} {Euler and the pentagonal number theorem},}\ } (\bibinfo
  {year} {2005})\BibitemShut {NoStop}%
\bibitem [{\citenamefont {Lind}(1996)}]{Lind96}%
  \BibitemOpen
  \bibfield  {author} {\bibinfo {author} {\bibfnamefont {D.~A.}\ \bibnamefont
  {Lind}},\ }\bibfield  {title} {\enquote {\bibinfo {title} {A zeta function
  for {$Z^d$}-actions},}\ }in\ \href
  {https://doi.org/10.1017/CBO9780511662812.019} {\emph {\bibinfo {booktitle}
  {{Ergodic Theory of {$Z^d$} Actions}}}},\ \bibinfo {editor} {edited by\
  \bibinfo {editor} {\bibfnamefont {M.}~\bibnamefont {Pollicott}}\ and\
  \bibinfo {editor} {\bibfnamefont {K.}~\bibnamefont {Schmidt}}}\ (\bibinfo
  {publisher} {Cambridge Univ. Press},\ \bibinfo {year} {1996})\ pp.\ \bibinfo
  {pages} {433--450}\BibitemShut {NoStop}%
\bibitem [{\citenamefont {{Wikipedia
  contributors}}(2025{\natexlab{d}})}]{wikiDedekindEta}%
  \BibitemOpen
  \bibfield  {author} {\bibinfo {author} {\bibnamefont {{Wikipedia
  contributors}}},\ }\href
  {https://en.wikipedia.org/wiki/Dedekind_eta_function} {\enquote {\bibinfo
  {title} {Dedekind eta function --- {Wikipedia, The Free Encyclopedia}},}\ }
  (\bibinfo {year} {2025}{\natexlab{d}})\BibitemShut {NoStop}%
\bibitem [{\citenamefont {Weisstein}(2026)}]{WolframDedekindEta}%
  \BibitemOpen
  \bibfield  {author} {\bibinfo {author} {\bibfnamefont {E.~W.}\ \bibnamefont
  {Weisstein}},\ }\href
  {https://mathworld.wolfram.com/DedekindEtaFunction.html} {\enquote {\bibinfo
  {title} {{Dedekind eta function}},}\ } (\bibinfo {year} {2026}),\ \bibinfo
  {note} {{From MathWorld--A Wolfram Resource}}\BibitemShut {NoStop}%
\bibitem [{\citenamefont {Gel'fand}\ and\ \citenamefont
  {Yaglom}(1960)}]{GelYag60}%
  \BibitemOpen
  \bibfield  {author} {\bibinfo {author} {\bibfnamefont {I.~M.}\ \bibnamefont
  {Gel'fand}}\ and\ \bibinfo {author} {\bibfnamefont {A.~M.}\ \bibnamefont
  {Yaglom}},\ }\bibfield  {title} {\enquote {\bibinfo {title} {Integration in
  functional spaces and its applications in quantum physics},}\ }\href
  {https://doi.org/10.1063/1.1703636} {\bibfield  {journal} {\bibinfo
  {journal} {J. Math. Phys.}\ }\textbf {\bibinfo {volume} {1}},\ \bibinfo
  {pages} {48--69} (\bibinfo {year} {1960})}\BibitemShut {NoStop}%
\bibitem [{\citenamefont {Ludewig}(2018)}]{Ludewig16}%
  \BibitemOpen
  \bibfield  {author} {\bibinfo {author} {\bibfnamefont {M.}~\bibnamefont
  {Ludewig}},\ }\bibfield  {title} {\enquote {\bibinfo {title} {Heat kernel
  asymptotics, path integrals and infinite-dimensional determinants},}\ }\href
  {https://doi.org/10.1016/j.geomphys.2018.04.012} {\bibfield  {journal}
  {\bibinfo  {journal} {J. Geom. Phys.}\ }\textbf {\bibinfo {volume} {131}},\
  \bibinfo {pages} {66--88} (\bibinfo {year} {2018})}\BibitemShut {NoStop}%
\bibitem [{\citenamefont {Barreto}(2008)}]{Barreto08}%
  \BibitemOpen
  \bibfield  {author} {\bibinfo {author} {\bibfnamefont {E.}~\bibnamefont
  {Barreto}},\ }\bibfield  {title} {\enquote {\bibinfo {title} {Shadowing},}\
  }\href {https://doi.org/10.4249/scholarpedia.2243} {\bibfield  {journal}
  {\bibinfo  {journal} {Scholarpedia}\ }\textbf {\bibinfo {volume} {3}},\
  \bibinfo {pages} {2243} (\bibinfo {year} {2008})}\BibitemShut {NoStop}%
\bibitem [{\citenamefont {Palmer}(2009)}]{Palmer09}%
  \BibitemOpen
  \bibfield  {author} {\bibinfo {author} {\bibfnamefont {K.}~\bibnamefont
  {Palmer}},\ }\bibfield  {title} {\enquote {\bibinfo {title} {Shadowing lemma
  for flows},}\ }\href {https://doi.org/10.4249/scholarpedia.7918} {\bibfield
  {journal} {\bibinfo  {journal} {Scholarpedia}\ }\textbf {\bibinfo {volume}
  {4}},\ \bibinfo {pages} {7918} (\bibinfo {year} {2009})}\BibitemShut
  {NoStop}%
\bibitem [{\citenamefont {{Wikipedia
  contributors}}(2023{\natexlab{b}})}]{wikiLyapTime}%
  \BibitemOpen
  \bibfield  {author} {\bibinfo {author} {\bibnamefont {{Wikipedia
  contributors}}},\ }\href {https://en.wikipedia.org/wiki/Lyapunov_time}
  {\enquote {\bibinfo {title} {Lyapunov time --- {Wikipedia, The Free
  Encyclopedia}},}\ } (\bibinfo {year} {2023}{\natexlab{b}})\BibitemShut
  {NoStop}%
\bibitem [{\citenamefont {Ding}\ and\ \citenamefont
  {Cvitanovi{\'c}}(2016)}]{DingCvit14}%
  \BibitemOpen
  \bibfield  {author} {\bibinfo {author} {\bibfnamefont {X.}~\bibnamefont
  {Ding}}\ and\ \bibinfo {author} {\bibfnamefont {P.}~\bibnamefont
  {Cvitanovi{\'c}}},\ }\bibfield  {title} {\enquote {\bibinfo {title} {Periodic
  eigendecomposition and its application in {Kuramoto-Sivashinsky} system},}\
  }\href {https://doi.org/10.1137/15M1037299} {\bibfield  {journal} {\bibinfo
  {journal} {{SIAM} J. Appl. Dyn. Syst.}\ }\textbf {\bibinfo {volume} {15}},\
  \bibinfo {pages} {1434--1454} (\bibinfo {year} {2016})}\BibitemShut {NoStop}%
\bibitem [{\citenamefont {Mestel}\ and\ \citenamefont
  {Percival}(1987)}]{varcyc}%
  \BibitemOpen
  \bibfield  {author} {\bibinfo {author} {\bibfnamefont {B.~D.}\ \bibnamefont
  {Mestel}}\ and\ \bibinfo {author} {\bibfnamefont {I.}~\bibnamefont
  {Percival}},\ }\bibfield  {title} {\enquote {\bibinfo {title} {{Newton}
  method for highly unstable orbits},}\ }\href
  {https://doi.org/10.1016/0167-2789(87)90072-8} {\bibfield  {journal}
  {\bibinfo  {journal} {Physica D}\ }\textbf {\bibinfo {volume} {24}},\
  \bibinfo {pages} {172} (\bibinfo {year} {1987})}\BibitemShut {NoStop}%
\bibitem [{\citenamefont {Glaser}(1970)}]{Glaser70}%
  \BibitemOpen
  \bibfield  {author} {\bibinfo {author} {\bibfnamefont {J.~I.}\ \bibnamefont
  {Glaser}},\ }\bibfield  {title} {\enquote {\bibinfo {title} {Numerical
  solution of waveguide scattering problems by finite-difference {Green's}
  functions},}\ }\href {https://doi.org/10.1109/TMTT.1970.1127265} {\bibfield
  {journal} {\bibinfo  {journal} {IEEE Trans. Microwave Theory Tech.}\ }\textbf
  {\bibinfo {volume} {18}},\ \bibinfo {pages} {436--443} (\bibinfo {year}
  {1970})}\BibitemShut {NoStop}%
\bibitem [{\citenamefont {Buzbee}, \citenamefont {Golub},\ and\ \citenamefont
  {Nielson}(1970)}]{BuGoNi70}%
  \BibitemOpen
  \bibfield  {author} {\bibinfo {author} {\bibfnamefont {B.~L.}\ \bibnamefont
  {Buzbee}}, \bibinfo {author} {\bibfnamefont {G.~H.}\ \bibnamefont {Golub}},\
  and\ \bibinfo {author} {\bibfnamefont {C.~W.}\ \bibnamefont {Nielson}},\
  }\bibfield  {title} {\enquote {\bibinfo {title} {On direct methods for
  solving {Poisson}'s equations},}\ }\href {https://doi.org/10.1137/0707049}
  {\bibfield  {journal} {\bibinfo  {journal} {SIAM J. Numer. Anal.}\ }\textbf
  {\bibinfo {volume} {7}},\ \bibinfo {pages} {627--656} (\bibinfo {year}
  {1970})}\BibitemShut {NoStop}%
\bibitem [{\citenamefont {Wood}(1971)}]{Wood71}%
  \BibitemOpen
  \bibfield  {author} {\bibinfo {author} {\bibfnamefont {W.~L.}\ \bibnamefont
  {Wood}},\ }\bibfield  {title} {\enquote {\bibinfo {title} {Periodicity
  effects on the iterative solution of elliptic difference equations},}\ }\href
  {https://doi.org/10.1137/0708041} {\bibfield  {journal} {\bibinfo  {journal}
  {SIAM J. Numer. Anal.}\ }\textbf {\bibinfo {volume} {8}},\ \bibinfo {pages}
  {439--464} (\bibinfo {year} {1971})}\BibitemShut {NoStop}%
\bibitem [{\citenamefont {Katsura}\ \emph {et~al.}(1971)\citenamefont
  {Katsura}, \citenamefont {Morita}, \citenamefont {Inawashiro}, \citenamefont
  {Horiguchi},\ and\ \citenamefont {Abe}}]{KMIHA71}%
  \BibitemOpen
  \bibfield  {author} {\bibinfo {author} {\bibfnamefont {S.}~\bibnamefont
  {Katsura}}, \bibinfo {author} {\bibfnamefont {T.}~\bibnamefont {Morita}},
  \bibinfo {author} {\bibfnamefont {S.}~\bibnamefont {Inawashiro}}, \bibinfo
  {author} {\bibfnamefont {T.}~\bibnamefont {Horiguchi}},\ and\ \bibinfo
  {author} {\bibfnamefont {Y.}~\bibnamefont {Abe}},\ }\bibfield  {title}
  {\enquote {\bibinfo {title} {Lattice {Green's} function. {Introduction}},}\
  }\href {https://doi.org/10.1063/1.1665662} {\bibfield  {journal} {\bibinfo
  {journal} {J. Math. Phys.}\ }\textbf {\bibinfo {volume} {12}},\ \bibinfo
  {pages} {892--895} (\bibinfo {year} {1971})}\BibitemShut {NoStop}%
\bibitem [{\citenamefont {Katsura}\ and\ \citenamefont
  {Inawashiro}(1971)}]{KatIna71}%
  \BibitemOpen
  \bibfield  {author} {\bibinfo {author} {\bibfnamefont {S.}~\bibnamefont
  {Katsura}}\ and\ \bibinfo {author} {\bibfnamefont {S.}~\bibnamefont
  {Inawashiro}},\ }\bibfield  {title} {\enquote {\bibinfo {title} {Lattice
  {Green's} functions for the rectangular and the square lattices at arbitrary
  points},}\ }\href {https://doi.org/10.1063/1.1665785} {\bibfield  {journal}
  {\bibinfo  {journal} {J. Math. Phys.}\ }\textbf {\bibinfo {volume} {12}},\
  \bibinfo {pages} {1622--1630} (\bibinfo {year} {1971})}\BibitemShut {NoStop}%
\bibitem [{\citenamefont {Katsura}, \citenamefont {Inawashiro},\ and\
  \citenamefont {Abe}(1971)}]{KaInAb71}%
  \BibitemOpen
  \bibfield  {author} {\bibinfo {author} {\bibfnamefont {S.}~\bibnamefont
  {Katsura}}, \bibinfo {author} {\bibfnamefont {S.}~\bibnamefont
  {Inawashiro}},\ and\ \bibinfo {author} {\bibfnamefont {Y.}~\bibnamefont
  {Abe}},\ }\bibfield  {title} {\enquote {\bibinfo {title} {Lattice {Green's}
  function for the simple cubic lattice in terms of a {Mellin-Barnes} type
  integral},}\ }\href {https://doi.org/10.1063/1.1665663} {\bibfield  {journal}
  {\bibinfo  {journal} {J. Math. Phys.}\ }\textbf {\bibinfo {volume} {12}},\
  \bibinfo {pages} {895--899} (\bibinfo {year} {1971})}\BibitemShut {NoStop}%
\bibitem [{\citenamefont {Morita}(1971)}]{Morita71}%
  \BibitemOpen
  \bibfield  {author} {\bibinfo {author} {\bibfnamefont {T.}~\bibnamefont
  {Morita}},\ }\bibfield  {title} {\enquote {\bibinfo {title} {Useful procedure
  for computing the lattice {Green's} function - square, tetragonal, and bcc
  lattices},}\ }\href {https://doi.org/10.1063/1.1665800} {\bibfield  {journal}
  {\bibinfo  {journal} {J. Math. Phys.}\ }\textbf {\bibinfo {volume} {12}},\
  \bibinfo {pages} {1744--1747} (\bibinfo {year} {1971})}\BibitemShut {NoStop}%
\bibitem [{\citenamefont {Morita}\ and\ \citenamefont
  {Horiguchi}(1971)}]{MorHor71}%
  \BibitemOpen
  \bibfield  {author} {\bibinfo {author} {\bibfnamefont {T.}~\bibnamefont
  {Morita}}\ and\ \bibinfo {author} {\bibfnamefont {T.}~\bibnamefont
  {Horiguchi}},\ }\bibfield  {title} {\enquote {\bibinfo {title} {Calculation
  of the lattice {Green's} function for the bcc, fcc, and rectangular
  lattices},}\ }\href {https://doi.org/10.1063/1.1665693} {\bibfield  {journal}
  {\bibinfo  {journal} {J. Math. Phys.}\ }\textbf {\bibinfo {volume} {12}},\
  \bibinfo {pages} {986--992} (\bibinfo {year} {1971})}\BibitemShut {NoStop}%
\bibitem [{\citenamefont {Horiguchi}(1971)}]{Horiguchi71}%
  \BibitemOpen
  \bibfield  {author} {\bibinfo {author} {\bibfnamefont {T.}~\bibnamefont
  {Horiguchi}},\ }\bibfield  {title} {\enquote {\bibinfo {title} {Lattice
  {Green's} function for the simple cubic lattice},}\ }\href
  {https://doi.org/10.1143/JPSJ.30.1261} {\bibfield  {journal} {\bibinfo
  {journal} {J. Phys. Soc. Jpn.}\ }\textbf {\bibinfo {volume} {30}},\ \bibinfo
  {pages} {1261--1272} (\bibinfo {year} {1971})}\BibitemShut {NoStop}%
\bibitem [{\citenamefont {Fiedler}(1973)}]{Fiedler73}%
  \BibitemOpen
  \bibfield  {author} {\bibinfo {author} {\bibfnamefont {M.}~\bibnamefont
  {Fiedler}},\ }\bibfield  {title} {\enquote {\bibinfo {title} {Algebraic
  connectivity of graphs},}\ }\href {http://dml.cz/dmlcz/101168} {\bibfield
  {journal} {\bibinfo  {journal} {Czech. Math. J}\ }\textbf {\bibinfo {volume}
  {23}},\ \bibinfo {pages} {298--305} (\bibinfo {year} {1973})}\BibitemShut
  {NoStop}%
\bibitem [{\citenamefont {Horiguchi}\ and\ \citenamefont
  {Morita}(1975)}]{HorMor75}%
  \BibitemOpen
  \bibfield  {author} {\bibinfo {author} {\bibfnamefont {T.}~\bibnamefont
  {Horiguchi}}\ and\ \bibinfo {author} {\bibfnamefont {T.}~\bibnamefont
  {Morita}},\ }\bibfield  {title} {\enquote {\bibinfo {title} {Note on the
  lattice {Green's} function for the simple cubic lattice},}\ }\href
  {https://doi.org/10.1088/0022-3719/8/11/002} {\bibfield  {journal} {\bibinfo
  {journal} {J. Phys. C}\ }\textbf {\bibinfo {volume} {8}},\ \bibinfo {pages}
  {L232} (\bibinfo {year} {1975})}\BibitemShut {NoStop}%
\bibitem [{\citenamefont {Anderson}\ and\ \citenamefont
  {Morley}(1985)}]{AndMor85}%
  \BibitemOpen
  \bibfield  {author} {\bibinfo {author} {\bibfnamefont {W.~N.}\ \bibnamefont
  {Anderson}}\ and\ \bibinfo {author} {\bibfnamefont {T.~D.}\ \bibnamefont
  {Morley}},\ }\bibfield  {title} {\enquote {\bibinfo {title} {Eigenvalues of
  the {Laplacian} of a graph},}\ }\href
  {https://doi.org/10.1080/03081088508817681} {\bibfield  {journal} {\bibinfo
  {journal} {Lin. Multilin. Algebra}\ }\textbf {\bibinfo {volume} {18}},\
  \bibinfo {pages} {141--145} (\bibinfo {year} {1985})}\BibitemShut {NoStop}%
\bibitem [{\citenamefont {Chen}(1987)}]{ChenM87}%
  \BibitemOpen
  \bibfield  {author} {\bibinfo {author} {\bibfnamefont {M.}~\bibnamefont
  {Chen}},\ }\bibfield  {title} {\enquote {\bibinfo {title} {On the solution of
  circulant linear systems},}\ }\href {https://doi.org/10.1137/0724044}
  {\bibfield  {journal} {\bibinfo  {journal} {SIAM J. Numer. Anal.}\ }\textbf
  {\bibinfo {volume} {24}},\ \bibinfo {pages} {668--683} (\bibinfo {year}
  {1987})}\BibitemShut {NoStop}%
\bibitem [{\citenamefont {Chung}\ and\ \citenamefont {Yau}(2000)}]{ChuYau00}%
  \BibitemOpen
  \bibfield  {author} {\bibinfo {author} {\bibfnamefont {F.}~\bibnamefont
  {Chung}}\ and\ \bibinfo {author} {\bibfnamefont {S.-T.}\ \bibnamefont
  {Yau}},\ }\bibfield  {title} {\enquote {\bibinfo {title} {Discrete {Green's}
  functions},}\ }\href {https://doi.org/10.1006/jcta.2000.3094} {\bibfield
  {journal} {\bibinfo  {journal} {J. Combin. Theory A}\ }\textbf {\bibinfo
  {volume} {91}},\ \bibinfo {pages} {19--214} (\bibinfo {year}
  {2000})}\BibitemShut {NoStop}%
\bibitem [{\citenamefont {de~la Llave}(2000)}]{dlLlave00}%
  \BibitemOpen
  \bibfield  {author} {\bibinfo {author} {\bibfnamefont {R.}~\bibnamefont
  {de~la Llave}},\ }\bibfield  {title} {\enquote {\bibinfo {title} {Variational
  methods for quasiperiodic solutions of partial differential equations},}\
  }in\ \href {https://doi.org/10.1142/4557} {\emph {\bibinfo {booktitle}
  {{Hamiltonian Systems and Celestial Mechanics (HAMSYS-98)}}}},\ \bibinfo
  {editor} {edited by\ \bibinfo {editor} {\bibfnamefont {J.}~\bibnamefont
  {Delgado}}, \bibinfo {editor} {\bibfnamefont {E.~A.}\ \bibnamefont
  {Lacomba}}, \bibinfo {editor} {\bibfnamefont {E.}~\bibnamefont
  {P{\'e}rez-Chavela}},\ and\ \bibinfo {editor} {\bibfnamefont
  {J.}~\bibnamefont {Llibre}}}\ (\bibinfo  {publisher} {World Scientific},\
  \bibinfo {address} {Singapore},\ \bibinfo {year} {2000})\BibitemShut
  {NoStop}%
\bibitem [{\citenamefont {Asad}(2007)}]{Asad07}%
  \BibitemOpen
  \bibfield  {author} {\bibinfo {author} {\bibfnamefont {J.~H.}\ \bibnamefont
  {Asad}},\ }\bibfield  {title} {\enquote {\bibinfo {title} {Differential
  equation approach for one- and two-dimensional lattice {Green}'s function},}\
  }\href {https://doi.org/10.1142/S021798490701244X} {\bibfield  {journal}
  {\bibinfo  {journal} {Mod. Phys. Lett. B}\ }\textbf {\bibinfo {volume}
  {21}},\ \bibinfo {pages} {139--154} (\bibinfo {year} {2007})}\BibitemShut
  {NoStop}%
\bibitem [{\citenamefont {Bhat}\ and\ \citenamefont {Osting}(2010)}]{BhaOst10}%
  \BibitemOpen
  \bibfield  {author} {\bibinfo {author} {\bibfnamefont {H.~S.}\ \bibnamefont
  {Bhat}}\ and\ \bibinfo {author} {\bibfnamefont {B.}~\bibnamefont {Osting}},\
  }\bibfield  {title} {\enquote {\bibinfo {title} {Diffraction on the
  two-dimensional square lattice},}\ }\href {https://doi.org/10.1137/080735345}
  {\bibfield  {journal} {\bibinfo  {journal} {SIAM J. Appl. Math.}\ }\textbf
  {\bibinfo {volume} {70}},\ \bibinfo {pages} {1389--1406} (\bibinfo {year}
  {2010})}\BibitemShut {NoStop}%
\bibitem [{\citenamefont {Stewart}\ and\ \citenamefont
  {G{\"o}kaydin}(2019)}]{SteGok19}%
  \BibitemOpen
  \bibfield  {author} {\bibinfo {author} {\bibfnamefont {I.}~\bibnamefont
  {Stewart}}\ and\ \bibinfo {author} {\bibfnamefont {D.}~\bibnamefont
  {G{\"o}kaydin}},\ }\bibfield  {title} {\enquote {\bibinfo {title} {Symmetries
  of quotient networks for doubly periodic patterns on the square lattice},}\
  }\href {https://doi.org/10.1142/s021812741930026x} {\bibfield  {journal}
  {\bibinfo  {journal} {Int. J. Bifurcat. Chaos}\ }\textbf {\bibinfo {volume}
  {29}},\ \bibinfo {pages} {1930026} (\bibinfo {year} {2019})}\BibitemShut
  {NoStop}%
\bibitem [{\citenamefont {{Wikipedia
  contributors}}(2025{\natexlab{e}})}]{wikiModBessel}%
  \BibitemOpen
  \bibfield  {author} {\bibinfo {author} {\bibnamefont {{Wikipedia
  contributors}}},\ }\href
  {https://en.wikipedia.org/wiki/Bessel_function#Modified_Bessel_functions}
  {\enquote {\bibinfo {title} {Modified {Bessel} functions --- {Wikipedia, The
  Free Encyclopedia}},}\ } (\bibinfo {year} {2025}{\natexlab{e}})\BibitemShut
  {NoStop}%
\bibitem [{\citenamefont {Artuso}, \citenamefont {Aurell},\ and\ \citenamefont
  {Cvitanovi\'c}(1990{\natexlab{b}})}]{AACII}%
  \BibitemOpen
  \bibfield  {author} {\bibinfo {author} {\bibfnamefont {R.}~\bibnamefont
  {Artuso}}, \bibinfo {author} {\bibfnamefont {E.}~\bibnamefont {Aurell}},\
  and\ \bibinfo {author} {\bibfnamefont {P.}~\bibnamefont {Cvitanovi\'c}},\
  }\bibfield  {title} {\enquote {\bibinfo {title} {Recycling of strange sets:
  {II. Applications}},}\ }\href {https://doi.org/10.1088/0951-7715/3/2/006}
  {\bibfield  {journal} {\bibinfo  {journal} {Nonlinearity}\ }\textbf {\bibinfo
  {volume} {3}},\ \bibinfo {pages} {361--386} (\bibinfo {year}
  {1990}{\natexlab{b}})}\BibitemShut {NoStop}%
\bibitem [{\citenamefont {Artuso}, \citenamefont {Rugh},\ and\ \citenamefont
  {Cvitanovi{\'c}}(2025)}]{CBconverg}%
  \BibitemOpen
  \bibfield  {author} {\bibinfo {author} {\bibfnamefont {R.}~\bibnamefont
  {Artuso}}, \bibinfo {author} {\bibfnamefont {H.~H.}\ \bibnamefont {Rugh}},\
  and\ \bibinfo {author} {\bibfnamefont {P.}~\bibnamefont {Cvitanovi{\'c}}},\
  }\bibfield  {title} {\enquote {\bibinfo {title} {Why does it work?}}\ }in\
  \href {https://ChaosBook.org/paper.shtml#converg} {\emph {\bibinfo
  {booktitle} {{Chaos: Classical and Quantum}}}},\ \bibinfo {editor} {edited
  by\ \bibinfo {editor} {\bibfnamefont {P.}~\bibnamefont {Cvitanovi{\'c}}},
  \bibinfo {editor} {\bibfnamefont {R.}~\bibnamefont {Artuso}}, \bibinfo
  {editor} {\bibfnamefont {R.}~\bibnamefont {Mainieri}}, \bibinfo {editor}
  {\bibfnamefont {G.}~\bibnamefont {Tanner}},\ and\ \bibinfo {editor}
  {\bibfnamefont {G.}~\bibnamefont {Vattay}}}\ (\bibinfo  {publisher} {Niels
  Bohr Inst.},\ \bibinfo {address} {Copenhagen},\ \bibinfo {year}
  {2025})\BibitemShut {NoStop}%
\bibitem [{\citenamefont {Gutzwiller}(1971)}]{gutzwiller71}%
  \BibitemOpen
  \bibfield  {author} {\bibinfo {author} {\bibfnamefont {M.~C.}\ \bibnamefont
  {Gutzwiller}},\ }\bibfield  {title} {\enquote {\bibinfo {title} {Periodic
  orbits and classical quantization conditions},}\ }\href
  {https://doi.org/10.1063/1.1665596} {\bibfield  {journal} {\bibinfo
  {journal} {J. Math. Phys.}\ }\textbf {\bibinfo {volume} {12}},\ \bibinfo
  {pages} {343--358} (\bibinfo {year} {1971})}\BibitemShut {NoStop}%
\bibitem [{\citenamefont {Chandler}\ and\ \citenamefont
  {Kerswell}(2013)}]{ChaKer12}%
  \BibitemOpen
  \bibfield  {author} {\bibinfo {author} {\bibfnamefont {G.~J.}\ \bibnamefont
  {Chandler}}\ and\ \bibinfo {author} {\bibfnamefont {R.~R.}\ \bibnamefont
  {Kerswell}},\ }\bibfield  {title} {\enquote {\bibinfo {title} {Invariant
  recurrent solutions embedded in a turbulent two-dimensional {Kolmogorov}
  flow},}\ }\href {https://doi.org/10.1017/jfm.2013.122} {\bibfield  {journal}
  {\bibinfo  {journal} {J. Fluid Mech.}\ }\textbf {\bibinfo {volume} {722}},\
  \bibinfo {pages} {554--595} (\bibinfo {year} {2013})}\BibitemShut {NoStop}%
\bibitem [{\citenamefont {Christiansen}, \citenamefont {Cvitanovi{\'c}},\ and\
  \citenamefont {Putkaradze}(1997{\natexlab{a}})}]{Christiansen97}%
  \BibitemOpen
  \bibfield  {author} {\bibinfo {author} {\bibfnamefont {F.}~\bibnamefont
  {Christiansen}}, \bibinfo {author} {\bibfnamefont {P.}~\bibnamefont
  {Cvitanovi{\'c}}},\ and\ \bibinfo {author} {\bibfnamefont {V.}~\bibnamefont
  {Putkaradze}},\ }\bibfield  {title} {\enquote {\bibinfo {title}
  {{Spatiotemporal} chaos in terms of unstable recurrent patterns},}\ }\href
  {https://doi.org/10.1088/0951-7715/10/1/004} {\bibfield  {journal} {\bibinfo
  {journal} {Nonlinearity}\ }\textbf {\bibinfo {volume} {10}},\ \bibinfo
  {pages} {55--70} (\bibinfo {year} {1997}{\natexlab{a}})}\BibitemShut
  {NoStop}%
\bibitem [{\citenamefont {Cvitanovi{\'c}}, \citenamefont {Davidchack},\ and\
  \citenamefont {Siminos}(2010)}]{SCD07}%
  \BibitemOpen
  \bibfield  {author} {\bibinfo {author} {\bibfnamefont {P.}~\bibnamefont
  {Cvitanovi{\'c}}}, \bibinfo {author} {\bibfnamefont {R.~L.}\ \bibnamefont
  {Davidchack}},\ and\ \bibinfo {author} {\bibfnamefont {E.}~\bibnamefont
  {Siminos}},\ }\bibfield  {title} {\enquote {\bibinfo {title} {On the state
  space geometry of the {Kuramoto-Sivashinsky} flow in a periodic domain},}\
  }\href {https://doi.org/10.1137/070705623} {\bibfield  {journal} {\bibinfo
  {journal} {SIAM J. Appl. Dyn. Syst.}\ }\textbf {\bibinfo {volume} {9}},\
  \bibinfo {pages} {1--33} (\bibinfo {year} {2010})}\BibitemShut {NoStop}%
\bibitem [{\citenamefont {Cvitanovi{\'c}}\ and\ \citenamefont
  {Gibson}(2010)}]{CviGib10}%
  \BibitemOpen
  \bibfield  {author} {\bibinfo {author} {\bibfnamefont {P.}~\bibnamefont
  {Cvitanovi{\'c}}}\ and\ \bibinfo {author} {\bibfnamefont {J.~F.}\
  \bibnamefont {Gibson}},\ }\bibfield  {title} {\enquote {\bibinfo {title}
  {Geometry of turbulence in wall-bounded shear flows: {Periodic} orbits},}\
  }\href {https://doi.org/10.1088/0031-8949/2010/T142/014007} {\bibfield
  {journal} {\bibinfo  {journal} {Phys. Scr. T}\ }\textbf {\bibinfo {volume}
  {142}},\ \bibinfo {pages} {014007} (\bibinfo {year} {2010})}\BibitemShut
  {NoStop}%
\bibitem [{\citenamefont {Willis}, \citenamefont {Cvitanovi{\'c}},\ and\
  \citenamefont {Avila}(2013)}]{ACHKW11}%
  \BibitemOpen
  \bibfield  {author} {\bibinfo {author} {\bibfnamefont {A.~P.}\ \bibnamefont
  {Willis}}, \bibinfo {author} {\bibfnamefont {P.}~\bibnamefont
  {Cvitanovi{\'c}}},\ and\ \bibinfo {author} {\bibfnamefont {M.}~\bibnamefont
  {Avila}},\ }\bibfield  {title} {\enquote {\bibinfo {title} {Revealing the
  state space of turbulent pipe flow by symmetry reduction},}\ }\href
  {https://doi.org/10.1017/jfm.2013.75} {\bibfield  {journal} {\bibinfo
  {journal} {J. Fluid Mech.}\ }\textbf {\bibinfo {volume} {721}},\ \bibinfo
  {pages} {514--540} (\bibinfo {year} {2013})}\BibitemShut {NoStop}%
\bibitem [{\citenamefont {Kreilos}\ and\ \citenamefont
  {Eckhardt}(2012)}]{KreEck12}%
  \BibitemOpen
  \bibfield  {author} {\bibinfo {author} {\bibfnamefont {T.}~\bibnamefont
  {Kreilos}}\ and\ \bibinfo {author} {\bibfnamefont {B.}~\bibnamefont
  {Eckhardt}},\ }\bibfield  {title} {\enquote {\bibinfo {title} {Periodic
  orbits near onset of chaos in plane {Couette} flow},}\ }\href
  {https://doi.org/10.1063/1.4757227} {\bibfield  {journal} {\bibinfo
  {journal} {Chaos}\ }\textbf {\bibinfo {volume} {22}},\ \bibinfo {pages}
  {047505} (\bibinfo {year} {2012})}\BibitemShut {NoStop}%
\bibitem [{\citenamefont {Suri}\ \emph {et~al.}(2020)\citenamefont {Suri},
  \citenamefont {Kageorge}, \citenamefont {Grigoriev},\ and\ \citenamefont
  {Schatz}}]{Suri20}%
  \BibitemOpen
  \bibfield  {author} {\bibinfo {author} {\bibfnamefont {B.}~\bibnamefont
  {Suri}}, \bibinfo {author} {\bibfnamefont {L.}~\bibnamefont {Kageorge}},
  \bibinfo {author} {\bibfnamefont {R.~O.}\ \bibnamefont {Grigoriev}},\ and\
  \bibinfo {author} {\bibfnamefont {M.~F.}\ \bibnamefont {Schatz}},\ }\bibfield
   {title} {\enquote {\bibinfo {title} {Capturing turbulent dynamics and
  statistics in experiments with unstable periodic orbits},}\ }\href
  {https://doi.org/10.1103/PhysRevLett.125.064501} {\bibfield  {journal}
  {\bibinfo  {journal} {Phys. Rev. Lett.}\ }\textbf {\bibinfo {volume} {125}},\
  \bibinfo {pages} {064501} (\bibinfo {year} {2020})}\BibitemShut {NoStop}%
\bibitem [{\citenamefont {Crowley}\ \emph
  {et~al.}(2022{\natexlab{b}})\citenamefont {Crowley}, \citenamefont
  {Pughe-Sanford}, \citenamefont {Toler}, \citenamefont {Grigoriev},\ and\
  \citenamefont {Schatz}}]{CPTGS22}%
  \BibitemOpen
  \bibfield  {author} {\bibinfo {author} {\bibfnamefont {C.~J.}\ \bibnamefont
  {Crowley}}, \bibinfo {author} {\bibfnamefont {J.~L.}\ \bibnamefont
  {Pughe-Sanford}}, \bibinfo {author} {\bibfnamefont {W.}~\bibnamefont
  {Toler}}, \bibinfo {author} {\bibfnamefont {R.~O.}\ \bibnamefont
  {Grigoriev}},\ and\ \bibinfo {author} {\bibfnamefont {M.~F.}\ \bibnamefont
  {Schatz}},\ }\bibfield  {title} {\enquote {\bibinfo {title} {Observing a
  dynamical skeleton of turbulence in {Taylor-Couette} flow experiments},}\
  }\href {https://doi.org/10.1098/rsta.2022.0137} {\bibfield  {journal}
  {\bibinfo  {journal} {Philos. Trans. R. Soc. Math. Phys. Eng. Sci.}\ }\textbf
  {\bibinfo {volume} {381}},\ \bibinfo {pages} {20220137} (\bibinfo {year}
  {2022}{\natexlab{b}})}\BibitemShut {NoStop}%
\bibitem [{\citenamefont {Gibson}\ and\ \citenamefont
  {Cvitanovi{\'c}}(2015)}]{GibsonMovies}%
  \BibitemOpen
  \bibfield  {author} {\bibinfo {author} {\bibfnamefont {J.~F.}\ \bibnamefont
  {Gibson}}\ and\ \bibinfo {author} {\bibfnamefont {P.}~\bibnamefont
  {Cvitanovi{\'c}}},\ }\href {https://ChaosBook.org/tutorials} {\enquote
  {\bibinfo {title} {Movies of plane {Couette}},}\ }\bibinfo {type} {Tech.
  Rep.}\ (\bibinfo  {institution} {Georgia Inst. of Technology},\ \bibinfo
  {year} {2015})\BibitemShut {NoStop}%
\bibitem [{\citenamefont {Gudorf}(2020)}]{GudorfThesis}%
  \BibitemOpen
  \bibfield  {author} {\bibinfo {author} {\bibfnamefont {M.~N.}\ \bibnamefont
  {Gudorf}},\ }\emph {\bibinfo {title} {{Spatiotemporal Tiling of the
  Kuramoto-Sivashinsky Equation}}},\ \href
  {https://ChaosBook.org/projects/theses.html} {Ph.D. thesis},\ \bibinfo
  {school} {School of Physics, Georgia Inst. of Technology}, \bibinfo {address}
  {Atlanta} (\bibinfo {year} {2020})\BibitemShut {NoStop}%
\bibitem [{\citenamefont {Gibson}(2019)}]{channelflow}%
  \BibitemOpen
  \bibfield  {author} {\bibinfo {author} {\bibfnamefont {J.~F.}\ \bibnamefont
  {Gibson}},\ }\href {http://Channelflow.org} {\enquote {\bibinfo {title}
  {{Channelflow}: {A} spectral {Navier-Stokes} simulator in {C}++},}\ }\bibinfo
  {type} {Tech. Rep.}\ (\bibinfo  {institution} {U. New Hampshire},\ \bibinfo
  {year} {2019})\ \bibinfo {note} {{\tt {Channelflow.org}}}\BibitemShut
  {NoStop}%
\bibitem [{\citenamefont {Willis}(2014)}]{openpipeflow}%
  \BibitemOpen
  \bibfield  {author} {\bibinfo {author} {\bibfnamefont {A.~P.}\ \bibnamefont
  {Willis}},\ }\href {http://Openpipeflow.org} {\enquote {\bibinfo {title}
  {{Openpipeflow}: {Pipe} flow code for incompressible flow},}\ }\bibinfo
  {type} {Tech. Rep.}\ (\bibinfo  {institution} {U. Sheffield},\ \bibinfo
  {year} {2014})\ \bibinfo {note} {{\tt {Openpipeflow.org}}}\BibitemShut
  {NoStop}%
\bibitem [{\citenamefont {Ruelle}(1972)}]{Ruelle72}%
  \BibitemOpen
  \bibfield  {author} {\bibinfo {author} {\bibfnamefont {D.}~\bibnamefont
  {Ruelle}},\ }\bibfield  {title} {\enquote {\bibinfo {title} {Statistical
  mechanics on a compact set with {$Z^\nu $} action satisfying expansiveness
  and specification},}\ }\href
  {https://doi.org/10.1090/S0002-9904-1972-13078-7} {\bibfield  {journal}
  {\bibinfo  {journal} {Bull. Amer. Math. Soc.}\ }\textbf {\bibinfo {volume}
  {78}},\ \bibinfo {pages} {988--991} (\bibinfo {year} {1972})}\BibitemShut
  {NoStop}%
\bibitem [{\citenamefont {Cvitanovi{\'c}}(2026{\natexlab{b}})}]{CBremark20-2}%
  \BibitemOpen
  \bibfield  {author} {\bibinfo {author} {\bibfnamefont {P.}~\bibnamefont
  {Cvitanovi{\'c}}},\ }\bibfield  {title} {\enquote {\bibinfo {title} {Remark
  20.2 {‘Pressure’}},}\ }in\ \href
  {https://ChaosBook.org/chapters/ChaosBook.pdf#rmark.20.2} {\emph {\bibinfo
  {booktitle} {{Chaos: Classical and Quantum}}}},\ \bibinfo {editor} {edited
  by\ \bibinfo {editor} {\bibfnamefont {P.}~\bibnamefont {Cvitanovi{\'c}}},
  \bibinfo {editor} {\bibfnamefont {R.}~\bibnamefont {Artuso}}, \bibinfo
  {editor} {\bibfnamefont {R.}~\bibnamefont {Mainieri}}, \bibinfo {editor}
  {\bibfnamefont {G.}~\bibnamefont {Tanner}},\ and\ \bibinfo {editor}
  {\bibfnamefont {G.}~\bibnamefont {Vattay}}}\ (\bibinfo  {publisher} {Niels
  Bohr Inst.},\ \bibinfo {address} {Copenhagen},\ \bibinfo {year}
  {2026})\BibitemShut {NoStop}%
\bibitem [{\citenamefont {Cvitanovi\'c}, \citenamefont {Gaspard},\ and\
  \citenamefont {Schreiber}(1992)}]{CGS92}%
  \BibitemOpen
  \bibfield  {author} {\bibinfo {author} {\bibfnamefont {P.}~\bibnamefont
  {Cvitanovi\'c}}, \bibinfo {author} {\bibfnamefont {P.}~\bibnamefont
  {Gaspard}},\ and\ \bibinfo {author} {\bibfnamefont {T.}~\bibnamefont
  {Schreiber}},\ }\bibfield  {title} {\enquote {\bibinfo {title} {Investigation
  of the {Lorentz} gas in terms of periodic orbits},}\ }\href
  {https://doi.org/10.1063/1.165902} {\bibfield  {journal} {\bibinfo  {journal}
  {Chaos}\ }\textbf {\bibinfo {volume} {2}},\ \bibinfo {pages} {85--90}
  (\bibinfo {year} {1992})}\BibitemShut {NoStop}%
\bibitem [{\citenamefont {Cvitanovi{\'c}}(1995)}]{Cvitanovic1995109}%
  \BibitemOpen
  \bibfield  {author} {\bibinfo {author} {\bibfnamefont {P.}~\bibnamefont
  {Cvitanovi{\'c}}},\ }\bibfield  {title} {\enquote {\bibinfo {title}
  {Dynamical averaging in terms of periodic orbits},}\ }\href
  {https://doi.org/10.1016/0167-2789(94)00256-P} {\bibfield  {journal}
  {\bibinfo  {journal} {Physica D}\ }\textbf {\bibinfo {volume} {83}},\
  \bibinfo {pages} {109--123} (\bibinfo {year} {1995})}\BibitemShut {NoStop}%
\bibitem [{\citenamefont {Cvitanovi{\'c}}(2025{\natexlab{d}})}]{aver_dasbuch}%
  \BibitemOpen
  \bibfield  {author} {\bibinfo {author} {\bibfnamefont {P.}~\bibnamefont
  {Cvitanovi{\'c}}},\ }\bibfield  {title} {\enquote {\bibinfo {title}
  {Averaging},}\ }in\ \href {https://ChaosBook.org/paper.shtml#average} {\emph
  {\bibinfo {booktitle} {{Chaos: Classical and Quantum}}}},\ \bibinfo {editor}
  {edited by\ \bibinfo {editor} {\bibfnamefont {P.}~\bibnamefont
  {Cvitanovi{\'c}}}, \bibinfo {editor} {\bibfnamefont {R.}~\bibnamefont
  {Artuso}}, \bibinfo {editor} {\bibfnamefont {R.}~\bibnamefont {Mainieri}},
  \bibinfo {editor} {\bibfnamefont {G.}~\bibnamefont {Tanner}},\ and\ \bibinfo
  {editor} {\bibfnamefont {G.}~\bibnamefont {Vattay}}}\ (\bibinfo  {publisher}
  {Niels Bohr Inst.},\ \bibinfo {address} {Copenhagen},\ \bibinfo {year}
  {2025})\BibitemShut {NoStop}%
\bibitem [{\citenamefont {Rugh}(2019)}]{Rugh19}%
  \BibitemOpen
  \bibfield  {author} {\bibinfo {author} {\bibfnamefont {H.~H.}\ \bibnamefont
  {Rugh}},\ }\bibfield  {title} {\enquote {\bibinfo {title} {Un interview de
  {David Ruelle}},}\ }\href
  {https://smf.emath.fr/publications/la-gazette-des-mathematiciens-161-juillet-2019}
  {\bibfield  {journal} {\bibinfo  {journal} {Gaz. Math.}\ }\textbf {\bibinfo
  {volume} {161}},\ \bibinfo {pages} {55--59} (\bibinfo {year}
  {2019})}\BibitemShut {NoStop}%
\bibitem [{\citenamefont {Rugh}(2020)}]{Rugh19a}%
  \BibitemOpen
  \bibfield  {author} {\bibinfo {author} {\bibfnamefont {H.~H.}\ \bibnamefont
  {Rugh}},\ }\bibfield  {title} {\enquote {\bibinfo {title} {An interview with
  {David Ruelle}},}\ }\href {https://doi.org/10.4171/news/115/6} {\bibfield
  {journal} {\bibinfo  {journal} {{EMS} Newsletter}\ }\textbf {\bibinfo
  {volume} {2020-3}},\ \bibinfo {pages} {21--24} (\bibinfo {year}
  {2020})}\BibitemShut {NoStop}%
\bibitem [{\citenamefont {Dahlqvist}(2023)}]{ChaosBookLeastAct}%
  \BibitemOpen
  \bibfield  {author} {\bibinfo {author} {\bibfnamefont {P.}~\bibnamefont
  {Dahlqvist}},\ }\bibfield  {title} {\enquote {\bibinfo {title} {Least action
  method},}\ }in\ \href
  {https://chaosbook.org/chapters/ChaosBook.pdf#section.34.3} {\emph {\bibinfo
  {booktitle} {{Chaos: Classical and Quantum}}}},\ \bibinfo {editor} {edited
  by\ \bibinfo {editor} {\bibfnamefont {P.}~\bibnamefont {Cvitanovi\'c}},
  \bibinfo {editor} {\bibfnamefont {R.}~\bibnamefont {Artuso}}, \bibinfo
  {editor} {\bibfnamefont {R.}~\bibnamefont {Mainieri}}, \bibinfo {editor}
  {\bibfnamefont {G.}~\bibnamefont {Tanner}},\ and\ \bibinfo {editor}
  {\bibfnamefont {G.}~\bibnamefont {Vattay}}}\ (\bibinfo  {publisher} {Niels
  Bohr Institute},\ \bibinfo {year} {2023})\ \bibinfo {edition} {14th}\
  ed.\BibitemShut {Stop}%
\bibitem [{\citenamefont {Lan}\ and\ \citenamefont
  {Cvitanovi{\'c}}(2004)}]{lanVar1}%
  \BibitemOpen
  \bibfield  {author} {\bibinfo {author} {\bibfnamefont {Y.}~\bibnamefont
  {Lan}}\ and\ \bibinfo {author} {\bibfnamefont {P.}~\bibnamefont
  {Cvitanovi{\'c}}},\ }\bibfield  {title} {\enquote {\bibinfo {title}
  {Variational method for finding periodic orbits in a general flow},}\ }\href
  {https://doi.org/10.1103/PhysRevE.69.016217} {\bibfield  {journal} {\bibinfo
  {journal} {Phys. Rev. E}\ }\textbf {\bibinfo {volume} {69}},\ \bibinfo
  {pages} {016217} (\bibinfo {year} {2004})}\BibitemShut {NoStop}%
\bibitem [{\citenamefont {Gudorf}(2021)}]{orbithunter}%
  \BibitemOpen
  \bibfield  {author} {\bibinfo {author} {\bibfnamefont {M.~N.}\ \bibnamefont
  {Gudorf}},\ }\href {https://orbithunter.readthedocs.io} {\enquote {\bibinfo
  {title} {{Orbithunter: Framework for Nonlinear Dynamics and Chaos}},}\
  }\bibinfo {type} {Tech. Rep.}\ (\bibinfo  {institution} {School of Physics,
  Georgia Inst. of Technology},\ \bibinfo {year} {2021})\BibitemShut {NoStop}%
\bibitem [{\citenamefont {Lakshmi}\ \emph {et~al.}(2021)\citenamefont
  {Lakshmi}, \citenamefont {Fantuzzi}, \citenamefont {Chernyshenko},\ and\
  \citenamefont {Lasagna}}]{LFCL21}%
  \BibitemOpen
  \bibfield  {author} {\bibinfo {author} {\bibfnamefont {M.~V.}\ \bibnamefont
  {Lakshmi}}, \bibinfo {author} {\bibfnamefont {G.}~\bibnamefont {Fantuzzi}},
  \bibinfo {author} {\bibfnamefont {S.~I.}\ \bibnamefont {Chernyshenko}},\ and\
  \bibinfo {author} {\bibfnamefont {D.}~\bibnamefont {Lasagna}},\ }\bibfield
  {title} {\enquote {\bibinfo {title} {Finding unstable periodic orbits: {A}
  hybrid approach with polynomial optimization},}\ }\href
  {https://doi.org/10.1016/j.physd.2021.133009} {\bibfield  {journal} {\bibinfo
   {journal} {Physica D}\ }\textbf {\bibinfo {volume} {427}},\ \bibinfo {pages}
  {133009} (\bibinfo {year} {2021})}\BibitemShut {NoStop}%
\bibitem [{\citenamefont {Azimi}, \citenamefont {Ashtari},\ and\ \citenamefont
  {Schneider}(2022)}]{AzAsSc22}%
  \BibitemOpen
  \bibfield  {author} {\bibinfo {author} {\bibfnamefont {S.}~\bibnamefont
  {Azimi}}, \bibinfo {author} {\bibfnamefont {O.}~\bibnamefont {Ashtari}},\
  and\ \bibinfo {author} {\bibfnamefont {T.~M.}\ \bibnamefont {Schneider}},\
  }\bibfield  {title} {\enquote {\bibinfo {title} {Constructing periodic orbits
  of high-dimensional chaotic systems by an adjoint-based variational
  method},}\ }\href {https://doi.org/10.1103/physreve.105.014217} {\bibfield
  {journal} {\bibinfo  {journal} {Phys. Rev. E}\ }\textbf {\bibinfo {volume}
  {105}},\ \bibinfo {pages} {014217} (\bibinfo {year} {2022})}\BibitemShut
  {NoStop}%
\bibitem [{\citenamefont {Parker}\ and\ \citenamefont
  {Schneider}(2022)}]{ParSch22}%
  \BibitemOpen
  \bibfield  {author} {\bibinfo {author} {\bibfnamefont {J.~P.}\ \bibnamefont
  {Parker}}\ and\ \bibinfo {author} {\bibfnamefont {T.~M.}\ \bibnamefont
  {Schneider}},\ }\bibfield  {title} {\enquote {\bibinfo {title} {Variational
  methods for finding periodic orbits in the incompressible {Navier-Stokes}
  equations},}\ }\href {https://doi.org/10.1017/jfm.2022.299} {\bibfield
  {journal} {\bibinfo  {journal} {J. Fluid. Mech.}\ }\textbf {\bibinfo {volume}
  {941}},\ \bibinfo {pages} {A17} (\bibinfo {year} {2022})}\BibitemShut
  {NoStop}%
\bibitem [{\citenamefont {Wang}\ and\ \citenamefont {Lan}(2022)}]{WanLan22}%
  \BibitemOpen
  \bibfield  {author} {\bibinfo {author} {\bibfnamefont {D.}~\bibnamefont
  {Wang}}\ and\ \bibinfo {author} {\bibfnamefont {Y.}~\bibnamefont {Lan}},\
  }\bibfield  {title} {\enquote {\bibinfo {title} {A reduced variational
  approach for searching cycles in high-dimensional systems},}\ }\href
  {https://doi.org/10.1007/s11071-022-08130-x} {\bibfield  {journal} {\bibinfo
  {journal} {Nonlinear Dynam.}\ }\textbf {\bibinfo {volume} {111}},\ \bibinfo
  {pages} {5579--5592} (\bibinfo {year} {2022})}\BibitemShut {NoStop}%
\bibitem [{\citenamefont {Page}\ \emph {et~al.}(2024)\citenamefont {Page},
  \citenamefont {Norgaard}, \citenamefont {Brenner},\ and\ \citenamefont
  {Kerswell}}]{PNBK22}%
  \BibitemOpen
  \bibfield  {author} {\bibinfo {author} {\bibfnamefont {J.}~\bibnamefont
  {Page}}, \bibinfo {author} {\bibfnamefont {P.}~\bibnamefont {Norgaard}},
  \bibinfo {author} {\bibfnamefont {M.~P.}\ \bibnamefont {Brenner}},\ and\
  \bibinfo {author} {\bibfnamefont {R.~R.}\ \bibnamefont {Kerswell}},\
  }\bibfield  {title} {\enquote {\bibinfo {title} {Recurrent flow patterns as a
  basis for two-dimensional turbulence: {Predicting} statistics from
  structures},}\ }\href
  {https://doi.org/https://doi.org/10.1073/pnas.2320007121} {\bibfield
  {journal} {\bibinfo  {journal} {Proc. Natl. Acad. Sci.}\ }\textbf {\bibinfo
  {volume} {121}},\ \bibinfo {pages} {23} (\bibinfo {year} {2024})}\BibitemShut
  {NoStop}%
\bibitem [{\citenamefont {Christiansen}, \citenamefont {Cvitanovi{\'c}},\ and\
  \citenamefont {Putkaradze}(1997{\natexlab{b}})}]{Christiansen96}%
  \BibitemOpen
  \bibfield  {author} {\bibinfo {author} {\bibfnamefont {F.}~\bibnamefont
  {Christiansen}}, \bibinfo {author} {\bibfnamefont {P.}~\bibnamefont
  {Cvitanovi{\'c}}},\ and\ \bibinfo {author} {\bibfnamefont {V.}~\bibnamefont
  {Putkaradze}},\ }\bibfield  {title} {\enquote {\bibinfo {title} {Hopf's last
  hope: {Spatiotemporal} chaos in terms of unstable recurrent patterns},}\
  }\href {https://doi.org/10.1088/0951-7715/10/1/004} {\bibfield  {journal}
  {\bibinfo  {journal} {Nonlinearity}\ }\textbf {\bibinfo {volume} {10}},\
  \bibinfo {pages} {55--70} (\bibinfo {year} {1997}{\natexlab{b}})}\BibitemShut
  {NoStop}%
\bibitem [{\citenamefont {Viswanath}(2007)}]{Visw07b}%
  \BibitemOpen
  \bibfield  {author} {\bibinfo {author} {\bibfnamefont {D.}~\bibnamefont
  {Viswanath}},\ }\bibfield  {title} {\enquote {\bibinfo {title} {Recurrent
  motions within plane {Couette} turbulence},}\ }\href
  {https://doi.org/10.1017/S0022112007005459} {\bibfield  {journal} {\bibinfo
  {journal} {J. Fluid Mech.}\ }\textbf {\bibinfo {volume} {580}},\ \bibinfo
  {pages} {339--358} (\bibinfo {year} {2007})}\BibitemShut {NoStop}%
\bibitem [{\citenamefont {Ding}\ \emph {et~al.}(2016)\citenamefont {Ding},
  \citenamefont {Chat{\'e}}, \citenamefont {Cvitanovi\'c}, \citenamefont
  {Siminos},\ and\ \citenamefont {Takeuchi}}]{DCTSCD14}%
  \BibitemOpen
  \bibfield  {author} {\bibinfo {author} {\bibfnamefont {X.}~\bibnamefont
  {Ding}}, \bibinfo {author} {\bibfnamefont {H.}~\bibnamefont {Chat{\'e}}},
  \bibinfo {author} {\bibfnamefont {P.}~\bibnamefont {Cvitanovi\'c}}, \bibinfo
  {author} {\bibfnamefont {E.}~\bibnamefont {Siminos}},\ and\ \bibinfo {author}
  {\bibfnamefont {K.~A.}\ \bibnamefont {Takeuchi}},\ }\bibfield  {title}
  {\enquote {\bibinfo {title} {Estimating the dimension of the inertial
  manifold from unstable periodic orbits},}\ }\href
  {https://doi.org/10.1103/PhysRevLett.117.024101} {\bibfield  {journal}
  {\bibinfo  {journal} {Phys. Rev. Lett.}\ }\textbf {\bibinfo {volume} {117}},\
  \bibinfo {pages} {024101} (\bibinfo {year} {2016})}\BibitemShut {NoStop}%
\bibitem [{\citenamefont {Williams}, \citenamefont {Gudorf},\ and\
  \citenamefont {Orlov}(2025)}]{WiGuOr25}%
  \BibitemOpen
  \bibfield  {author} {\bibinfo {author} {\bibfnamefont {S.~D.~V.}\
  \bibnamefont {Williams}}, \bibinfo {author} {\bibfnamefont {M.~N.}\
  \bibnamefont {Gudorf}},\ and\ \bibinfo {author} {\bibfnamefont {D.~M.}\
  \bibnamefont {Orlov}},\ }\bibfield  {title} {\enquote {\bibinfo {title}
  {Understanding plasma turbulence through exact coherent structures},}\ }\href
  {https://doi.org/10.1063/5.0282368} {\bibfield  {journal} {\bibinfo
  {journal} {Physics of Plasmas}\ }\textbf {\bibinfo {volume} {32}},\ \bibinfo
  {pages} {092302} (\bibinfo {year} {2025})}\BibitemShut {NoStop}%
\bibitem [{\citenamefont {Axenides}, \citenamefont {Floratos},\ and\
  \citenamefont {Nicolis}(2024)}]{AxFlNi24}%
  \BibitemOpen
  \bibfield  {author} {\bibinfo {author} {\bibfnamefont {M.}~\bibnamefont
  {Axenides}}, \bibinfo {author} {\bibfnamefont {E.}~\bibnamefont {Floratos}},\
  and\ \bibinfo {author} {\bibfnamefont {S.}~\bibnamefont {Nicolis}},\ }\href
  {https://doi.org/10.48550/ARXIV.2401.08521} {\enquote {\bibinfo {title}
  {Exponential mixing of all orders for {Arnol'd} cat map lattices},}\ }
  (\bibinfo {year} {2024})\BibitemShut {NoStop}%
\bibitem [{\citenamefont {Politi}\ and\ \citenamefont
  {Torcini}(1992{\natexlab{b}})}]{PolTor92}%
  \BibitemOpen
  \bibfield  {author} {\bibinfo {author} {\bibfnamefont {A.}~\bibnamefont
  {Politi}}\ and\ \bibinfo {author} {\bibfnamefont {A.}~\bibnamefont
  {Torcini}},\ }\bibfield  {title} {\enquote {\bibinfo {title} {Periodic orbits
  in coupled {H{\'e}non} maps: {Lyapunov} and multifractal analysis},}\ }\href
  {https://doi.org/10.1063/1.165871} {\bibfield  {journal} {\bibinfo  {journal}
  {Chaos}\ }\textbf {\bibinfo {volume} {2}},\ \bibinfo {pages} {293--300}
  (\bibinfo {year} {1992}{\natexlab{b}})}\BibitemShut {NoStop}%
\bibitem [{\citenamefont {Baake}, \citenamefont {Hermisson},\ and\
  \citenamefont {Pleasants}(1997)}]{BaHePl97}%
  \BibitemOpen
  \bibfield  {author} {\bibinfo {author} {\bibfnamefont {M.}~\bibnamefont
  {Baake}}, \bibinfo {author} {\bibfnamefont {J.}~\bibnamefont {Hermisson}},\
  and\ \bibinfo {author} {\bibfnamefont {A.~B.}\ \bibnamefont {Pleasants}},\
  }\bibfield  {title} {\enquote {\bibinfo {title} {The torus parametrization of
  quasiperiodic {LI}-classes},}\ }\href
  {https://doi.org/10.1088/0305-4470/30/9/016} {\bibfield  {journal} {\bibinfo
  {journal} {J. Phys. A}\ }\textbf {\bibinfo {volume} {30}},\ \bibinfo {pages}
  {3029--3056} (\bibinfo {year} {1997})}\BibitemShut {NoStop}%
\bibitem [{\citenamefont
  {Cvitanovi{\'c}}(2022{\natexlab{c}})}]{CvitanovicYT02}%
  \BibitemOpen
  \bibfield  {author} {\bibinfo {author} {\bibfnamefont {P.}~\bibnamefont
  {Cvitanovi{\'c}}},\ }\href {https://youtu.be/Ztt1v8uGCUE} {\enquote {\bibinfo
  {title} {A chaotic field theory: {Fundamental} fact},}\ } (\bibinfo {year}
  {2022}{\natexlab{c}})\BibitemShut {NoStop}%
\bibitem [{\citenamefont {Isola}(1990)}]{Isola90}%
  \BibitemOpen
  \bibfield  {author} {\bibinfo {author} {\bibfnamefont {S.}~\bibnamefont
  {Isola}},\ }\bibfield  {title} {\enquote {\bibinfo {title}
  {{$\zeta$}-functions and distribution of periodic orbits of toral
  automorphisms},}\ }\href {https://doi.org/10.1209/0295-5075/11/6/006}
  {\bibfield  {journal} {\bibinfo  {journal} {Europhys. Lett.}\ }\textbf
  {\bibinfo {volume} {11}},\ \bibinfo {pages} {517--522} (\bibinfo {year}
  {1990})}\BibitemShut {NoStop}%
\bibitem [{\citenamefont {Muresan}(2009)}]{Muresan09}%
  \BibitemOpen
  \bibfield  {author} {\bibinfo {author} {\bibfnamefont {M.}~\bibnamefont
  {Muresan}},\ }\href {https://doi.org/10.1007/978-0-387-78933-0} {\emph
  {\bibinfo {title} {{A Concrete Approach to Classical Analysis}}}}\ (\bibinfo
  {publisher} {Springer},\ \bibinfo {address} {New York},\ \bibinfo {year}
  {2009})\BibitemShut {NoStop}%
\end{thebibliography}%
